# Euclid

## Mapping the geometry of the dark Universe

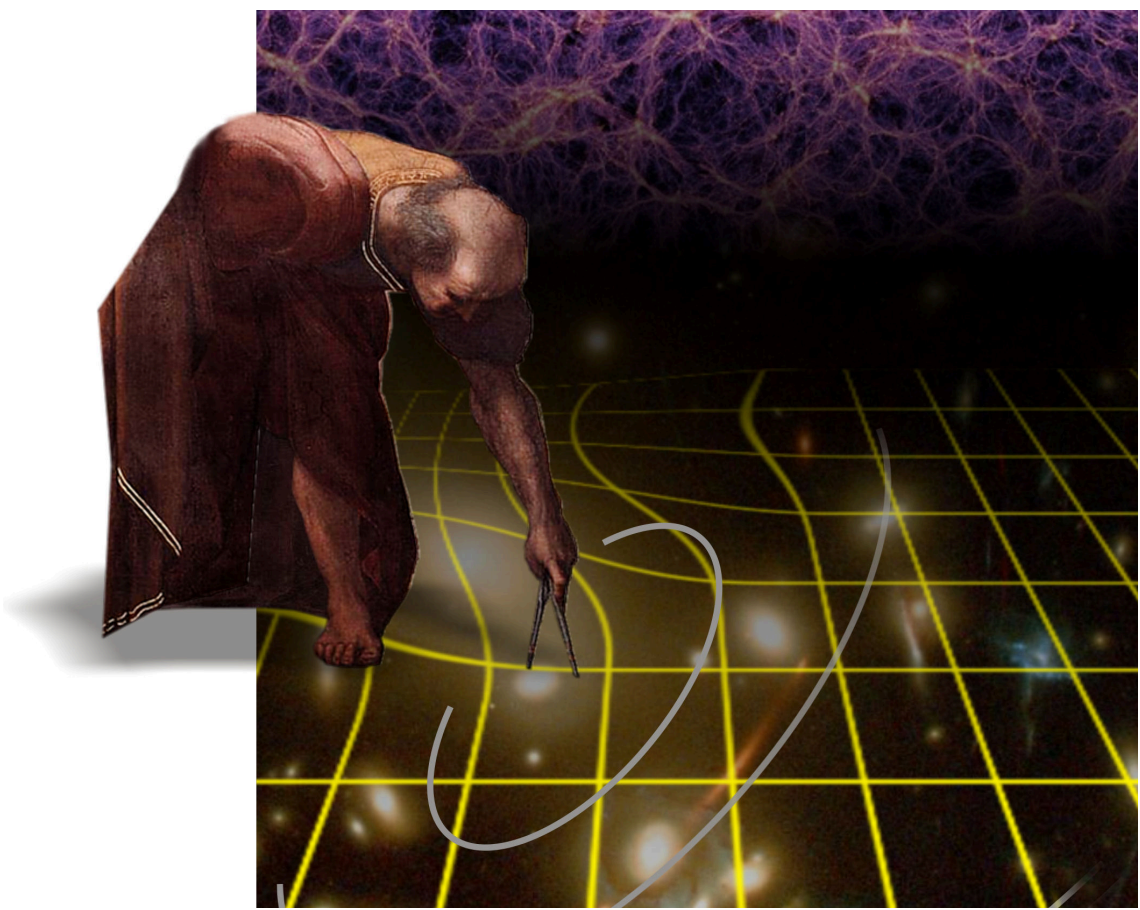

**Assessment Study Report**

**European Space Agency**

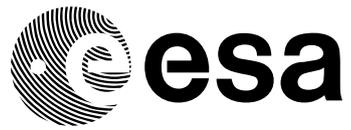

# Euclid

## Mapping the geometry of the dark Universe

## Assessment Study Report

1 December 2009

**On the front cover**: composite of a fragment from Raphael's fresco "The School of Athens" in the Stanza della Segnatura of the Vatican Palace depicting the greek mathematician Euclid of Alexandria, a simulation of the cosmic web by Springel et al, and an image of Abell 1689; the composition is made by Remy van Haarlem (ESA/ESTEC).

# Euclid Mission Summary

| Main Scientific Objectives |
| --- |
| ***Understand the nature of Dark Energy and Dark Matter by:*** |
| • Measuring the DE equation of state parameters $w_0$ and $w_a$ to a precision of 2% and 10%, respectively, using both expansion history and structure growth. |
| • Measuring the growth factor exponent γ with a precision of 2%, enabling to distinguish General Relativity from the modified-gravity theories |
| • Testing the Cold Dark Matter paradigm for structure formation, and measure the sum of the neutrino masses to a precision better than 0.04eV when combined with Planck. |
| • Improving by a factor of 20 the determination of the initial condition parameters compared to Planck alone. |

| SURVEYS | | |
| --- | --- | --- |
| | Area (deg2) | Description |
| Wide Survey | 20,000 | Entire extragalactic sky with galactic latitude b>30 deg Shear measurements for 40 galaxies/arcmin$^2$ Spectroscopic measurements for 70 million galaxies |
| Deep Survey | 40 | In at least 2 patches of > 10 deg$^2$ 2 magnitudes deeper than wide survey |

| PAYLOAD | | | | | |
| --- | --- | --- | --- | --- | --- |
| Telescope | 1.2 m Korsch | | | | |
| Instrument | Imaging Instrument | | | | Spectrometer |
| Field-of-View | 0.48 deg$^2$ | | | | 0.48 deg$^2$ |
| Capability | Visual Imaging | NIR Imaging Photometry | | | NIR slitless spectrometry $\lambda/\Delta\lambda \sim 500$ |
| Wavelength range | R+I+Z (550–920nm) | Y (920-1146nm), | J (1146-1372 nm) | H (1372-2000nm) | 1000-2000 nm |
| Sensitivity | 24.5 mag (10σ extended source) | 24 mag 5σ point source | 24 mag 5σ point source | 24 mag 5σ point source | $4 \times 10^{-16}$ erg cm$^{-2}$ s$^{-1}$ For emission line 7σ at 1600 nm for unresolved source. |
| Detector Technology | 36 arrays 4k×4k CCD | 18 arrays 2k×2k NIR Array | | | 8 arrays 2k×2k NIR Array |
| Pixel Size | 0.1 arcsec | 0.3 arcsec | | | 0.5 arcsec |

| SPACECRAFT | |
| --- | --- |
| Launcher | Soyuz ST-2.1 B from Kourou |
| Orbit | Sun Earth Lagrange point 2 (L2) |
| Pointing | 35 mas relative pointing error over one dither exposure with VIS 100 mas absolute measurement accuracy |
| Stabilization | Step and stare |
| Lifetime | 5 years nominal |
| Operations | 4 hours per day contact, 850 Gbit/day in K band |

| MASS and POWER BREAKDOWN | | |
| --- | --- | --- |
| | *Mass (kg)* | *Power (W)* |
| Payload Module | 855 | 350 |
| Service Module | 691 | 595 |
| Propellant | 150 | |
| Adapter / Harness and PDCU losses | 100 | 58 |
| Margin (20%) | 309 | 201 |
| **Total** | **2105** | **1204** |

# Foreword

We provide here a brief summary of the historical events behind the creation of Euclid, including details of the various reviews and proposals undertaken before this report. In response to the ESA Cosmic Vision 2020-2025 Call for M-Class missions, two dark energy related proposals were submitted, which shared the same objective, namely to study dark energy and other aspects of fundamental cosmology, and provide unique legacy science. The Dark Universe Explorer (DUNE) mission intended to measure the effects of weak gravitational lensing. The other proposal, the Spectroscopic All Sky Cosmic Explorer (SPACE), aimed at measuring the baryonic acoustic oscillations and redshift-space distortion patterns in the Universe.

A Concept Advisory Team (CAT) was appointed whose mandate was to recommend to ESA which mission concept should be taken for a CV assessment study. Chaired by Prof. M. Longair (Cavendish Laboratory) and composed of members from the DUNE and SPACE consortia and independent European scientists, the CAT decided to have a combined mission which can simultaneously measure weak lensing *and* baryonic acoustic oscillations (BAOs). The combination of the two methods provides an advantage over any individual method as it allows scientist to probe the true, underlying explanation for the accelerating Universe (e.g. dark energy or modified gravity) while ensuring that systematic uncertainties are well controlled. Within the boundary conditions of a 1.2m primary mirror and a limited amount of near-infrared detector arrays, the science concept implied a survey of 20,000 deg$^2$ extragalactic sky with instruments capable of (1) diffraction limited imaging and (2) near-infrared imaging to perform photometry in three bands, and (3) medium resolution slit-spectroscopy with digital micro-mirror devices (DMD) in the infrared.

In spring 2008, an ESA internal pre-assessment study was conducted in the ESTEC Concurrent Design Facility (CDF) to identify the mission drivers, feasibility points and key enabling technologies for the upcoming industrial and payload studies. Following the Invitation to Tender for a 1 year assessment study of Euclid, tenders were accepted from prime contractors TAS-Italy (Torino) and Astrium GmbH (Friedrichshafen). These studies started in September 2008. Two instrument consortia were accepted to ESA's call for Declaration of Interest for payload studies. The consortium for the visual and near-infrared imaging photometer for the weak-lensing experiment is led by A. Réfrégier from CEA Saclay. The consortium providing the NIR spectrometer for the BAO experiment is led by A. Cimatti from Bologna University. These studies started in October 2008 for 10-months duration. In parallel, ESA has initiated a technological evaluation of the DMD, an essential component of the spectrometer as originally proposed for SPACE. The studies progressed in parallel under the scientific supervision of the Euclid Science Study Team (ESST).

Phase 1 of the industry study ended in January 2008, and concluded that the optics design that satisfies both instruments was very demanding and also not compatible with a continuous scanning observing mode. The initial baseline review was unable to conclude on a feasible design, and an extension to Phase 1 was requested assuming a step and stare observing strategy.

At this point, during the first months of 2009, a discussion began at higher management level on a possible merging with the NASA Joint Dark Energy Mission. The studies were redirected to consider design options compatible with the expected merged mission. But due to internal reasons, NASA could not proceed with the planned joint ESA/NASA AO on a timescale which is in line with the Cosmic Visions schedule. ESA concluded that further formal agreements with NASA, as well as possible AO, should stop while ESA concludes the Cosmic Vision down selection, and the US community finishes their Decadal Review (due mid-2010). Future discussions are possible but at the moment, ESA is addressing dark energy via a European mission.

Consequently, the Euclid concept was reinforced, but with important programmatic changes. To avoid effects of lost time the industry and consortia were provided with an ESA-imposed optics design solution, and a recommendation to set as baseline a simpler slitless design. The slit spectroscopy design is kept as an option acknowledging the larger scientific potential over a slitless design and the different timescales for the environmental tests of the DMD component. To stay in line with the schedule of M-Class Cosmic Vision studies, Phase 2 was limited to studies for accommodation & folding of this optics design, and followed the Phase 1 system design aspects with no major trade-offs. Approximately 10 weeks of study were left available to the industry and instrument consortia teams instead of a nominal 6 months.

*The Euclid Study Team, November 2009*

# Authorship and Acknowledgements

**Euclid Science Study Team**
| | |
|---|---|
| A. Cimatti | University of Bologna, Italy |
| R. Laureijs | ESA/ESTEC, The Netherlands (Chair) |
| B. Leibundgut | ESO Garching, Germany |
| S. Lilly | ETH Zurich, Switzerland |
| R. Nichol | Univ. Portsmouth, United Kingdom |
| A. Réfrégier | CEA Saclay, France |
| P. Rosati | ESO Garching, Germany |
| M. Steinmetz | Astrophysical Institut Potsdam, Germany |
| N. Thatte | Univ. Oxford, United Kingdom |
| E. Valentijn | Univ. Groningen, The Netherlands |

**ESA Study Team at ESTEC**
| | |
|---|---|
| L. Duvet | Study Payload Manager |
| R. Laureijs | Study Scientist |
| D. Lumb | Study Manager |
| L. Boisnard | Study Manager (until March 2009) |
| G. Saavedra Criado | Systems Engineer |

**ESA Planning and Coordination Office at ESA HO and ESTEC**
M. Coradini, P. Escoubet, F. Favata and T. Prusti

**ESA editorial support**:
| | |
|---|---|
| A. Salama | ESAC, Spain |

## Contributions from:

Lauro Moscardini (U. di Bologna), Claudia Quercellini (U. Roma Tor Vergata), Mario Radovich (INAF – Oss. Capodimonte), Anna Romano (U. Roma La Sapienza), Roberto Scaramella (INAF – Oss. Roma), Andrea Zacchei (INAF - Oss. Trieste), **Spain:** Ricard Casas (IEEC Barcelona), Martin Crocce (IEEC Barcelona), Pablo Fosalba (IEEC Barcelona), Enrique Gaztanaga (IEEC Barcelona), Ignacio Sevilla (CIEMAT Madrid), Eusebio Sanchez (CIEMAT Madrid), Juan Garcia-Bellido (UAM Madrid), **Switzerland:** Rongmon Bordoloi (ETH Zurich)**,** Julien Carron (ETh Zurich)**,** Frederic Courbin (EPFL Lausanne)**,** Stephane Paltani (ISDC/U. Geneva)**,** Justin Read (U. Zurich/U. Leicester)**,** Robert Smith (U. Zurich), **UK:** Filipe Abdalla (UCL), Manda Banerji (UCL), David Bacon (U. Portsmouth), Kirk Donnacha (UCL), Ignacio Ferreras (MSSL-UCL), Gert Hutsi (UCL), Alan Heavens (IfA Edinburgh), Tom Kitching (IfA Edinburgh), Lindsay King (IoA Cambridge), Ofer Lahav (UCL), Katarina Markovic (UCL/Ludwigs-Maximilians U.), Richard Massey (IfA Eindinburgh), Michael Schneider (U. Durham), Fabrizio Sidoli (UCL), Jiayu Tang (UCL/IPMU), Shaun Thomas (UCL), Lisa Voigt (UCL), **USA:** Joel Berge (JPL), Benjamin Dobke (JPL), Richard Ellis (Caltech), David Johnston (Northwest. U.), Michael Seiffert (JPL), Ali Vanderveld (JPL), **Other:** Eduardo Cypriano (U. Sao Paulo), Hakon Dahle (U. Oslo), Ariel Goobar (U. Stockholm, Sweden), Konrad Kuijken (Leiden)

**Instrument: France:** Christophe Cara (CEA Saclay), Sandrine Cazaux (CEA Saclay), Arnaud Claret (CEA Saclay), Philippe Daniel-Thomas (CEA Saclay), Eric Doumayrou (CEA Saclay), Cydalise Dusmesnil (IAS Orsay), Jean-Jacques Fourmond (IAS Orsay), Jean-Claude Lecleech (IAS Orsay), G. Morinaud (IAS Orsay), Samuel Ronayette (CEA Saclay), Zihong Sun (CEA Saclay), Tristan VanDenBerghe (IAS Orsay), **Germany:** Reiner Hofmann (MPE), Rory Holmes (MPIA), Reinhard Katterloher (MPE), Oliver Krause (MPIA), **Italy:** Marco Frailis (INAF Oss. Trieste), Pasquale Cerulli Irelli (INAF IFSI), Stefano Gallozzi (INAF Oss. Roma), Michele Maris (INAF Oss. Trieste), Renato Orfei (INAF IFSI), Fabio Pasian (INAF Oss. Trieste), Andrea Zacchei (INAF Oss. Trieste), **Spain:** Laia Cardiel (IFAE Barcelona), Francesc Madrid (IEEC Barcelona), Santiago Serrano (IEEC Barcelona), **Switzerland:** Adrian Glauser (ETH Zurich), Francois Wildi (ISDC/U. Geneva), **UK:** Eli Atad-Ettedgui (UK-ATC), Ian Bryson (UK-ATC), Richard Cole (MSSL), Jason Gow (Open U), Phil Guttridge (MSSL) **,** Andrew Holland (Open U)**,** Neil Murray (Open U)**,** Kerrin Rees (MSSL), Phil Thomas (MSSL), Dave Walton (MSSL), **USA:** Parker Fagrelius (JPL), Kirk Gilmore (Stanford)**,** Steven Kahn (Stanford)**,** Andrew Rasmussen (Stanford)**,** Suresh Seshadri (JPL)**,** Roger Smith (Caltech)**,** Harry Teplitz (IPAC/Caltech)

## Euclid Near Infrared Spectrograph (ENIS) Consortium

**PI**: Andrea Cimatti (University of Bologna – Department of Astronomy). **Program Manager:** L. Valenziano (INAF – IASFBO, Italy), C. Butler (Advisor, INAF – IASFBO, Italy). **Instrument Manager:** F.M. Zerbi (INAF – OA Brera, Italy).

**Board of Co-I's:** F. Bertoldi (Uni Bonn, Germany), O. Le Fevre (LAM, France), P. Lilje (Uni Oslo, Norway), R. Rebolo Lopez (IAC, Spain), M. Robberto (STScI, USA), H.J.A. Röttgering (Sterrewacht Leiden, Netherlands), R. Sharples (Uni Durham, UK), W. Zeilinger (Uni. Wien, Austria).

**Ground Segment Manager:** F. Pasian (INAF – OA Trieste, Italy). **General Science Managers:** G. Zamorani (INAF – OA Bologna, Italy), L. Guzzo (INAF – OA Brera, Italy). **Simulations Managers:** B. Garilli (INAF – IASFMI, Italy), P. Rosati (ESO, Germany).

**Science Topics Coordinators – BAO:** C.M. Baugh (Uni Durham, UK), Y. Wang (Uni Oklahoma, USA), W. Percival (ICC, Portsmouth, UK), **Clusters:** S. Borgani, A. Biviano (Uni Trieste, Italy), **Redshift-Space Distorsions:** L. Guzzo (INAF – OA Brera, Italy), **Galaxy biasing:** E. Branchini (Uni. Roma 3), **Galaxies:** G. Zamorani (INAF –OA Bologna, Italy), M. Cirasuolo (ROE, UK), R. Salvaterra (INAF – OA Brera, Italy), **AGN:** A. Martinez-Sansigre (MPIA Heidelberg, Germany), **Our Galaxy:** M.R. Zapatero Osorio (IAC, Spain).

**Compact Engineering Team:** F.M. Zerbi (Underline: Coordinator, INAF – OA Brera), C. Butler, A. De Rosa, G. Morgante, L. Nicastro, M. Trifoglio (INAF – IASFBO), B. Garilli, P. Franzetti (INAF – IASFMI), C. Bonoli, F. Bortoletto, E. Giro (INAF – OA Padova), E. Diolaiti, G. Zamorani (INAF – OA Bologna), F. Pasian, A. Zacchei (INAF – OA Trieste), L. Corcione, S. Ligori (INAF – OA Torino), P. Leutenegger (TAS, Milano), V. De Caprio, L. Guzzo, M. Riva, P. Spanò (INAF – OA Brera), C.M. Baugh, P. Clark, R. Content, N.E. Looker, R. Sharples, G. Talbot (Uni Durham), R. Grange, L. Martin, A. Origne, T. Pamplona, C. Surace, F. Zamkotsian (LAM), J. Walsh, P. Rosati (ESO – ST/ECF), M. Robberto (STScI), R. Rebolo Lopez (IAC).

**Science Advisory Team:** A. Cimatti (Underline: Coordinator, Uni Bologna), C.M. Baugh (Uni Durham), S. Borgani (Uni Trieste), M. Franx (Sterrewacht Leiden), M. Lehnert (Obs. de Paris Meudon / GEPI), P. Lilje (Uni. Oslo), A. Martinez-Sansigre (MPIA Heidelberg), R. Rebolo Lopez (IAC), Y. Wang (Uni Oklahoma), W. Zeilinger (Uni. Wien).

**General Collaborators (Science and Payload) – France:** E. Daddi (CEA), F. Hammer (Obs. de Paris Meudon), B. Milliard, C. Moreau, C. Schimd (LAM), C. Benoist, A. Bijaoui, C. Ferrari, F. Martel, S. Maurogordato, T. Regimbau, E. Sleza (OCA), T. Contini (Lab. d' Astroph. de Toulouse-Tarbes). **Germany:** M. Kümmel, H. Kuntschner (ESO-ST/ECF), E. Schinnerer, J. Steinacker, F. Walter (MPIA, Heidelberg), J. Kurk (MPE), M. Polder, C. Porciani, K. Reif (Uni Bonn), B. Ciardi (MPA). **Italy:** D. Bottini, M. Fumana, D. Maccagni, L. Paioro, M. Scodeggio (INAF-IASFMI), M. D'Alessandro, A. Renzini (INAF – OA Padova), A. Bianco, A. Iovino, E. Majerotto, E. Molinari, C. Carbone, L. Ciotti, G. Cosentino, I. Foppiani, F. Marulli, C. Nipoti, S. Pellegrini, L. Schreiber (Uni. Bologna), G. Bregoli, A. Buzzoni, S. Bardelli, S. Ettori, C. Gruppioni, M. Mignoli, L. Pozzetti, N. Roche, D. Vergani, E. Zucca (INAF – OA Bologna), S. Cristiani, G. De Lucia, M. Magliocchetti, E. Pian, E. Vanzella (INAF – OA Trieste), A. Bulgarelli, C. Burigana, F. Cuttaia, A. Gruppuso, N. Mandolesi, N. Masetti, F. Finelli, F. Gianotti, E. Palazzi, L. Stringhetti, M. Villa (INAF – IASFBO), A. Franceschini, G. Rodighiero (Uni Padova), M. Bersanelli, D. Maino, A. Mennella, M. Tomasi (Uni Milano), A. Zonca, B. Cappellini (INFN, Milano), E. Oliva (INAF – OA Arcetri), A. Ferrara (SNS Pisa). **Netherlands:** G. Verdoes (Uni Groningen), B. Braam, R. Vink (TNO). **Norway:** O. Elgaroy, H.K.K. Eriksen, F.K. Hansen, S.V. Haugan, D.F. Mota, M. Wold (Uni Oslo). **Romania:** L. Popa (Inst. for Space Sciences). **Spain:** M. Balcells, S. Iglesias-Roth, J.A. Rubino, M.R. Zapatero-Osorio (IAC), A. Perez Garrido, A. Díaz Sánchez, I. Villó Pérez (UPCT, U. Politecnica de Cartagena). **UK:** S. Blake, R. Bower, S. Cole, C.S. Frenk, J. Geach, S. Morris, M. Schneider, T. Shanks, I. Small, M. Ward (Uni. Durham), R. Kennicutt (IoA, Cambridge), I. Bryson, J.S. Dunlop, R.J. McLure (ROE), M.J. Jarvis (Uni Hertsfordshire).

**Others:** P. Astier (LPNHE, CNRS-IN2P3 and Univ. of Paris), A. Belikov (Groningen), A. Ealet (CPPM, Marseille), C. Maraston (Portsmouth), C. Lacy (Durham), S. Smartt (Queen's Univ. Belfast), S. Viti (UCL).

# 1 Executive Summary

Euclid is a high-precision survey mission to map the geometry of the Dark Universe with demonstrated feasibility. Euclid's Visible-NIR imaging and spectroscopy of the entire extragalactic sky will further produce extensive legacy science to the boundaries of the visible universe.

**Primary Science Objectives:** Over the last decades, a combination of observations has led to the emergence and confirmation of the concordance cosmological model. In this model, the Universe has evolved from a homogeneous state after the Big Bang, to a hierarchical assembly of galaxies, clusters and superclusters at our epoch. Remarkably, the energy density of the resulting Universe is dominated by two mysterious components. First, 76% of its energy density is in the form of Dark Energy, which is causing the Universe expansion to accelerate. The existence and energy scale of Dark Energy is in conflict with our knowledge of fundamental physics. A key question in this regard is whether it behaves like the cosmological constant ($\Lambda$) introduced by Einstein. Another 20% of the energy is in the form of dark matter, which exerts a gravitational attraction as normal matter, but does not emit light. While several candidates exist in particle physics, such as supersymmetric extensions of the Standard Model, the nature of dark matter is unknown. One possibility to explain one or both of these puzzles is that Einstein's General Relativity, and thus our understanding of gravity, needs to be revised on cosmological scales. Together, dark energy and dark matter pose some of the most important questions in fundamental physics today.

Euclid is a high-precision survey mission designed to answer these fundamental questions. To do so, Euclid will map the large-scale structure of the universe over the entire extragalactic sky out to redshifts of 2 (about 10 billion years ago), thus covering the period over which dark energy accelerated the universe expansion. The mission is optimised for two primary cosmological probes: Weak gravitational Lensing (WL) and Baryonic Acoustic Oscillations (BAO). Weak lensing is a method to map the dark matter and measure dark energy through the distortions of galaxy images by mass inhomogeneities along the line-of-sight. BAOs are wiggle patterns imprinted in the clustering of galaxies which provide a standard ruler to measure dark energy and the expansion in the universe. Surveyed in the same cosmic volume, these techniques not only provide systematic cross-checks but also a measurement of large scale structure via different physical fields (potential, density and velocity), which are required for testing Dark Energy and gravity on cosmological scales. Euclid will also make use of several secondary cosmological probes such as the Integrated Sachs Wolfe Effect (ISW), galaxy clusters and redshift space distortions to provide additional measurements of the cosmic geometry and structure growth.

WL and BAO require a high image quality for the shear measurements, near-infrared spectroscopic and imaging capabilities to measure galaxies at redshifts z>1, a very high degree of system stability to minimize systematic effects, and the ability to survey the entire extra-galactic sky. Such a combination of requirements cannot be met from the ground, and demands a wide-field Visible/NIR space mission. A central design driver for Euclid is the ability to provide tight control of systematic effects in space-based conditions and to measure WL and BAO simultaneously.

With its wide-field capability and high-precision design, Euclid will achieve the following science objectives in fundamental cosmology: (1) Euclid will measure the dark energy equation of state parameters $w_0$ and $w_a$ to a precision of 2% and 10% from the geometry and structure growth of the Universe. Euclid will thus achieve a Dark Energy Figure of Merit of 500 (1500) without (with) Planck Priors, thus improving by a factor of 50 (150) upon current knowledge. (2) Euclid will test the validity of General Relativity against modified gravity theories, and measure the growth factor exponent $\gamma$ to an accuracy of 2%. (3) Euclid will study the properties of dark matter by mapping its distribution, testing the Cold Dark Matter paradigm and measuring the sum of the neutrino masses to a few 0.01eV in combination with Planck. (4) Euclid will improve the constraints on the initial condition parameters by a factor of 2-30 compared to Planck alone. Euclid is therefore poised to uncover new physics by challenging all sectors of the cosmological model. The Euclid survey can thus be thought as the low-redshift, 3-dimensional analog and complement to the map of the high-redshift universe provided by CMB experiments.

**Legacy Science:** Beyond these breakthroughs in fundamental cosmology, the Euclid surveys will provide unique legacy science in various fields of astrophysics. In the area of galaxy evolution and formation, Euclid will deliver high quality morphologies, masses, and star-formation rates for billions of galaxies out to z~2, over the entire extra-galactic sky, with a resolution 4 times better and 3 NIR magnitude deeper than ground based survey. The Euclid deep survey will probe the 'dark ages' of galaxy formation as it is predicted to find thousands of galaxies at z>6, of which about 100 could be at z>10 i.e., probing the era of reionization of the Universe. These high redshift galaxies and quasars will be critical targets for JWST and E-ELT. Closer to home, Euclid will augment the Gaia survey of our Milky Way, taking it several magnitudes deeper. Below V=20, Euclid will provide complementary information to Gaia, adding infrared colours and spectra for every Gaia star it observes; hence breaking the age-metallicity degeneracy, which is critical for the chemical enrichment history of our Galaxy. Also, the all-sky coverage of Euclid will detect nearby extremely low surface brightness tidal streams of stars thus allowing us to probe the formation and evolution of our own Galaxy. Euclid will also provide key measurements of the mass function of galaxy clusters (esp. in combination with eROSITA, Planck and SZ telescopes), and of over $10^5$ strong lensing systems, and should find thousands of intermediate-redshift supernovae in the near-IR. Euclid could also undertake a programme to detect earth-mass planets in the habitable zone through the microlensing technique.

**Surveys:** Euclid's primary wide survey will cover 20,000 deg$^2$, i.e. the entire extragalactic sky, thus measuring shapes and redshifts of galaxies to redshift 2 as required for weak lensing and BAO. For weak lensing, Euclid will measure the shape of over 2 billion galaxies with a density of 30-40 resolved galaxies per arcmin$^2$ in one broad visible R+I+Z band (550-920 nm) down to AB mag 24.5 (10σ). The photometric redshifts for these galaxies will reach a precision of $\sigma_z/(1+z)$= 0.03-0.05. They will be derived from three additional Euclid NIR band (Y,J,H in the range 0.92-2.0 micron) reaching AB mag 24 (5σ) in each, complemented by ground based photometry in visible bands derived through engaged collaborations with ground based projects such as DES and Pan-STARRS. To measure the shear from the galaxy ellipticities a tight control is imposed on possible instrumental effects and will lead to the variance of the shear systematic errors to be less than $10^{-7}$. The BAO are determined from a spectroscopic survey with a redshift accuracy of $\sigma_z/(1+z)$ ≤0.001. The Euclid baseline is a slitless spectrometer with $\lambda/\Delta\lambda$=500, which will detect predominantly Hα emission line galaxies. The limiting line flux level is 4 $10^{-16}$ erg s$^{-1}$cm$^{-2}$ (point source 7σ at 1.6 micron), yielding 70 million galaxy redshifts with a success rate in excess of 35%. The option of a slit spectrometer survey based on DMD technology is discussed in Appendix A.

Euclid will also perform a deep survey, about 2 mag deeper than the wide survey and covering an area of about 40 deg$^2$. Although unique as a self standing survey, the deep survey will also monitor the stability of the spacecraft and payload through repeated visits of the same regions.

**Payload:** The baseline payload consists of a Korsch telescope with a primary mirror of 1.2 m diameter. The telescope is designed to provide a large field of view (0.5 deg$^2$) to an imaging instrument with a visible channel (VIS) and and a NIR imaging channel (NIP) and a NIR spectroscopic channel (NIS). VIS and NIP support the weak lensing probe whereas NIS is designed to perform the wide spectroscopic galaxy survey.

VIS will measure the shapes of galaxies with a resolution of 0.18 arcsec (PSF FWHM) with 0.1 arcsec pixels in one wide visible band (R+I+Z). NIP contains three NIR bands (Y, J, H), employing HgCdTe NIR detector with 0.3 arcsec pixels. NIS, the spectroscopic channel, operates in the wavelength range 1.0-2.0 micron at a spectral resolution $\lambda/\Delta\lambda \sim 500$, employing 0.5 arcsec pixels.

The optional spectroscopic implementation is slit spectroscopy using digital micro-mirror devices (DMDs). Providing technical challenges can be resolved, such an instrument would deliver continuum spectra down to H(AB)=22 mag corresponding to 150 million spectra distributed over 20,000 deg2 assuming a selection rate of 30%. In this study report we provide an overview of the differences compared to the baseline.

To accomplish the surveys within the nominal mission lifetime of 5 years, each instrument has a large field of view and the system design is optimized for a sky survey with fast attitude slews to support a step-and-stare tiling mode. To meet the survey depth and sensitivity, the telescope will have a well baffled design and is cooled to minimize background noise. For the NIR detectors, on-board on-the-ramp processing will be performed, i.e. combining image frames to lower the noise. The NIR related optics and detectors are cooled down to ~100 K.

**Mission:** The spacecraft will be placed in a large L2 orbit which will ensure stable thermal and observing conditions. The satellite will be launched on a Soyouz ST-2.1B rocket from Kourou. The nominal mission duration is 5 years and the observations will be done in step-and-stare mode. A fine guidance system will provide a relative pointing accuracy of 35mas over a 450s exposure. Image dithering will be achieved at spacecraft level to fill detector gaps, provide sub-pixel information, and allow correction for cosmic rays. The survey speed with a relatively large number of detectors of 36 4k×4k CCDs and 26 2k×2k NIR detector arrays, in combination with the dithered exposures yields a data rate of 850 Gbits/day. To accommodate such a rate the K-band transmission is required for data transfer from the spacecraft to the Cebreros ground station.

**Ground segment data handling:** The large data volume and high-precision requirements of Euclid will be handled with a data handling architecture based on existing space-based and ground-based projects. Dedicated teams from the PI-led instrument consortia will process the data from their instruments during all phases of the mission through Instrument Operations Centres (IOCs) and several Science Data Centres (SDCs). The IOCs will be responsible for the first level standard data processing (calibration, removal of the instrumental effects, etc) and for requesting to the SOC corrective action in operations. The SDCs will be in charge of second and third level data products and the development of simulation pipelines. The data handling system includes a common archive, the Euclid Mission Archive (EMA) which will support the sharing of data within the project, the reporting of quality controls and a built in redundancy in the key processing tasks. The calibrated and qualified data of the EMA will form the Euclid Legacy Archive (ELA) which will be delivered to the astronomical community at large.

**Management**: For the space segment, ESA will provide the spacecraft and telescope through a selected industrial contractor. The instruments will be procured by nationally funded instrument consortia. A PFM development philosophy for the payload will be adopted in order to achieve a launch date before the end of 2018. The main lead item for the development schedule is the development of the instruments which will be started early in the definition phase.

Science operations and data reduction will be carried out by ESA through the MOC and SOC and by the two Instrument Consortia who are leading respective IOCs and SDCs, under the scientific coordination by the Euclid Science Team. The structure of the science ground segment allows flexibility to include the active participation of the astronomical community in a wide range of activities, ranging from hardware development to the generation of a large number of diverse scientific products. The Euclid Science Team has the scientific supervision of the generation of the raw, calibrated and advanced science data products. These products will be made available to the astronomical community at large on an annual release basis with a one year proprietary period after a calibration period of one year.

# Table of Contents

# 2 Euclid Science Objectives

## 2.1 Introduction

### 2.1.1 Cosmology today

Over the last decade, a combination of diverse observations has led to the emergence and confirmation of the ΛCDM concordance model for cosmology. In its framework, the Universe has evolved from a nearly homogeneous state immediately after the Big Bang to its current highly inhomogeneous state, through a hierarchical assembly of galaxies, clusters and superclusters (Figure 2.1a). This single model self-consistently explains the anisotropy fluctuation spectrum of the Cosmic Microwave Background (CMB), the abundances of the light elements from primordial nucleosynthesis, the large-scale structure of the Universe today, and the redshift-distance relation for supernovae.

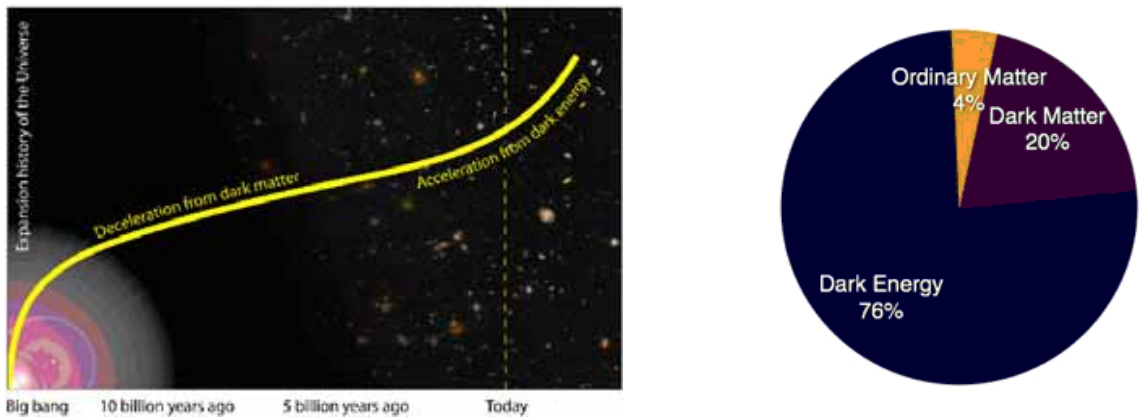

*Figure 2.1: a. (left panel): The Universe evolves from a homogeneous state after the big bang through cooling and expansion. The small initial inhomogeneities grow through gravity to produce the large-scale structures that we see today. b. (right panel): The mass-energy budget at our cosmological epoch. The Universe appears to be dominated by mysterious ingredients, dark energy and dark matter, whose nature poses some of the most pressing questions in fundamental physics.*

Against this remarkable observational success, the ΛCDM model demands that the global mass-energy density of the Universe today is dominated by two mysterious components. First, about 76% of the overall mass-energy density is in the form of dark energy, which causes the Universe to accelerate in its expansion at the current epoch. Another 20% is in the form of Dark Matter, which exerts a gravitational attraction in the same way as normal matter, but does not emit light. Only 4% of the universe is made up of the ordinary baryonic matter composed of protons, neutrons and electrons, out of which stars, planets and ourselves are made (see Figure 2.1b). While particle physics suggests a number of candidates for the dark matter, e.g. in Super-Symmetric extensions of the Standard Model, its nature is unknown. Dark Energy is difficult to reconcile with fundamental physics. Current observations are consistent with dark energy being in the simplest form of a cosmological constant, described by a constant pressure-to-density ratio ($w$) that is constant and equal to -1. Such an equation of state is what would be expected for the *energy of vacuum*; however an energy scale up to 120 orders of magnitude higher is obtained from general quantum field theory predictions. This and other fundamental physical difficulties could be explained if dark energy is a dynamical scalar field, for which $w$ varies with time (Section 2.2.1). Alternatively, the observed acceleration could instead indicate that Einstein's General Relativity (GR), and thus our understanding of gravity, needs to be revised on cosmological scales. Together, dark energy and dark matter pose some of the most important questions in fundamental physics.

The above different explanations can be discriminated by measuring: (*i*) the geometry of the Universe through the relation between redshift and distance and (*ii*) the growth rate of structures as they collapse under gravitational attraction while the Universe expands. The two most important properties of the dark energy for cosmological observations are the total amount and its equation of state. Both of these are best measured using observations of the geometry and growth rate of the Universe. Furthermore, the combination of these two observations provides a cross-check on the dark energy-dominated model and may instead signal the need for a new theory of gravity. The properties of dark matter are best probed by measuring the statistical properties of density fluctuations and the density profiles of bound objects.

## 2.1.2 Euclid science concept

The goal of the Euclid mission is to study the dark sector of the Universe to answer the following key questions in fundamental cosmology:

1. Is the Dark Energy merely a cosmological constant, as introduced by Einstein, or is it a scalar field that evolves dynamically with the expansion of the universe?

2. Alternatively, is the Dark Energy instead a manifestation of a breakdown of General Relativity on the largest scales?

3. What are the nature and properties of dark matter?

4. What are the initial conditions in the Early Universe, which seeded the formation of cosmic structure?

**High Precision Cosmology:** Euclid is a high precision survey mission that is optimised for two independent cosmological probes: Weak Gravitational Lensing from a high-resolution imaging survey and Galaxy Clustering from a massive spectroscopic redshift survey. For both probes, Euclid is designed to provide large statistics and tight control of systematic effects.

- Weak gravitational lensing (WL) involves the small distortions in the shapes of distant galaxies due to the bending of light by the intervening matter distributed along the line of sight. By correlating the shapes of large numbers of individual galaxies, these small systematic distortions can be measured. By doing this for galaxies at different distances, we can characterize the intervening mass distribution at different distances, and hence at different cosmic epochs, yielding information on the shape and growth of the power spectrum of density fluctuations. The cosmological information on the geometry of the Universe comes from the dependence of the lensing effect on the angular diameter distance and on tracking the angular size of features in the density fluctuation spectrum with redshift (Fig. 2.2a).

- Galaxy clustering is quantified through galaxy-galaxy correlations – or equivalently by the Fourier-space power spectrum P(k). The Baryonic Acoustic Oscillations (BAO) are a fossil of the early universe giving rise to a characteristic feature at comoving separations of ~150 Mpc in the galaxy correlation function. This scale is accurately known from the CMB, turning it into a uniquely powerful universal standard rod. By measuring its position as a function of redshift, one directly probes the expansion history and thus the equation of state of dark energy (Fig. 2.2b). At the same time, the anisotropy of clustering in redshift space provides a quantitative measurement of the growth rate of structure. This is produced by galaxy coherent flows towards overdensities, tracing the cosmic growth rate at different epochs.

These are two powerful independent probes of both geometry and growth, and their combination makes Euclid unique in terms of its ability for cross-checks and cross-calibrate of the systematic effects of each probe (e.g. photo-z calibration for WL and galaxy bias for BAO, see Sect. 2.4).

**All-Sky Map of the Dark and Visible Universe at 0<z<2:** In order to answer the above questions, we need to map the geometry and the growth of structure in the Universe as a function of cosmological time, or redshift (z), over the broad span of the expansion history during which the Dark Energy emerged to dominance, i.e. 0<z<2. Euclid is designed to do this by means of systematic surveys of galaxies over a large fraction of the observable Universe.

To achieve the required precision on parameters such as the equation of state *w(z)*, the observable extra-galactic sky (i.e. $2\pi$ sr, or 20,000 deg$^2$) must be covered. Cosmic structures should be mapped through all the available fields, namely the gravitational potential plus the density and peculiar velocity fields traced by galaxies. This ensures sufficient information to check for systematic effects and to test GR through relative inconsistencies. Such unprecedented map of the Universe spanning three-quarters of its life will uniquely complement the single snapshot at *z*~1100 provided by CMB experiments as WMAP and Planck.

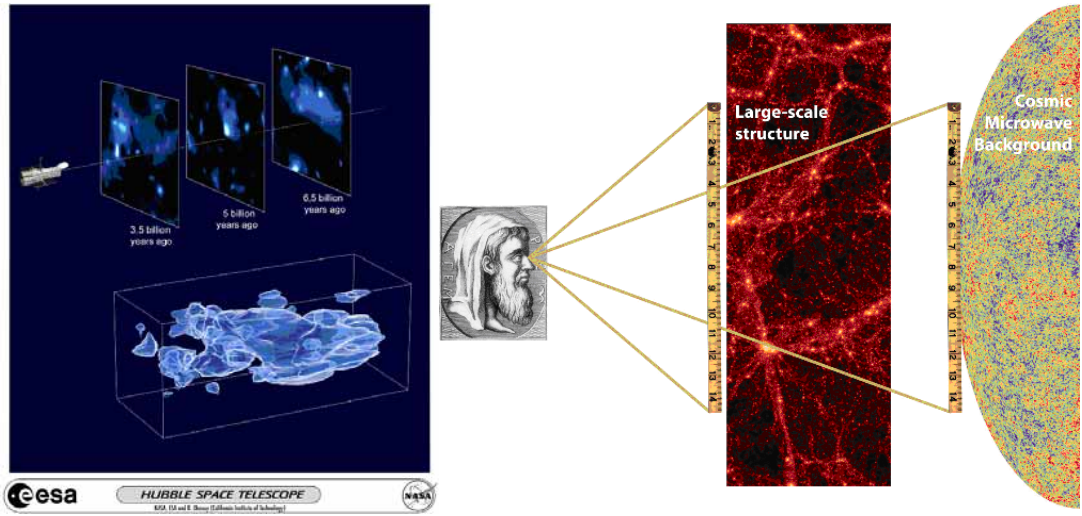

*Figure 2.2: Illustration of the two primary cosmological probes of Euclid: weak gravitational lensing and Baryonic Acoustic Oscillations. a. (left panel) The distribution of dark matter measured from the HST COSMOS survey using weak gravitational lensing. This survey covers an area similar to 8 times the size of the Moon. Euclid will produce higher resolution 3-D dark matter maps over the entire sky. Credit: NASA, ESA, and R. Massey. b. (right panel) Galaxy clustering as a probe of the geometry of the universe. The same acoustic features (BAO) seen in the CMB can be observed in the distribution of galaxies, providing a standard cosmological ruler.*

In addition, Euclid's unprecedented imaging and spectroscopic surveys will give us the ability to use several other complementary cosmological probes, including, galaxy clusters counts and the Integrated Sachs Wolfe effect. These probes will provide complementary constrains on dark energy and other cosmological parameters as well as additional cross-checks on systematics. Such a combined approach has been recommended by both the ESA-ESO Working Group on Fundamental Cosmology (Peacock et al. 2006) and a similar panel, the US DETF (Dark Energy Task Force; Albrecht et al. 2006).

**Legacy Science:** The Euclid surveys will carry an unprecedented legacy value for many areas of astronomy: galaxy evolution, the search for high-redshift objects, strong lensing. Additional surveys may also be carried out by Euclid to allow the study of galactic structure and the search for extra-solar planets with microlensing. The mission and the associated data management structure will guarantee the exploitation of such unique legacy potential by the wider community.

## 2.1.3 The Case for Going to Space

To achieve the above cosmological objectives, measurements of the shapes and redshifts of galaxies need to be obtained over very large volumes and with very high precision. Small statistical and systematic errors require the stable observing conditions in space free of atmospheric effects. Specifically, the observations carried out by Euclid in space will afford the following major improvements over ground based observations (see section 2.4 for more detailed discussions):

**Imaging with small and stable point spread function**: space based observations will provide a point spread function (PSF, i.e. the size of a point source) which is more than 3 times smaller than ground based PSF's which are dominated by atmospheric seeing. In addition, the stability of a source image observed in space is an order of magnitude better than that on ground, due to the high thermal stability in space, and the absence of windshake, atmospheric seeing, airmass, etc

**Deep NIR photometry with minimum background:** Space is exempted from the absorption from the atmosphere and will thus yield NIR photometry over the entire extra-galactic sky three magnitudes deeper than what can be achieved from the ground. This provides a small scatter and outlier rate for the photometric redshifts, which are required for weak lensing measurements.

**Deep NIR spectroscopy with minimum background:** Space observations provide deep NIR spectroscopy and thus redshifts for galaxies in the redshift range $0.5 < z < 2$ over the entire extra-galactic sky. Such volumes cannot be achieved from the ground in this wavelength range due to absorptions and emission lines of the atmosphere.

## 2.2    Primary Science Objectives

### 2.2.1  Dark energy

During 1980's, the Cold Dark Matter (CDM) model emerged as a scenario to reconcile the theoretical prejudice for a Universe with flat geometry ($\Omega = 1$) with the hierarchical nature of large-scale structures and other astrophysical observations. However, it became evident that the flat CDM model was not able to explain a number of observations. For instance it did not provide enough power at large scales to justify the observed clustering of galaxies and galaxy clusters, while at the same time had problems with galaxy peculiar velocities on small scales. These instead could be explained by a low-density ($\Omega_m < 0.5$) solution pointing to an open geometry. This contradicted the CMB anisotropy upper limits. In 1998, two teams using supernovae to measure luminosity distances made the amazing discovery that the expansion rate of the Universe is accelerating (Riess et al, 1998, Perlmutter et al, 1999). Combined with the existing data and, soon after, with the first quantitative measurement of the CMB anisotropies by the BOOMERANG experiment (de Bernardis et al. 2000), a fully coherent picture emerged: the Universe is indeed spatially flat, but matter makes up only 24% of the critical energy density today, with the remaining fraction provided by a "dark energy", so named to reflect the fact that it does not interact emit or absorb electromagnetic radiation, and does not behave like matter. Over the last decade, experimental evidence from many different observations has confirmed this as the basic make up of our Universe, most notably the high-precision all-sky CMB maps by the WMAP satellite (Spergel et al., 2003).

Dark energy is considered to be one of the biggest puzzles facing physics today. It provides the majority of the energy density in the Universe, but the most natural candidate (postulated by Einstein over 80 years ago) the cosmological constant, has a major problem. Using quantum field theory we naively expect that zero-point fluctuations should set the vacuum energy scale at 40 to 120 orders of magnitude higher than what is observed. Finding a vacuum energy which is extraordinarily small, but not zero, is the fine-tuning problem. While some speculative mechanisms have been proposed to explain this, it is widely regarded as un-accounted for in our current physical understanding.

Another conceptual problem with the concordance model is known as the coincidence problem. This relates to the relative densities of dark energy and matter. The density of matter ($\rho_m$) drops as the Universe expands whereas the density of dark energy due to a cosmological constant ($\rho_\Lambda$) remains fixed. The coincidence is that we appear to live in a special time in cosmic history where these two energy densities are comparable. This is uncomfortable due to a temporal version of the Copernican principle – why should we live at any special time in the history the Universe?

Dark energy models can be treated as fluids parameterised with an energy density ($\rho$) and pressure ($p$). The acceleration of the expansion rate, expressed with the cosmic scale factor $a(t)$, is given by,

$$\ddot{a} = -\frac{4\pi G a}{3}\left(\rho + 3p/c^2\right)$$

,

where $c$ is the speed of light. The scale factor $a$ is linked to the observational redshift $z$ by the Universe scale factor $a = 1/(1+z)$. Both $\rho$ and $p$ can evolve with time so it is convenient to work with the equation of state parameter ($w$) of dark energy, which is the ratio of pressure to density,

$$w \equiv p/\rho c^2.$$

To achieve accelerated expansion we need to satisfy the condition $p < -\rho c^2/3$, which requires a $w < -1/3$. A cosmological constant corresponds to $w = -1$ and can be compared to the equation of state of (pressureless) matter with $w = 0$. Dynamic models, including quintessence, allow the equation of state to vary over time see e.g. Copeland et al. (2006).

In principle, dynamical models can have any evolution of $w$ (see Figure 2.3b for example of current constraints). For example, quintessence models can be expressed in quantum field theory terms as an evolving scalar field, characterized by some potential, and any $w$ evolution can be recreated by selecting an appropriate form of the potential. However, the simplest non-trivial parameterisation of $w$ that allows for time variation is (Chevallier & Polarski, 2001; Linder, 2003; hereafter referred to as "CPL"), $w(a) = w_0 + (1-a)w_a$.

This captures a wide range of models that can be separated in the $w_0$-$w_a$ plane. These include *thawing* ($w$ slowly deviating from -1 over time) and *freezing* ($w$ asymptotically approaching -1 over time) models, as well as *k-essence* (interacting dark energy-dark matter), *tachyonic*, *SUSY scalar fields* and *Chaplygin gas* (see Section 2.3.6). The CPL parameterization is equivalent to a more general formulation,

$$w(a) = w_p + (a_p - a)w_a,$$

which is essentially a Taylor expansion about a specific redshift ($z_p$) set by some 'pivot scale' $a_p$ instead of today ($z = 0$; $a = 1$). With this form, the errors on $w_p$ and $w_a$ become uncorrelated and the error on $w_p$ corresponds to the error we would measure for a constant $w$. Following Albrecht et al. (2005) it is convenient to define a Figure of Merit (FoM$_{DE}$) as to quantify a given experiment's ability to measure the dark energy equation of state

$$FoM_{DE} = \frac{1}{\Delta w_p \Delta w_a}.$$

This FOM is widely used. By combining all current surveys its value is FoM$_{DE} \sim 10$ (Komatsu et al. 2009). Euclid will improve this figure by nearly two orders of magnitude (see section 2.3.5), reaching a value of ~1500 for the Euclid+Planck combination.

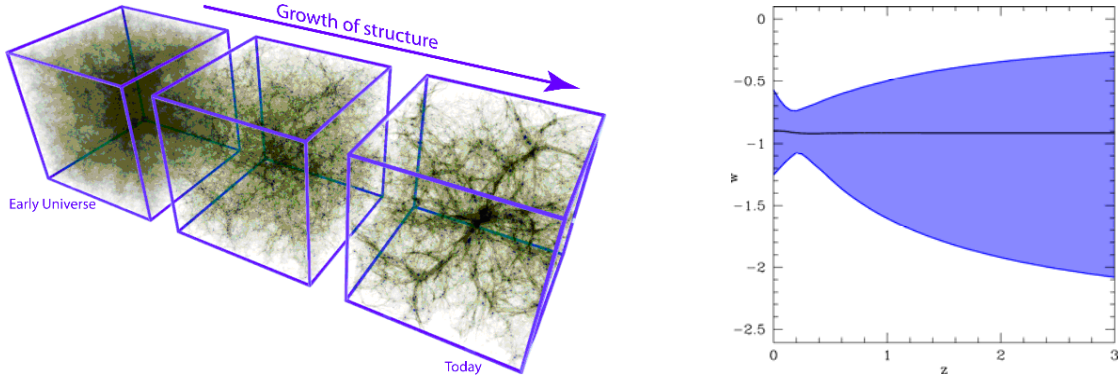

*Figure 2.3: a. (Left) Growth of structure in a numerical simulation (Credit: V. Springel). b. (Right) Constraints on w(z) from supernova data (SNLS 1st year) and CMB data (WMAP 3yr).*

Given a specific equation of state we can evaluate the geometry of the Universe or 'Hubble parameter' ($H(a) = \dot{a}/a$). This in turn sets the (comoving) distance to an object at a given redshift, which for a flat cosmology is given by,

$$D(z) = \int_0^z \frac{c}{H(z')} dz'.$$

As well as the geometry, dark energy also affects the growth of structure (illustrated in Figure 2.3a). The expansion of the Universe acts as a damping term for the growth of density fluctuation so as the expansion rate increases, the growth of structure slows down. Importantly, however, the growth rate $f(a)$ depends also on the gravity theory, which makes it a fundamental probe of modified gravity (see next section). $f(a)$ is well approximated by the phenomenological relation (Wang & Steinhardt, 1998; Amendola & Quercellini, 2004)

$$f(a) \cong \Omega_m(a)^{\gamma_m},$$

where $\gamma_m$ is the matter growth index $\gamma_m = 0.55 + 0.05[1 + w(a = 0.5)]$ (Linder, 2005), an expression valid for smooth dark energy models like quintessence. In the GR framework, measurements of $f(a)$ represent a fundamental consistency test of our cosmological assumptions, while carrying additional information on $w(a)$ beyond the geometric probes of the expansion rate.

The scientific dark energy targets for Euclid are to

*(i) Measure w(a), reaching a precision of 2% on $\Delta w_p$ and 10% on $\Delta w_a$ (corresponding to FoM$_{DE}$ = 500) for Euclid alone (ii) Test whether there are deviations from w = -1, indicating a dynamical dark energy. (iii) Measure the cosmic geometry and growth rate in several redshift bins from z = 0.5 to z = 2.*

## 2.2.2  Test of gravity

The standard approach to dark energy, discussed in the previous section, deals with accelerated expansion within the framework of a nearly isotropic and homogeneous solution of Einstein's theory of gravity. An alternative approach to understanding the apparent acceleration of the Universe is to ask whether there is something wrong with either the theory itself, or else the way we use it. For instance, the real Universe is certainly not perfectly homogeneous and isotropic, and since GR is a non-linear theory, corrections could become important and can lead to the similar observational signatures as accelerated expansion. This class of explanations for the dark energy, often dubbed "backreaction", is expected to generically lead to a violation of the Copernican test described at the end of this section. But first we discuss the more common second alternative where the equations of GR are modified.

**Modified Gravity:** Although we know that gravity is the main long-range force that shapes our Universe, we have very little direct information on it at scales beyond the solar system. The gravitational interaction has been studied exclusively in the laboratory and in systems of a size comparable to the solar system. On galactic and extragalactic scales there is little direct observational evidence that the interactions of ordinary matter, dark matter and dark energy are indeed described by standard GR. Even models that radically depart from canonical gravity, such as Bekenstein's Tensor-Vector-Scalar (TeVeS) theory, cannot be convincingly ruled out. The problem is further exacerbated on cosmological scales. And even if we had precise knowledge of gravity today we still could not exclude significant deviations in the past. The exploration of gravity beyond the "here" and "now" is one of the most exciting goals of cosmology and Euclid is well posed to make unprecedented advances (see Figure 2.4).

Recently, a vast array of alternative theories of gravity capable of explaining the acceleration has been proposed. These theories include: a) the addition of an extra gravitational scalar field, as in scalar-tensor theories, or in f(R) or f(G) modifications of the Hilbert-Einstein Lagrangian; b) models motivated by superstring/M-theory with extra spatial dimensions which lead to theories, in which gravity is modified on large-scales such as popular Dvali-Gabadadze-Porrati (DGP) model; c) models with more complicated corrections to gravity due to vector fields or bi-metric (such as TeVeS) and multi-metric theories where gravity and matter respond to different metrics. Some of these concepts have a long history; extra-dimensional Kaluza-Klein theories were proposed in the 1920s and the Brans-Dicke scalar-tensor model in the 1960s. This reminds us how difficult - and how urgent - it is to make progress in testing gravity on large scales, in addition to the problem of cosmic acceleration.

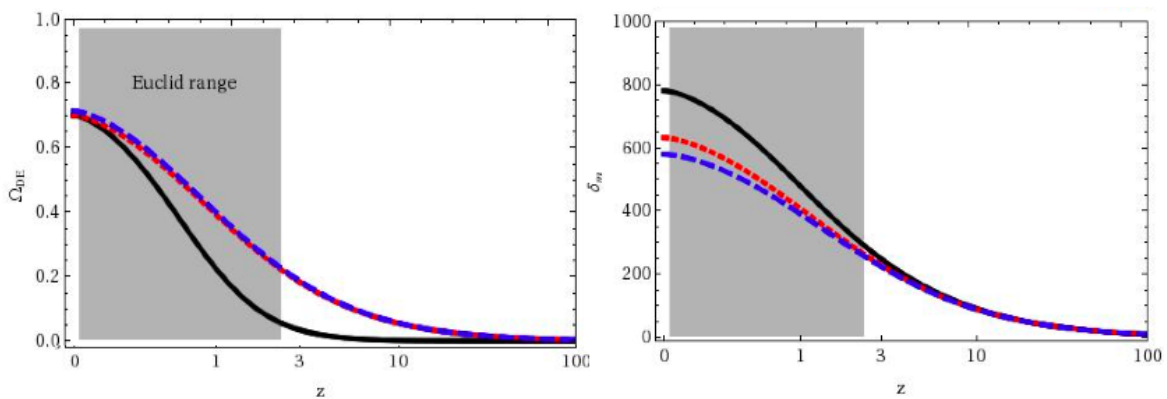

*Figure 2.4: a. (left) behaviour of $\Omega_{DE}(z)$ for ΛCDM (black line) and two models with identical background expansion (DGP, red, and a quintessence model with the same w(z), blue). b. (right): Growth factor $\delta_m$ (in arbitrary units) for the same models (here $\delta_m$ is the integral of the growth rate f(a) defined in the text). Although DGP and the quintessence model cannot be distinguished by measuring w(a), they lead to a different growth rate.*

As discussed in the previous section, in GR there is a clear relation between the expansion rate *H(z)* and the growth of structure, implying for ΛCDM a growth index $\gamma_m$=0.55. A deviation of the measured $\gamma_m$ from this value would point towards a large-scale modification of gravity. In the weak gravitational field limit appropriate for cosmology, a very general picture for cosmic dynamics requires the existence of two space-time dependent gravitational potentials: Φ, which measures spatial curvature and Ψ which measures time-dilation (akin to Newton's gravitational potential). In standard gravity these two potentials are identical and linked to the matter fluctuations by the relativistic version of the classical Poisson equation. In contrast, in modified gravity models, or in models in which dark energy clusters or interacts, Φ and Ψ need not be equal (their difference being called *anisotropic stress*) and the relation with matter fluctuations must be generalized. Any theory of gravity predicts a specific form of the anisotropic stress and the Poisson equation. Measuring Φ and Ψ independently means therefore mapping the large-scale geometry and dynamics of our Universe, as they govern the evolution of particles and fields. However, massive particles respond mostly to Ψ alone, while massless particles, like photons, respond to both. Therefore, to test each of them independently, we need to combine observations probing matter perturbations with observations of how light rays respond to gravity, as in gravitational lensing. Euclid will thus probe directly the relationship between Φ and Ψ through its unique combination of measurements of galaxy clustering (including redshift distortions) and galaxy shapes.

Finally, note that knowledge of the global dynamics alone (e.g. from Supernova luminosity distances) cannot distinguish between GR and modified gravity models, as any expansion history can be accounted for within GR with an appropriate $w(z)$.

**Testing the Cosmological Principle:** With a multiple-probe approach Euclid will also be able to test the Cosmological Principle – that the Universe is homogeneous and isotropic on large scales. This principle is a basic cornerstone of modern cosmology and testing it on large scales is vital. By combining measurements of the expansion history $H(z)$ and the distance $D(z)$ to a given redshift we can extract the local spatial curvature at that redshift (Clarkson et al. 2009). In back-reaction theories where the universe is strongly perturbed or where we live in a special place (e.g. in the centre of a giant void), the curvature would vary strongly as a function of redshift. Such strong deviations from the Friedmann-Lemaitre-Robertson-Walker metric, which predicts a constant curvature κ, would be visible to Euclid.

In summary, to make progress in understanding gravity and testing the cosmological principle we need:

*(i) To measure the growth rate of fluctuations, $f_g(z)$ in several bins between z = 0 and z = 2 with an error of less than 10%, or the growth rate parameter γ to a precision of a few percent of better, (ii) To extract separate constraints on the two relativistic potentials Φ and Ψ by combining probes of weak lensing, galaxy clustering and redshift space distortions and (iii) To measure independently the expansion rate and the angular diameter distances in several redshift bins to test the cosmological principle.*

## 2.2.3 Dark matter

Over 80% of the matter in the Universe is in a dark, non-baryonic form known as dark matter. Due to this dominance, dark matter drives structure formation and as it undergoes gravitational collapse, baryonic matter follows. The behaviour of stars, galaxies and gas therefore depends a great deal on the underlying gravitational potential created by the dark matter field. Dark matter interacts gravitationally in the same way as normal baryonic matter, however it does not interact through the electromagnetic force. Observations must therefore rely on inferring its presence through the gravitational effect it has on light (gravitational lensing) or baryonic matter (galaxy clustering, rotation curves etc). The presence of dark matter has been detected on all scales: on galaxy scales a non-luminous matter component is needed to explain observed radial velocity profiles, on cluster scales the velocity dispersion of galaxies implies a large non-luminous matter component and on cosmological scales the total matter contribution to the energy budget of the Universe is approximately 6 times larger than the baryonic matter content alone. To explain all of these independent observations, our concordance model calls for dark matter that is cold (not relativistic) and non-interacting (i.e. only responds to gravity). The Cold Dark Matter (CDM) paradigm underpins our cosmological standard model, ΛCDM, and yet the nature of the cold dark matter is entirely unknown.

To explain dark matter two competing theories were initially proposed: dark matter could be a macroscopic population of compact objects or dark matter could be a new sub-atomic particle. However microlensing observations of Milky Way halo objects have now ruled out a population of dark compact objects convincingly. Within the sub-atomic description there are many potential dark matter candidates that come from non-standard theories of particle physics; for example in minimal supersymmetric theories the neutralino, gravitino and axino could be stable long-lived particles (Steffen, 2008). Experimental upper limits exist from direct detection experiments on the interaction cross-section of dark matter with baryonic matter. However the self-interaction cross-section (dark matter-dark matter collisions) and kinetic temperature can only be determined from astronomical observations. If dark matter has a non-negligible self-interaction cross-section dark matter-dark matter collisions will occur in the highest density regions of the Universe, at the centers of galaxies and galaxy clusters. If dark matter particles are fast moving (not cold but "warm" or "hot"), then the dark matter density in the centers of galaxies and clusters, and the abundance of small galaxies will be reduced relative to the cold dark matter paradigm. This is because the dark matter will tend to stream out of the small high-density regions. Dark matter in general forms *halos* over a wide range of scales from galactic to cluster sized clumps (see Figure 2.5 for an illustration). Measuring the sub-structure or "clumpiness" and the global properties of galactic and cluster scale dark matter halos can help to determine dark matter properties.

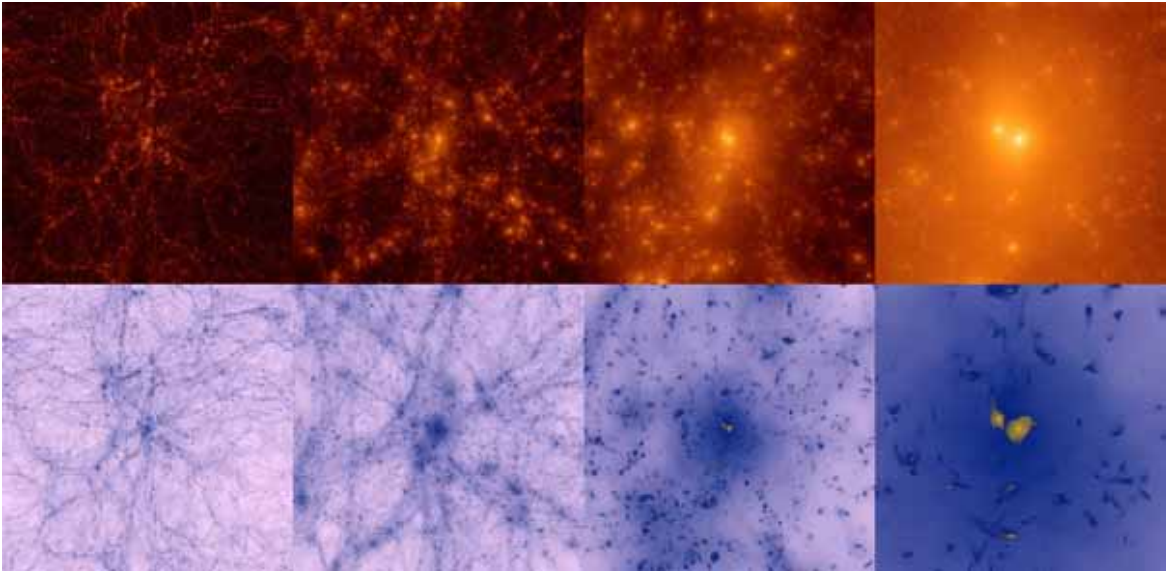

*Figure 2.5: The top panels show the dark matter distribution in a concordance cosmology simulation and the bottom panel shows the distribution of gas (baryonic). We see that gas follows the matter closely. Moving from left to right, the panels zoom in by a factor of 4 into a central cluster. On the far left we see that dark matter on cosmic scales forms an intricate web (sometimes referred to as the cosmic web). On the far right we see a zoom in a bound halo, which is where galaxies reside. (c.f. Pichon, Rasera and Teyssier – Horizon Project)*

Euclid will constrain the dark matter particle mass using gravitational lensing and galaxy clustering (Section 2.3.7) using the number of dark matter haloes and substructure over a wide range of masses. By measuring the large halo properties of dark matter around thousands of galaxy clusters over a wide range of masses Euclid will constrain the "clumpiness" and "cuspiness" of the cluster scale dark matter distribution. Gravitational lensing and galaxy clustering can also be used in 3D to construct the cosmic web of dark matter, these techniques are complementary – gravitational lensing probes the dark matter distribution directly and galaxy clustering maps the baryonic distribution. On large cosmic scales the dark matter properties are encapsulated in its power spectrum, which is the Fourier transform of the auto-correlation of the dark matter distribution. Figure 2.6 shows how galaxies and other probes can be combined to measure the power spectrum and compare it to model predictions.

Massive neutrinos are a known, hot (relativistic) component of dark matter. As a result they damp the formation of structure by providing an effective pressure as they stream out of areas of high density in the early Universe; this effect leaves a distinctive signature on the dark matter power spectrum. Neutrino oscillation experiments have determined the mass difference (squared) between neutrino species, which proves that neutrinos have non-zero mass, but they cannot constrain the total (sum) of the neutrino masses to the accuracy required to place constraints on particle physics models. The constraints on the mass squared also leave an ambiguity in the hierarchy of the individual neutrino masses. Since neutrinos are very light and contribute only a fraction of the dark matter, it is necessary to measure a large range of scales to achieve the required sensitivity. Euclid will use gravitational lensing and galaxy clustering to produce an unprecedented measurement on the detailed shape of the power spectrum, to constrain the neutrino mass and number to a high accuracy, while being also sensitive to individual neutrino mass states and constrain the neutrino hierarchy (Section 2.3.7). This measurement will also contain memory of initial conditions from the Big Bang (see Sections 2.2.4, 2.3.5 and 2.3.7).

The scientific dark matter targets for Euclid therefore are

*(i) To detect dark matter halos between a mass scale of > $10^{15} M_\odot$ to $10^8 M_\odot$. (ii) To determine the neutrino mass to the percent level, approximately 0.03 eV and place percent-level constraints on the number of neutrino species and the neutrino hierarchy. (iii) To determine the dark matter mass profile on cluster and galactic scales. (iv) To reconstruct an all-sky three-dimensional dark matter map to z=2.*

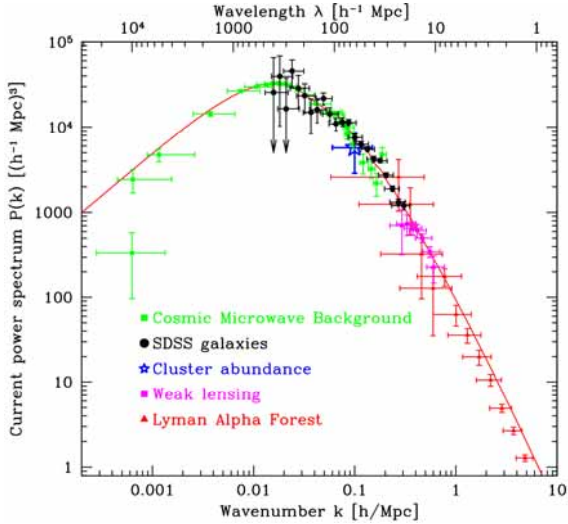

*Figure 2.6: Current measurements of the dark matter power spectrum P(k) from the CMB as well as from weak lensing (cosmic shear), galaxy clustering, cluster abundances and the Lyman alpha forest (c.f. Tegmark et al., 2004).*

## 2.2.4 Cosmic Initial Conditions

In our standard cosmological model, the initial primordial fluctuations are caused by a mechanism called inflation (Starobinsky, 1980; Guth, 1981), which can be described by a rapid expansion occurring approximately $10^{-36}$ seconds after the Big Bang, see Figure 2.7. During inflation quantum fluctuations are inflated to macroscopic scales. The major success of inflation is that it can explain several conundrums in the Big Bang theory, including the observed flatness and extreme homogeneity of the Universe; these are a result of the early inflationary period causing the entire observable Universe to originate from a very small and causally connected volume.

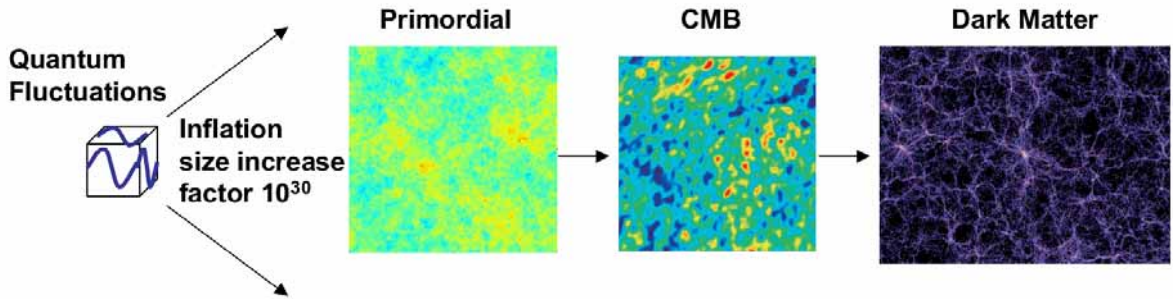

*Figure 2.7: The evolution of structure is seeded by quantum fluctuations amplified by inflation. These seeds of structure grow to create the CMB anisotropies after approximately 100,000 years and eventually the dark matter distribution of today after 13.6 billion years. The CMB probes the Universe when it was a few 100,000 years old, while Euclid will observe the Universe from a few billion years old to the present. By combining Euclid and CMB data we will map the evolution of structure over orders of magnitude in time.*

Most inflation models require a new scalar field. This field is subject to quantum fluctuations and if the potential is suitably chosen, the field gives rise to exponential expansion. The slow-roll inflation models are the simplest of these models that lead to inflation, in these models the field rolls slowly down the potential. There is currently no strong reason to favour a particular form of the inflationary potential and future observations from Euclid and Planck will help discriminate between models. The power spectrum of the quantum fluctuations that are produced during the inflationary period is linked to evolution of the field. It is often written in terms of a scalar spectral index *n* that can depend on scale,

$$P(k)_{init} \propto k^{n(k)}.$$

For slow-roll models, $n$ is close to 1 (corresponding to a scale-invariant spectrum of perturbations), but there are small deviations due to the field evolution. The spectral index can also be expressed using a second parameter, the running spectral index $\alpha$, defined by $n(k)=n(k_0)+1/2\alpha\ln(k/k_0)$, where $k_0$ is the pivot scale. Determining the value of the spectral index and its running is therefore of considerable interest because it is a direct probe of the initial conditions and the inflationary period.

These primordial fluctuations subsequently evolve through gravitational instability to become cosmic structure observed today in the Universe – dark matter accumulates around the early small overdensities and forms early structures that the baryonic matter flows into as described in Section 2.2.3. Hence the matter power spectrum $P(k)$ is a processed version of the initial power spectrum, and so is a powerful probe of the dynamics of inflation and ultimately of inflationary models. In fact the dark matter power spectrum can be used to measure a number of inflationary parameters: the slope of the power spectrum can help us understand the spectral index and the running spectral index; the amplitude of the power spectrum $\sigma_8$ is directly related to the amplitude of the primordial fluctuations.

In addition to scale-invariance, simple inflation models usually predict that the primordial fluctuations are very nearly Gaussian in the early Universe, hence detecting any non-Gaussianity is a signature that the inflationary period was considerably more complex. In the case of Gaussian fluctuations, the dark matter power spectrum contains the full statistical information about the perturbations; however in the presence of non-Gaussianities, more information can be extracted. The simplest quantity to investigate is $f_{NL}$, which parameterises a quadratic ($\chi^2$) admixture to the initial perturbations. This leads for example to a non-zero bispectrum. Euclid will measure the non-Gaussianity parameter $f_{NL}$ to high precision using both gravitational lensing and galaxy clustering (see Section 2.3.5 for details).

The CMB is an exquisite probe of the initial conditions – both primordial parameters and non-Gaussianity. The photons observed in the CMB are a tracer of the state of the Universe after only a few hundred thousand year, hence the fluctuations in the CMB closely trace the primordial fluctuations. However the constraints from the CMB have finite information available since the CMB photons come mostly from a single time (recombination) and from large scales which results in parameter degeneracies. Euclid will observe the Universe from when it was a few billion years old to the present day, and on much smaller scales, thus measuring the primordial power spectrum through the evolved dark matter power spectrum. Euclid and the CMB will therefore have perfect synergy – probing the Universe over many orders of magnitude in time and scale, and, in combination, reach high accuracies.

In summary, to make progress towards understanding the origin of structure we need:

*(i) To measure the matter power spectrum over a large range of scales in order to extract values for the parameters $\sigma_8$ and $n$ to less than 1% and improve constraints on $\sigma_8$ and $n$ by over a factor 30 and 2 respectively compared to Planck alone. (ii) For models allowing for running spectral index and a tensor to scalar ratio, to improve constraints on $n$ and $\alpha$ with respect to Planck alone by a factor 2. (iii) To measure the non-Gaussianity parameter $f_{NL}$ to $\pm 10$.*

## 2.2.5  Summary of Primary Science Objectives

We have outlined the main four sectors of the $\Lambda$CDM concordance model, which have become the standard model of cosmology. We summarise these targets in Table 2.1.

In section 2.3 we identify the cosmology probes that Euclid will use to meet these objectives and how these experiments can be designed to have the statistical power that is necessary. This will focus mainly on weak gravitational lensing and galaxy clustering. To meet this statistical potential, Euclid will need to overcome a number of challenges so as to control systematic errors. Some of the main difficulties are discussed in section 2.4 along with mitigation strategies for Euclid.

# 2.3  Science objective by measuring galaxy shapes and positions

By designing a survey able to reach galaxies up to a redshift of 2, we will be able to precisely track the evolution of large-scale structure back to when the Universe was one third of its current age. In this section, we demonstrate the wealth of information provided by high precision measurements of galaxy shapes and positions.

*Table 2.1: Summary of Euclid's primary science objectives*

| Sector | Euclid Targets | |
|---|---|---|
| Dark Energy | (i) | Euclid *alone* to measure $w_p$ and $w_a$ to 2% and 10% (FoM$_{DE}$ = 500) |
| | (ii) | Look for deviations from $w$ = -1, indicating a dynamical dark energy. |
| | (iii) | Measure the cosmic expansion history to better than 10% for several redshift bins from $z = 0.5$ to $z = 2$. |
| Test of Gravity | (i) | Measure the growth index, $\gamma_m$, to a precision better than 2%. |
| | (ii) | Measure the growth rate to better than 5% for several redshift bins between $z = 0.5$ and $z = 2$ |
| | (iii) | Separately constrain the two relativistic potentials $\Phi$ and $\Psi$ |
| | (iv) | Test the cosmological principle |
| Dark Matter | (i) | Detect dark matter halos between a mass scale of >$10^{15}$ to $10^8$ M$_\odot$ |
| | (ii) | Accuracy of a few hundredths of an eV on the sum of neutrino masses, the number of neutrino species and the neutrino hierarchy. |
| | (iii) | Measure the dark matter mass profile on cluster and galactic scales. |
| Initial Conditions | (i) | Measure the matter power spectrum on a large range of scales in order to extract values for the parameters $\sigma_8$ and $n$ to 1%; improve constraints on $\sigma_8$ and $n$ by over a factor 30 and 2 respectively compared to Planck alone |
| | (ii) | For extended models, improve constraints on $n$ and $\alpha$ with respect to Planck alone by a factor 2. |
| | (iii) | Measure the non-Gaussianity parameter $f_{NL}$ to ± 10. |

## 2.3.1 Expansion and Growth Histories through Gravitational Lensing

Light is gravitationally deflected by the curvature of space-time around massive objects, which can distort the images of background objects such as galaxies. The most subtle effect, weak lensing, will make an intrinsically circular galaxy appear as an ellipse (Blandford et al 1991, Bartelmann & Schneider 2000, Réfrégier 2003). Lines of sight that pass closer to more massive foreground bodies undergo a stronger deflection, this can produce higher order distortions like flexion (Goldberg & Natarajan 2002) (which turns circular galaxies into banana-shaped arcs), and strong lensing, where a single source will give rise to multiple images and giant arcs. Each of these lensing regimes (examples shown in Figure 2.8) requires high image quality and resolution, which is best achieved by a space-based telescope. They all provide unique measures of the dark matter and are directly sensitive to all mass (dark matter, hot gas, stars etc), irrespective of the dynamical or thermal state of the lensing material.

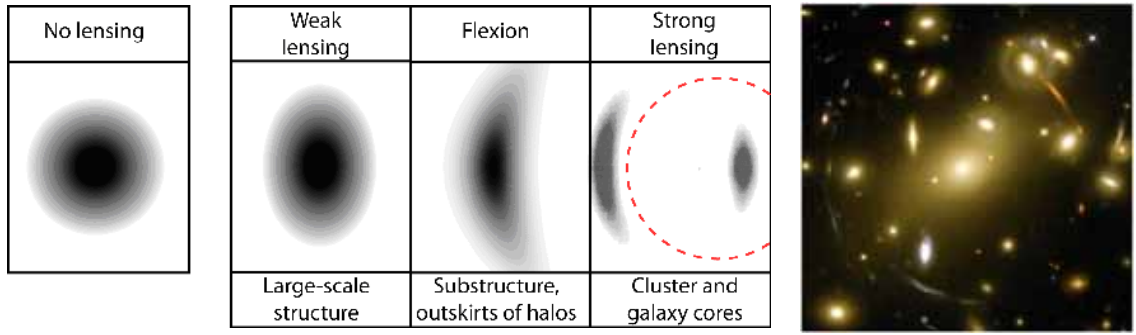

*Figure 2.8: a. (Left) Illustrations of the effect of a lensing mass on a circularly symmetric image. Weak lensing elliptically distorts the image, flexion provides an arc-ness and strong lensing creates large arcs and multiple images. b. (Right) Galaxy cluster Abell 1689, strongly lensed arcs can be seen in around the cluster. Every background galaxy is weakly lensed.*

Euclid has been optimised for weak lensing. By measuring the coherent pattern in galaxy ellipticities that gravitational lensing by large-scale structure imprints, we are able to extract information on the shape of the matter power spectrum, broad-band power, neutrino effects etc.; the growth factor as a function of redshift; and the redshift-distance relations, which are a direct probe of the geometry and expansion history of the Universe. This can be seen by considering the integrated foreground mass, $\kappa$, which determines the strength of the lensing signal to a galaxy at redshift, $z_s$,

$$\kappa = \frac{3H_0^2\Omega_m}{2c^2} \int_0^{\chi_s} d\chi \frac{D(\chi)D(\chi_s-\chi)}{\chi_s}(1+z)\delta(\chi),$$

where $\delta$ is mass over density, $z$ is redshift, $\chi$ is the comoving radial distance, $D$ is the angular diameter distance, $\chi_s$ is the co-moving radial distance to the source galaxies at redshift $z_s$ (i.e. those that are lensed), $\Omega_m$ is the matter density of the Universe, $H_0$ is the present-day Hubble constant and $c$ is the speed of light. Weak lensing is therefore a very precise probe of cosmology and offers us a broad handle on all cosmological sectors with the growth of mass perturbations and background geometry simultaneously contributing to signal.

A large number of statistical methods have been developed for capturing this information. In this report we focus mainly on the power spectrum (two-point correlation statistic – e.g. see Figure 2.9) and the bispectrum (three-point statistic) for which the Euclid weak lensing survey has been optimised. The redshift information of the galaxies can be used by either dividing the galaxies into redshift bins and cross-correlating the shapes of galaxies in different bins (Hu 1999, Jain & Taylor 2003, Bernstein & Jain 2004) or by using a full 3D treatment of the lensing field (Heavens, Kitching & Taylor 2006). Additional tests, including the shear ratio test of cosmic expansion history (Taylor et al 2007) and galaxy-galaxy lensing (e.g. Parker et al 2007), will add to Euclid's performance.

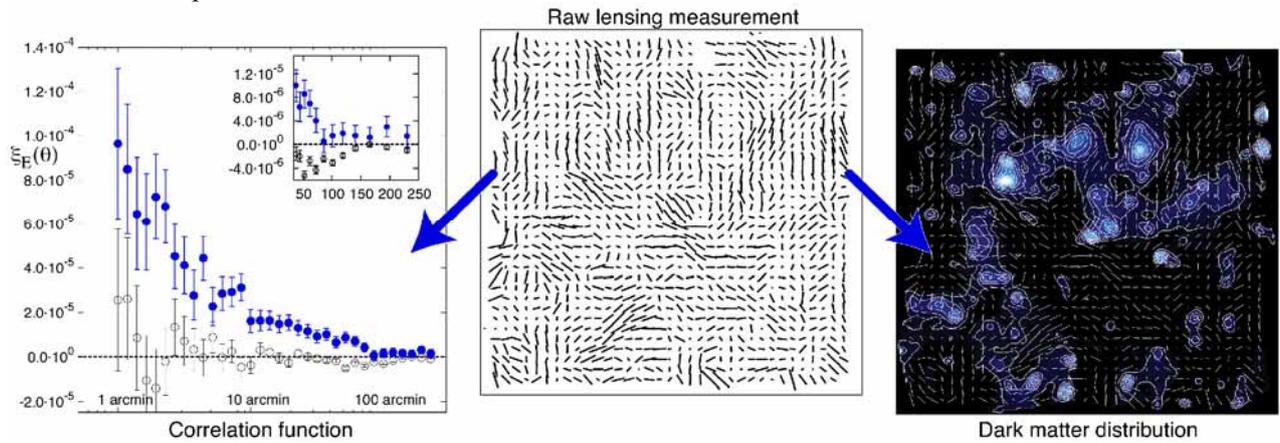

*Figure 2.9: (Middle panel) Raw gravitational lensing data obtained from HST observations. Each line shows the average shape of about 200 galaxies. Lines of sight without coherent lensing produce circular average galaxies, represented by a dot; the line length indicates the magnitude of the shear, and the orientation shows the major axis. (Right panel) Foreground dark matter lenses are mapped by filtering the observed shear field for the circular patterns. (Left panel) The clumpiness of dark matter on different physical scales can be quantified statistically via the correlation between shear along different lines of sight (data from the Canada_France-Hawaii telescope). In both cases, an independent measurement of systematics can be made via the B-mode curl signal, which is not produced by lensing and acts as a control experiment. This is not shown for the map, but is the open symbols in the left panel; it should be consistent with zero but is not quite because of residual systematic effects due to the Earth's atmosphere.*

As a stand-alone method the lensing constraints from Euclid will improve upon the accuracy of CMB and large scale structure experiments in many cosmological sectors (e.g. dark energy, modified gravity, dark matter). In combination with other cosmological probes, particularly galaxy clustering, lensing constraints will break many parameter degeneracies. The remarkable ability of weak lensing to illuminate the nature of the Universe has been noted by many reports including the ESO-ESA Working Group on Fundamental Cosmology (Peacock et al., 2006) and the Dark Energy Task Force (Albrecht et al., 2006).

Measurements of cosmic shear were first enabled by the development of large area digital imaging detectors a decade ago (Bacon, Réfrégier & Ellis 2000, Kaiser et al 2000, Wittman et al 2000, Van Waerbeke et al 2000). The technique has since evolved rapidly and now all current and planned wide-area optical surveys include a major weak lensing component that typically drives the optical requirements and survey area. Recent studies have measured cosmic shear from both the ground over tens of square degrees (Jarvis et al 2006, Hetterscheidt et al 2007, Fu et al 2008) and from space over degree-scales (Massey et al 2007, Schrabback et al 2007). Despite the relatively small-scale surveys available today, weak lensing has been used to measure many physical phenomena. Weak lensing has provided precise constraints on modified gravity (e.g. Thomas et al., 2008; Dore et al., 2007) where constraints have ruled out DGP models at 2-sigma. The neutrino mass (e.g. Ichiki et al., 2009; Tereno et al., 20009) has been constrained to $m_v < 0.54$eV using weak lensing measurements. Notably weak lensing has provided the most direct evidence for dark matter in both clusters (e.g. the "bullet cluster", Clowe et al., 2004) and on cosmological scales (e.g. Massey et al, 2007) where dark matter has been mapped in 3D using gravitational lensing (e.g. Taylor et al., 2006; Simon et al., 2009).

Future weak lensing missions will need to control a number of observational and astrophysical systematic effects to reach their full potential. Only a dedicated space-based mission can mitigate all these problems, in ways that are simply impossible from the ground. The main systematic effects are discussed in section 2.4.1 along with details of how Euclid will control their impact. In particular, deep near infrared photometry is needed for accurate redshift measurements of distant galaxies ($z > 1$) which cannot be achieved from the ground. Space also allows us to perform measurements with a small and stable Point Spread Function (PSF) which is not possible from the ground because the PSF properties are dominated by atmospheric effects.

**Optimization and Sensitivity:** The optimal survey configuration depends on the area of the survey ($A$), the number density of galaxies with measureable shapes ($n_g$), and the median redshift of the galaxies ($z_m$). Cosmic shear surveys are also sensitive to accuracy with which galaxy redshifts can be measured ($\sigma(z)$) and calibrated. For a fixed observing time and a median galaxy redshift greater than z~0.8, the dark energy equation of state parameters $w_0$ and $w_a$ are determined more accurately by increasing the survey area instead of increasing its depth (Amara & Réfrégier 2007). Only after the entire extragalactic sky has been covered is statistical power improved by increasing survey depth. This optimization assumes that systematic errors are always kept below statistical errors. At the level of precision required by Euclid, a number of potential sources of systematic contamination must be controlled (i) high quality galaxy images with well known, stable PSFs are needed for accurate PSF deconvolution (ii) well-understood photometric redshifts are needed for measuring the evolution of cosmic structure (iii) precise photometric redshifts are needed for removal of any intrinsic shape alignments of galaxies. All of these are addressed by an optical and NIR survey performed in the stable and low background environment of space, as discussed further in Section 2.4.

## 2.3.2 Expansion and Growth Histories through Galaxy Clustering

The power spectrum $P(k)$ or the two points correlation function of the galaxy distribution, and its evolution with time, encapsulates information on the cosmological model and the values of its fundamental parameters. On large scales, the distribution of galaxies follows the distribution of seed density fluctuations created in the early Universe, and encodes the physics of this era. The global shape of the power spectrum therefore depends on the baryon, dark matter and neutrino densities. On smaller scales, we use the clustering pattern at different redshifts as a standard ruler to probe the expansion history of the Universe. Baryon Acoustic Oscillations (BAOs) provide a standard ruler that can measure dark energy with the lowest systematic error of any known probe (Albrecht et al. 2009). With careful modelling of galaxy properties, we can expand on this test, and use the full 3D power spectrum. The velocities of galaxies produced by the growth of density fluctuations will affect the power spectrum by introducing anisotropy in the observed clustering. These redshift-space distortions are sensitive to the growth rate on cosmic scales and thus can test the fascinating hypothesis that dark energy is a modification of the gravity theory. Euclid will perform high precision measurements of the full power spectrum $P(k)$, the BAO feature and redshift distortions for redshifts out to z~2, providing the expansion and growth histories of the Universe over the time interval when dark energy becomes dynamically important.

**Geometry and Expansion History using Baryonic Acoustic Oscillations:** BAOs are the fossils of sound waves in the photon-baryon plasma that existed in the early Universe. As the Universe expanded and cooled, the photons and baryons decoupled, and the acoustic waves became frozen, imprinting their signature on both the CMB and the matter distribution. The resulting peaks in the CMB fluctuation spectrum were first detected by experiments in 2000 (Boomerang: de Bernardis et al. 2000) and were detected with high signal to noise from space by WMAP (Hinshaw et al. 2003). The corresponding signal in the power spectrum $P(k)$ is out of phase with the CMB peaks and has weaker amplitude, because the density of dark matter is higher than that of baryons by a large factor. This BAO signal has been seen in the galaxy distribution as a preferred co-moving separation of galaxies of ~150 Mpc (using the SDSS: Eisenstein et al. 2005), or, equivalently, as a series of oscillations in the galaxy power spectrum (using the 2dFGRS: Cole et al. 2005). Further analyses of the SDSS and 2dFGRS have now fully exploited the largest volume of the Universe currently observed covering ~1 Gpc[3] (Percival et al 2007; Cabre & Gaztanaga 2009; Percival et al. 2009). There are tentative signs that, by separating the radial and tangential BAO signals, competitive constraints can be derived on $w(z)$, without the need to appeal to external datasets (Gaztanaga et al. 2008a, 2008b; Sanchez et al. 2009). In fact, the measured size of the BAO feature is a projection of the sound horizon $s$ at the baryon-drag epoch (shortly after the decoupling of CMB photons from baryons). BAO therefore constrain $sH(z)$ in the radial direction, and $s/D(z)$ in the direction transverse to the line of sight, where $H(z)$ is the Hubble parameter, and $D(z)$ is the angular diameter distance.

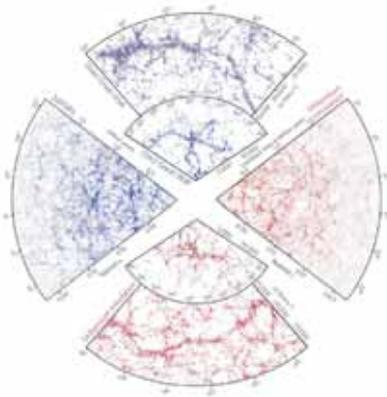
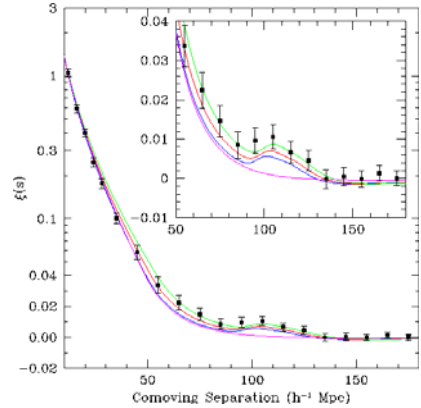

*Figure 2.10: a. (Left panel) The galaxy distribution in the largest surveys of the local Universe, compared to simulated distributions from the Millennium Run (Springel et al. 2005); b. (Right panel) The two-point correlation function of SDSS "luminous red galaxies", in which the BAO peak at ~$10^2$ $h^{-1}$ Mpc has been clearly detected (Eisenstein et al. 2005).*

Despite the relatively recent discovery of BAO in the local galaxy distribution, this technique has already demonstrated its ability to place competitive constraints on the values of cosmological parameters (e.g. Sanchez et al 2009; Percival et al 2009). The full power of this technique is realised when combined with CMB observations (e.g. Wang & Mukherjee 2006; Percival et al 2007; Komatsu et al. 2008; Sanchez et al 2009).

**The Full Galaxy Power Spectrum:** The Euclid galaxy redshift survey will provide an unprecedented precise measurement of galaxy clustering over the maximum possible range of separations given by the full sky, not just on the BAO scale. The full galaxy clustering signal (i.e. the complete power spectrum) can be used as our standard ruler, rather than extracting out just the BAO signal. This provides significantly more information, boosting the FoM of dark energy by a factor of a few beyond using BAO only constraints (see Table 2.2 below). This extra information comes at the price of an increased systematic burden: we now need to model the relationship between the full galaxy-clustering signal and the predicted linear matter power spectrum (Rassat et al. 2009). We therefore consider both approaches: the conservative BAO-only dark energy constraints, and constraints using the full power spectrum.

The use of the galaxy power spectrum $P(k)$ yields stronger constraints on dark energy than using BAO alone because $P(k)$ itself also includes information about dark energy (i.e. our standard ruler is itself interesting). For example, in quintessence models with appreciable densities of dark energy around the epoch of matter-radiation equality, the width and form of the turnover in the matter power spectrum is significantly different from that expected in ΛCDM (see e.g. Jennings et al. 2009). $P(k)$ provides constraints on the matter density, baryon density, and primordial matter power spectrum that are complementary to those from the CMB.

**Growth History of Structure from Redshift-Space Distortions:** As a sort of "test particles", galaxies partake in the overall growth of cosmic structures, which are slowly built up by gravity starting from initial fluctuations. This induces "peculiar" velocities, i.e. galaxy motions that deviate from the smooth Hubble flow. On large scales these velocities take the form of coherent bulk flows out of voids towards filaments and superclusters, whilst on small scales, the motions are randomised inside gravitationally bound structures. As a result, the observed galaxy distribution in redshift space is distorted, introducing a measurable anisotropy in the clustering signal. The "strength" of the distortion field is given by the product $f\sigma_8$ of the growth rate $f$, as defined in Sect. 2.2.1: $f \equiv \mathrm{dlog}(G)/\mathrm{dlog}(a)$, with the *rms* amplitude of fluctuations in the matter, $s_8$ (Kaiser 1987). Since velocities are non-relativistic, we are only testing time-like metric fluctuations. This information is therefore fully complementary to weak-lensing measurements. The value of $f*s_8$ (or equivalently of the observed distortion parameter $\beta \equiv f/b$, where $b$ is the linear bias factor of the galaxies being analyzed, given on large scales as $b=\sigma_8(\mathrm{gal})/\sigma_8$ -- Kaiser 1987, Hamilton 1998), can be extracted by proper modeling the anisotropy of $P(k)$ or the correlation function. Recent work (e.g. Guzzo et al. 2008, Percival & White 2008; Wang 2008; White et al. 2009, McDonald & Seljak 2009) has demonstrated that redshift-space distor-tions measured at different redshifts from a large galaxy redshift survey provide us with a powerful method to trace *f(z)* back in time, which we have seen in Section 2.2.2 allows to test for modifications of GR.

**Optimizations and Sensitivity:** The statistical potential of a redshift survey that aims at measuring accurately cosmological parameters through the clustering power spectrum (or correlation function) and its anisotropy, depends essentially on: (1) Survey area; (2) Redshift range covered; (3) Galaxy number density as a function of redshift; (4) Galaxy bias as a function of redshift; (4) Redshift accuracy. To optimize the survey performances we used values and functional forms for these quantities stemming from our current knowledge of the galaxy distribution, galaxy formation models and the instrument performance. For the latter, we have considered and carefully compared the two options of a slitless (emission-line galaxy) survey and a DMD slit survey (see Appendix 1). Plots for the assumed redshift distribution of galaxy density and galaxy bias are shown in Section 2.5.

## 2.3.3  Experiment Design for Lensing and Clustering Science

Weak Lensing and galaxy clustering form an ideal combination of probes for detailed and powerful measurements of the dark sector with natural synergies in observing strategy: (i) both measurement galaxy properties in a similar wavelength range, (visible to near infrared (NIR)), (ii) there is no need for repeated observations, (iii) both favour surveys covering the widest possible area for a median redshift *z*>0.8. Given this the measurements still have key differences. Gravitational lensing relies on very high quality images of the galaxies, with redshift determined photometrically. For galaxy clustering studies, redshifts need to be determined spectroscopically. To meet these requirements, Euclid will perform an ultra wide survey of > 20,000 deg$^2$ with high quality images for galaxy shapes, optical and NIR photometry for photo-z, and NIR spectroscopy for spectroscopic-z.

**Imaging Components of the Wide Survey:** To reach percent level precision on equation of state parameters, Euclid will measure the shapes of 30 to 40 galaxies per arcmin$^2$ over the 20,000 deg$^2$ observable extragalactic sky. This leads to a total of roughly 3 billion galaxy measurements with a median redshift at *z*~1. The imaging instrument must have sufficient accuracy to be able to resolve small distant galaxies requiring a PSF that is stable and small (0.18 to 0.23 arcsec).

**Spectroscopy Component of Wide Survey:** To reach our objectives redshifts need to be measured to a precision of $\sigma(z) = 0.001(1+z)$. For Euclid we are aiming for galaxies in the range of at least 0.5<*z*<2.1. This will require spectroscopy in the NIR using Hα emission as tracer. With this redshift range in place, increasing the area maximizes Euclid's performance. This drives us toward a wide (20,000 deg$^2$) survey.

## 2.3.4  Euclid Additional Cosmological Probes

Here we specifically highlight the added benefit of: (i) galaxy clusters and (ii) integrated Sachs-Wolfe effect coming from the Euclid wide survey.

**Galaxy Clusters:** Galaxy clusters are the most massive objects in the Universe and bear imprints of three processes: a) the spectrum of initial fluctuations; b) the growth of these structures over time and c) the dynamics of the collapse of halos. This threefold dependency makes clusters an excellent probe of the growth of structure in the Universe. The quantities of interest are the redshift distribution of clusters and their spatial distribution. These quantities depend on the observed volume, which in turn is cosmology dependent. The expected redshift distribution of clusters above a given mass in different cosmologies is shown in Figure 2.11a. Here we assumed that all clusters above a mass limit of $5 \times 10^{13}$ h$^{-1}$ M$_\odot$ are detected. The estimate of the mass limit is based on the properties of the SDSS cluster sample (Koester et al. 2007, Rozo et al. 2007). We predict that Euclid will discover over half a million clusters. This number is most likely to be larger due to the up-scatter of low mass clusters into the sample. It is evident from the plot that galaxy cluster counts can be a sensitive probe of different models of dark energy and in particular modifications of gravity. The distribution of clusters can provide important complementary information (Majumdar & Mohr 2004; Lima & Hu 2004) and allows additional (self-) calibration of the cluster uncertainties. In addition, this provides the possibility of constraining non-Gaussian models through scale-dependent bias (Fedeli et al. 2009; Oguri 2009).

The precision of cosmological parameters derived from galaxy cluster surveys, however, is currently limited by the uncertainty of the cluster mass estimates. For the large number of clusters that will be detected in the Euclid wide survey, the shot-noise term is almost negligible compared to the sample variance. Hence the interpretation of galaxy cluster counts requires a clear understanding of the selection function, systematic errors, and in particular the scatter and statistics of the relevant mass-observable relations. The large sample and the complementary methods of detecting and observing clusters enable a self- and cross-calibration of scaling relations and scatter around them (Lima & Hu 2004, 2005, Cunha 2008).

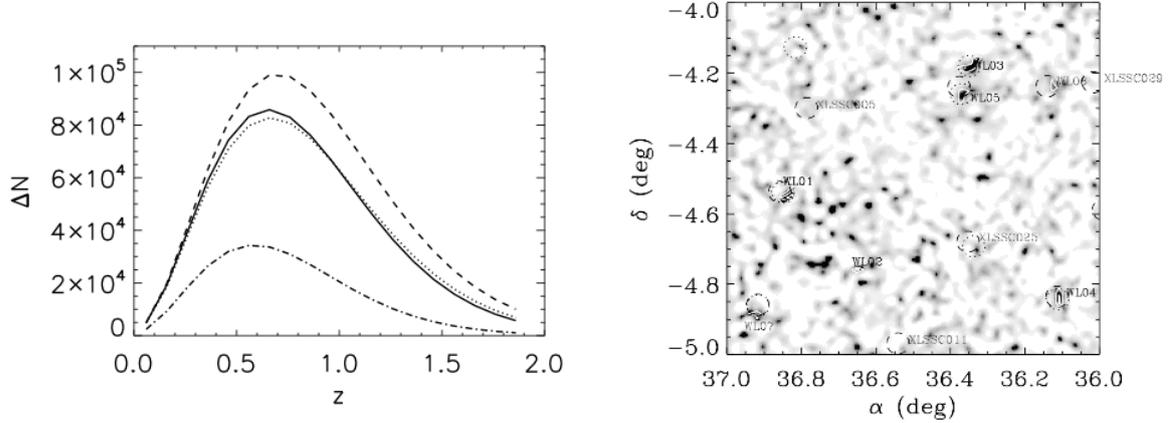

*Figure 2.11: a. (Left) The expected redshift distribution of galaxy clusters with mass larger than $5 \times 10^{13}$ $h^{-1}$ $M_{\odot}$ in different cosmologies, over 20,000 $deg^2$ and in bins of width $\Delta z$=0.1, for a survey as outlined in Section 2.3.3. Lines show a $\Lambda CDM$ model (solid), a w=-0.9 model (dotted) and a modified gravity model ($\gamma$=0.68; dashed).The dot-dashed line is for a $\Lambda CDM$ model with a mass limit of $8 \times 10^{13}$ $h^{-1}$ $M_{\odot}$. The prediction assumes a mass-observable relation with a 10% systematic bias in log(Mass). b (Right) Weak Lensing Shear map for a CFHTLS field containing X-ray selected galaxy clusters (Berge et al. 2008).*

For a Euclid type experiment clusters can be selected in mainly three ways: a) by targeting galaxy over-densities, either in projection only or by combining with redshift space information, with methods like the brightest cluster galaxy (maxBCG, Koester et al. 2007) or the optical richness by selecting galaxies along the red sequence (Rozo et al. 2007, 2009), b) strong lensing (Smith et al. 2003); c) weak lensing (Wittman et al, 2000, 2006; Sheldon et al. 2007, Berge et al. 2008) and peak statistics. Identifying clusters of galaxies by a collection of galaxies is sensitive to the galaxy formation process, which is not well understood. However, empirical optical richness – mass relations have been successfully used to determine cosmological parameters (Gladders et al. 2007, Rozo et al. 2009). Identification and follow-up observations of clusters detected in weak lensing surveys have been employed in combination with X-ray selected clusters (Pedersen and Dahle, 2006, Berge et al. 2007). These studies show weak lensing mass estimates considerably improve the calibration of the X-ray mass – temperature relation.

Future surveys like eRosita using X-rays, or Planck using Sunyaev-Zel'dovich selection, will discover thousands of clusters. Euclid will provide mass proxies via the stacking of clusters in bins of the observables, like X-ray or SZ flux, and estimating the masses via weak lensing shear. Although the method is noisy for individual clusters, it has the advantage that it does not depend on uncertain cluster gas or galaxy formation physics. Furthermore, Euclid's spectroscopy will directly give a redshift determination for a vast number of cluster galaxies identified in these cluster surveys. This will in turn allow one to carry out dynamical measurements of cluster masses, either through the virial theorem (e.g. Girardi et al. 1998), or through the method of "caustics" (e.g. Diaferio 1999). This has important consequences for cosmological applications of the Euclid cluster samples. Tests based both on observational data and on numerical simulations show that a robust determination of cluster velocity dispersion, $\sigma_v$, to be used for virial mass estimates, requires measuring redshifts for at least 50 member galaxies (e.g., Biviano et al. 2006). In Figure 2.12 we show the redshift dependence of the limiting value of $M_{200}$[1] of the clusters for which 50 redshifts can be measured, for both slitless H$\alpha$ and slit surveys. The slitless H$\alpha$ survey will provide about 120 such clusters at z>0.5, a number that increases to about 4000 for the slit survey, of which about 10% will be at z>0.8.

---

[1] Here $M_{200}$ is defined as the mass within the radius encompassing a density of 200 times the cosmic critical density.

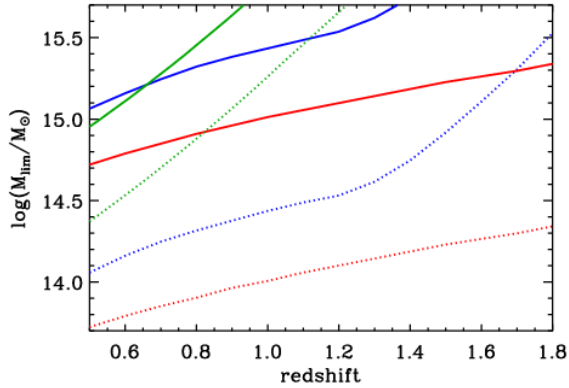

*Figure 2.12: The redshift dependence of the minimum value of the cluster mass $M_{200}$ for which at least 5 (dotted lines) and 50 (solid lines) redshifts are measured, for the slitless Halpha survey (blue lines), and the DMD slit survey (red lines). These results are based on using the relation between $M_{200}$ and richness by Popesso et al. (2007), with richness computed by integrating the luminosity functions by Lin et al. (2003) and Iglesias-Paramo et al. (2002), down to the surveys' sensitivity limits, using a 1/3 sampling rate. The green curves show the limiting cluster mass ($H\_0=70$ km/s/Mpc) for Weak Lensing detections with S/N=7 (solid line) and 3 (dotted line)*

**Integrated Sachs Wolfe Effect:** The integrated Sachs-Wolfe effect (Sachs & Wolfe 1967, ISW) is a secondary anisotropy of the CMB and a direct signature of dark energy. The effect is caused by the decaying gravitational potentials of low redshift large-scale structure. This may occur through cosmic curvature, in the presence of dark energy at late times, or in alternative models of gravity. Assuming general relativity is correct, and that the Universe is spatially flat, then a detection of the ISW effect is a direct signature of the presence of dark energy at late times. Furthermore, it has recently been shown that the ISW spectrum will also present a signature of redshift distortions (Rassat, 2009), increasing the amount of information which can then be extracted from the ISW signal as well as providing an extra handle on the growth through the distortion parameter β (see section above on redshift distortions). In addition, the ISW effect also traces the presence of dark energy inhomogeneities on large scales (Weller & Lewis 03, Bean & Dore 03). The ISW effect is a weak signal and can be detected only through correlations of large-scale structure with the CMB (Crittenden & Turok, 1996). It has been detected with several galaxy survey data sets and WMAP at 2-2.5σ significance; the significance increases to 4.5 σ when all current surveys are used in combination (Giannantonio et al. 2008).

Signal-to-noise analysis have shown that the ideal survey for detecting ISW in cross-correlation studies should cover at least half of the sky, with a median redshift distribution around 0.9 and detect more than 10 galaxies per squared arcmin (Douspis et al 2008). The wide survey of Euclid, optimised for the weak lensing science case, satisfies these requirements. Tomographic analysis with Euclid and Planck will provide a signal-to-noise of the ISW of more than 5σ. Combining ISW for Euclid with Planck data, gives constraints of order 0.1 and 0.6 on $w_0$ and $w_a$ (Douspis et al. 2008). As the ISW signal is sensitive to the derivative of the growth factor, its combinations with other Euclid probes (Lensing, galaxy clustering, redshift distortions) will allow to put strong constraints on alternative models of gravity.

### 2.3.5 Combined Analysis and Model discrimination

The combined power of the Euclid probes will allow a simultaneous measurement of all sectors of the cosmological model to very high precision using the low redshift Universe. This is the natural extension of high precision measurements currently being performed by Planck, which measures the Universe in its very early high redshift state. The combination of these two regimes will allow us to track the evolution of the Universe to remarkable accuracy. To illustrate this we begin with a brief synopsis of the Euclid probe performance using an eight parameter cosmological model, with fiducial values {$w_p$: -0.95, $w_a$: 0.0, $\Omega_m$: 0.25, $\Omega_\Lambda$: 0.75, $\Omega_b$: 0.0445, $\sigma_8$: 0.8, $n_s$: 1.0, $h$: 0.7}. This model assumes Einstein gravity and allows for a curved Universe containing dominant dark energy and dark matter components.

The expected cosmological performance for the Euclid weak lensing probe was calculated using the weak lensing tomography technique (Hu et al 1999; Amara & Réfrégier, 2007), in which the galaxies are divided into a number of redshift bins. We also performed calculations using 3D cosmic shear (Heavens et al., 2006; Kitching et al., 2007) and found excellent agreement between these different predictions. To forecast the dark energy constraints from galaxy clustering for the Euclid galaxy redshift survey, we use two methods: (1) The BAO "wiggles only" method developed by Seo & Eisenstein (2007), which only uses the baryon acoustic features from the reconstructed matter density field, and discards other cosmological information;

(2) The P(k) method developed by Seo & Eisenstein (2003), which utilises the measured galaxy power spectrum in its entirety, with a cutoff at intermediate scales to avoid non-linear effects (Rassat et al. 2009).

*Table 2.2: Fisher matrix errors and Figure of Merit predictions. Expected errors on cosmological parameters for Euclid compared and combined with CMB experiments. The Euclid values include the constraints from weak lensing tomography, galaxy clustering, cluster counts and the ISW effect. Euclid alone is able to meet our dark energy and initial conditions science objectives, with many of the cosmological parameters being measured at the sub-percent level.*

| | Dark Energy | | Densities | | | Initial Conditions | | Hubble | DE FoM[2] |
|---|---|---|---|---|---|---|---|---|---|
| | $\Delta w_p$ | $\Delta w_a$ | $\Delta\Omega_m$ | $\Delta\Omega_\Lambda$ | $\Delta\Omega_b$ | $\Delta\sigma_8$ | $\Delta n_s$ | $\Delta h$ | |
| Current +WMAP[3] | 0.13 | - | 0.01 | 0.015 | 0.0015 | 0.026 | 0.013 | 0.013 | **~10** |
| Planck | - | - | 0.008 | - | 0.0007 | 0.05 | 0.005 | 0.007 | **-** |
| Euclid Req. | 0.018 | 0.15 | 0.004 | 0.012 | 0.006 | 0.004 | 0.007 | 0.022 | **400** |
| Euclid Goal | 0.016 | 0.13 | 0.003 | 0.012 | 0.005 | 0.003 | 0.006 | 0.020 | **500** |
| Euclid +Planck | 0.010 | 0.066 | 0.0008 | 0.003 | 0.0004 | 0.0015 | 0.003 | 0.002 | **1500** |
| **Factor gain on Current** | **13** | **> 15** | **13** | **5** | **4** | **17** | **4** | **7** | **150** |

The exquisite accuracy that will be achieved by the Euclid cosmological probes is shown in Table 2.2. It will provide improvement of many orders of magnitude over current constraints, and will represent a fundamentally new level of precision. For example current dark energy constraints, of order 10% on the equation of state (e.g. Komatsu et al 2009), require many assumptions – e.g. on the curvature of the Universe – and require the combination of a large number of heterogeneous data sets and techniques. Euclid will constrain the dark energy equation of state to 1% alone, with no additional assumptions. This is a result of the power of the individual probes and the in-built complimentarily of these probes simultaneously measured by Euclid.

We also see from Table 2.2 that Planck alone is not able to constrain dark energy parameters. The Planck all-sky CMB map will measure the geometric distance to the surface of last scattering to extreme accuracy; however this constraint is subject to large degeneracies (Bond & Efstathiou 1999). Hence Planck accurately probes the matter dominated epoch but cannot constrain the dark energy equation of state to high accuracy.

Planck combined with Euclid will provide significant improvement of the dark energy equation of state parameters. The complementarity of the Euclid probes with the CMB constraints results in percent and sub-percent accuracy in all cosmological sectors. Table 2.2 shows that the relative densities of the constituents of the Universe will be constrained to within 0.1%, that the initial conditions of the Universe will be constrained to 0.1% and that the constraint on the evolution of the equation of state of dark energy is reduced by a factor of 2. The resulting FoM for dark energy is over 10 times the expected FoM for any pre-Euclid experiment (Albrecht et al., 2009) that will have FoM's of approximately 100. This powerful synergy is a consequence of the complementary nature of the low-redshift Euclid probe and the high redshift CMB Planck experiment.

The results presented here are consistent with the findings of the ESA-ESO working group on fundamental cosmology (Peacock et al., 2006), the NASA dark energy task force (Albrecht et al., 2006) as well as numerous articles available on the predicted constraints obtainable for the Euclid cosmological probes.

In Section 2.4 we will extend this investigation to discuss the impact of systematic effects on these statistical constraints and how they can be contained and mitigated. The raw statistical power of Euclid is matched by the level at which systematic effects can be controlled.

---

[2] Note that Fisher matrix FoM predictions are not more accurate than 10%. We have therefore rounded the numbers to aid the reader.

[3] Results taken from five year WMAP results in Komatsu et al 2009.

### 2.3.6 Dark Energy and Modified Gravity Constraints

The combined predicted constraints for a Euclid weak lensing experiment, BAO observations, redshift-space galaxy power spectrum and combined Euclid constraints are shown in Figure 2.13 along with the expected results combining Euclid with Planck. We see how the multiple probes constrain the dark energy and cosmological parameters in complementary ways leading to sub-percent accuracy in the combined error on dark energy. We also show points in the parameter space where some of the proposed dark energy models are situated, including ΛCDM. Any deviation from $w_0 = -1$ or $w_a = 0$ would immediately rule out Einstein's Cosmological Constant, while distinguishing between a freezing and thawing model (see Section 2.2). In addition, detection of a phantom model ($w_0 < -1$), would signal non-standard physics or interacting dark matter/dark energy models, with the precise location indicating which is viable.

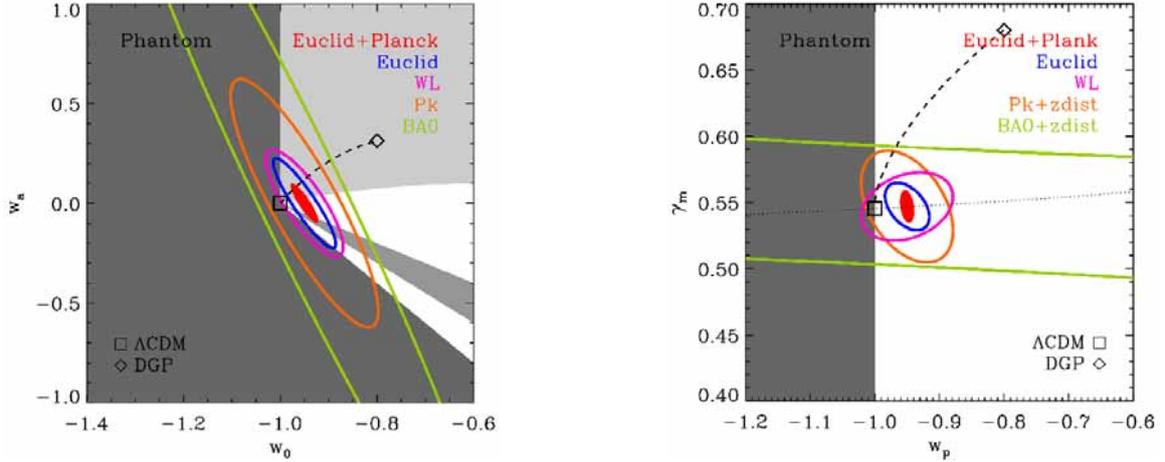

*Figure 2.13: a. (left) Predicted constraints from Euclid on the dark energy $w_0$-$w_a$ plane. The grey areas show different region relevant for DE theory. The darkest grey region $w_0 < -1$ is the Phantom zone, while the others show 'thaw' and 'freeze' models (middle grey and lightest grey). The outer (green) ellipses show the constraints from BAO, orange shows the galaxy power spectrum, $P(k)$, purple weak lensing alone, and inner blue ellipse the combined Euclid probes. The inner red ellipse is the combined Eulcid and Planck constraints. The square denotes ΛCDM and diamond DGP in parameter space, with the dotted line connecting them showing where extended DGP models lie. b. (right) Similar constraints in the growth index, $\gamma_m$, and $w_p$.*

An alternative explanation for the acceleration of the Universe is a deviation from Einstein gravity on large scales. These models also lead to predictions in the $w_0$-$w_a$ plane. Figure 2.13a shows the DGP model (diamond), which is some 10-sigma away from ΛCDM in the $w_p$ direction. Some modified gravity theories can mimic the expansion history of dark energy; however modified gravity models can also be distinguished from dark energy models by their effect on the evolution of structure. In Figure 2.13b we show the expected constraints on the structure growth parameter $\gamma_m$ in combination with the dark energy equation of state parameter at its pivot redshift, $w_p$. The individual probes form a complementary set of constraints that in combination constrain the structure growth parameter to sub-percent accuracy. Such a level of accuracy will distinguish between a number of modified gravity models. For example the string theory inspired DGP braneworld model is over 20-sigma from ΛCDM in the $\gamma_m$-$w_p$ plane, and can be clearly distinguished from Einstein gravity. Other modified gravity models, such as f(R), f(G) and multi-metric theories will also make predictions which can be tested and either ruled out or point to new physics on large scales.

Figure 2.14(a) shows the growth rate of matter perturbations, $f(z) = \Omega_m^\gamma$, as a function of redshift, which can be measured from redshift-space distortions by the distortion parameter, $\beta(z) = f(z)/b(z)$, if we know the evolution of galaxy bias, $b(z)$. Assuming the bias evolution is known from the weak lensing survey or from measurements of higher-order moments of the galaxy density field (Verde et al., 2002), we have simulated the Euclid spectroscopic galaxy survey and measured the growth rate from redshift-space distortions as a function of redshift (data points and errors). The assumed ΛCDM model is shown, as well as models based on coupling dark matter and dark energy and the flat DGP model. We conclude that the growth index, $\gamma_m$, of matter perturbations measured by Euclid can clearly distinguish between competing models for the dark energy.

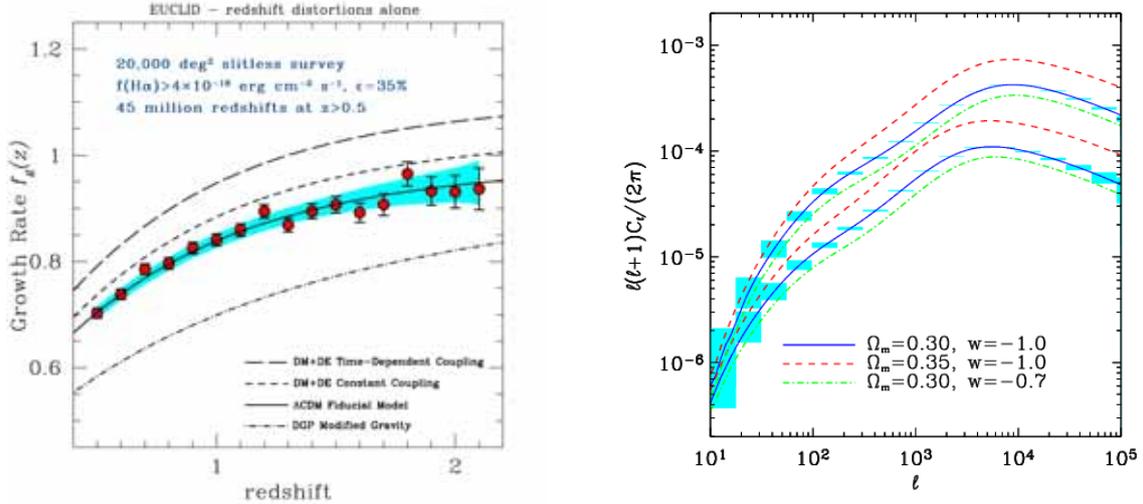

*Figure 2.14: a. (left) The growth rate of matter perturbations as a function of redshift. Data points and errors are from a simulation of the spectroscopic redshift survey. The assumed ΛCDM model, coupled dark matter/dark energy modes and DGP are also shown. b. (right): The predicted cosmic shear angular power spectrum at z=0.5 (bottom curves) and z=1 (top curves) for a number of cosmological models*

The primary signal for weak lensing is the Cosmic Shear power spectrum. Figure 2.14(b) shows the expected precision with which Euclid will be able to measure the shear power spectrum for galaxies separated into two redshift bins, at $z=0.5$ and $z=1$. Note that there will be a cross-power spectrum between the two redshift bins (not plotted). Figure 2.14(b) also shows how the shear power spectrum changes for different cosmological models, where we have varied the matter density parameter and a constant equation of state, $w$. We see the effect of an increase in $\Omega_m$ in an increase in the power, as the lenses have more mass, and a shift in the peak as we alter the matter-radiation equality redshift. Changing $w$ has a much smaller effect; decreasing $w$ suppresses the shear power spectrum, since structure grows faster, and so has lower amplitude at higher redshift, and is further away so shifted to smaller angular scales. The shear power spectrum is the low-redshift analogue of the well-known CMB power spectrum. While the CMB only probes effectively the single redshift $z=1100$, cosmic shear probes the entire redshift range out from $z=0$ to the maximum redshift of the galaxy sample.

### 2.3.7 Dark Matter Constraints

A key science output of Euclid will be an all-sky three-dimensional dark matter map, see Figure 2.15. This will map the dark matter structure directly and in three dimensions covering the evolution of the Universe out to a redshift of 2. Dark matter reconstruction in 3D has already been used to detect multiple dark matter haloes along the line of sight (e.g. Taylor et al., 2006; Simon et al., 2009). By simultaneously mapping the galaxy distribution Euclid will probe the growth of structure and galaxy environment over orders of magnitude in time. The Euclid local Universe dark matter map will be complementary to all-sky high redshift CMB maps, for example from the on-going Planck experiment. The cross correlations of such experiments will be a unique scientific tool enabling dark matter and dark energy evolution to be mapped over the majority of the history of the Universe.

By measuring the averaged mass profiles of hundreds of thousands of galaxy groups and clusters over a wide range of masses and redshifts, the characteristic clumpiness and cuspiness of dark matter structures can be tightly constrained as a function of redshift and mass. Constraining the evolution of cluster mass profiles will place limits on the overall dark matter content of the Universe as well as the energetic and mass properties of the dark matter particles. This technique has already been applied to a sample of 30,000 groups and clusters in the SDSS (Mandelbaum et al. 2006) to rule out a simple mass model - the isothermal density profile so characteristic of galaxies has been ruled out at cluster scales at over 3-sigma. Euclid will decrease statistical errors on, for example, NFW mass and concentrations by a factor of 20 for dark matter halos with masses between $10^{13}$ M$_\odot$ to $10^{15}$ M$_\odot$ - this corresponds to statistical errors of less than 1%. Most constraining will be the combination of weak lensing (that probes the outer part of clusters) and strong lensing that probes the inner core (e.g. Bardeau et al., 2007) such measurements could constrain the mass to light ratio and mass profile of hundreds of thousands of clusters with masses between $10^{12}$ M$_\odot$ to $10^{15}$ M$_\odot$.

Euclid will also constrain the shape of dark matter haloes. Evans & Bridle (2008) used SDSS to measure the average cluster halo ellipticity and ruled out spherical dark matter haloes at 99.6% confidence. Using flexion constraints Euclid should improve on these constraints by a factor of 100 (Hawken et al 2009). Euclid will also constrain the triaxiality of dark matter haloes using gravitational lensing (e.g. Corless et al., 2007).

Euclid will be extremely sensitive to any dark matter particle with a mass in the range of a few hundred to thousand eV. A direct effect of dark matter is to change the relative abundance of sub-structure in galaxy clusters. For dark matter with a particle mass of 3 KeV the majority of substructure below a mass of $10^9$ M$_\odot$ will be diffused. Using a weak lensing flexion variance (Bacon et al., 2009) such a signature could be distinguished from ΛCDM at a 10-sigma confidence. Markovic et al. (2009) have shown that Euclid, using cosmic shear power spectra, plus Planck could detect particle with a mass of 1 KeV.

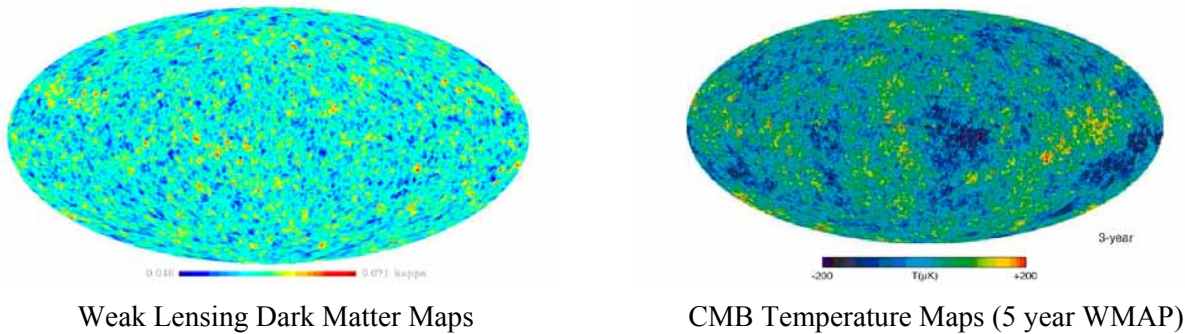

Weak Lensing Dark Matter Maps      CMB Temperature Maps (5 year WMAP)

*Figure 2.15: a. (left) All sky mass map from weak lensing (Teyssier et al., 2008) for a Euclid survey, based on a 70 billion particle N-Body simulation. b. (right) This can be compared with the all sky temperature maps of the CMB, such as the WMAP 5 Year all sky CMB temperature map at a resolution of ~0.2 deg (Hinshaw et al., 2008). Euclid will produce a 3D maps at redshifts between 0 and 2 at arcminute scales.*

Massive neutrinos are a natural (hot) dark matter component. Euclid will constrain both the total mass of neutrino and the number of massive neutrinos to 0.03eV and 0.1 respectively (Kitching et al., 2007). In comparison with current and future particle physics experiments Euclid provides a complementary measure of the neutrino mass. In addition to these constraints Euclid will measure the mass of individual neutrino species (de Barnardis et al., 2009; Elgaroy & Lahav, 2005), and will decisively distinguish between the normal and inverted neutrino hierarchies.

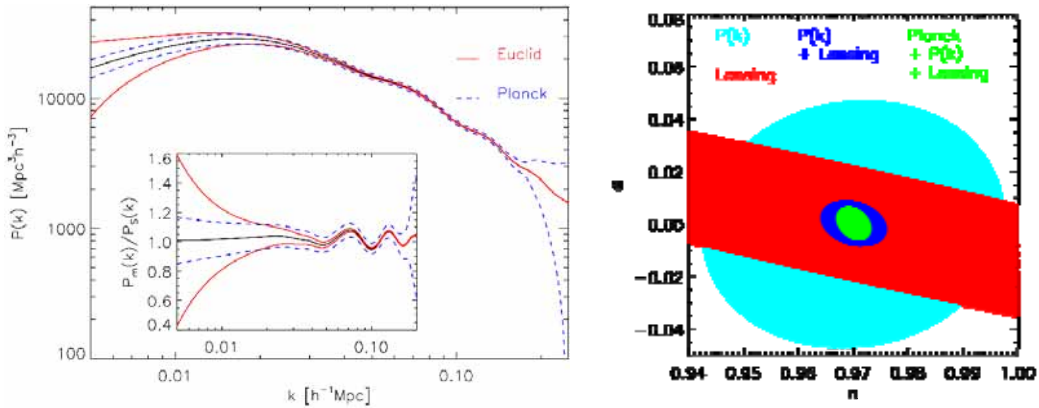

*Figure 2.16: a. (Left) Euclid and Planck both probe the power spectrum over a complementary range of scales. b. (Right) For extended models allowing a running spectral index (α) and tensor perturbations, we show that the stand alone Euclid probes can constrain α.and n to sub-percent level. Constraints on n increase by a factor of 2 when combined with Planck data.*

## 2.3.8 Initial Conditions Constraints

The Euclid matter power spectrum, as well as put constraints on dark energy and structure evolution from redshift 0 to 2, will provide a high precision measurement of the primordial power spectrum (PPP), which is necessary to understand the origin of structure and to discriminate between different inflationary scenarios. The combination of galaxy clustering and cosmic shear from Euclid and the Planck CMB data will provide a measurement of the shape of the PPP, where degeneracies with other cosmological parameters can be broken. Figure 2.16a illustrates the complementary range of scales probed by Planck and Euclid, for a particular choice of cosmological parameter values; this figure can be compared to Figure 2.6 in section 2.2.3 which examines current constraints on the PPP. Euclid will determine with great accuracy the parameters describing the shape of the PPP: the amplitude ($\sigma_8$), the spectral index ($n_s$) and the running ($\alpha$) (see Table 2.2, and Figure 2.16b). Combining constraints from cosmic shear, galaxy clustering and the CMB will help discriminate between families of single field inflationary models.

Euclid will also constrain the non-Gaussianity of the primordial fluctuations, as parameterised by $f_{NL}$, in three different ways: from weak-lensing studies, from the large-scale clustering of galaxies, and from cluster counts. Combining these datasets and using prior information on the cosmological parameters from the Planck satellite will give unprecedented constraints on $f_{NL}$. Forecasts based on the Fisher matrix formalism show that Euclid will be able to measure $f_{NL}$ to better than $\Delta f_{NL}=\pm10$ (Carbone, Verde & Matarrese 2008).

## 2.4 Precision Cosmology from Space

We have seen how the Euclid probes have the statistical potential to meet our science goals. One of the main design drivers of Euclid is to control of systematic errors so that they remain subdominant to the statistical errors. Here we describe the main source of systematic errors and the ways that they will be mitigated with Euclid's space-based observations.

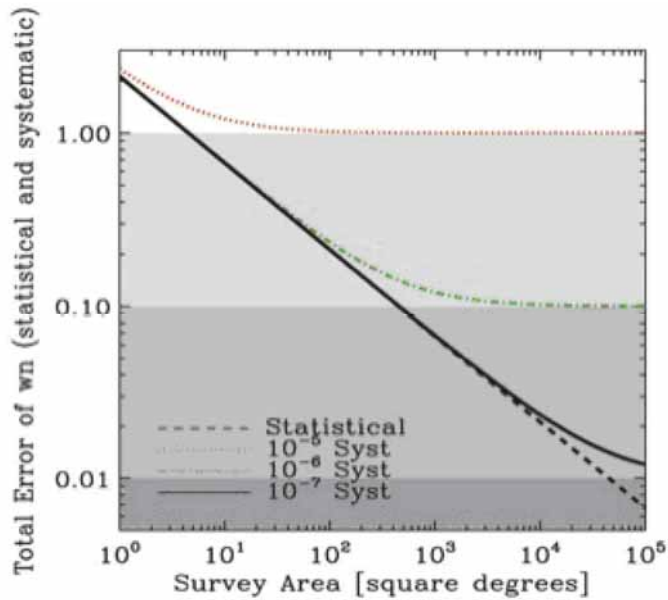

*Figure 2.17: Advantages of space based observations in order to reach Euclid's cosmological objectives. The total error on the equation of state decreases statistically as the area of a survey is increased. However systematic effects limit the achievable dark energy constraint. For Euclid to achieve 2% on the dark energy equation of state requires an area of 20,000 square degrees and shape systematic levels with a variance of $10^{-7}$ (Cf. Amara & Réfrégier 2008). Such a systematic precision can only be achieved with the stability and accuracy of space-based observations.*

### 2.4.1 Weak Lensing

Euclid's imaging survey must contain 30-40(requirement-goal) galaxies per $amin^2$ useful for lensing, with a median redshift greater than $z_m = 0.8$, with photometric redshift errors better than $0.05(1+z)$ (see Section 3 for details on these requirements, and Amara & Réfrégier 2007, Kitching et al 2009). Such a survey has a high-level requirement on the variance of the systematic contribution to the shear power spectrum of $\sigma^2_{sys}<10^{-7}$ (Figure 2.17; Amara and Réfrégier 2008). This translates into a limit that the shear systematic error needs to be $\delta\gamma \leq 3\times10^{-3}$ and the mean redshift calibration error must be $\delta\bar{z} \leq 0.002(1+z)$; see Appendix 2 for detailed descriptions of performance predictions.

**Galaxy Counts for Shear Measurement:** The EIC has used two independent image simulation pipelines to evaluate the expected galaxy number counts for Euclid. Both image simulation pipelines produce realistic galaxies using a Shapelet decomposition (Réfrégier 2003, Massey et al 2004; Melchior et al. 2007) of the Hubble Space Telescope (HST) Ultra Deep Field (UDF; Beckwith et al., 2006). The two image simulations pipelines (described in Meneghetti et al. 2008) have undergone detailed cross-comparisons with each other and the HST COSMOS survey (see Meneghetti et al 2009 for details of these comparisons). Simulations from one of the pipelines have also been used by the worldwide weak lensing community in the Shear TEsting Program (STEP; Massey et al 2007). For the Euclid specific configuration, the simulations use accurate models of the total Euclid throughput, noise characteristics, PSF shape and the average zodiacal sky background. The image simulations have then been analysed using three independent shear measurement algorithms (RRG, Rhodes, Réfrégier & Groth 2000; *lens*fit, Miller et al., 2007, Kitching et al., 2008; im2shape, Bridle, 2004). The results are summarised in Figure 2.18 which shows the number of galaxies useful for lensing for two examples of cuts on the data: (i) simple cuts based on SExtractor parameters keeping galaxies with S/N >10 and FWHM size greater than 1.25 times that of the PSF, (ii) cuts based on shear measurement pipelines such that the error on measured ellipticity is δe < 0.1. We see that, for both of these, the galaxy number counts meet our requirement.

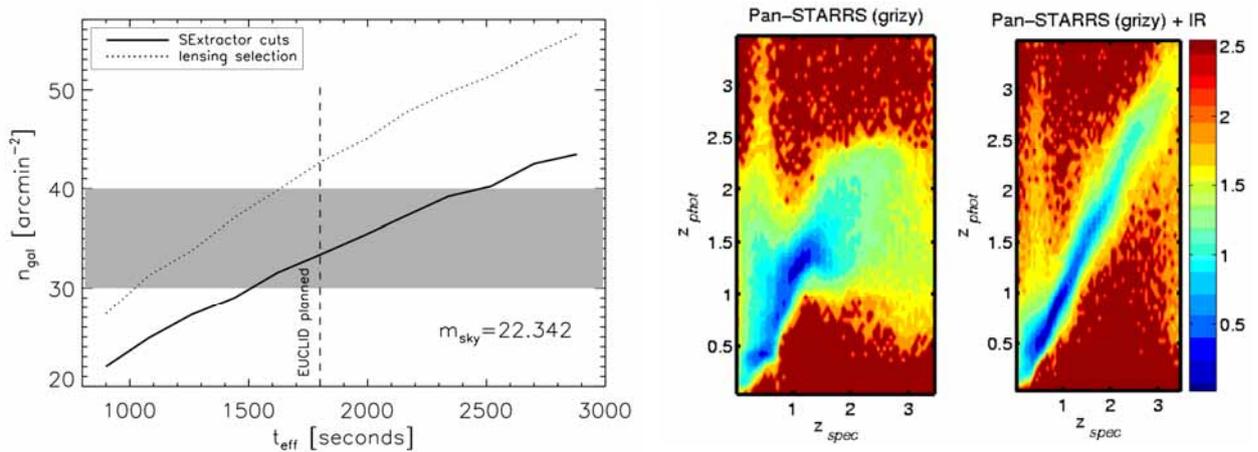

*Figure 2.18: a. (Left) The expected number counts of galaxies useful for lensing as a function of exposure time. The solid line is made using a simple cut on SExtractor detection with S/N>10 and FHWM[gal]>1.25FWHM[PSF], the dashed line is from the shape measurement pipelines that sum the lensing weight assigned to each galaxy, with a cut in ellipticity error of 0.1. We see that we are able to reach our requirements of 30-40 gal/amin². b. (Right) Shows the redshift measurement for PanSTARRS with and without the Euclid NIR bands (c.f. Abdalla et al 2007). We find that with DES, PanSTARRS-2 and a fortiori PanSTARRS-4 and LSST we will be able to meet out requirements of δz = 0.05(1+z).*

**Photo-z Performance:** Photometric redshift estimates are needed for two key steps in the weak lensing analysis. The first is to divide galaxies into redshift slices and the second (discussed below) to characterize the distribution of galaxies in each slice. Using a set of photo-z simulations (described in detail in Abdalla et al 2008 and Bordoloi et al 2009) we showed that our requirement of $\sigma(z) < 0.05(1+z)$ could be achieved by combining deep NIR photometry from Euclid (with magnitude limits of $Y_{AB}=24$, $J_{AB}=24$, $H_{AB}=24$) with DES or PanSTARRS-2. Clearly surveys that go deeper (such as PanSTARRS-4 and LSST) would further improve our photo-z performance. We also simulated the sky coverage and complementary for DES and PanSTARRS (Paulin-Henriksson et al 2009). Figure 2.19 shows the sky coverage from the two sights (Hawaii and La Silla Chile) and shows that we are able to cover the Euclid survey by combining these two surveys.

**Photo-z Calibration:** The mean of the galaxy redshift distribution needs to be known to better than $\delta\bar{z} \leq 0.002(1+z)$ and can be measured in a number of different ways. The simplest and most direct is to measure the spectroscopic redshift of a random subsample of galaxies. For this direct approach, $10^4$-$10^5$ spectra would be needed per bin with high completeness and with a spatial sampling not alleviate cosmic variance. This places the calibration burden on the spectroscopy. An alternative approach is to extract the information from the photo-z estimates. Many of the current photo-z methods output the full probability distribution of the redshift of a galaxy, which can be used to reconstruct the distribution of galaxies in the bin (Bordoloi, et al 2009). The left panel of Figure 2.20a shows the performance of this method for a set of PanSTARRS configurations. We find that for a PanSTARRS2-like survey this technique is able to meet our requirements. It should be made clear however that this method does not do away with the need for a spectroscopic subsample. The spectroscopic subsample is used to calibrate the photo-z probability estimates. This relaxes slightly the required number of spectra, but most importantly it relaxes our requirement on completeness and spatial sampling.

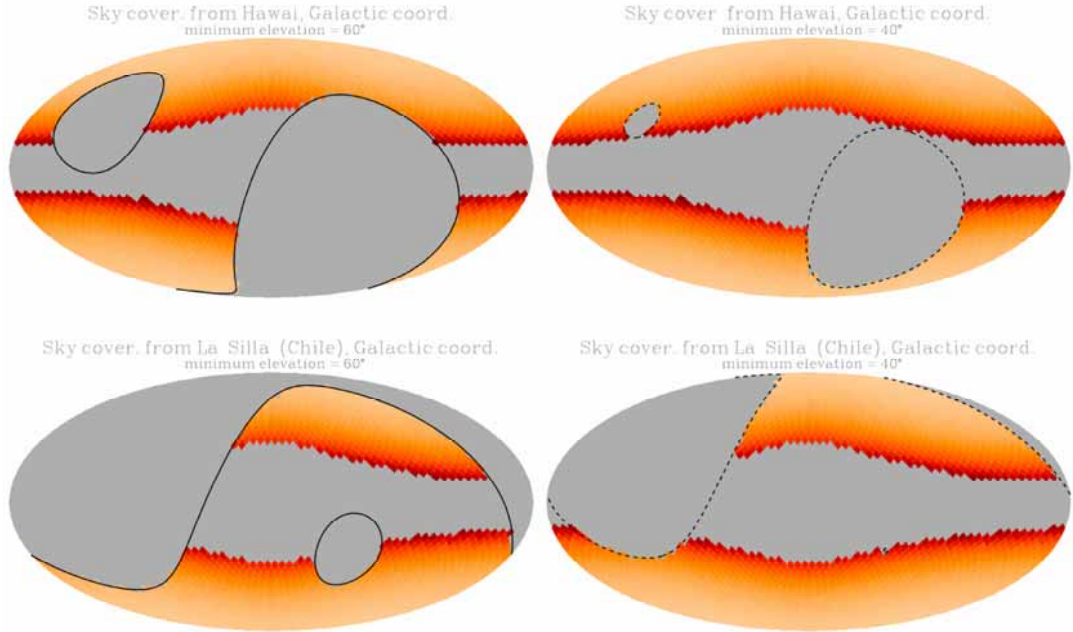

*Figure 2.19: The Sky coverage achievable from Hawaii i.e. at a latitude of +19.83° - PanSTARRS (top panels) and La Silla i.e. at a latitude of −29.25° - DES (bottom panels), for minimum elevations of 60° (solid lines, right panels) and 40° (dotted lines, left panels) in Galactic coordinates. The colour scale shows the density of stars within a magnitude 18 ≤ I ≤ 20 (the lightest and the darkest corresponds to 0.3 stars/arcmin² and more than 3 stars/arcmin², respectively), while the grey regions are excluded. We see that PanSTARRS and DES together can fill in the Euclid survey.*

The most unambiguously reliable way to obtain the necessary spectroscopic subsample is from space. JWST will give the infrared spectroscopy in a low background regime. At $H_{AB}$ = 24, an equivalent depth to the requirement in the optical, a 3600s exposure with JWST/NIRSpec gives S/N = 80 at 1.6 microns for a point source at R = 100 over the full spectral range 1-5 micron, with the S/N degrading by a factor of about two for realistic galaxies. At this depth a multiplex of about 100 with NIRSpec is achievable, and 1000 pointings would require ~2000 hours including overheads. A DMD spectrometer on Euclid would require about 15 hours to secure a redshift at $H_{AB}$ = 24. With a multiplex gain of 1500 per pointing, 100 pointings are needed corresponding to roughly 1500 hours of integration. Spectroscopy from the ground is also possible at $I_{AB}$ = 24.5. The VVDS Ultra-deep project has already obtained spectra using two grisms over the whole 3600-9300A range and demonstrates a success rate in redshift determination of over 80%. At $I_{AB}$ = 24.5, this success rate can be achieved with 10 hours integration, per grism, with 1000 slits per pointing. A sample of 100,000 galaxies could therefore be achieved with 100 pointings, requiring roughly 2000h of observing time on the ESO VLT with current instrumentation on a single telescope. Future upgrades of VIMOS would significantly reduce this requirement.

**PSF Characterisation:** Correcting for the Point Spread Function (PSF) of the instrument is a crucial step in weak lensing since this can have a large effect on the measured ellipticity of galaxies. To do this we measure the PSF from the images of stars around the galaxy (see Figure 2.20b). In Paulin-Henriksson (2008) it was shown how errors in the PSF measurement propagate into errors on shear measurement. To reach the Euclid requirements we need to measure the PSF using more than 50 bright stars per galaxy. This PSF correction is able to correct of the PSF variation on large scales, however Amara et al, (2009) show that the variability on small scales (smaller than the scale containing 50 stars – roughly 50 arcmin²) needs to be controlled in hardware. This means that the variability needs to be constant both spatially and temporally. This is a unique advantage of space observations.

Beyond the PSF spatial variability we have also investigated the impact of the shape of the PSF itself. To do this the concept of complexity and sparsity was introduced into this problem in Paulin-Henriksson et al. (2009). We find that the PSF must be simple (low complexity) and described by a small number of degrees of freedom. These findings have been translated into a set of requirements in section 3. One important property of the PSF that we have highlighted is wavelength dependence. Due to a number of chromatic effects we can expect that Euclid's PSF will depend on the wavelength of the source, this is important because in taking images we integrate the light over a wavelength range within the photometric band. This effect must be tackled in two key steps. The first is the wavelength dependence of the instrument must be modeled and measured. The second step is that it must be corrected using the spectrum (Spectral Energy Distribution – SED) of the galaxy. Cyrpiano et al. (2009) have considered three contributions to the PSF; (i) optical diffraction, (ii) detector modulation transfer function (MTF) and (iii) space craft jitter. This work finds that stars cover a range of spectral properties and thus are able to model the wavelength of the instrument. To correct for wavelength dependence we need to estimate the SED of the galaxy, which can be done using the same routines used for photometric redshift measurements. We find that the requirements on photometry coming from wavelength dependence are weaker than those from photometric redshifts.

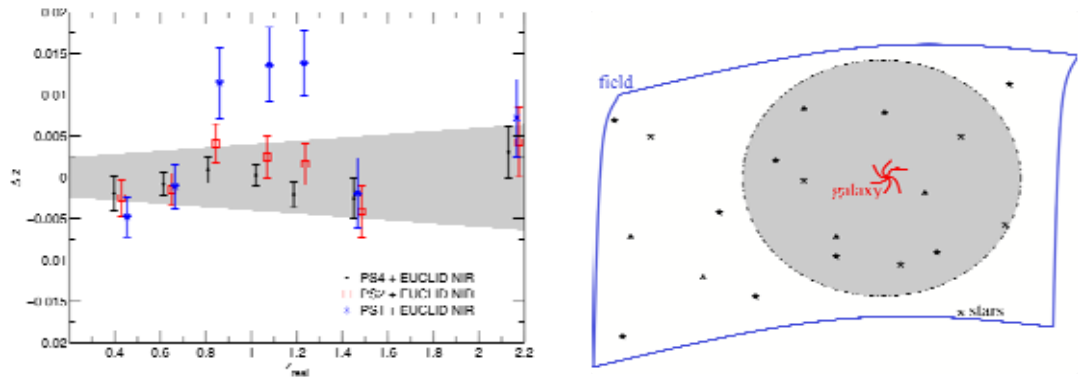

*Figure 2.20: (Left) The precision with which we are able to calibrate the mean redshifts of the tomographic bins for PanSTARRS-1, 2 and 4. To meet our requirements (shown by the grey area) we need to combine Euclid with PanSTARRS-2; see also Bordoloi et al( 2009). (Right) An illustration of the requirement on the number of stars needed to estimate the PSF at each galaxy position to sufficient accuracy. For each galaxy we require that 50 stars are needed for PSF characterisation.*

**Detector Performance:** A full analysis technique for studying CCD Charge Transfer Efficiency (CTE) effects on galaxy shape measurements has been presented by Rhodes et al, (2009). This analysis used a combination of laboratory tests on irradiated CCDs and a software model developed to mimic the process of CCD readout. The model has been tested against and calibrated upon data from the Hubble Space Telescope (Massey et al. 2009) and laboratory-controlled data (Dawson et al. 2008, Rhodes et al. 2009). The model's parameters include the density and characteristic release time of multiple charge trap species. Several CCDs have been irradiating in the LBNL 88-Inch Cyclotron with a (large) known flux of 12.5 MeV protons, and the image degradation was measured in subsequent First Pixel Response/Extended Pixel Edge Response (FPR/EPER) tests and behind point sources created by $^{55}$Fe radiation. The effect on galaxy shapes was demonstrated to be linearly dependent on trap density (by testing both the increased degradation over time and as a function of on-chip distance from the readout register), so the effect at any given point in the mission could be deduced by rescaling the trap density measured in the irradiated CCD.

The software to model a CCD readout can also be used (in an inverse mode) to mitigate the adverse effects on weak lensing measurements. This has been demonstrated to reduce the effects of CTE by a factor of 30 on HST data (Massey et al 2009). Further improvement may be with minimal overhead during the mission to better calibrate the readout model using on-board EPER/FPR and pocket pumping tests. The conclusion is that a coherent solution has been demonstrated in which a combination of radiation-tolerant hardware and software post-processing can reduce the effect of CTE on galaxy shapes to 10% of the total shape measurement error budget (Rhodes et al 2009).

**Shape Measurement Method:** Members of the EIC are involved in code development in two ways, (i) in house simulations such the one described in Meneghetti et al. (2009) and (ii) through the The GRavitational lEnsing Accuracy Testing (GREAT) challenges (Bridle et al., 2009; Kitching et al, in prep.), which are an international effort also designed to bring in expertise from the machine learning and computer science community. The targets for the GREAT simulations are designed to match the requirement of $\sigma^2_{sys} < 10^{-7}$ needed for Euclid. In the six months of the first competition we have improved by a factor of two with some methods reaching the accuracy necessary to achieve our targets in some observing conditions. Shape measurement is of key importance for Euclid and it is likely that the final Euclid analysis will need to be performed with multiple independent codes. The independent pipelines will enable cross checks and comparisons to be made. Within Euclid we already have a number of independent methods (*lens*fit, im2Shape, Shapelets) and more are likely to be developed before Euclid launches.

**Intrinsic Alignment:** Intrinsic alignments are the primary weak lensing astrophysical systematic and have two effects: (i) on small scales galaxies can become aligned due to alignment of angular momentum vectors or from tidal alignments. Removing close pairs of galaxies from any correlation should however mitigate this effect to a large degree (e.g. Heymans & Heavens, 2003). (ii) the lensing signal from background galaxies can become aligned with a foreground tidal field, which in turn aligns foreground galaxies - this is known as the shear-intrinsic effect and is more difficult to remove. There are a number of methods that have been developed to either remove the shear-intrinsic alignment signal from the data or use extra information to pin down the nature of this effect. Joachimi & Schneider (2008, 2009) introduced a nulling method that transforms the shear signal in a purely geometric way such that intrinsic alignments are largely down-weighted, however this also affects the achievable error on cosmological parameters from the shear signal. Joachimi & Bridle (2009) consider the complete set of galaxy shape and position information that will be available from the Euclid survey and marginalize over intrinsic alignments and the galaxy bias without substantial prior information.

**Theoretical Uncertainties:** The theoretical predictions that will be used in the Euclid analysis require further developments as we move forward. Examples of this include the predictions for non-linear growth of structure, for which N-Body simulations have been used. Today we rely on fitting function to the simulations (Peacock and Dodds 1996, Smith et al 2004), however Euclid will likely require direct simulation predictions. This is not seen as critical since N-Body simulations are becoming routine. The current challenges for the simulations are to study high-order effects using, for instance, full ray-tracing simulations to calculate the lensing signals and to include a broader range of physical processes such as baryonic physics and exotic dark energy models. Much of this work is already underway. Many ray-tracing studies have been performed in lensing (including Wambsganss et al 1998, Meneghetti et al 2001, Barber et al 2004, White et al 2004, Teyssier et al 2008 and many others). Simulations with baryons have been performed to study the impact on lensing (e.g. Teyssier et at 2009).

## 2.4.2 Galaxy distribution

For direct observations of the galaxy distribution, one of the key advantages of space is the ability to obtain galaxy redshifts over the full extragalactic sky out to z=2, within a relatively short time. This allows an order-of-magnitude improvement over existing and ongoing surveys from the ground, in terms of the cosmological constraints possible from statistical measurements of the galaxy distribution. To show this, Figure 2.21 compares the expected constraints on the Hubble parameter $H(z)$ and the angular diameter distance $D(z)$ from BAO measurements (see section 2.3.2 for details), and on $f\sigma_8$ from redshift-space distortion measurements (see section 2.3.2 for details), from the key ongoing ground-based redshift surveys, the Baryon Oscillation Spectroscopic Survey (BOSS) and the Wiggle-z survey, against the constraints expected from slitless and DMD Euclid surveys.

**Catastrophic redshift errors:** In the slitless spectroscopic survey there will be a fraction of catastrophic redshift errors. These are mainly due to either a wrong identification of a true emission line or the association of a line (and, therefore, redshift), to the wrong object in the case of confusion from overlapping spectra. From our extensive simulations we estimate that this fraction is expected to be in the range 10-15%. The effect of these catastrophic errors is that the BAO signal is degraded, without biasing the result. This effect can be mitigated through a number of steps such as: i) use of photometric redshifts to eliminate a large fraction of these errors from the sample to be analyzed; ii) estimate of the true distribution of the galaxy redshifts that are wrongly assigned to $z_{obs}$, knowing this distribution means that it can be modeled in the analysis. We already have an estimate of this distribution from the simulations; for Euclid this will be estimated by comparing, even in limited regions of the sky, our redshift determinations with those from other existing spectroscopic surveys.

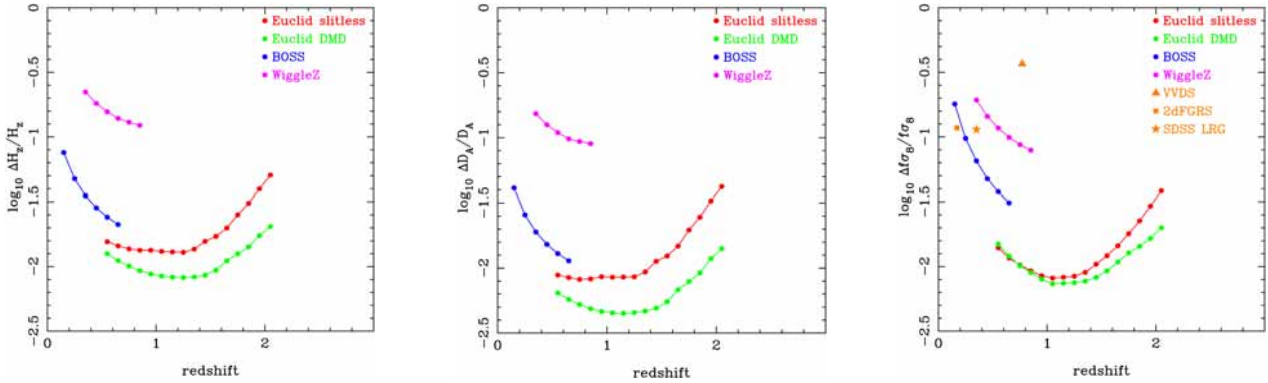

*Figure 2.21: (left and middle panels) forecast of the fractional accuracies in H(z) (left) and $D_A(z)$ (middle) expected from the Euclid slitless survey (flux>4x10^-16 erg s^-1 cm^-2, 35% efficiency), the Euclid optional DMD multi-slit survey (H<22, 36% sampling), and two ongoing/planned BAO surveys, WiggleZ and BOSS; (right panel) comparison of the expected accuracies of the slitless and DMD surveys to those from current and soon-to-come projects - the accuracy is expressed as the fractional error on the quantity f * σ8(mass), corresponding to the observable product β * σ8(gal).*

**Non-linear density field growth:** On very large scales, where density fluctuations are small, the evolution of the fluctuations is understood: the density field can be decomposed into Fourier modes and in the regime of small amplitude modes the density waves on different scales evolve independently of one another, according to linear perturbation theory. The power spectrum of fluctuations, the expectation value of the square of the mode amplitude, retains its shape in the linear regime, growing in amplitude with time due to gravitational instability. Eventually, the fluctuation amplitude on a given scale will approach unity. In this limit, called the nonlinear regime, the evolution becomes more complex, as the different Fourier modes become coupled.

Fortunately, cosmologists have a well developed and established tool for modeling the growth of fluctuations deep into the nonlinear regime, N-body simulation of gravitational collapse (e.g. Davis et al., 1985). This allows the departure from linear perturbation theory to be followed with high accuracy for density fluctuations in the collisionless dark matter. The nonlinear growth of fluctuations can have an impact on the appearance of the power spectrum or two-point correlation function even on scales as large as those associated with the BAO.

These effects manifest themselves as a shift in the observed BAO scale compared with the CMB-calibrated prediction, along with a damping of the higher harmonic BAO in the power spectrum or, equivalently, a change in the shape of the BAO spike in the correlation function (e.g. Seo & Eisenstein 2005; Springel et al. 2005; Angulo et al. 2008). This systematic is straight forward to model using large-volume N-body simulations. A series of papers have investigated the implications of the nonlinear growth of density perturbations on the appearance of the BAO (Huff et al. 2007; Angulo et al. 2008; Seo et al. 2008; Smith et al. 2008; Takahashi et al. 2008; Nishimichi et al. 2009). As a result, the effect is well calibrated and understood.

Motivated by the N-body simulation results, two general approaches have been proposed to deal with nonlinear effects: modeling based on extensions to linear perturbation theory (e.g. Jeong & Komatsu 2006, 2009; Smith et al. 2008; Sanchez, Baugh & Angulo 2008) and reconstruction of the linear theory spectrum (Eisenstein et al. 2007; Seo et al. 2008; Padmanabhan, White & Cohn 2009; Noh, White & Padmanabhan 2009).

The perturbation theory approach has been revolutionised by the renormalised perturbation theory (RPT) devised by Crocce & Scoccimarro (2006). The RPT approach takes into account mode coupling and gives a description of the nonlinear two point clustering of the dark matter which agrees remarkably well with the results from N-body simulations (Crocce & Scoccimarro 2008). A model based on RPT has been successfully applied to the BAO signal measured in the luminous red galaxy catalogue in the Sloan Digital Sky Survey by Sanchez et al. (2009).

The reconstruction approach can potentially yield the tightest constraints on the BAO scale (Seo et al. 2008), by using phase information to reconstruct the linear clustering spectrum from an evolved galaxy distribution. This minimizes the impact of nonlinear growth on the constraining power of BAO as a dark energy probe (Eisenstein et al. 2007; Seo et al. 2008). Figure 2.22 shows a prediction for the effect of reconstruction on the BAO peak in the correlation function. Seo et al. (2008) argue that the shifts of the BAO peak can be predicted numerically, and can be substantially reduced (to less than 0.1% between z= 0.3 and 1.5) using a simple reconstruction method. Padmanabhan, White & Cohn (2009) showed that reconstruction indeed reduces the mode-coupling term in the power spectrum, thus reducing the bias in the estimated BAO scale when the reconstructed power spectrum is used.

In addition to nonlinear growth in the underlying density perturbations, when a galaxy's position is inferred from its redshift, the non-linear peculiar velocity on top of the Hubble flow can distort the pattern of galaxy clustering. Distortions due to the small scale, random motions of galaxies inside virialised structures can have an impact on the power spectrum, although these can be reduced using a "Fingers-of-God" compression technique, in which identified galaxy groups are replaced by a single object with the centre of mass velocity of the group (e.g. Tegmark et al., 2004).

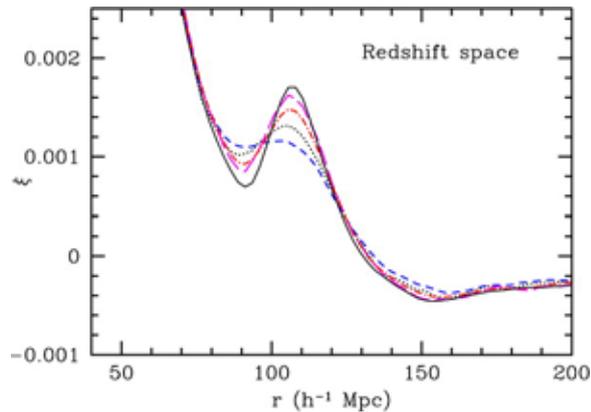

*Figure 2.22: Redshift-space matter correlation function after reconstruction by the linear-theory density-velocity relation, with the density field Gaussian-filtered (Eisenstein et al. 2007). The black solid line shows the correlation function at z = 49. The blue short-dashed line shows the redshift-space correlation function at z = 0.3, where the acoustic peak has been smeared out. The black dotted line shows the real-space correlation function for comparison. The red dot-dashed line shows the effects of reconstruction for a 10 Mpc/h Gaussian filtering; the magenta long-dashed line is the result when one compresses the "Fingers of God" prior to the reconstruction.*

**Galaxy bias:** The difference between galaxy and underlying mass density distributions is termed "galaxy bias". Galaxy formation is a local process. If we look on sufficiently large scales and assume a Gaussian distribution for the matter fluctuations, the galaxy over-density field is drawn from a simple scaling of the matter over-density field (Cole & Kaiser 1989). On large-scales we can therefore measure the relative clustering of the mass from the galaxy distribution after including a single nuisance parameter for the overall amplitude of clustering. On small scales and at late-times, non-linear processes affect the galaxies and mass in the Universe in different ways and we cannot use galaxy positions to easily infer the clustering of the mass. When predicting our ability to use the clustering of galaxies measured by Euclid to constrain cosmological parameters in Section 2.3.5, we included a small-scale cut-off beyond which there is assumed to be no information (this was varied between $k$=0.14h Mpc$^{-1}$ at $z$=0.5 to k=0.2 hMpc$^{-1}$ at $z$=2).

Because BAO are a large-scale phenomena (the co-moving sound horizon at the baryon-drag epoch is approximately 150 Mpc) their positions in the galaxy power spectrum are unaffected by the large-scale bias, and are only weakly dependent on the small-scale bias. They can also be extracted from the power spectrum in a way that cleanly separates them from the primary effects of galaxy bias (Sanchez et al. 2008), which is why they are considered such a robust probe of cosmic expansion. However, to use the full power spectrum or correlation function as a standard ruler, we do need to model galaxy bias. For Euclid we will do this using a suite of numerical simulations, coupled with analytic theory such as the halo-model (e.g. Peacock & Smith, 2000). We will also consider non-parametric methods to measure galaxy bias from counts in cells together with simple models to parameterize our ignorance about bias, allowing us to marginalize over this uncertainty (e.g. Smith et al. 2007). Another possibility is to estimate the bias parameter directly from the data by computing the bispectrum as it was done for the 2dF galaxy redshift survey (Verde et al. 2002). Euclid will measure redshifts for galaxies with a range of properties, which are expected to have different small-scale bias (e.g. Cresswell & Percival, 2009). We will therefore be able to test the importance of galaxy bias by checking that we recover the same cosmological model constraints from different galaxy populations, and using clustering measurements on different scales. Finally, it is worth commenting that the existence of galaxy bias extends the breadth of cosmological measurements made by Euclid, allowing constraints to be set on primordial non-Gaussian fluctuations in the gravitational potential (Dalal et al., 2008), and can also boost the signal recovered from redshift-space distortion measurements (McDonald & Seljak, 2009). Measurements of large and small-scale galaxy bias will also set tight constraints on galaxy formation models (see Section 2.5 - additional science section).

**Non-linear redshift-space distortions:** Peculiar velocities arising during the non-linear regime can affect and bias the estimate of the growth rate from redshift distortions. The parameter $\beta=f$/bias, or equivalently the combination $\beta\sigma_8(\text{galaxy})=f\sigma_8(\text{mass})$ can only be measured from the observed redshift distortions when density fluctuations are in their linear phase of growth. In reality, the anisotropies of galaxy two-point statistics in redshift space (correlation function and power spectrum) result from the combined effect of linear, coherent motions on large scales and highly non-linear, random motions within virialized structures ("Fingers of God"). The small-scale nonlinear effects can in general be successfully modeled, to reproduce the complete distortion pattern and disentangle the linear contribution (e.g. Hamilton et al., 1997, Guzzo et al., 2008). In this framework the effect of the small velocity bias of galaxies measured in numerical experiments (e.g. Faltenbacher et al., 2005) can be marginalized over without affecting the estimate of the growth rate. Building on existing experience in Euclid we have performed new extensive tests on larger N-body simulations to compare the statistical and systematic errors reachable using different estimators (Branchini et al., 2010), including the effect of redshift errors. These tests demonstrate that Euclid will be able to deliver unbiased estimates of $\beta$ to percent accuracy in each of its several redshift slices. This is achieved by the huge volume and good galaxy sampling of Euclid within each of its redshift sub-samples, which allow an exquisite clustering signal on scales of approximately 100 $h^{-1}$ Mpc to be recovered. This makes non-linear effects and their accurate modeling less crucial than for current or forthcoming galaxy surveys.

## 2.4.3 Advantages from space based observations

Using the advantage of a stable space based platform Euclid will mitigate or remove each weak lensing systematic to unprecedented accuracy. Only from space, with this level of control can the science objectives of Euclid be achieved. Every improvement translates into larger science performance for Euclid. For the weak lensing probe this is encapsulated in Figure 2.17. This shows how the error on the dark energy parameter $w$ varies as a function of survey area and shape systematic level. Only with 20,000 deg$^2$ and a variance for shear systematics $\sigma_{sys}^2 < 10^{-7}$ can an accuracy on $w$ of 2% be reached. This can only be reached with the PSF stability and resolution of space based observations. On the ground, shape measurements the PSF stability is limited by atmospheric seeing, windshake, and flexure. The spatial resolution available from the ground is limited for the same reasons. In addition, the statistical errors and outlier rates of photometric redshifts can only be achieved using the deep NIR photometry of Euclid, which is unobservable from the ground due to high background emission from the atmosphere. Figure 2.20 shows photometric redshift accuracies obtained by combining the on-board Euclid bands with photometry from ground based surveys available by the time Euclid will be operational.

For galaxy redshift surveys, Euclid will yield medium resolution ($R$~500) spectra of about 70 million galaxies in the same survey area in a NIR wavelength range which is difficult to access from ground for faint galaxies between redshift $z$=1 to 2 due to atmospheric emission. The high number of atmospheric sky lines requires a higher spectral resolution from ground than that needed to reach the same redshift accuracy from space. Figure 2.21 shows the improvement in distance scale measures of Euclid using the BAO technique compared to ground based surveys.

# 2.5 Additional Science with Euclid

Beyond the exquisite cosmological measurements it will make, Euclid will produce a massive legacy of deep images and spectra over at least half of the entire sky. This will be a unique resource to the astronomical community and impact all areas of astronomy. For example, Euclid's spatial resolution of 0.2 arcseconds should be compared with the median seeing of 1.43 arcseconds for SDSS and the expected seeing of 0.8 arcseconds for PanSTARRS, i.e., factors of 50 and 16 respectively in terms of spatial concentration, and comparable to the resolution of HST-WFPC2. Moreover, Euclid will push the Gaia photometry 5.5 magnitudes deeper and be >1000 times larger than the deep VISTA surveys. Euclid spectra will sample about 50% of all objects to H(AB)~19, which is unobtainable from the ground. With Euclid, the majority of the new sources identified by future imaging observatories, from radio to X-rays, will be readily associated to a known redshift, out to z~2. This adds immediately an enormous power to the science return of these other projects, as it eliminates the inevitable time-consuming phase of redshift follow-up. Euclid will be a discovery machine on an unprecedented scale, perhaps the major feeder for more detailed studies both with ground-based facilities (ALMA/ELT/SKA) and satellites (JWST).

The impact on modern astronomy of such legacy science from these massive surveys of the sky is now clear. Recently, Madrid & Macchetto (2009) performed an analysis of the productivity of all major observatories in the world and ranked the Sloan Digital Sky Survey (SDSS) first because of the sheer number of citations received from thousands of papers. Likewise, the HST was ranked third primarily because of the Hubble Deep Field and other ACS surveys. The SDSS, although predominately designed around galaxy large-scale structure and high redshift quasars, provided breakthrough discoveries on brown dwarfs, white dwarfs, the structure of Milky Way, gravitational lensing and galaxies 100-fold less luminous than any known before. The SDSS showed that both spectroscopy and imaging have enormous serendipity potential, and, in particular, that unprecedented imaging, as Euclid promises to deliver, holds a near-inexhaustible set of discovery opportunities for the whole astronomical community.

## 2.5.1 Galaxy Formation and Evolution

A key area of legacy science with Euclid will be galaxy evolution. In detail, determining the physical processes that regulate the growth of galaxies and their link with massive black holes is one of the outstanding problems of modern astrophysics. In the standard model of galaxy formation, galaxies coalesce from gas that cools inside dark matter halos (White & Rees 1978). Its special challenge derives from the diversity of physical processes that could be important. Some of these processes are internal to the galaxy – disk instabilities, bar formation, ejection/injection from stars and black holes – whereas, others are external – such as merging, interaction, and tidal stripping, to cite only a few examples. The relative mix and dominance of these processes as a function of time and environment are thought to regulate the properties of galaxies we observe in the local Universe.

**Resolved multi-wavelength images and spectra for millions of galaxies:** The last decade has shown that the combination of wide-field imaging and spectroscopy can provide breakthroughs in our understanding of galaxy evolution, e.g., the SDSS has shown the power of precise information on the numerous physical properties of galaxies (star formation rates, AGN activity, morphologies, stellar masses, etc). At high redshift, Euclid will provide a tremendous leap forward in this respect, as the Wide and Deep surveys will break completely new ground by achieving orders of magnitude gain in the combined parameter space of area, depth, wavelength coverage and spatial resolution. The Wide survey will span a transverse size of ~1 Gpc (comoving) at z~0.1, up to more than 20 Gpc (comoving) at z~6, with a total volume in excess of ~1200 $Gpc^3$ – returning images for more than a billion galaxies. This data will simultaneously control the shot noise (statistical precision) and cosmic variance (systematic uncertainty) of any measurements made by Euclid.

In the near-IR, the Y, J and H wavelength coverage down to AB~24 will probe the rest-frame optical light of high-z galaxies and enable photometric redshift measurements to an accuracy of $\Delta z/(1+z) \sim 0.05$, and reliable conversion of the observed galaxy properties to key physical parameters like stellar mass and star formation rate – out to epochs well beyond the peak of the integrated star formation rate in the universe (Lilly et al 1995; Madau et al 1996). The exquisite <0.2 arcsecond resolution of the AB~24.5 optical images will provide detailed morphological and structural information, e.g., bulges, disks, bars, spiral arms, lopsidedness, out to z~2, and fundamental structural information such as size, concentration, at any redshifts. At AB ~ 26, the Deep survey will fill in the unexplored niche of area and depth between Euclid's Wide Imaging Survey and the extraordinarily deep, but narrow-beam surveys of the JWST, and will unveil galaxies at high redshifts that will remain undetected in ground-based surveys.

At the same time, Euclid-NIS will identify and measure the redshifts of star-forming galaxies (predominantly based on Hα) out to z~2, with a sensitivity to star formation rate of SFR ≈ 5, 20, 120 solar masses a year for z = 0.5, 1, 2 respectively. As shown in Figure 2.23, we expect 70 million redshifts (even accounting for a spectroscopic redshift success rate of ε ≈ 30 – 50 %). We note that the current samples of star forming galaxies in the range 0.5<z<2 are only a few tens of thousands (going up to ≈10^5 at z~1 with the ongoing ESO VLT VIPERS survey). Thus, the E-NIS Wide spectroscopic survey will give over a 100-fold improvement in the size of samples available and will allow us to probe galaxy evolution in very narrow bins of mass, type, star-formation and environment; a necessary step forward to pin-down the main mechanisms of galaxy mass assembly. In addition, the Deep Euclid survey will provide ≈ 46,000 Hα emitters per $deg^2$ to a flux limit of $F(H\alpha) \approx 5 \times 10^{-17}$ erg $s^{-1}$ $cm^{-2}$. Both the Wide and Deep spectroscopic surveys will also include the brighter, but rarer population of AGNs (i.e. ≈ 1% of the whole galaxy population), based on the detection of other emission lines over a wider redshift range (e.g. [O II]3727 at 1.7 <$z$< 4.4, [O III]5007 1 <$z$< 3, CIV 1549 at 5.5 <$z$< 12, Lyα at 7 <$z$< 15).

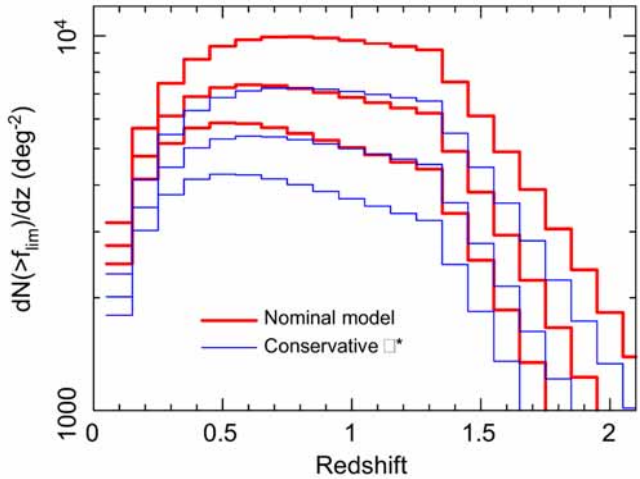

*Figure 2.23 The expected redshift distribution for Hα emitters for 3 different flux limits: 5, 4, and $3 \times 10^{-16}$ erg $m^{-2}s^{-1}$ for the lower, middle and upper pairs of red and blue curves, respectively. The two sets of blue and red curves encompass the current uncertainty on the luminosity function at $z<2$. The blue lines represent a conservative estimate which was adopted in our predictions for the slitless survey (adapted form Geach et al. 2009).*

The synergy generated by the availability of both deep optical/IR imaging and spectroscopy for ~70 million galaxies will be a revolutionary aspect of Euclid. The combination of the IR colors and an accurate redshift are vital to derive reliable stellar population parameters such as the current star formation rate, dust attenuation, the slope of the initial mass function, the absolute luminosity, and stellar mass. Euclid data on galaxies up to z~1 will help break well-known degeneracies between some of these parameters. This unparalleled combination will be essential in establishing the co-evolution of galactic morphology, star-formation and AGN activity as a function of mass and environment. A comprehensive morphological analysis of the survey can be carried out down to the spectroscopic limit of the surveys with a resolution around 1.5 kpc at z~1. This will make headway by enabling the quantification of the redshift evolution of (a) the morphology-density relation, in connection with the merger rates, an essential step to understanding structure formation; and (b) fundamental galactic scaling relations connecting star formation and dynamical assembly histories of galaxies, which bridge the gap between the dark-matter-driven merger rates and the poorly known baryon physics controlling star formation.

**Statistical studies of galaxy properties:** A clear advantage of Euclid over previous and planned surveys of galaxies will be the sheer number of galaxies available with morphological information and photometric (>$10^9$) or spectroscopic ($7 \times 10^7$) redshifts. This massive database will be important for many areas of galaxy evolution studies including: *i)* The co-evolution of the multi-variate distribution functions (e.g., luminosity, stellar mass, morphology, etc.) of star-forming galaxies and AGNs; *ii)* The cosmic evolution of the star formation density and activity at 0.5<z<2, for different Hubble types, based on rest-frame optical spectral energy distributions and on the Hα luminosity; *iii)* The merger rate of different galaxy populations at $z < 2$; *iv)* The identification of the rarest and most massive early-type galaxies at z>1.5, to constrain the evolution of the exponential ends of their luminosity and mass functions; a key test for galaxy formation models; *v)* The evolution of physical diagnostics derived from emission lines ratios – including dust extinction measurements, gas metallicities, ionization processes, and AGN activity (at suitable redshifts, e.g. z>1 for Hβ and [OIII]5007, and line fluxes, as Hα/Hβ≥2.8). The large volume probed by Euclid will furthermore make it possible, for the first time, to map the small- and large-scale galactic environment at high redshifts, and to perform, at those early epochs, a statistically significant analysis of the effect of environment on galaxy and AGN properties.

**The most luminous objects in the very early universe:** The first generations of galaxies were assembled at a redshift of roughly z~7-10+, just 500-800 Myr after recombination, contemporaneous with the re-ionization of the universe. We know little about galaxies in this period. Despite great effort with HST and large ground-based telescopes, a handful of galaxies have been reliably detected at z>7, compared with ~1000 galaxies detected to date at z~6, just 200-400 Myr later, near the end of the re-ionization epoch. For example, the current state-of-the-art is the ground-breaking images delivered by the new Wide Field Camera-3, recently installed on the HST, which provided only 5 candidate z~8 Ly-break galaxies (see Figure 2.24 for details), which is just too small a number for a quantitative study of the galaxy population at such early epochs.

Euclid will dramatically change this situation with its Deep Survey, whose four-color imaging will yield galaxy populations statistics by itself. Using the Lyman-dropout technique (Steidel et al 1990) in the near-IR, the Deep survey will detect thousands of the most luminous objects in the early Universe at $z$>6: for star formation rates of order and above 30 $M_\odot$/yr, about $10^4$ star-forming galaxies at $z$~8 and up to $10^3$ at z~10.

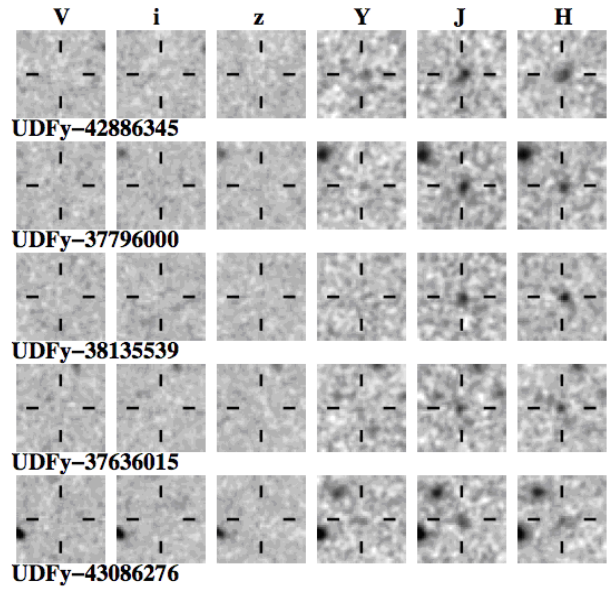

*Figure 2.24 : Candidate galaxies at z~8 from WFP3 observation of the Ultra Deep Field (Taken from Bouwens et al (2009)*

In synergy, the large sky coverage of the Deep spectroscopic survey will be vital for "blind" spectroscopic searches for Lyα emission (arising from the most luminous star-forming regions), and could detect ≈ 10 − 100 sources at $z \geq 6$ per deg$^2$. Moreover, it could find ≈ 1 − 20 galaxies at $z \geq 7$ (per deg$^2$) and possibly up to 5 galaxies at $z \geq 8$ (assuming the present Lyα emitter luminosity function with no redshift evolution and a line limiting flux of 5×10$^{-17}$ erg s$^{-1}$ cm$^{-2}$; so far only one such source has been detected at z~7, see Iye et al. 2007). A fraction (~10%) of these Lyα emitters could show the He II λ1640 emission line due to Population III stars (Scannapieco et al. 2003, Schaerer 2003). The Deep spectroscopic survey should also detect ~ 50 spatially extended Lyα nebulae (Lyα "blobs"), with luminosities of L(Lyα) ~ 5×10$^{43}$ erg s$^{-1}$ at $z \geq 7$ (see e.g. Ouchi et al. 2009). These nebulae are thought to be objects with large cooling clouds accreting onto a massive halo or young galaxies experiencing intense outflows produced by starbursts or nuclear accretion.

High redshift quasars (QSOs) may also play a crucial role in re-ionizing the universe at epochs beyond $z$~7. To date, no quasars have been found above $z$=6.4, and presently planned ground-based programs will struggle to push this beyond $z$~7 because of the need for deep, wide area NIR imaging and spectra (see e.g. Venemans et al. 2007). The optical and NIR Wide Imaging Survey of Euclid will identify likely QSOs by their colors out to redshifts of $z$~12, if they exist. Many of these objects may be difficult to find by spectroscopic techniques, as their Lyα lines are often weak or even absent (e.g. Fan et al 2008). The likely steep luminosity function of these objects implies that maximising the imaging depth over the wide survey is paramount: even a factor of two in flux limit improvement will increase the number of QSOs by a factor of ten. Such searches, and their follow-up, may well hold the key to the rate of black hole growth at z>8 and hence answer questions about the masses of likely "seed" black holes, a fundamental open question in galaxy formation. The Deep Euclid imaging survey should add up to a thousand QSOs at $z$>7, depending on the (unknown) details of the luminosity function at such early epochs. Finally, Euclid spectra will be ideal for directly detecting, without any pre-selection, the broad emission lines of quasar spectra, especially in the absence of the bright OH lines in the atmosphere (e.g. White et al. 2003). We estimate that E-NIS could detect QSOs with L(bol)≥10$^{47}$ erg/s through the direct detection of Lyα and CIV (assuming L(Lyα)/L(bol)=1% and L(C IV)/L(bol)=0.25%). Therefore, in the wide spectroscopic survey, we expect ~10 - 30 QSOs per square degree at 6≤ $z$ ≤7, and up to ~30 per square degree at $z$>7, with the possibility of ten of these objects at $z$>9 depending on QSO luminosity function evolution at $z$>6.

*Euclid's detection of these high redshift sources (Ly-break galaxies, Lyα emitters and quasars) will pro-vide crucial targets for the E-ELT and the JWST, which will achieve greater depths but on much smaller areas, thus probing a fully complementary part of the luminosity functions.*

## 2.5.2 Structure Formation

A fundamental issue in astrophysics is how galaxies trace the underlying dark matter in the Universe, as a function of both scale and redshift. All of the Euclid surveys will provide unprecedented access to the details of this relationship through a number of complementary probes.

**The relationship between mass and light:** By linking the weak lensing maps of the total mass density with the stellar masses and luminosities of the more than a billion galaxies detected in the imaging surveys (out to $z{\sim}3$), Euclid will provide a direct mapping between the total mass density and the distribution of galaxies and their properties (stellar mass, luminosity, morphological type, etc), i.e., it will directly determine the cosmo-logical "bias", which is at the core of understanding galaxy formation and much of cosmology. Moreover, the imaging surveys will unveil a very large number of strong lensing systems: about $10^5$ galaxy-galaxy lenses, $10^3$ galaxy-quasar lenses and several thousand strong lensing arcs in clusters. Several tens of galaxy-galaxy lenses will be *double* Einstein rings (Gavazzi et al. 2008), which are exceptionally powerful probes of the cosmological model as they simultaneously probe several redshifts. These strong gravitational lenses will provide an unparalleled sample of intrinsically extremely faint, but highly magnified, objects at very high redshifts, which will then be exciting targets for spectroscopic studies with the JWST and E-ELT.

**Intrinsic Alignments:** In addition to measuring the cosmic shear, the shapes of galaxies measured in the Euclid imaging data will reveal possible alignments of galaxy ellipticities induced by tidal interactions. The presence of such "intrinsic alignments" is a source of contamination for weak lensing measurements, and will be carefully corrected for in Euclid (see Section 2.4.1). Such alignments would however provide crucial information on the hierarchical assembly of structure in the universe. Recent measurements of intrinsic ellipticity correlations (Okumura et al. 2009, Brainerd et al. 2009) show the presence of a strong alignment that varies among different galaxy types. On large scales, the mean galaxy ellipticity alignment should behave linearly with the tidal field, and observations of the intrinsic alignments will provide a basic test for the *assumed* relationship between the observed galaxy density and the tidal field.

**Halo statistics:** The Euclid surveys will be critical for detailed studies of the clustering of different types of galaxies over a wide range of scale. In recent years, such studies have highlighted the existence of features in the (two-point) correlation function of galaxies consistent with the emerging "halo model" description for the distribution of galaxies in the Universe. For example, Zehavi et al. (2005) showed evidence for a "kink" in the correlation function consistent with the expected transition from pairs of galaxies residing within the same dark matter halo, to pairs of galaxies sharing different halos. The halo model therefore provides an in-tuitive way of understanding how galaxies trace the underlying dark matter, and provides significantly more flexibility than simply assuming galaxies are linear, "biased" tracers of the dark matter. The model naturally explains scale-dependent features (as seen by Zehavi et al. 2005) as well as incorporating the concept of central and satellite galaxies within a halo, which can have different evolutionary paths dependent on their relative location in their host halos, e.g., see the recent work of Skibba et al. (2009) which used the halo model, in conjunction with the mark correlation function, to study the color and morphology evolution of satellite and central galaxies separately.

This methodology will be revolutionized by the size and scale of the Euclid samples of objects, providing accurate measurements of a hierarchy of possible correlation functions, e.g., the three-point function (Kulkarni et al. 2005) and mark correlation functions (see Skibba et al. 2009 and references therein). The richness of this dataset of correlation functions will provide detailed constraints on how different galaxies (or quasars) trace the underlying dark matter halos and how this mapping changes with cosmic time and the build-up of the dark matter halos. Such measurements are ideally suited to statistical comparisons of the observational data with the next generation of detailed galaxy evolution simulations which create mock galaxy catalogues by populating dark matter halos extracted from state-of-the-art numerical simulations, e.g., the latest semi-analytical methodology discussed by Baugh et al. (2005), Croton et al. (2006) and Cai et al. (2008), or the more empirical methods employed by Conroy & Wechsler (2008).

*Therefore, Euclid provides the opportunity to truly understand the details of "galaxy bias", which has been the Achilles' Heel of cosmology for decades.*

**Clusters of Galaxies:** Clusters of galaxies are the largest scale signposts of structure formation at early epochs, and the end products of the hierarchical formation of structure in the Universe. Present imaging studies have revealed reliable galaxy cluster detections to z~1.5. Comparison of the Wide NIR imaging survey to semi-analytic models (Somerville et al 2009) shows that Euclid will find all the massive, and well-populated, galaxy clusters out to z~4 if they exist. More precisely, Euclid's near-infrared imaging will allow the identification, through their red sequence, of hundreds of galaxy clusters at z~2-3 with $M>10^{14}$ M$_\odot$ (comparable with the Virgo Cluster) and thousands of structures with $M > 1$-3 x $10^{13}$ M$_\odot$ (mass scale of the massive groups). The latter are the likely environments in which the peak of QSO activity at z~2 takes place, and hold the empirical key to understanding the heyday of QSO activity.

The Euclid Wide spectroscopic survey will open up an additional discovery space by returning ~10,000 clusters, each with more than 20 redshifts, at z≥0.5 for a slitless Hα survey, and ~900 in a slitless continuum survey (z<0.8; see Figure 2.25). A cluster with ≥20 concordant redshifts is easily detected against the background, guaranteeing clean and complete cluster samples. Twenty redshifts are sufficient for a reliable determination of the velocity dispersion, $\sigma_v$. We also expect more than 100 clusters with ≥50 concordant redshifts, which will be used to calibrate the effect of substructures and of using different cluster galaxy populations (star-forming vs. passively-evolving) in the $\sigma_v$ estimate.

Such wealth of Euclid data on high redshift galaxy clusters (dynamic measurements, weak lensing masses, deep high-resolution optical and NIR imaging of galaxy members) will allow for a detailed calibration of the cluster mass function, and a quantitative comparison of total, baryonic and stellar masses within clusters (using different proxies and approaches). Furthermore, the Euclid clusters can be "stacked" into a single composite cluster with ~3×10⁵ galaxies at z~0.5-0.8 (after rescaling the individual clusters to a characteristic scale, e.g., $M_{200}$). This will constrain the evolution of cluster dynamics using both the traditional Jeans analysis as well as the "caustic" technique for the determination of the cluster mass profile, $M(r)$. With the Euclid cluster sample it will be possible to confirm or reject the standard NFW model (Navarro, Frenk & White 1996) for the inner slope of clusters' dark matter halo profiles, and to constrain their concentration parameter to an unprecedented accuracy, i.e. ±3%, or ~30 times better than current estimates based on the largest spectroscopic database available for cluster galaxies at z~0.5. The redshift evolution of $M(r)$ will be a crucial test for theoretical predictions of the evolution of cosmic structures.

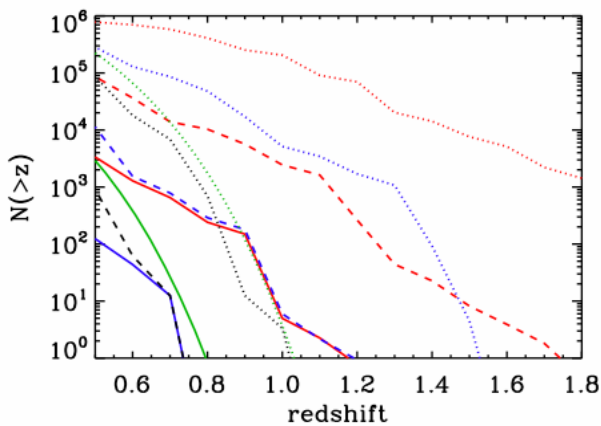

*Figure 2.25: The expected number of clusters with more than a given number of galaxy spectroscopic redshifts (N) within $r_{200}$ as a function of redshift (the dotted, dashed and solid lines are for n≥5, 20, 50, respectively). Black, blue and red lines are for a slitless continuum survey, the slitless Hα survey, and the DMD-slit survey. The green lines refer to the cumulative number for weak lensing selection of cluster with S/N=3 (dotted) and S/N=7 (solid).*

Finally, both the wide imaging and spectroscopic surveys will advance our understanding of the evolution of galaxies in the cluster potentials. In addition to a z~0.5 "stacked" spectroscopic cluster sample (~700 times larger than currently available), the availability of an Hα-flux (rather than [OII]-flux) selected sample of star forming cluster galaxies will largely minimize biases introduced by dust extinction. Furthermore, Euclid's optical morphologies, and deep NIR galaxy images for robust stellar mass estimates, will, combined with the spectroscopic diagnostics, map the evolution of star-formation rates and the growth of supermassive black holes in cluster galaxies, as a function of galactic structural properties and cluster-centric distance, over a wide span of galaxy- and cluster-mass scales. The large size of the data-set, the availability of morphologies, colors, and SFR for spectroscopically-confirmed cluster members, over the full range of galaxy densities, will allow to disentangle local effects of evolution from effects related to the global properties of the clusters, and to identify the physical processes responsible for differential galaxy evolution in high-density environments.

The next generation of wide-area Sunyaev-Zeldovich (e.g., SPT, CCAT) and X-ray (e.g., eROSITA, WFXT) surveys will detect a huge number of distant clusters (from ≈5000 for eROSITA to >$10^5$ for WFXT) only up to $z\sim2$. Capitalizing on the large area covered down to AB~24 in the NIR, the Euclid Wide Imaging Survey will be uniquely positioned to detect bound structures at higher redshifts, and down to one order of magnitude smaller scales. In the $10^{14}$ M$_\odot$ 'massive groups', the combination of relatively high mass densities and low velocity dispersions is expected to maximize the strength of tides and to trigger strong galactic evolution. Furthermore, E-NIS will provide the absolutely mandatory spectroscopic follow-up of all clusters detected by these future S-Z and X-ray clusters surveys, thereby substantially augmenting their scientific return.

## 2.5.3 Structure and Evolution of the Milky Way Galaxy

By returning a vast catalogue of stars in the Milky Way and nearby galaxies, the Euclid surveys will enable a treasure of legacy science based on structural and stellar population studies with unprecedented depth, wavelength coverage, and spatial resolution. Euclid will massively augment the Gaia survey of our Milky Way, taking it several magnitudes deeper. Especially when combined with other deep surveys by ground-based surveys (SDSS, PanSTARRS, SkyMapper), Euclid will provide 4-D information on the positions and velocities of hundreds of millions of medium and high latitude stars, highly constraining the integrals of motion when locked into the Gaia reference frame. Moreover, the 4 optical/IR colours will provide photometric distances (i.e., a $5^{th}$ dimension) allowing us to trace large-scale Galactic streams and structure to much greater distances than Gaia, and accessing a wide and more representative range of the HR diagram (rather than relying on giant tracers). Below V=20, the spatial sampling of Euclid will provide complementary information to the Gaia astrometry, photometry and spectroscopy, adding infrared colours, and providing infrared spectra for every single Gaia star it observes; hence breaking the age-metallicity degeneracy, which is critical for the chemical enrichment history of our Galaxy.

The full sky coverage of Euclid is expected to lead to the discovery of more Milky Way satellite galaxies if they exist. Finding these nearby and extremely low surface brightness objects depends critically on wide-field continuous photometric coverage. The most recently found satellites were from the SDSS, i.e., 12 objects detected over 8000 deg$^2$ down to a r ~ 22.5. Euclid will make a substantial leap in the definition of both the faintest-end of the luminosity function, and the radial distribution of dwarfs, putting stringent constraints on the nature of dark matter, the epoch of reionization, and the properties of star formation in small galaxies. The combination of Euclid and Gaia will give European scientists the premier facilities for current-epoch local galactic structure studies.

The spectroscopic wide survey data will also allow us to obtain the first nearly complete census of L and T ultracool dwarfs in the vicinity of the Sun, with the potential to extend the L-T substellar census to unprecedentedly cool temperatures (below 600-700 K) and low luminosities (e.g. "Y" dwarfs), corresponding to planetary-mass objects of less than ~15 times the mass of Jupiter. Determining the space density of L-T-Y objects will set strong constraints on the slope of the field mass function at sub-stellar masses. Furthermore, the T- and Y-type substellar bodies have temperatures comparable to the gas giant exoplanets and the giant planets of the solar system, so the spectroscopy provided by E-NIS will uncover the properties of the planetary atmospheres.

Young star-clusters are full of free-floating objects with masses as low as 3-5 Jupiter masses (e.g. Caballero et al 2007), with spectral energy distributions peaking at 1-2 mm. It is unclear how many of these were born as planets, and became unbound, and how many arose directly from cloud fragmentation. Their low masses permit their ejection from star clusters, leading to the prediction of about a billion free-floating 'super-Jupiters' in the Milky Way. For relatively young ages (~ ten million years), the Euclid Wide Imaging Survey could find all such systems present within ~1kpc from the Sun. Discovering these objects is crucial for understanding the efficiency of (retained) planet formation.

## 2.5.4 Supernovae

The details of the optimal observing strategy for the Euclid Deep survey will be defined in the Definition Phase but it is already clear that the cosmological weak lensing measurements require often, and regular, observations of the deep fields to monitor the stability of the point-spread function. This time sequence of imaging and spectroscopic data should therefore provide an interesting database of transients like Type Ia Supernovae (SNe Ia), which are now proven cosmological distance indicators. For example, recent results based on ~250 distant SNe Ia from the SuperNova Legacy Survey (SNLS), combined with WMAP-5 data and BAO data, constrained w to better than 5% statistical accuracy, or ~7% when systematic uncertainties are included (Sullivan et al in prep).

The statistical uncertainty on w from SNe Ia is now reduced to the level where systematic effects are dominant. Therefore, to make any major future cosmological advances with Supernovae, one needs to control the systematic uncertainties. The Euclid Deep survey provides a unique combination of area, stability and depth to achieve this goal, particularly in the NIR where the sky background is significantly lower than from the ground and where the effects of extinction by dust are greatly reduced. Therefore, the Euclid Deep survey should allow us to create a larger and, more importantly, better-controlled SNIa sample for cosmology.

By the time Euclid is launched, ground-based SNe surveys are expected to have collected ~1000 well-measured SNe with redshifts up to z~1. Perhaps a few hundred of these will have NIR measurements (at z<0.5), which will provide the strongest control on systematics. In addition a smaller number (perhaps 50) at higher redshifts will have been studied from space using HST. Euclid, via the repeat imaging of 10 deg$^2$ patches of the deep survey, could detect and measure light-curves for 1000-2000 SNe Ia at z<0.7 (in the Euclid observed J band) during the five-year mission. This is the redshift range where the acceleration of the Universe is strongest. In addition, a further 1000-2000 SNeIa, up to z~1, will be detected just at peak brightness. The power of such a Euclid SN survey would be greatly enhanced if coordinated with a ground-based campaign obtaining spectroscopic redshifts and optical light curves.

**Supernova Colours:** The key aspect of any Euclid SN survey will be the availability of deep and stable NIR colours for SNe. This data is extremely difficult to obtain from the ground and is vital for advancing this area of research. Present SNe Ia surveys demonstrate that SNe Ia colour variations are not only due to Milky Way-like dust. Collecting SNe colours over a large wavelength range is the key to characterising and separating intrinsic and dust-induced colour variations of supernovae. The NIR imaging data alone will allow us to measure distances in the rest-frame I-band, where the Hubble diagram has a scatter of only 0.13 (Freedman et al, 2009).

Overall, we expect the J-band photometry will be the most sensitive for SNe, with the Y and H bands providing additional colour information. The R+I+Z filter will provide useful information (e.g., morphology of SN host galaxies and position of the SN within its host) but will require calibration for precision light-curve photometry. If we combine with ground-based data, we could compare standard distances and rest-frame I-band distances to the same supernovae, and use the large wavelength coverage to study colour variations and in turn tighten distance measurements. This would yield a high quality Hubble diagram in the rest-frame I-band with thousands of events to z ~0.7.

**SNe Ia Spectroscopy:** Spectroscopy is needed to measure both the redshift (from the host galaxy) and to determine the SN Type. Euclid itself, through the deep spectroscopic survey, could provide spectra for the brightest SNe as well as redshifts for most of the host galaxies below z~0.5. Large telescopes (including VLT, JWST and E-ELT) could be used to gain spectra for the remaining objects, while in the case of DMD spectroscopy, Euclid could easily provide spectra for all the detected SNe and host galaxy redshifts to a limit of H<24. However, we stress that repeated slitless spectroscopic visits to the deep field (e.g. with a cadence of ~5days), correlated with photometry, would provide a unique time series of SN spectra for free, which could then provide accurate typing information and a SN redshift. The possibility of generating SN photometric "light-curve" measurements from the slitless spectroscopy could also be investigated, but is untested and would require accurate positioning, dithering capability and high sensitivity.

**Supernovae from massive stars:** In addition to cosmological measurements with SNe Ia, we will also detect two other interesting classes of supernovae. For many years, several methods to derive the distance to Type II SNe (or "core collapse" SNe) have been proposed based on the expanding photosphere effect which relates the SN luminosity to the geometry of the photosphere. Now tight Hubble diagrams are being constructed with the current scatter on the distance of 10%. This is somewhat larger than currently derived with type Ia SNe, but it is considered to be subject to lower systematics because the luminosity is directly related to the geometry of the source and metallicity is not expected to have a role as the hydrogen shell is optically thick. We expect Euclid to detect between 3000-6000 Type II SNe out to z~0.5, thus providing a complementary cosmological test compared to Type Ia SNe and other cosmological probes.

Secondly, the discovery of ultra-bright Type II-n SNe, which peak at $M_B \approx$ -23, opens-up the possibility of detecting large numbers of such SNe at high redshift. Their physical nature is currently debated but it is clear that an extremely massive progenitor star (60-150 solar masses) needs to be invoked. The wide Euclid imaging survey will be sensitive to such SNe out to $z \approx$ 2-3, and as they are bright in the near UV continuum (with exceptionally bright Ly$\alpha$ emission), they could be detectable to $z \approx$ 5 in the Deep fields. Due to their long light-curve durations (a few hundred days), the time dilation at $z \approx$ 2-5 would mean that even a low cadence would detect these in significant numbers (possibly a few thousand in the $z \approx$ 2-3 range and maybe 100 at $z \approx$ 5). Such extreme objects would be ideal targets for JWST and ELT.

Finally we note that with several thousand SNe discovered by Euclid imaging, accurate measurements of SN rates as a function of redshift will be possible. Such measurements impact many areas of astrophysics by constraining the star formation history and chemical evolution of the Universe, and by constraining models of the progenitors of SN explosions of all types.

**Other Transients**: Besides supernovae, Euclid will detect other transient objects through both the 10% overlap between adjacent images in the Wide survey, and the sequence of exposures within any single pointing. Statistical studies of the variability of AGNs and strongly gravitationally-lensed QSOs will be possible, and at moderate Galactic latitudes, variable stars of every type will be detected (e.g. eclipsing, accreting, eruptive, ellipsoidal, pulsating, occulting and interacting stars). All of these objects will have multi-colour photometry into the infrared, and many will have infrared spectroscopy. Closer to home, the deep Euclid images near the ecliptic plane will identify a myriad of objects within the Solar system (asteroids, comets and trans-Neptunian objects). For such studies, Euclid's infrared and high spatial resolution optical capabilities are unique, and complementary to many of the ground-based facilities (such as PanSTARRS) addressing this area of science.

## 2.5.5 Extra Euclid Survey Science

The legacy science return of Euclid can be further enhanced through extensions of the Euclid mission to undertake additional surveys. Here we highlight two specific examples.

**Survey of the galactic plane:** The standard Euclid wide surveys will identify a large numbers of faint stars. An extended mission to survey the Galactic Plane would however add large populations of young stellar objects, variables and rare and unusual objects. More broadly, it would provide, through deep infrared imaging (coupled with high spatial resolution optical imaging and, for moderate crowding conditions, with infrared slitless spectroscopy) the most detailed structural studies ever of the thin and thick disk of our Galaxy. A survey of the Galactic Plane would trace the spiral arms, the extent of the Galactic bar, star formation rates, star streams and associations, and would reach obscured young thin-disk regions in the disk plane and bulge inaccessible to the Gaia mission.

**Planet Detection:** Another possible extension to the Euclid mission is the search for extrasolar planets through their microlensing signal. This signal arises from a temporary magnification of a galactic Bulge source star by the gravitational potential of an intervening lens star passing near the line of sight, with an impact parameter smaller than the Einstein ring radius $R_E$. A planet orbiting the lens star generates a caustic structure in the source plane. The source star transiting or passing next to one of these caustics will have an altered magnification compared to a single lens, showing a brief flash or a dip in the observed light curve. The duration of such planet lensing anomalies scales with the square root of its mass, lasting typically one hour for Mars, few hours for an Earth, up to 2-3 days for a Jupiter. A 3-month extension of the Euclid's mission would be sufficient to undertake a unique microlensing survey of the Galactic Bulge, reaching detection limits which would include all planets similar to those in our solar system except Mercury.

The high angular resolution of Euclid's imaging capabilities, and the uninterrupted visibility and NIR sensitivity, will be ideal for such an experiment. They will provide detections of microlensing events using as sources G and K Bulge dwarfs stars – therefore detecting planets down to 0.1-1 $M_\oplus$ from orbits of 0.5 AU. Such a space-based microlensing survey is the only way to gain a comprehensive census and understanding of the nature of planetary systems and their host stars, which is required to understand planet formation and habitability.

## 2.6    Synergy with other missions

Euclid will provide, for the first time, a homogeneous optical and NIR imaging and spectroscopic survey of the entire extragalactic sky. Previous "all-sky" optical surveys have required different instruments, at different hemispheres, leading to major concerns about the calibration of such data-sets on the latest scales probed. Because of the all-sky nature of Euclid, it provides natural synergy with other present, and planned, space-based all-sky surveys, including;

- Planck/WMAP (radio), which provides detailed cross-correlations of these two datasets looking for the CMB lensing signal, ISW effect, Rees-Sciama effect, SZ and CMB foregrounds,
- eROSITA (X-ray) which enhances studies of the cluster mass function and details of the intra-cluster medium,
- WISE (mid-IR), providing additional infrared fluxes on the brightest Euclid sources, which can help in the accurate determination of stellar masses and dust.

Euclid also has key synergies with wide-angle ground-based surveys including

- HyperSuprimeCam (HSC) and Pan-STARRS(1,2,4), which will survey the northern hemisphere and provide deep, optical color information (photo-z's) and time-domain information over the coming decade. In the southern hemisphere, DES, Skymapper, and a number of important ESO surveys (VST/VISTA) will also provide deep optical/IR data soon with the prospect of LSST being available (in the south) by the time Euclid is launched. All of these surveys will clearly benefit from the rich optical/NIR imaging and spectroscopic data from Euclid and will help in the definition of Euclid photometric redshifts.
- Spectroscopic surveys including the SDSS-III BOSS, HETDEX and possibly the "all-sky" missions like BigBOSS. In the southern hemisphere, there are several large area spectroscopic surveys via AAT (WiggleZ) and ESO.
- Radio facilities like ALMA, LOFAR and SKA will be enhanced by Euclid provide redshifts and optical data on radio detected sources (galaxies and quasars). This will revel the details of obscured AGNs and possible feedback processes at play.

Euclid has synergy with large-mirror, smaller-area facilities like JWST, ELT, VLTs, etc. These can provide detailed spectral information for the faint populations discovered by Euclid, as well as their spectroscopic redshifts, important for calibrating Euclid's photo-z's at faint magnitudes.

# 3 Scientific Requirements

In this Chapter, we show how the science objectives for Euclid (see summary table in Section 2.2.5) translate into science requirements. The resulting top level science requirements for the weak lensing and galaxy distribution measurements are described below (Section 3.1), along with the applicable boundary conditions (Section 3.2). The flow down to mission, survey and payload requirements is described in Sections 3.3 and 3.4. The top level requirements and the resulting experiment are based on extensive simulations performed by the instrument consortia. The simulations demonstrating the feasibility of the experiment are described in Appendices 1 and 2.

## 3.1 Top level scientific requirements and boundary conditions

### 3.1.1 Weak lensing top level science requirements

The Weak Lensing survey involves the measurement of (1) the shapes of galaxies and (2) the corresponding redshift of each galaxy for a given survey depth and area. The galaxy shear is measured in the visible from diffraction limited and well sampled images from space. To determine the redshifts of galaxies, multi-band photometry is employed to obtain the photometric redshift or "photo-z". The Euclid science objectives (Chapter 2) are achieved if the following top level requirements are met:

***Survey geometry and depth:*** the galaxy statistics should encompass (1) a survey area of at least 20 000 deg$^2$, equivalent to the entire useful extragalactic sky, and (2) a galaxy number density which is larger than 30(requirement)-40(goal) galaxies per square arcmin with a median redshift $z_m$>0.8.

***Systematics:*** Measuring the galaxy shear from the galaxies' apparent ellipticities and orientations imposes strong control of possible instrumental effects causing distortions in the galaxy images. The variance due to systematic errors in the shear measurement has to be less than $\sigma_\gamma^2 = 10^{-7}$ (see Amara & Réfrégier 2008 for details)

***Photometric redshifts accuracy***: the redshift from photometric measurements should have an accuracy better than 5%(1+z) (requirement, 3%(1+z) goal) with low catastrophic failure rate to build redshift bins. The error in the mean of the n(z) distribution of each bin must be less than $z= 0.002(1+z)$, this is achievable with a sub-sample of about $10^5$ spectra for the direct calibration approach (see also Section 2.4.1 and Bordoloi et al. 2009 for a less demanding approach).

We summarize the main requirements for the WL survey in Table 3.1. We will explain in more detail the choice of these requirements in Section 3.2.

The minimum number of galaxies per square arcmin demands a minimum sensitivity of the system which to first order depends on the size of the primary mirror, the integration time, and system throughput. The minimum survey area together with the total duration of the mission imposes a minimum field of view and maximum integration time.

The required maximum variance due to systematic errors is a factor 50 more stringent than what can be achieved presently from the ground. It imposes strong requirements on the image quality: the shape of the point spread function (PSF) must be well-understood, image distortions over the field of view have to be minimized, and the galaxies need to be well-sampled according to the size of the PSF. The determination of the photometric redshifts imposes wavelength coverage in the NIR band.

### 3.1.2 Wide Field Spectroscopic survey top level science requirements

By measuring the characteristic scale in the galaxy power spectrum (or correlation function) as a function of redshift both in the tangential and redshift directions, one directly probes the expansion history *H(z)* and thus the equation of state of dark energy *w(z)*. At the same time, the statistical distortion of the clustering pattern is a direct consequence of the growth of structure. The Euclid wide-field spectroscopic survey is to measure the galaxy power spectrum *P(k)* for different redshift bins, BAOs and growth factor, and exploit them to place stringent constraints on the Dark Energy Equation of State (EOS) and cosmological parameters in synergy with the WL experiment. The top level requirements are:

***Redshift Accuracy:*** the accuracy of the spectroscopic redshift of each detected galaxy must be $\sigma_{\Delta z} \leq 0.001(1+z)$.

***Survey Area***: to meet the science objectives the galaxy statistics should encompass an area of $\geq 20{,}000$ deg$^2$, equivalent to the entire useful extragalactic sky.

***Statistics and depth***: the number of galaxy redshifts for the spectroscopic survey has to be at least $7 \times 10^7$.

<u>Baseline: Slitless spectroscopy:</u> the minimum number of galaxy redshifts for the spectroscopic survey is $7 \times 10^7$. This can be achieved reaching H$\alpha$ emission line limiting fluxes in the range of $4 \times 10^{-16}$ erg s$^{-1}$ cm$^{-2}$, $7\sigma$ at 1.6 micron for an unresolved source, and with a redshift success rate $\geq 35\%$. The sensitivity to continuum flux expressed as limiting magnitude is H$<19.5$ (AB).

<u>Optional: DMD-based spectroscopy:</u> the number of galaxy redshifts for the DMD spectroscopic survey is in excess of $1.5 \times 10^8$ galaxies for a limiting magnitude of H$<22.0$ (AB).

The expected redshift distributions of measured redshifts imply a median redshift of $z \sim 1.1$ and $z \sim 0.9$, with an upper quartile at $z \sim 1.35$ and $z \sim 1.15$ for the slitless and DMD-based surveys, respectively.

To cover the redshift range $0.5<z<2.0$, the spectroscopic measurements should cover the infrared at wavelengths in excess of $\sim$1 micron. We summarize the main requirements of the spectroscopic survey in Table 1.

### 3.1.3  The Euclid Deep field survey

It has been recognized that the Euclid surveying capabilities will be unique, and can provide in a relatively short period of time a large survey area of high quality images in the visible, deep NIR photometric images and NIR spectroscopy. A deep survey of several tens of square degrees, some 2 magnitudes deeper than the Euclid wide survey would be unprecedented in terms of area, wavelength range, and depth, and can neither be done from ground nor with other (planned) space missions. The top level requirements for the deep survey are:

***Survey depth:*** the deep survey must be 2 magnitudes deeper, than the wide survey.

***Survey area***: the total area must be several tens of square degrees ($\sim$40 deg$^2$)

The wide survey not only provides a large amount of additional science information, but it is also necessary for calibration purposes. The survey depth will be achieved by repeating about 40 times the observations of the same fields using the same wide survey observing modes. By carefully choosing the time intervals between the repeats, the deep survey data will be used to monitor the stability of the payload and spacecraft. In addition, a large number of spectra can be used for the calibration of the weak lensing photometric redshift determination.

### 3.1.4  The need for a space mission in L2

The top level science requirements imply the following important features: high image quality, accessibility to near infrared wavelengths, and a homogeneous survey of the extragalactic sky with a minimum of sources of systematic effects. In the following we will argue that a space mission in L2 is the best option to meet these conditions.

To meet the required redshift accuracies for the WL and spectroscopic surveys, the photometry and spectroscopy must be performed in the near-infrared at wavelengths beyond $\sim$1 micron. The top level science requirements aim at observing galaxies with redshifts in the range at $0<z<2$ with a median at $z\sim1$. At these redshifts the galaxy spectral energy distributions must be probed in the near infrared. The prime spectroscopic diagnostic emission line (H$\alpha$) for slitless spectroscopy must be observed in the infrared between 1.0 and 2.0 micron in order to achieve the required redshift determination of galaxies at $0.5 < z < 2.1$.

The near infrared measurements have to be carried out from space to meet the required accuracies. In the wavelength range between 0.8 and 2 micron, about 30% is invisible from ground and for the remaining fraction bright night sky lines completely dominate the background. At low spectral resolution (R$\sim$500), the filling factor of the sky lines is virtually 100%. These emission lines, variable in intensity on time scales of minutes, add background noise, and make cosmic line and redshift determination of faint sources from the ground extremely difficult, no matter how long one integrates. The absence of the Earth's atmosphere is the unique advantage of infrared spectroscopy and imaging photometry from space, and cannot be duplicated using current or future ground-based telescopes.

The weak lensing requirement to limit the variance in shear systematics implies a high image quality and resolution. This can only be met in space, because observing from space provides a constant diffraction limited resolution, a well-controlled and stable environment, and avoids sources of systematic errors caused by the Earth's atmosphere and thermal variations, which seriously limits similar observations from ground.

With the expected conditions in space the Euclid survey will produce homogeneous visible and near infrared images of the entire extra-galactic sky (> 20,000 deg$^2$) at a diffraction limited spatial resolution not possible from ground, and near-infrared (NIR) images in one or more bands of the same area.

It has been decided to use as baseline a large amplitude free-insertion libration orbit around the Sun-Earth second Lagrange point (SEL2 or L2). This type of orbit is the same as the one used by Herschel/Planck, which is an ESA cornerstone mission. This orbit has the great advantage that there are no disturbances by the Earth magnetic field or Moon, no thermal perturbations, and no gravity gradient as is the case for Earth bound orbits. The Euclid wide survey can be performed completely without occultation or illumination of the payload by Sun, Earth or Moon, which is advantageous for obtaining a homogeneous survey. Finally, L2 has a benign radiation environment, which can be a limiting factor in terms of exposure time and total lifetime of the detectors selected.

## 3.1.5 Other Requirements

*Technological readiness*: The mission should comply with the technical constraint that all subsystems should have a technology readiness level (TRL) larger than 5.

*Telescope aperture*: The required sensitivity, wavelength coverage, and spectral resolution for Euclid are modest compared to other astronomy space missions. In order to meet the sensitivity and angular resolution a 1 meter class telescope is sufficient. An early assessment showed that an aperture of maximum 1.2m diameter is sufficient to meet the science objectives.

*Mission duration*: To meet the science objectives a nominal mission duration of maximum 5 years is assumed. This maximum guaranteed lifetime requirement directly sets an upper limit on the survey speed, which is a trade off between the main aperture size, instantaneous field of view, and duty cycle of an experiment.

*Daily telemetry rate*: A constraining condition is the daily telemetry rate for the science data which are collected by the experiments. For K-band communications, the maximum daily downlink telemetry rate is 0.85 Tbits/day for a daily telemetry communication period of 4 hours. Due to the large information content of the Euclid data, no specific scientific data processing will be done on board to extract the science content. In terms of on-board data processing, it is foreseen to perform the so called up-the-ramp stacking of the data (thereby removing glitches and cosmic rays) for the near-infrared images, and lossless data compression of all data. The data rate constrains the number of detectors and hence the maximum field of view which can be fitted for a given pixel size. It also limits the maximum number of independent exposures (e.g. by using filters or by dithering) which can be collected during one day.

*Table 3.1: Essential requirements to meet the science objectives*

| Category | Item | Requirement |
|---|---|---|
| WL Survey Geometry | Survey Area | 20 000 deg$^2$ extragalactic, contiguous |
| | Galaxy distribution | 30 (required) - 40 (goal) galaxies/arcmin$^2$ usable for WL with a median redshift $z_m$>0.8 |
| WL Systematics | Shear measurement | shear systematics variance $\sigma_{sys}^2$ <10$^{-7}$ |
| WL Photometric redshifts | Statistics | $\sigma(z)/(1+z)$<0.05, 0.03 (requirement, target) with low catastrophic failure rate to build redshift bins |
| | Calibration | Error in the mean of the n(z) distribution of each bin <0.002, achievable with a subsample of 10$^5$ spectra |

| Category | Item | Requirement |
|----------|------|-------------|
| Spectroscopic Survey | Redshift Accuracy | $\sigma_z < 0.001$ |
|  | Survey Area | at least 20 000 deg2 |
|  | number of redshifts | Number = $7 \times 10^7$ galaxies as minimum |
|  | Limiting magnitude or H$\alpha$ flux | Baseline: Slitless spectroscopy<br>Emission line flux $> 4\times10^{-16}$ erg cm$^{-2}$ s$^{-1}$ (7$\sigma$ at 1.6 micron, unresolved source)<br>Continuum magnitude H(AB) = 19.5 mag<br><br>Option: Slit (DMD) spectroscopy; H(AB)=22 mag |
|  | Redshift Distributions | Median redshift of z~1.1 and z=0.9, with an upper quartile at *z*=1.35 and *z*~1.15 for slitless and DMD based surveys, respectively |

## 3.2 Derivation of the weak lensing and spectroscopic survey science requirements

### 3.2.1 Measuring the galaxy shear

We have discussed in section 2.4.1 how our science requirements translate into top level science requirements and beyond. We summarise this discussion here.

Amara and Réfrégier (2007), amongst others, have shown that maximal dark energy constraints are achieved with a wide field survey that covers the entire observable extra-galactic sky. This is because the dark energy FoM scales linearly with total number of galaxies, to first order, and a wide survey (with a median redshift greater than 0.8) maximizes that total number of lensed galaxies. Further more, error predictions show that to reach percent level precision Euclid must contain at least 30 galaxies/arcmin$^2$ (with a goal of 40) over an area of 20,000 square degrees.

Furthermore to reach its full potential, galaxy redshifts need to be measured to $\delta z /(1+z)< 0.05$(required)-0.03(goal) (see Abdalla et al 2008, Kitching et al 2008). This can be achieved by adding deep NIR photometry ($Y_{AB}$ = 24, $J_{AB}$ = 24 and $H_{AB}$ =24) to planned ground based surveys.

To keep systematic effects subdominant to statistical errors Amara and Réfrégier (2008) have shown that the variance of the power spectrum error must be smaller than $\sigma^2_{sys} < 10^{-7}$. This is the top level systematics requirements that can be translated to lower levels.

- The mean redshit of the galaxies in a redshift bin need to be know to better than $\delta\bar{z} < 0.002(1 + z)$ (See Abdalla et al 2008, Bordoloi et al 2008). This in turn translated in to a conservative requirement of $10^5$ spectroscopic subsample.
- Paulin-Henriksson et al (2007, 2008) have shown that errors in the PSF estimation propagate to errors in the shear. This leads to a number of requirements: (i) PSF needs to be measured over 50 bright stars which places a stability requirement on scales smaller than 50 arcmin, (ii) The PSF must have a size smaller than the smallest galaxy to be used and (iii) the ellipticity of the PSF must be smaller than 5%.
- Image simulations show that with a small PSF (less than 0.2 arcsec) and a depth of 24.5 in a broad r+i+z filter (discussed in Meneghetti et al 2009 and section 2.4.1) will provide us with 30 to 40 galaxies/arcmin$^2$ with a median redshift greater than 0.9.

### 3.2.2 Galaxy Redshift distribution

**Emission line flux limit and number of redshifts for the slitless spectroscopic survey:** The cosmological predictions discussed extensively elsewhere in this document have been obtained assuming : 1) a wavelength range of $1 - 2$ μm, corresponding to a target redshift range of $0.5 < z < 2.0$ for H$\alpha$ emitters; 2) our best model for the counts and d$n$/d$z$ of H$\alpha$ emitters (Geach et al. 2009; see Section 2.2.5); and 3) the currently estimated, and possibly conservative, global (i.e., integrated over all Hα fluxes above the limit) success rate of ~ 35% resulting from our extensive simulations. The results of these simulations show that a competitive value for DETF FoM for the slitless spectroscopic survey is obtained for an Hα limiting flux of $4 \times 10^{-16}$ erg cm$^{-2}$ s$^{-1}$, 7σ at 1.6 micron and unresolved source. This flux limit corresponds to a total number of measured redshifts in the range $7 \times 10^7$. These results produce the following requirements:

*The slitless spectroscopic survey shall measure redshifts of Hα emitters over 20,000 deg$^2$ down to an Hα flux of 4 x 10$^{-16}$ erg cm$^{-2}$ s$^{-1}$ over the redshift range 0.5 < z < 2.0.*

Note that, given its characteristics, a slitless spectroscopic survey will have a success rate in measuring redshifts that is a function of the emission line flux. As such, a slitless survey can not be simply characterized by a single emission line flux limit. Indeed, our simulations show that we will be able to measure accurate redshifts, although with a low success rate, also for galaxies below the nominal flux limit given above. Given the steep log N – log S number counts of Hα emitters in this flux range, the number of galaxies with a measured redshift below the nominal flux limit will not be negligible. At this stage these galaxies are not included in our derivation of the cosmological figures of merit.

**H-band magnitude limit and number of redshifts for the DMD spectroscopic survey:** Cosmological forecasts have been produced assuming: 1) a wavelength range of 0.9 – 1.7 μm (consistent with the original DMD spectrograph design, when it was the E-NIS baseline); a limiting magnitude of $H_{AB} = 22.0$; and 3) our best model for the counts and $dn/dz$ of galaxies as a function of H band magnitude (see Appendix A). The results of these simulations show that highly competitive figures of merit for the DMD spectroscopic survey can be obtained with a total number of measured redshifts of the order of $1.5 \times 10^8$. With a global (i.e., integrated over all H band magnitudes) average success rate of ~90% (Appendix 1), these numbers produce the following requirements:

*The DMD spectroscopic survey shall measure redshifts of galaxies over 20,000 deg$^2$ down to a limiting magnitude of $H_{AB}$ < 22.0 over the redshift range 0.4 < z < 2.0.*

Given an area of 20,000 deg$^2$, a surface density of ~24,000 galaxies deg$^{-2}$ at H<22.0 and z > 0.4, a sampling rate of ~ 33%, and a measurement success rate of ~90%, the total number of measured redshifts will be ~ 1.6 x $10^8$.

**Redshift accuracy:** Extensive simulations show that the statistical reconstruction of large-scale structure and the accurate measurement of the BAO fluctuations in the power spectrum analysis requires a redshift accuracy of $\sigma_{\Delta z} \leq 0.001(1+z)$. In case of larger values of $\sigma_{\Delta z}$, the BAO FoM will decrease significantly. A similar accuracy is required also for the redshift distortion analysis. This requirement applies to both the slitless and the DMD spectroscopic surveys. Simulations show that both surveys, with the baseline spectral resolution, can meet this requirement.

## 3.3 Survey Requirements

### 3.3.1 Simultaneous WL and Spectroscopic surveys

The top level requirement cover the entire extragalactic sky (more than 20 000 deg$^2$) for both the weak lensing as well for the wide spectroscopic survey experiments, put stringent requirements on how the survey is conducted within nominal mission time of 5 years. In order to meet the top level science requirements and the boundary conditions the weak lensing and spectroscopic redshift surveys have to be carried out simultaneously. As a consequence, the field of views of the instruments are harmonized such that the same area sizes can be covered in the same amount of time. The areas do not have to cover the same sky. Considering the required pixel-size (see Section 4.4) and the maximum amount of detectors available, it has been decided to adopt a FoV of ~0.48 deg$^2$ for all instruments.

The exposure times and duty cycles of the instruments have to be consistent such that the available time per pointing is used optimally.

The observing strategies and observing modes of the different instruments need to be harmonized such that the measurement of one instrument does not invalidate the measurement of another instrument.

In addition, feasibility studies have led to a step-and-stare observing strategy, as opposed to a continuous scanning survey strategy, because it was not possible to operate the infrared detectors in time delayed integration (TDI) mode such as used for e.g. Gaia.

## 3.3.2 Wide survey requirements

In Table 3.2 the survey requirements for the weak lensing and the spectroscopic surveys are listed. For Euclid, the sky is subdivided in patches, which consists of fields, and the fields are made out of (dithered) exposures. The 20 000 deg$^2$ area covers the two disconnected galactic caps with galactic latitude |b|>30 deg. The survey has to consist of homogeneously collected patches of a given minimum size (larger than 20×20 deg$^2$), which is the useful scale relevant for weak lensing. With a field of view of ~0.48 deg$^2$, a patch has to be made out of ~900 fields, which can be achieved in less than a month time.

The weak lensing experiment requires that the image is dithered with 4 dithers. One dither corresponds to one stable exposure of the detector arrays. The collection of dithered images has several advantages. Dithering will fill the gaps between the detector arrays, mitigates the effect of clusters of bad pixels, improves the PSF calibration from fields stars and galaxy shape measurement from sub-pixel information, provides a distortion map, enables cross correlations between exposures and corrects for cosmic rays.

In addition to dithering, the weak lensing experiment requires an overlap between fields to ensure astrometric and photometric solutions when coadding fields, and to perform shape measurement cross-checks.

In order to minimize the confusion due to overlapping slitless spectra, the spectra have to be rotated, or alternatively, different parts of the spectral range have to be selected, see Section 3.4.3 for more detail. The minimum amount of overlap between NIS fields is to be decided.

The depth of the survey in terms of limiting magnitude in AB magnitudes or in terms of emission line flux, has been derived from the top level science requirements (Section 3.2).

*Table 3.2: Euclid wide survey requirements for imaging and spectroscopy*

| Weak lensing wide survey requirements | | |
|---|---|---|
| Duration | | Less than 4.5 year |
| Survey Strategy | Area | 20000 sq degrees, \|b\| > 30º |
| | Contiguous patches | >20º×20º |
| | Overlap | 2.5% on each side of an image |
| | Dithers | ≥ 3-4 dithers covering detector gaps |
| Depth | Shape measurement channel | R+I+Z$_{AB}$ > 24.5 (10σ extended source) |
| | Photometric channel | Y$_{AB}$ > 24 (5σ point source) |
| | | J$_{AB}$ > 24 (5σ point source) |
| | | H$_{AB}$ > 24 (5σ point source) |

| Spectroscopic Wide Survey | |
|---|---|
| Duration | Less than 4.5 year |
| Survey Strategy | Area | 20000 sq degrees, \|b\| > 30º |
| | Overlap | TBD |
| | Geometry | Contiguous patches |
| | Dithers | Not required |
| Depth for slitless spectroscopy | Line flux 4×10$^{-16}$ erg cm$^{-2}$ s$^{-1}$ (7σ, at 1.6 micron, unresolved source), and continuum magnitude limit H(AB) = 19.5 mag (TBC) |
| Slitless spectroscopy success rate | >35% |
| Depth for slit spectroscopy | H(AB)=22.0 (5σ continuum) |
| Slit spectroscopy sampling rate | >33% |

## 3.3.3 Deep survey requirements

With the capabilities of Euclid, the top level deep survey requirements can be fulfilled in a much shorter observing time than the wide survey. Deep survey areas of several tens of square degrees can be obtained in a few months observing time. Since the deep survey observations will be used to monitor the stability of the system, it is necessary that the observing mode is identical to that of the wide survey observing mode. This condition is fulfilled by just repeating a wide survey measurement on the same sky position until the required gain in depth (2 mag deeper) has been achieved.

As a goal, the frequency or "cadence" of the deep survey repeated measurements should be twice per month. The feasibility of the frequency depends on the limitations imposed by the wide survey strategy based on the viewing constraints, and the timescales of the thermo-elastic disturbances induced by the changes in solar aspect angle necessary to move from a wide survey to a deep survey field. Both limitations can be mitigated by choosing the deep survey areas close to the ecliptic poles, which have good visibility throughout the year.

The deep survey will also be used to obtain a large number of redshifts which will be applied to calibrate the photometric redshift method. Slitless spectroscopy, which is Euclid's baseline, can provide a large number of galaxies for this calibration, but additional ground or space based spectroscopic redshifts are still needed to complement the sample to arrive at the required number of 100,000 spectra down to H(AB) = 24 mag. In the optional case where DMD based slitless spectrometry will be available, all calibration spectra can be obtained from the Euclid deep survey.

The deep survey requirements are listed in Table 3.3.

*Table 3.3: Euclid deep survey requirements for imaging and spectroscopy*

| Deep Survey | | |
|---|---|---|
| Duration | The total fractioned time is less than 4 months | |
| Survey Strategy | Area | > 40 deg$^2$ |
| | Geometry | More than 2 patches larger than 10 deg$^2$ |
| | Visits | Stack built from wide survey like sub images evenly timed over the course of the mission |
| Imaging Depth | All imaging channels | 2 mag deeper than wide survey |
| Spectroscopic Survey Depth for slitless spectroscopy | Magnitude and emission line flux limit | Emission line flux $4 \times 10^{-17}$ erg s$^{-1}$ cm$^{-2}$ (7$\sigma$ at 1.6 micron, unresolved source); H(AB)=21.5 mag |
| Spectroscopic Survey Depth for slit spectroscopy | Magnitude limit | H(AB) < 24.0 mag corresponding to R+I+Z < 24.5 mag |

## 3.3.4 Survey completeness requirements

The wide survey scanning strategy needs to be well optimized even with a nominal mission time of 5 years. From an operational point of view it is important to assess the required completeness of the survey due to the viewing constraints and due to the fact that it is not possible to trade survey depth for survey area once the field observing mode has been fixed.

The completeness requirements are given at different scales in area. To meet the science objectives, the total wide survey area to be collected must be more than 95% of the required area of 20,000 deg$^2$. The wide survey shall be contained in two contiguous areas at the two hemispheres centred on the galactic poles with no holes due to missed patches. For the wide survey planning, the sky has been divided up in tiles or "patches" of ~440 deg2, which are built from individual Euclid fields of 0.5 deg$^2$.

In case a nominal patch of 440 deg$^2$ cannot be completed in the allocated amount of observing time at a given epoch, the "lost" fields are required to be situated at the edge of the patch. This means that holes inside a patch should be avoided and that a failed field inside a patch has to be rescheduled. The extra time necessary for the rescheduling is then compensated by the unobserved fields at the edge of a patch. In case of non-responsive detectors, VIS and NIP can tolerate one non responsive detector in the field without changing the scanning strategy. In case of one (or more) non-responsive detectors for NIS or more than one non-responsive detector for VIS or NIP, the scanning strategy or the field observing strategy has to be revised. Systematic failures at the smallest scales - of pixels or clusters of pixels - are tolerated as long as the field observation meets the dithering requirements which already should take care of the gaps between the detectors in the field of view.

## 3.4   Instrument requirements

### 3.4.1   The weak lensing experiment: Visible shape measurement channel

The weak lensing measurements consist of the deconvolution of galaxy shapes from the Point Spread Function (PSF). This process involves measuring the PSF from the stars and using this to correct nearby galaxy shapes and putting stringent requirements on the imaging quality of the instrument. In Table 3 are listed the derived instrument requirements for the weak lensing experiment.

The galaxy shapes are measured from visible images taken in one broad red band. Galaxy shapes for gravitational lensing are best measured in the red part of the visible spectrum.

The visible shape measurement involves the determination of the ellipticity of a galaxy. This imposes several requirements on the properties of the system point spread function (PSF), which is parameterized by its full width at half maximum (FWHM) and ellipticity. The size of the PSF in terms of FWHM has to be correctly sampled with respect to the size of the detector pixels. With a choice of 0.1 arcsec pixels, the PSF must be in the range 0.18-0.23 arcsec FWHM at a reference wavelength of 800nm.

Control of systematics is of crucial importance for the Euclid mission. It is assumed that there are sufficient calibration stars in the field to calibrate the shape of the system PSF. Stars with brightness R~18 mag will be used, which occur at a frequency of about 1 per arcmin[2]. We require that the system PSF has a low intrinsic ellipticity, low wings, and high stability with small variations on timescales ~3 days to ensure that the calibration stars in the field are sufficient for calibration. In addition, due to the broad wavelength range of the visible band, the ellipticity and FWHM of the PSF as a function of wavelength must be calibrated (see section 2.4.1). The system PSF also must have a small number of degrees of freedom for its dynamic variation. By dynamic we mean the part that cannot be calibrated from an image taken at a different time, such as a calibration run.

As a result of the tight requirements imposed on the PSF, the overall image quality must be high. To be able to relate the (bright) calibration stars with the galaxies at the detection limit we need to ensure our knowledge of the detector response over a sufficiently large dynamic range of 1 to 1000. Geometrical image distortions on scales of 1 arcmin should be minimized and if they do occur it must be possible to calibrate them out. An upper limit is imposed on the number of glitches and dead pixels per exposure. In addition the straylight level should be well below the sky background.

### 3.4.2   The weak lensing experiment: Near infrared photometry

Accurate determination of photometric redshifts (photo-z) with z<2 require photometry in near infrared bands at wavelengths less than 2 micron. The Euclid weak lensing experiment will have 4 bands onboard for the photo-z determination: 1 band from the visible shape measurement and 3 bands in the NIR. These will be complemented by ground based photometry. The combination of DES and Pan-STARRS2 surveys, in the south and north respectively, will be sufficient to achieve the photometric redshift performance requirements, while deeper surveys such as LSST and Pan-STARRS4 would further improve the photometric redshifts (see Section 2.4.1 and Abdalla et al. 2008). Collaborations with the DES and Pan-STARRS ground based projects have been established to define the synergy and collaborative scheme between Euclid and these surveys.

Euclid will have 3 bands (designated Y, J, and H) covering the NIR range from 0.9-2.0 micron. This selection of filters is difficult to obtain from the ground, and therefore provides a unique synergy to ground based surveys, which will cover most of the visible spectrum. To maximize the photometric results in the NIR, a sampling of about 1 detector pixel (or slightly higher) per system PSF will be applied. For the photo-z measurement it is important that the photometry is uniform within a field and among areas on the sky. As a consequence we put a strong requirement on the relative photometric accuracy within a field: after full data processing this accuracy should be better than 0.5%. At a later stage the data can be scaled to adjust the photometry to a consistent absolute calibration based on standard stars or other data sets. To minimize systematics from external light sources, the straylight level should be well below the sky background level.

*Table 3.4: Instrument requirements for the imaging channels.*

| | | |
|---|---|---|
| **Visible Shape Measurement Channel** | | |
| Spectral band | 1 broad red band | R+I+Z (550–920nm) |
| PSF | Size (FWHM 800nm) | Azimuthal average between 0.18 and 0.23 arcsec (not including pixelisation) |
| | Sampling (at 800 nm) | CCD pixelsize = 0.1 arcsec |
| | Complexity | 'Low wings' (see text) |
| | Ellipticity | Less than 5% target on the full FoV;< 10% as a maximum |
| | Stability | Ellipticity and FWHM rms variation less than 0.02% over 50 arcmin$^2$ (corresponding to 50 calibration stars) |
| | Chromatic | Little wavelength dependence |
| Image Quality | Cosmetics | < few % of bad pixels per exposure |
| | Linearity | Instrument calibratable for S/N 1-1000 |
| | Distortion on 1" scale | < 1% anywhere in the FoV calibrated at 0.1% level |
| | Diffuse straylight | Stray light level less than 20% of zodiacal background at the ecliptic poles |
| **NIR Photometric Channel** | | |
| Spectral bands | 3 bands | Y (920-1146nm), J(1146-1372nm), H(1372-2000nm) |
| Photometric accuracy | Final accuracy after data processing | ≤0.5% |
| PSF | Size | System: 0.30 to 0.36 arcsec FWHM in J band |
| | Sampling | Less or equal to 1 pixel per FWHM in J band |
| Image quality | Diffuse straylight | Stray light level less than 20% of zodiacal background at the ecliptic poles |

## 3.4.3 The spectroscopic redshift survey instrument

For the spectroscopic redshift survey instrument, the instrument requirements are driven by the decision which kind of spectrometer is available fore the mission. The baseline for Euclid is a slitless spectrometer, which will be constrained by the sky background noise and the level of confusion of the spectra.

Investigations concerning the usage of a high performance slit spectrometer are ongoing, but slit spectroscopy is for the present study optional. We will mention the instrument requirements for such an instrument without going into further detail (see also Appendix 1).

From the general capabilities of a slitless instrument, it is recognized that the survey comprises the detection of predominantly emission line galaxies. The Hα line will be the main spectral feature for the determination of the redshift. Given the high number of detectable spectra in the field of view, and the size of the spectra covering the full wavelength range, a large number of spectra will be contaminated by spectra from other galaxies. Based on the top-level science requirement on the number density, the depth of such an emission line survey must be $4 \times 10^{-16}$ erg cm$^{-2}$ s$^{-1}$ (7σ point source) in the line at 1.6 micron. The same survey will have a sensitivity to the continuum flux equivalent to a limiting magnitude of H(AB)=19.5 mag. The success rate (i.e. the fraction of galaxies that are detected out of the number that can be detected) should be in excess of 35%. This high fraction can be achieved by measuring the same field of view at different rotation angles, such that spectra overlap differently at different angles. Alternatively, different parts of the spectral range can be selected in different exposures of the same field through the use of filters, which make shorter spectra for each exposure, thereby reducing the amount of confusion among spectra in the field of view.

To meet the required redshift range the spectrometer should cover a wavelength range of 1.0 to 2.0 micron nominally, with goal a larger range. The nominal range corresponds to 0.5 <z< 2.1, thus any extensions in the wavelength will provide larger redshift coverage. To achieve the required redshift accuracy the spectral resolution should be R=500, as constant as possible over the wavelength range with a resolution element of 2 detector pixels resolution.

It is necessary to obtain a NIR image of the same field as covered by the slitless spectrograph, with a depth which is sufficiently deep to always allow association between an emission line detected in the dispersed image with a counterpart in the field image. The NIR image will provide positions of the objects, provide the zero-point in the wavelength scale, and remove ambiguities with zero order spectra contamination. Equally important is the fact that the NIR image will give the object sizes and orientations enabling the correct definition of the best extraction aperture of the spectra, as well as the flagging of contaminated spectra.

Within the Euclid concept, the NIR photometry channel from the weak lensing experiment can be used to obtain the field image. An image exposure with the H-band would provide a depth of H(AB) < 24 mag, with is sufficient to meet the needs of the slitless spectroscopy.

The imaging quality of the spectrometer is important to minimize the loss of sensitivity. For the imaging mode the size of the PSF at 80% encircled energy must be less than 1 arcsec. In case of spectroscopic mode the size of the PSF at 80% encircled energy must be better than 1 detector pixel in cross dispersion mode.

*Table 3.5a: Instrument requirements for the slitless spectrometer.*

| Spectroscopic Channel Requirements: Slitless Spectroscopy Baseline | | |
|---|---|---|
| Wavelength Range | 1.0-2.0 micron baseline <br> <1.0 and >2.0 micron goal | Spectroscopy Mode |
| Spectral Resolution | 500±20 | Variations over wavelength range are small |
| Wavelength accuracy | 0.5 wavelength resolution element or 1 detector pixel | For all identifiable spectra on the array |
| Depth | $< 4 \times 10^{-16}$ erg cm$^{-2}$ s$^{-1}$ (7σ point source at 1.6 micron) | For emission line galaxies |
| | H(AB) = 19.5 mag (5σ per spectral element) | Continuum |
| Minimum detection efficiency or success rate | >35% | Fraction of all detectable emission line galaxies |
| Image quality | 80% Encircled Energy of point source: <br> (1) within 1 arcsec in image mode <br> (2) better than 1 pixel in cross dispersion in spectroscopic mode | System PSF and sampling |
| Observing Strategy | 4 (TBC) rotation angles, or 4 (TBC) sub-samples of the spectral apertures | Several spectral images per field of view |
| Level of straylight in the FoV | < 20% of the Zodiacal light at the ecliptic poles | |

*Table 3.5b: Instrument requirements for the DMD based slit spectrometer.*

| Spectroscopic Channel Requirements: <br> DMD based multi-object slit spectroscopy option | | |
|---|---|---|
| Wavelength Range | Imaging Mode <br> For target selection | 1.3-1.7 micron |
| | Spectroscopy Mode | 0.9-1.7 micron baseline <br> 0.8-1.8 micron goal |
| FoV | | X (TBD) fraction of visible channel FoV |
| Spectral Resolution | 400 | Wavelength range variation: min 300, max: 600 |
| | | Variation over FoV < 5-10% |
| Depth | Imaging Mode | H(AB)=22.5 (10σ point source, TBC) |
| | Spectroscopy Mode | H(AB)=22.0 (5σ continuum, TBC) |
| Multi slit positioning | Object Selection | No overlap of spectra on the detector array |
| | | Object centred in the slit |
| | Sampling Rate | > 33% of objects selected simultaneously |
| Observing Strategy | Collect one image with DMD in imaging mode, subsequently use this image to select objects to be observed and to set the DMD for observation in spectroscopy mode. | |
| Level of straylight in the FoV | < 10% of the Zodiacal light at the ecliptic poles | |

# 4    Payload

In this section we summarise the features of the model payload and the summary performance parameters. The instrument designs are challenging, but the heritage from Gaia and JWST ensures high confidence in the teams to deliver the performance required.

## 4.1    Telescope

The heart of the instrument payload is the telescope, which has to provide a compromise between excellent visible channel imaging quality and a simultaneous field for near-IR spectroscopy. A number of different designs have been produced by industry and institute consortia representing complex mission and system trade-offs made at an early stage of the study. Eventually an optical design solution was defined by ESA (D-SRE-PJ) and offered to both industry and instrument consortia as a common reference design.

This telescope is a modified Korsch with a primary diameter is 1.2 m, and central obscuration 0.4 m. The F/21 telescope has focal pupils and light paths for the VIS and NIP. VIS and NIP use a common M3 optic. The telescope has an afocal pupil and light path for the NIS which uses a different M3. Residual aberrations are corrected out by a simple two lens corrector system positioned in between the M2 and M3. The nominal off axis angle for VIŞ has been arranged at 0.72 deg. As a consequence, the third mirror is used off-axis and the dichroic is placed at the pupil image and separates VIS and NIP channels. The dichroic reflects the VIS beam and transmits the NIP beam.To ensure excellent coverage for field overlaps, the field of views (FoV's) for all three payloads is about 1.0×0.5 deg². Figure 4.1 displays the outline design for the telescope as supplied to industry and consortia teams, but before folding of design for optimum accommodation.

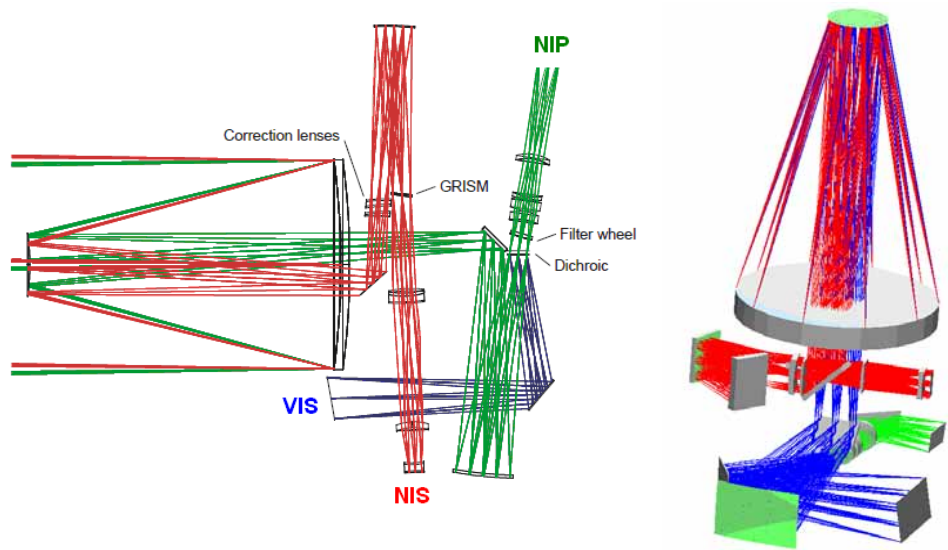

*Figure 4.1: Euclid Telescope design (blue-VIS, green-NIP, red-NIS); (left) complete system before folding (right) consortium concept after folding*

All three systems achieve almost diffraction limited optical performances across the whole image field. Spherical lenses of NIR materials with heritage are employed (namely fused silica, ZnSe, CaF2 and S-FTM16). All of the focal planes are flat, and all mirrors used are flat or simple conic aspheres.

NIS employs the common VIS/NIP M1 and M2 mirrors but combined with a M3 mirror that together with two spherical corrector lenses (CaF2 and S-FTM16) collimates the beam towards the disperser pupil plane.

The folding of this baseline design has been considered, trading off a number of issues such as: available space within launcher; necessary clearance of components; space for mechanical structures, mechanisms; optical performance; thermal design at spacecraft and payload level; mechanical stability; baffling, shielding and minimizing the number of folding elements to maximize throughput. Figure 4.2 shows two examples of the folding of the optical design representing different extremes of compactness achievable, pending the per-formance optimizations, and radiometric loss to additional folds. The EADS design (left) maximizes the compactness at the expense of additional mirrors.

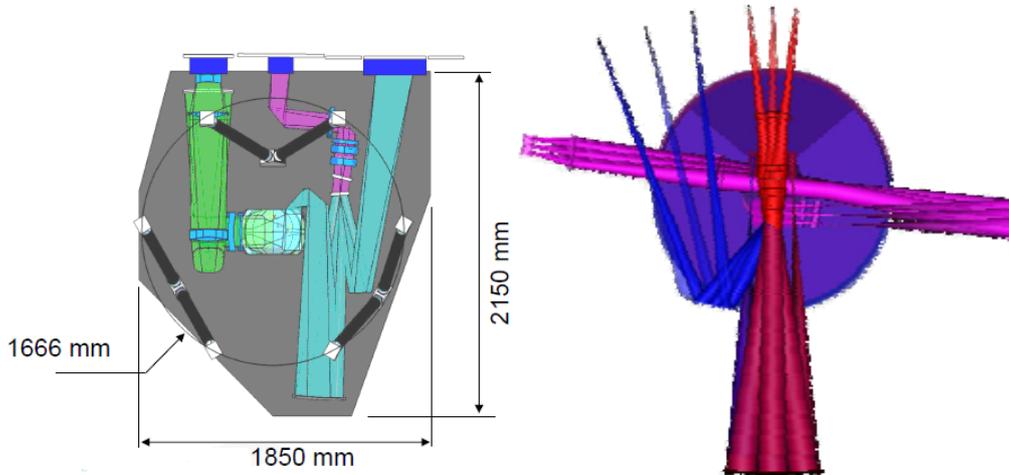

*Figure 4.2: Two examples of optical design folded to match accommodation in spacecraft PLM volume (EADS –left, Thales – right)*

To meet scientific performance objectives, such as an internal background well below the zodiacal sky background, the telescope and payload module have to operate at a reduced temperature. The maximum telescope temperature was determined to be ~240K, whilst the NIR detector temperatures should be ~100K to minimise dark current noise. Once it is determined that a cold temperature is needed for the payload, then AIV issues were considered to be of secondary importance, and payload temperatures will be allowed to passively cool to the minimum value, and local temperatures actively heated where necessary and/or for stability requirements. The eventual temperature defined will lead to a modification of the optics designs for specific cryogenic refractive constants, and at this point a re-optimisation of optical designs will be made.

For the M1/M2 sub-systems, industry has concentrated on two options that reflect their respective experience base (SiC and Zerodur). The 1.2m primary mirror is supported via 3 isostatic bipods and is on the upper side of PLM platform plate. In the SiC options, this mirror must be brazed due to its large size. A specific process for CVD + brazing is needed for good optical performance, but is under development with AMOS+Boostec under a GSTP contract (22072 /08/NL/NA). Alternatively the use of a light-weighted zerodur primary eases the thermal stability aspects of the primary figure control, and then a ceramic (eg $Si_3N_4$) strut structure is required. The radial gradients within the primary, the variation of an overall uniform temperature and the M1-M2 cavity temperature must be controlled to almost an order magnitude higher tolerance for SiC than Zerodur+ceramic options (and the CTE value itself depends upon chosen temperature with SiC preferring a lower temperature). The secondary mirror (0.37m) is supported by a spider. The spider is on the top of a truss made by hexapod or 12 struts. This truss also supports the baffle/thermal shield for the upper part of the payload. It will also support the cover during the launch. The secondary mirror is integrated on a mechanism for alignment and/or focus. This mechanism may be used one time after cooling down to correct any misalignment, and potentially for any relaxation and thermal changes to focus, especially in the case of large excursions of the Solar Aspect Angle (SAA) from a nominal 0 deg up to 30 deg or even 45 deg. Use of an all-SiC design allows an athermal homothetic approach to guarantee PSF stability and ellipticity. The Zerodur/ceramic approach requires more careful active thermal controlhile being less sensitive in principle to PLM power dissipation and SAA variations.

A Monte Carlo analysis has been performed on the ellipticity budget, using a Wave Front Error (WFE) repartition based on tolerances in the following Table 4.1: Manufacturing and positioning of the optical elements will be very challenging. The polishing accuracy required should be achievable but will require state-of-art polishing techniques. Mechanical tolerances are also stringent, and the distance between M1 and M2 must be maintained within a few microns The PSF is calculated at the waveband of 800 nm with a detector MTF of 40% for a set of statistical WFE vectors. For each WFE, the PSF was calculated and from this the associated parameters of FWHM and ellipticity.

The nominal system FWHM is typically 0.15 arcsec and does not enlarge quickly for random errors anywhere near as rapidly as ellipticity and Encircled Energy measures (Figure 4.3). In the visible, while the nominal design does not meet the Survey and Instrument requirements (see Section 3.3 and 3.4) on PSF FWHM, it meets the Top Level Scientific Requirements (Section 3.1) on shear error as evaluated from simulations. The main simulation tool that have been used, estimates the bias on the shear measurements, for a given system PSF and galaxy population. This is done by considering two steps in the shear measurement process: first the calibration of the PSF using a given number of stars with a fixed SNR; second the measurement of the shear by a fit of a galaxy shape convolved with a PSF.

*Table 4.1: Parameters used for tolerance of the VIS channel*

| Parameter | Value |
|---|---|
| Powered mirrors radius of curvature | ±0.5mm |
| Flat mirrors flatness | λ/5 at 633nm |
| M1 and M2 positioning | ±20μm and tip/tilt: ±0.001° |
| M3 positioning | ±30μm and tip/tilt: ±0.005° |
| First fold mirror positioning | ±100Nm and tip/tilt: ±0.001° |
| Detector positioning | ±50μm and tip/tilt: ±0.01° |
| Other elements positioning | ±60μm to ±100μm, ±0.01° |
| M1 and M3 surface error | 15 nm RMS (30nm WFE) |
| Other elements surface error | 10 nm RMS (20 nm WFE) |

In the NIP channel, a sensitive analysis of under-sampled system PSF to the intra-pixel response of the NIR detector on radiometric accuracy has been performed, based on 1.7 μm cut-off detector information. This study shows that Top Level radiometric accuracy requirements are achievable with the current design system.

When in orbit, ellipticity will be calibrated using stars within the field of view. Stability is specified to be of 0.02%, within a region containing ~50 stars of the required Signal to Noise and is very sensitive to focus, and a focus mechanism might be needed to avoid too large focus/ ellipticity change. An associated wave front sensor may be needed to measure the optical quality after cooling down.

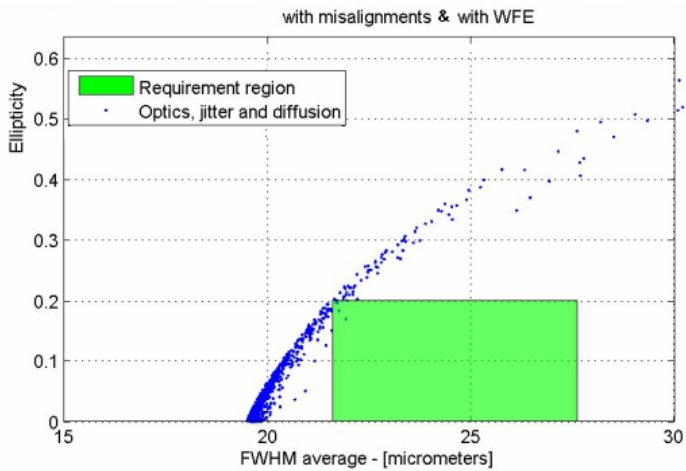

*Figure 4.3: FWHM and ellipticity budget for a range of random alignment errors and in-orbit effect*

In addition to requirement refinement, analysis of the capability to enlarge the system PSF has been investigated at system level. PSF could be enlarged by degrading the MTF of the detector but ~10% is needed to get a FWHM of 2.1 pixels. A defocus of ~160 nm might achieve a FWHM of 2 pixels, but the ellipticity increases then to 30%, and the PSF is quite distorted. Adding spherical aberration and defocus could enlarge the FWHM but again the ellipticity becomes larger than specification.

To meet the top level MRD specification a Line of Sight (LoS) variation must be obtained by a circular motion of 1.2 pixels PTV of LoS, homogeneous in all directions. This LoS variation might be done: (i) Either by the satellite: this impacts all 3 instruments and the impact is uniform within the whole FoV. (ii) Moving a folding mirror: this could be done for VIS and for NIP separately.

A refinement of the Instrument requirement is under analysis especially to check impact of relaxing PSF size on other system PSF contributors (PSF complexity, AOCS, detector pixel response). The investigation of PSF enlargement capability is considered as a backup technical solution.

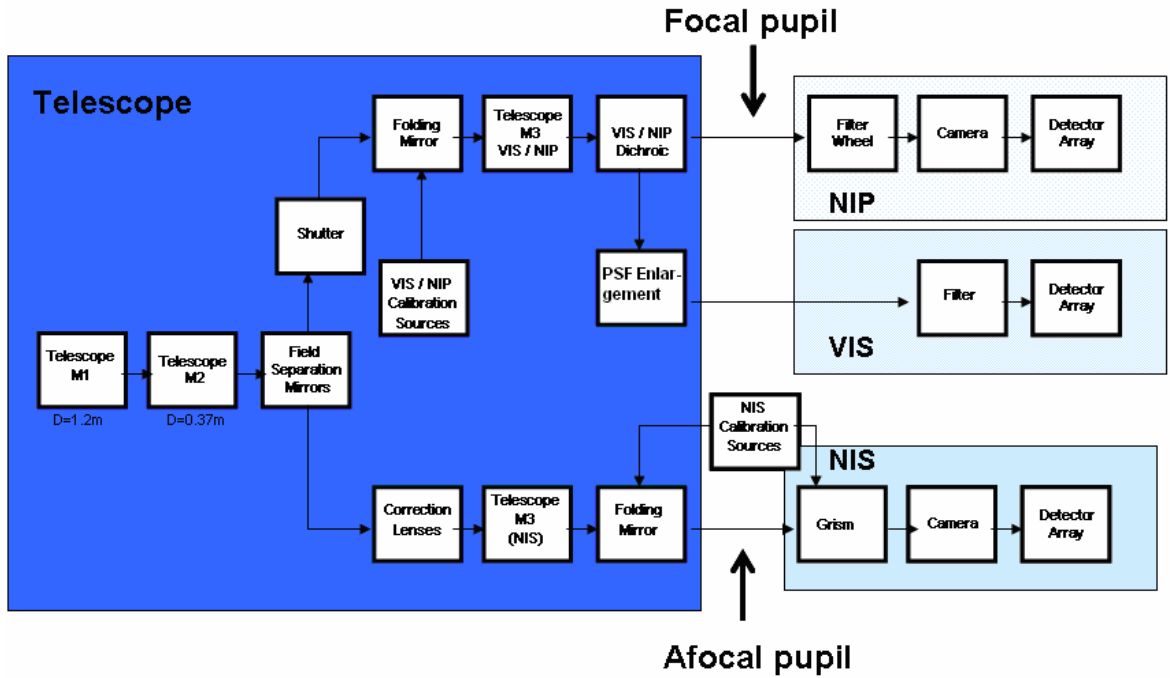

*Figure 4.4: Block diagram of payload elements (one possible version of instrument interfaces)*

Figure 4.4 provides a block diagram of the payload components up to the focal plane assemblies. The instruments can be mechanically separated to ensure an ease of integration for separate deliveries during AIV. This is also considered by industry and it allows optimizing the mass distribution of the payload module optical bench. Another possibility is followed by the EIC consortium and consists of a single weak lensing instrument with interfaces after M3 and including both the VIS and NIP channels.

## 4.2   Euclid Imaging Channels Instrument (VIS and NIP)

The Visible and Near Infrared Imaging Channels instrument provides shape measurements in the visible and photo-z information in the near infrared. It is optimised to fulfill the needs of its primary scientific goal, weak lensing. The interface to the payload module (PLM) is still TBD. As was defined in the PDD, Euclid Instrument Channels starts after the telescope M3. In the current concept of the EIC consortium VIS and NIP channels would be delivered integrated on a common composite support structure (COMA = Common Opto-Mechanical Assembly). Alternative solutions will be explored in Definition Phase. Three electronics boxes are associated to the instrument and integrated on the Payload, the Payload Data Handling Unit (PDHU) and the Payload Mechanism Control Unit (PMCU) and the NIP electronics CCU.

The COMA is conceived as an aluminum honeycomb and carbon fiber sheets bench holding the dichroic and the fold mirror of the visible path, and supporting additional functions such as a shutter for visible read out and a visible calibration unit. The COMA would provide the thermal and mechanical interface towards the payload and ensure optical baffling to the VIS/NIP channels. (Note the option of an all-SiC PLM could force a more expensive design solution on the consortium concept).

The fold mirror is a 290 mm x190 mm flat mirror with alignment capability through 3-balls joint for fine tuning of the focus on the VIS-FPA. The dichroic is a 120 mm in diameter plate that separate the visible wavelength range (reflection) and the NIR wavelength range (transmission). The calibration unit in the visible path consists of 3 visible LEDs at 600 nm, 750 nm and 900 nm with a lambertian diffuser in front of it and located below the VIS fold mirror. It allows illumination of the VIS Focal Plane Array with flatness better than 5% for calibration. The shutter mechanism is located in front of the dichroic and prevents trails in the images during readout.

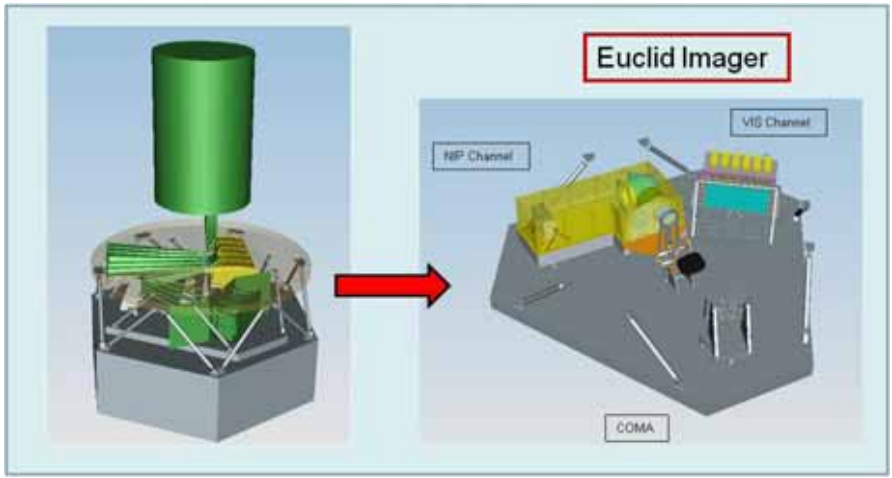

*Figure 4.5: Instrument team proposed folding of the telescope with Imaging Channels instrument volume allocation (left: green block) and I/F to payload, on the Right the Euclid Imaging Channels instrument*

## 4.2.1 Visible Channel description

The Visible Channel (VIS) is composed of 4×9 CCDs with 12 μm pixels (baseline e2v CCD203-82 CCD with optimized packaging for field of view gap) that covers the 0.5 deg$^2$ visible field of view (see Figure 4.7). The baseline CCDs are full frame types of 4096×4096 pixels. They are used in conjunction with the mechanical shutter in the optical path to prevent image smearing during readout. The CCDs are butted with minimum gaps between the sensitive areas but still require dithering pattern to provide maximum gap filling of the visible field of view in order to prevent specific loss of information at inter CCD spatial frequencies.

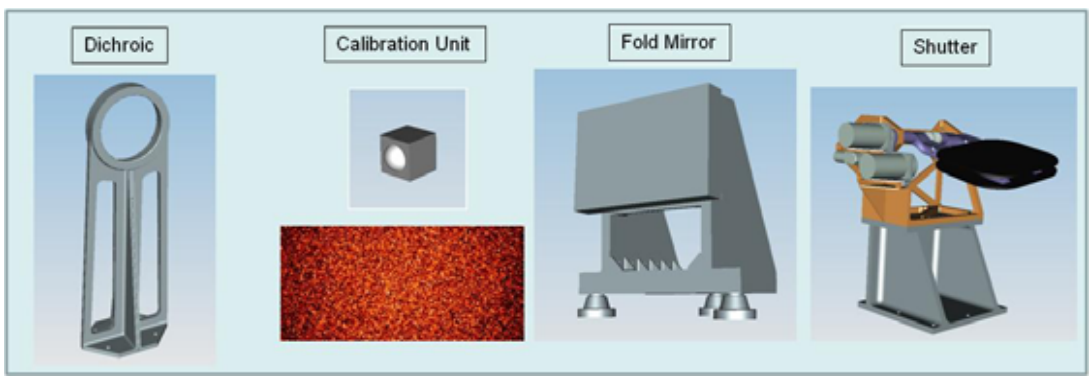

*Figure 4.6: Sub-elements of the VIS/NIP channel*

The architecture of the proposed electronics to control and read out the 36 CCDs has been developed from experience with previous projects, most notably Gaia. Unlike Gaia where the CCDs are all operated independently, the Euclid VIS CCDs are all operated in synchronism. This simplifies the sub-system design. Each CCD is served by a proximity electronics module (PEM). Each row of PEMs is served by an interconnect module (IM). Because an individual clock sequence generators for each CCD is not needed for Euclid, it is beneficial to combine the PEM & IM functions into one unit. This unit is called a Read Out Electronics (ROE) unit. In principle, one clock sequence generator could be used to drive the entire Euclid VIS array but due to redundancy considerations, an ROE will be provided per three CCDs. This architecture has been arrived at after several trade-off iterations. There are many drivers including the system level grounding plan, the AIV & spares philosophy and the optimization of functions per block. Our architecture eliminates the need for any data buffer memory in each ROE. This is a major simplification that reduces cost, power and mass. To preserve the redundancy and grounding concept, each ROE is provided with its own Power Supply Unit (PSU). The performance of the PSU is critical to overall instrument performance because supply line noise can easily be the factor that limits the achievable signal to noise ratio. Considering the "step and stare" observing strategy, the CCDs observe for a relatively long time and are then read out quickly. Very high speed readout is not required because a substantial amount of time is available during each spacecraft re-pointing. The ROE units digitize with high resolution (200kHz / 16-bit) and provide low noise, high precision video processing in order to sample accurately the telescope PSF.

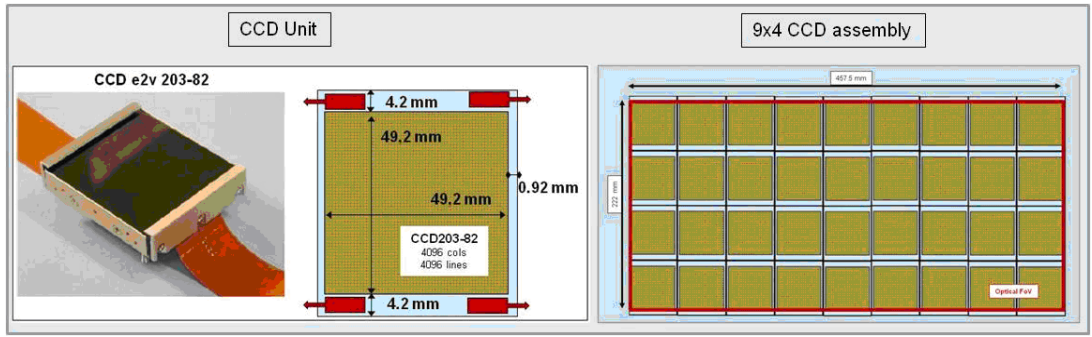

*Figure 4.7: CCD block unit and associated gaps (left), 4x9 CCD focal plane assembly (right)*

The VIS units composed of CCDs + ROE + PSU are all integrated into a Focal Plane Assembly thermal mechanical structure. This design will benefit from experience of large CCD focal plane assembly from the MegaCam and Gaia instruments. The 4×9 CCDs are integrated on a SiC structure. The ROE and PSU may be held by an aluminum structure that provide harness interface from the electronics boxes to the CCDs. The CCD plate and the electronics structure are mechanically and thermally decoupled as their thermal and mechanical requirements are different (electronics box operating between 240 K and 300 K, CCD operating around 160K). They are only electrically linked by the CCDs harness. The CCD plate and electronics structure have separate interfaces to the COMA bench. The CCD plate is isolated from the bench by insulating material pads and from the radiating electronics structure by radiative screen. 4 additional CCDs may be added with no major change to the concept to implement FGS sensors on the visible focal plane, but the interface definition to spacecraft AOCS and SVM subsystem must be elaborated.

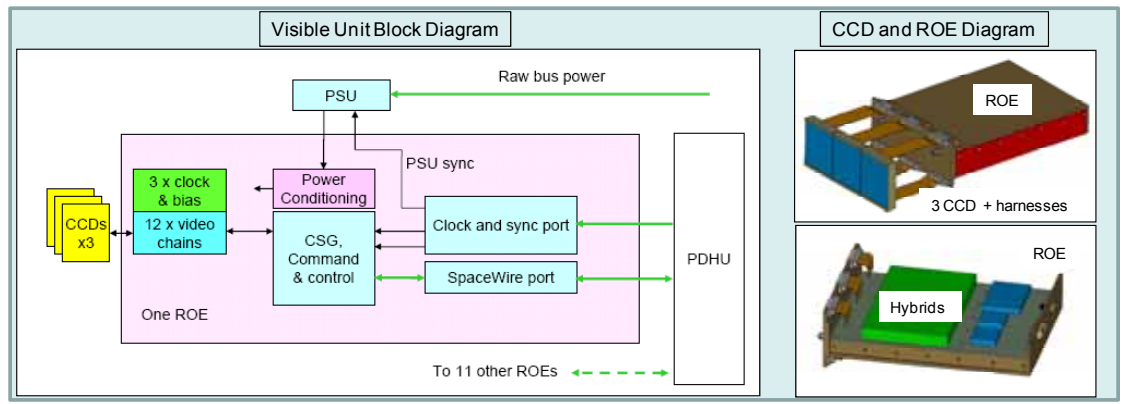

*Figure 4.8: CCD and ROE block unit and diagram – consortium implementation*

In order to minimize the risk associated with the ROE architecture, an early breadboard program has already started at MSSL. It has already been demonstrated that the radiation-qualified electronics to support three CCDs will fit on the ROE board size using only surface mount devices (i.e. without any hybrid microcircuits) at engineering model quality. The necessary performance at room temperatures has already been achieved and the board is not being used for PSF characterisation and radiation testing. The development of such a board is ongoing, to verify that there are no cross-talk problems between the channels and hence further reduce risk to the eventual flight system performance.

The CCD plate and Electronics structures are both isolated from the PLM bench and thermally linked to a radiator that is directly in view of cold space.

Digitized CCD data and asynchronous commands are communicated by a bi-directional SpaceWire link between each ROE unit and the Payload Data Handling Unit. A common clock, to synchronise the readout of all the CCDs, is fed from the PDHU to all the ROE units.

The total number of CCDs and Proximity Electronics Module (PEMs) to be procured for VIS (36 for science + 4 for FGS ) is quite large compared to other space programmes but still less compared with GAIA (more than 100 CCD). There will be technical trade-offs to be completed as early as possible in the PEM design phase related to costs and programmatic risks between different solutions. From the GAIA experience, it can be concluded that suitable technical solutions exist that can be implemented with reasonable low programmatic risks against the required model delivery dates. Also the EIC consortium is already bread-boarding a viable PEM solution.

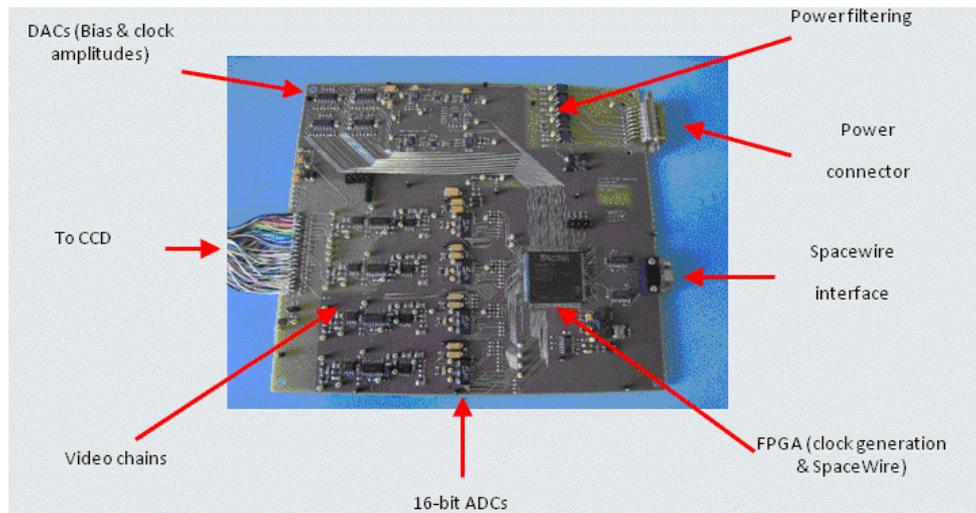

*Figure 4.9: The design Evaluation Model read-out electronics*

The CCD procurement rate on GAIA is ~6 detectors per month when ramped to full production rate, while the Euclid procurement is classed by E2V as requiring no more fabrication effort than that for a typical procurement order placed by a large ground-based observatory. Any identified development areas in the CCD design established in Definition Phase should not impact on detailed PEM driving requirements, but nevertheless CCD development batches would need to be initiated early, and the cost/schedule risk by leaving this to National Agencies needs to be considered. Overall the detector procurement has been designated as a nationally funded item, and yet the schedule requires a commitment to begin procurement before the mission selection to implementation.

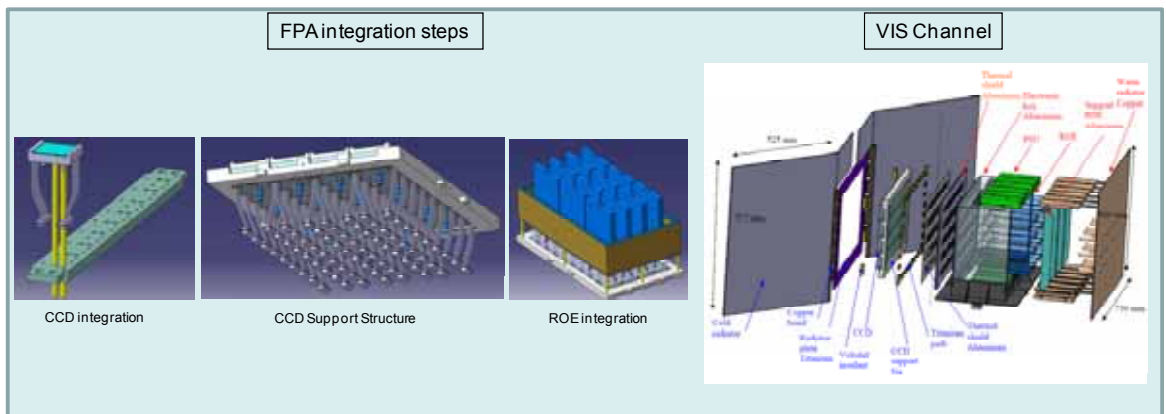

*Figure 4.10: Thermal Mechanical design of VIS channel*

For the PEMs, there must be close interaction between the EIC consortium and potential PLM contractors even during Definition Phase because:

- The technical trade-offs should be carried out in an approach to consider design to cost, design for test and design to production rate; weighting, the non recurring and recurring costs, against physical features and performances
- The long lead items procurement should start at the Preliminary Design Review at the latest – some may have to be ordered before, e.g. once the detailed review of requirements is completed.
- The PEMs flight production must be organised in batches, each batch manufacturing being started within ~2 weeks (pending the batch size) after the previous one, and progressing in parallel in order to meet implied schedule constraints
- The mechanical and thermal vacuum facilities have to be sized according to the size of a whole batch for acceptance tests that include electrical and functional tests, vibration tests, then performance tests in thermal vacuum, thermal cycling, and final electrical and functional tests.

## 4.2.2  NIP Channel description

The NIP channel provides photometry on the same field as the visible channel in 3 bands (Y: 920 – 1146 nm, J: 1146 – 1372 nm, H: 1372 –2000 (goal 2500) nm. The optical base plate of the instrument is attached via three Inconel bi-pods to the bench structure. The NIP channel is composed of a structural box containing a Filter Wheel Assembly (FWA) with 3 IR filters and a diffuser/shutter, a focal reducer, and a NIR FPA. The focal reducer is a 4-lens (ZnSe, fused Si, CaF2, SF57HHT) block that includes a fold mirror due to instrument volume constraints. The VIS shutter can be used as a calibration source for the NIP instrument. Tungsten lamps, mounted on the dichroic holder or connected via optical fibers, are used to illuminate the shutter's diffuse back surface. The calibration setup will fully illuminate the pupil, located at the dichroic, and provides flat fielding calibration of the NIP focal plane through all the optical elements (dichroic, filter and lenses) with a reasonable representation of the telescope beam. The VIS shutter (~150 K), can also be used as a dark calibration target for the NIP instrument without the need for additional mechanisms. As a backup option an additional position in the filter wheel could house a diffuser/shutter for calibration.

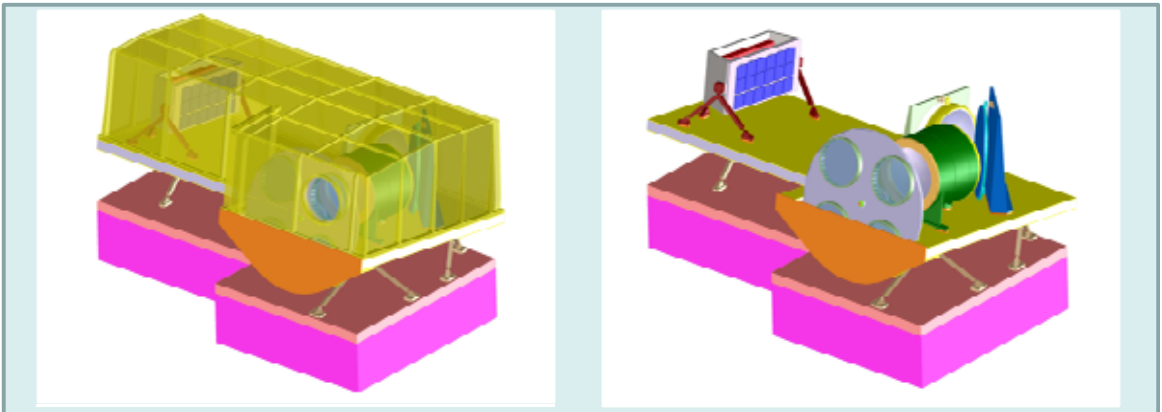

*Figure 4.11: NIP channel elements implementation.*

The FWA is made up of a 410 mm (diameter) Al disk, a dedicated duplex bearing system and actuator. The 3 infrared filters will be housed in filter mounts that will protect the filters from launch vibration, compensate for the differential thermal expansion during cool down (ambient to ~150 K) and hold the filters steady during science exposures.

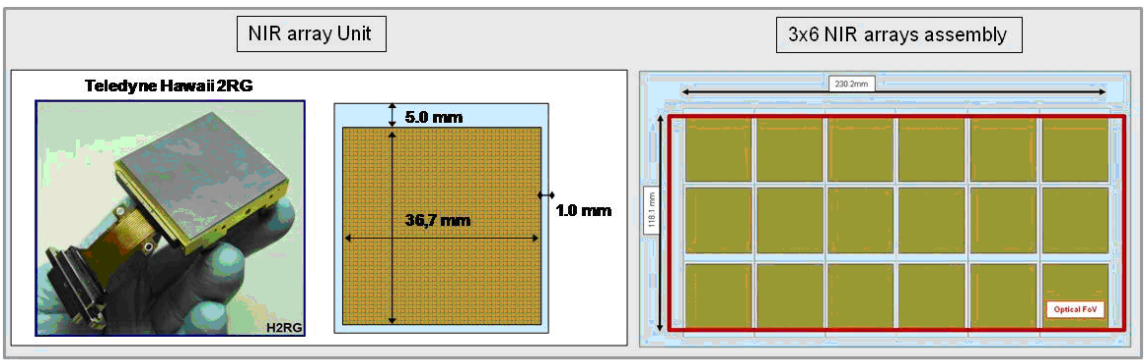

*Figure 4.12: NIR array block unit and associated gaps (left), 3x6 NIR focal plane assembly (right)*

The NIP FPA is composed of 18 (3×6) Teledyne Hawaii 2RG IR arrays that cover the 0.5 square-deg NIR field of view (see Figure 4.12). . The baseline NIR arrays have 2048×2048, 18 μm pixels, with a cut-off wavelength of 2.5 μm. For gap filling, the NIP will have the same dither as the visible channel because dithering is achieved at spacecraft level. The detectors at 100 - 120K (TBC) are mounted in SiC housing at 150K thermally decoupled from the base plate by bipods made from titanium. The temperature of the detectors and SIDECARs (System for Image Digitalization, Enhancement, Control and Retrieval) ASICs (Application Specific Integrated Circuit) is regulated via specific thermal control loops.

The analog-to-digital conversion is performed by the SIDECAR electronics, also provided by Teledyne. The 18 H2RG arrays are connected via a custom flex harness to 2 (TBC) SIDECAR chips mounted inside the FPA housing close to the arrays. The main task of the SIDECAR is to provide the detector readout and analog-digital conversion. The basic SIDECAR architecture can be divided into the following major blocks: analog bias generator, A/D converter, digital control and timing generation, data memory and processing, and digital data interface. An additional CCU electronics unit devoted to the processing of the NIP raw data (slope generation, glitch and saturation detection) and interfacing to the PDHU is associated to the channel but integrated on the payload structure.

## 4.2.3 Thermal architecture

The thermal architecture assumes that the PLM structure is cooled down passively by the satellite to ~150K. The instrument is isolated from the radiative environment by MLI and dedicated radiators for the NIP structure (150 K) the NIR FPA (100 K), the VIS FPA (150 K) and the VIS electronics (240−300 K) are directly located on the cold side of the spacecraft. The NIP, the VIS FPA and the VIS electronics are thermally decoupled from the PLM bench for accurate thermal control.

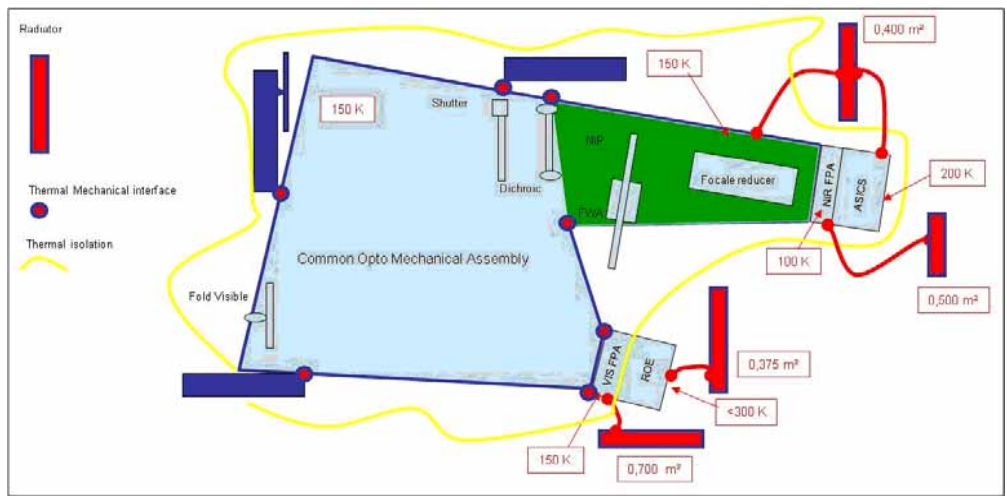

*Figure 4.13: Euclid Imaging Channels temperature architecture.*

## 4.2.4 Electronics architecture

The overall electronics architecture is described in Figure 4.14. Low level analog electronics are enclosed in VIS ROE and NIP ROE respectively for the VIS and NIP instruments. These electronics shall be close to the detectors in order to limit the length of the harness between the two stages ensuring high performance. Control electronics for instruments mechanisms, motor controls and position sensors are hosted by the PMCU. It also includes drive electronics to bias the calibration sources and temperature monitoring lines for focal planes temperatures acquisition. Digital electronics is mainly distributed into the PDHU (central data handling) and CCU units. The PDHU is in charge of the control of all the sub-systems of the instrument by distributing low level commands, collecting and monitoring housekeeping parameters. It is also in charge of collecting all the uncompressed data streams transferred from the subsystems and perform lossless compression in order to match instrument data rates and spacecraft telemetry rate. The CCU performs the processing of the NIP raw data. It performs data reduction from over-sampled detector data (digitized in NIP-ROE).

The command distribution is ensured by the PDHU. This unit receives time-tagged commands from the S/C which are then distributed to the destination sub-system: VIS-ROEs, CCU, NIP SIDECARS through CCU and PMCU. For standard electronics units (PDHU and PMCU) the secondary power supplies are derived from the S/C power bus by mean of DC/DC converters located inside the unit. For more critical electronic units (ROE) the power converters are located outside of the unit to avoid risk of electrical coupling and hence performance degradation. Therefore dedicated units will host these power converters for both VIS and NIP SIDECARS. The VIS PSU is either interfacing the PCDU by means of a redundant power and then distributed to individual unit or by means of 12 power lines directly connected to individual units. A primary principle is to suppress whenever possible the identified single point failure: the aim is to avoid the loss of the complete instrument due to a single failed function. When looking at the architecture, PDHU and PMCU appear to be main single point failures since the PDHU provides interface for both downstream data upstream commands and synchronization to the whole instrument and the PMCU controls the optical mechanisms which are mandatory for the instrument operation. Cold redundancy of these units is therefore foreseen. The PDHU interfaces both the VIS and the NIP instruments through dedicated high-speed data links (e.g. SpaceWire). On the spacecraft side it communicates with the SVM CDMU through a SpaceWire link (TBC), from which it receives telecommands and to which sends the housekeeping telemetry. On the other side science telemetry is send directly to the Mass Memory through a dedicated high-speed data link.

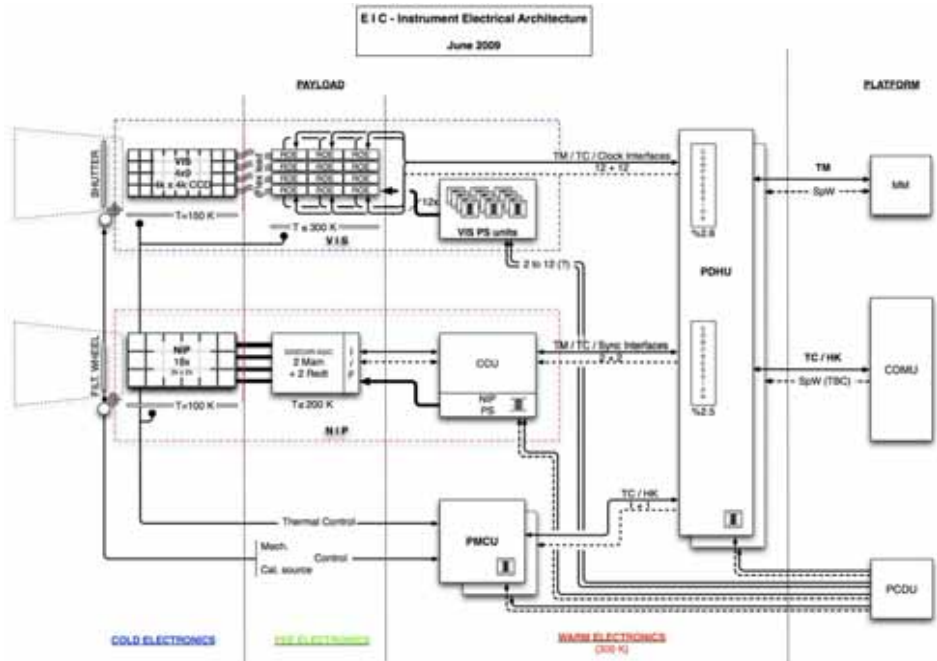

*Figure 4.14: Euclid Imaging Channels electrical architecture.*

## 4.2.5  Mass and power budget

The overall mass and power budgets (including 20% margin) are given in Tables 4.2 and 4.3 according to the consortium design,. The budget is hostage to detailed mass allocation for structures that can change with integration approach, but here is reported on worst case optical bench estimates.

## 4.2.6  Critical items

The main critical items associated with the Euclid VIS and NIP are linked to the need of calibration and stability of the performance over a long 5-years mission period, together with large field image and large amount of data to downlink.

Weak Lensing requires high accuracy in the shape measurement of faint galaxies and therefore system PSF degradation effects must be quantified and the impact on final performance is being assessed. The CCDs and NIR detectors performance stability over time are currently being investigated at MSSL, CEA and Open University especially for degradation of capability induced by radiation. This supplements a test activity funded by ESA within industry on CCD radiation effects. While expecting evaluations of performance with hardware-in-the-loop data from representative CCDs, we can highlight that HST CCDs are operated in a warmer regime than the Euclid ones are, and greater precision will be required for Euclid, but the concepts have been proven on real data. Recent work by EIC members have explored the higher accuracy regime required for Euclid in the framework of p-channel CCDs (Rhodes et al. 2009) and have concluded that the systematic errors introduced by radiation damage effects are acceptable.

Table 4.2: Euclid Imaging Channels mass budget

| System | Sub-system | Mass (incl. 20% margin) |
|---|---|---|
| *VIS* | | *114 kg* |
| | CCD+harness | 11 |
| | FPA Structure/Radiator | 54 |
| | ROE+harness | 15 |
| | PSU+harness | 4 |
| | Electronics box/Radiator | 30 |
| *NIP* | | *111 kg* |
| | NIP Structure | 54 |
| | FWA | 5 |
| | Focal Reducer | 27 |
| | NIP FPA | 10 |
| | Calibration Unit | 2 |
| | Electronics | 13 |
| *Dichroic /Shutter* | | *14kg* |
| *COMA* | | *93kg* |
| *PDHU* | | *16 kg* |
| *PCMU* | | *13 kg* |
| **Total** | | **361 kg** |

Table 4.3: Euclid Imaging Channels power budget

| System | Sub-system | Power (incl. 20% margin) |
|---|---|---|
| *VIS* | | *134 W* |
| | CCD | 6 |
| | ROE | 80 |
| | PSU | 48 |
| *NIP* | | *37 W* |
| | FWA | |
| | Focal Reducer | |
| | NIP FPA | 5 |
| | Electronics | 32 |
| *PDHU* | | *74* |
| *PCMU* | | *22* |
| **Total** | | **267 W** |

The NIP photometric budget requires a very low noise, demanding low internal thermal background (cold telescope) and low detector readout noise. The data handling algorithms to accommodate this sampling have been identified, and not found to be critical with respect to processing power and resources. However, this needs careful checking against evolution in Hawaii detector noise performance vs. temperature.

The large focal planes in VIS and NIP generate a large data volume to be stored and downloaded to ground. The lossless compression algorithm, telemetry capability and mass memory associated with these data volume requires further analysis. RICE lossless compression algorithms for the VIS and the NIP (TBC) have been evaluated on simulated image including radiation effects to confirm the 2.8 VIS and 2.5 NIP compression factor.

The specific image data processing associated with weak lensing science demands a close connection between science requirements and system performance evaluation during the definition phase. This ensures that the instrument performance remains commensurate with mission goals, especially in terms of trade-offs involving dithering, PSF sampling, wings profile, ellipticity and stability over time.

For NIS and NIP the same detectors (Hawaii 2RG and Sidecar proximity electronics) are used. Due to the high number of detectors required (18 for NIP + 8 for NIS) for the baseline, a common procurement is attractive, more especially as a long production run will be required, and ITAR issues may add schedule risks. Such procurement could be initiated centrally, since it should begin before phase B2. However, as with the CCD procurement this is foreseen to be a National Agency procurement, commensurate with recommendations from the SPRT report.

## 4.3 Near IR Spectrometer (NIS)

NIS observes adjacent non-overlapping areas of sky in each pointing. The baseline for NIS is a slitless spectrograph. An optional multi-slit solution based on Digital Micromirror Devices (DMD) has also been studied at system and subsystem level by the ENIS consortium and is described in Appendix 1. Industry was directed not to study this option as a consequence of the inability to secure an acceptable Technology Readiness Level in time before the conclusion of the Assessment Study.

The NIS slitless spectrograph design is based on top-level requirements listed in Table 4.4.

Table 4.4: Summary of Slitless spectrograph top-level requirements

| Spectrograph parameters | Values |
|---|---|
| FOV | 0.5 x 1.0 deg$^2$ = 0.5 deg$^2$ |
| Plate scale at detector | 0.44-0.45 as/px |
| Wavelength band | 1.0 to 2.0 μm (or 0.85-1.7 μm) |
| Spectral resolution | R = 500 (constant in range) |
| Mitigation of Spectral Confusion | Multiple Roll Angles or "Multifiler" Approach |
| Imaging-mode | Limited to astrometric mapping |

At design level the ENIS consortium kept open the possibility to change the spectral coverage from 1-2 um to 0.85-1.7 um. In the context of Euclid timescale, there is a solid perspective that this could provide a significant gain in the FoM of the spectroscopic survey cosmological results (BAO and growth factor).

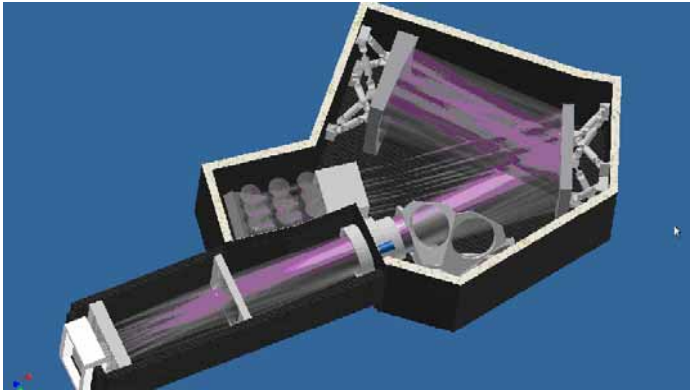

Figure 4.15: The Opto-mechanical design of the Slitless spectrograph

## 4.3.1 Optomechanical Design

NIS is based on an optical system made by a collimator forming a pupil at the dispersive element location and a camera for proper pixel-scale matching. The optical design appropriately folded and materialized in a suited mechanical configuration is shown in Figure 4.15 and the optical element layout in Fig 4.15. Mirror elements are in light-weighted SiC with JWST heritage, and the lenses derive from developments of many ground-based near-IR instruments, but with special care for radiation-hardened glass.

The dispersion for spectroscopy is obtained via a grism permanently inserted in the optical beam. The grism is designed to provide constant spectral resolution along the wavelength range of operation instead of the normally used constant dispersion.

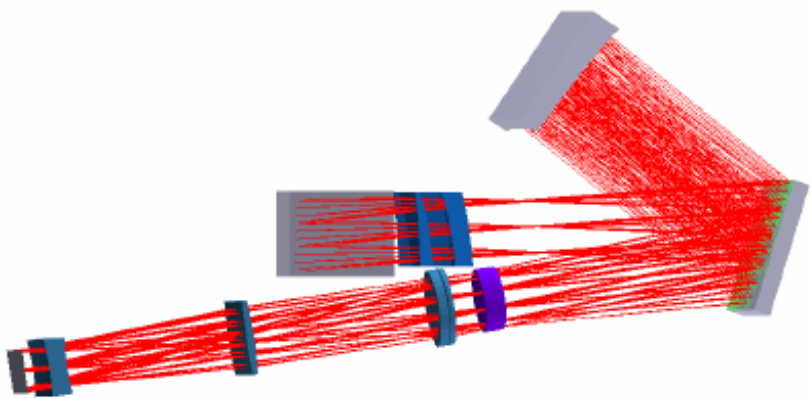

Figure 4.16: Euclid NIS optical design. Telescope beam enters from the center of the image, hits the pick-off mirror (in grey), pass through two corrector lenses (blue), then is directed to the collimator mirror M3 after passing onto a flat folding mirror. After collimation, light is sent to the disperser (violet) passing again onto the flat folding mirror. Finally light is refocused onto the NIR focal plane array by the camera optics, made of four lenses.

The disperser comprises 2 prisms (ZnSe and Infrasil) and a ruled grating (~17 l/mm). It is mounted in mechanism that allows rotation around its optical axis. This allows varying the orientation of the dispersion with respect to the focal plane coordinates, enabling the multiple roll-angles needed for spectra extraction. Close to the grism a filter wheel allows to insert a blank to collect dark measurements, a counter-dispersion grism for the astrometric mapping in imaging mode. NIS will be operated at a temperature expected between 120 and 150 K obtained and maintained via passive cooling.

The Filter wheel has been purposely designed and dimensioned with extra positions. This allows an implementation with minor impact on the design of the "multi-filter" option. In this option the dispersion grism is maintained in its position and filters to reduce the spectral range per exposure are accommodated on the filter wheel. A Scientific performance evaluation for the two cases is on-going. The current preliminary design can easily be oriented to one or the other solution.

### 4.3.2 Detector system

The Focal Plane of NIS is a single unit accommodating 8 chips (2048×2048 H2RG chips) closely packed and mutually aligned, mounted on a common cold reference plate, made of SiC. It is also connected via a flexi-cable to the proximity electronic ASICS–SIDECAR board. The detector housing is shown in Figure 4.17.

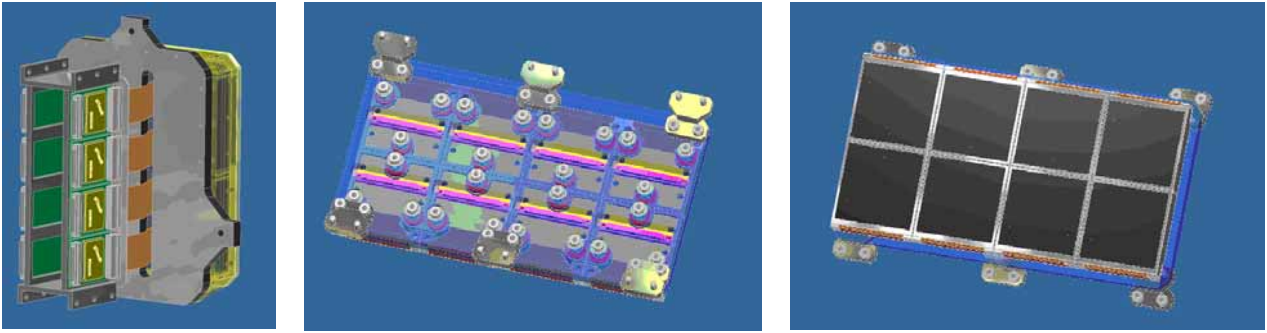

*Figure 4.17: Details of the Focal Plane Design. The eight chips forming the E-NIS focal plane are arranged in order to minimize the gap between chips. An unavoidable residual gap of about 2.9 mm (about 80 arcsec) remains and has been taken into account in the performance evaluation.*

### 4.3.3 Thermal Architecture

The thermal architecture of NIS is schematically represented in figure 4-18. NIS will be enclosed in a shielded environment thermalized to about 150 K, passively obtained via a radiator looking to the cold sky. The detector system needs to be operated at ~90K with very high stability in the timescale of an exposure. This is achieved via a dedicated passive radiator and an active controlled heater.

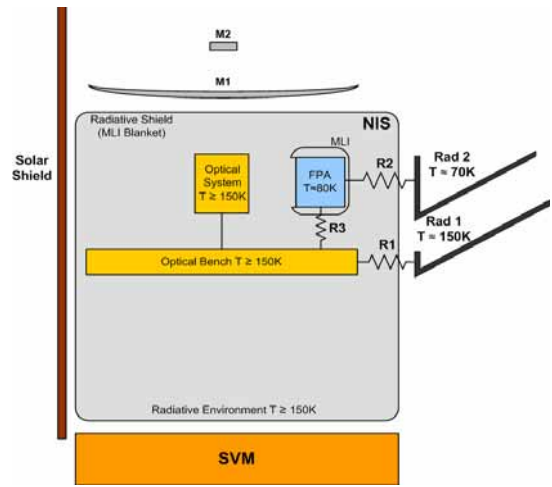

*Figure 4-18: Thermal architecture of the E-NIS Spectrograph.*

### 4.3.4 Electronics Architecture

The electronic block diagram for the spectrograph is sketched in the following Figure 4.19.

The focal plane assembly is segmented in four groups, each one having two H2RG detectors, two SIDECAR read-out ASICs and one I/O FPGA that handles the communication with the Data Acquisition and Processing Unit (DAPU). Scientific data are sent to the DAPU, where they are multiplexed and stored in the unit's memory, waiting for the CPU to run the deglitching preprocessing algorithm. The detectors are read-out using "up-to-the-ramp" integration. Each pixel is sampled at 16 bit and read-out continuously to allow the ramp reconstruction. The deglitching preprocessing algorithm analyses each pixel to determine if it has been hit by cosmic events. If no cosmic hits have been recorded the net integrated charge value is sent to the ICU. For pixels that have been hit by cosmic rays, all intermediate readings are analysed to preclude the outliers. In the ICU, science data coming from the DAPU are compressed by a lossless algorithm and packetized with CCSDS format. Finally data are sent to the S/C mass memory for storage and transmission to ground.

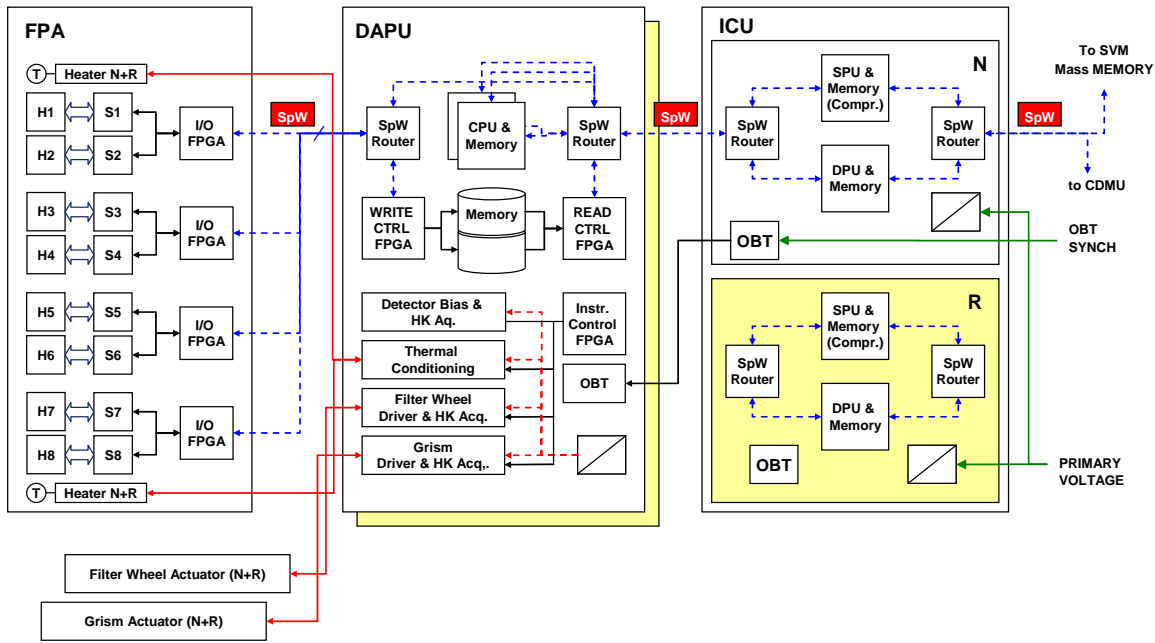

*Figure 4.19: NIS electronic layout*

In order to minimize the interconnections between the different units and implement the required redundancies, the adopted communication standard is SpaceWire, that allows high throughput rates and for which routing solutions are already available following space standards.

## 4.3.5  Summary of operations

The NIS spectrograph is characterized by one single main observing mode: the acquisition of a slitless spectroscopic image of the monitored field. However, slitless data reduction techniques require that each field is observed at a different orientation of the dispersion with respect to the image coordinates (roll angles) to disentangle confused spectra. Each spectroscopic image is then the association of 4 frames collected at different roll angles.

A possible alternative is to introduce passband filters in order to reduce the length of the spectrum and hence the confusion. This possibility, having negligible impact on the mission architecture, will be possibly studied in the further phases.

Following the need of dithering of the imagers, NIS will synchronize each roll angle with a dither. As a consequence each of the 4 individual frames collected, maps the targeted sky shifted by a dither step and at the given roll angle. In order to make the spectra post processing possible, each step must be reconstructed at a sub-pixel level. This is obtained via reducing relative pointing error in the spacecraft and via astrometric cross-mapping between the NIS and the NIP fields. This requires an additional auxiliary imaging observing mode. Therefore NIS will acquire (at most) a frame in imaging mode every pointing, before the implementation of the dithering sequence. This image will be used to re-construct the astrometric mapping between NIS and NIP by recognition of a number of bright sources. The astrometric matrix will be used in data reduction to achieve the sub-pixel precision needed for spectra extraction.

Performing frequent NIS imaging with a short integration time (tens of seconds) down to H<19-20 ("open exposure") is considered an interesting possibility, provided that it does not impact significantly on the "spectroscopic" integration time and S/N and it does not impact the flight hardware design.

### 4.3.6 Mass and power budget

The overall mass budget (including 20% margin) is given in Table 4.5. The allocation of functionality, and the definition of structural elements between fixations and supports also varies in all 3 solutions (consortium and 2 industries) and reflects the wide differences between the columns. The budget is hostage to detailed mass allocation for structures that can change with integration approach, but here is reported on worst case optical bench estimates.

*Table 4.5: Euclid Spectrometer Channels mass budget*

| System | Sub-system | Mass (incl. 20% margin) |
|--------|-----------|------------------|
| *NIS* | | *119 kg* |
| | Detectors, ASIC +harness | 8 |
| | Optics | 41 |
| | ICU+harness | 15 |
| | DAPU | 20 |
| | Thermal | 6 |
| | Structure | 29 |

*Table 4.6: Euclid Spectrometer Channels power budget*

| System | Sub-system | Power (incl 20% margin) |
|--------|-----------|------------------|
| *NIS* | | *83 W* |
| | Detectors, ASIC +harness | 5 |
| | DAPU | 46 |
| | ICU | 32 |

### 4.3.7 Critical items

Zero wavelength calibration requires precise (post facto) knowledge of target location. This can be achieved with a single full wavelength image exposure before spectra are accumulated, and/or with reference to the equivalent field observed (non-contemporaneously) with NIP. The cross registration therefore demands very high geometrical stability (<10 microns) over a number of pointings/dithers.

An allocation in the photometric budget assumes no more than 20% of zodiacal light background as a straylight component, but this depends critically on the assumption of detector response uniformity. Consequently the design for internal thermal background can be critical, and a more detailed trade-off for detector and optics temperature, cold stop provision etc. needs to be elaborated.

The proposed optical design is refractive and is based on glasses that have never been characterized at the temperature regime envisaged. Characterization of their cryogenic behaviour is mandatory in the early breadbording phase, and the supply of representative FM batch radiation hard glasses procured as Long Lead Item.

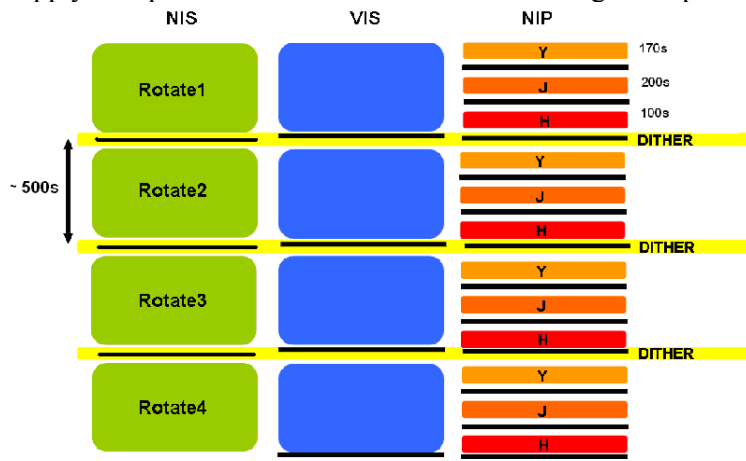

*Figure 4.20: Euclid observation sequence.*

## 4.4 Observation sequence

The instrument main observation sequence is composed of 4 dithering steps on the same fields. During each steps, there is 1 frame in the VIS, 3 frames in the NIP (1 in each band) and one rotation position of the NIS grism (or filter in the multi-filters configuration). This scheme implies the NIP FWA is rotating while the VIS and NIS are integrating. Between each frame, the shutter is closed, and dithering steps of ~100" are achieved at satellite level. At the end of the observation sequence, the satellite slews to the following field.

This sequence described in Figure 4.20, with approximately 500s observation time per dither step, and after accommodating dither step slews (<60s) and frame step slews (<230s) allows the observation of 36 fields per day. Different assumptions in the radiometric models lead to different analyses of a field duration. Comparison of the data suggests 500s is a comfortable functioning point for overall duty cycle calculations.

The associated instrument daily telemetry rates are given in Table 4.6. These calculations assume that the data processing for the NIR instruments with Up-the-Ramp sampling include recognition of, and correction for cosmic ray events.

*Table 4.6: Euclid Imaging Channels daily telemetry rate*

| | VIS | NIP | NIS | Total |
|---|---|---|---|---|
| # Detectors | 36 | 18 | 8 | |
| Pixels/detector | 16M | 4M | 4M | |
| Bits/pixel | 16 | 16 | 16 | |
| Raw Data/frame | 9.7 Gbit | 1.2 Gbit | 0.5 Gbit | |
| Frames/field | 4 | 12 | 5 | |
| Fields/day | 36 | 36 | 36 | |
| Raw data/day | 1397 Gbit | 522 Gbit | 97 Gbit | 2016 Gbit |
| Compression factor | 2.8 | 2.5 | 1.5 | |
| Data / day | 499 Gbit | 209 Gbit | 65 Gbit | 773 Gbit |
| Margin/OH | | | | 10% |
| **TOTAL** | | | | **849 G bit /day** |

notes:
[1]: 2.8 compression ratio for VIS lossless compression (in PDHU)
[2]: 2.5 compression ratio for NIP lossless compression (in PDHU)
[3]: 2 compression ratio for NIP lossless compression (in PDHU)
Note: 3% overhead is considered after CCSDS formatting.

## 4.5  Radiometric performance

Radiometric performance was estimated using standard parameters such as 97% mirror reflectivity, lens transmission, Hawaii array QE data, 75% grism efficiency etc.

*Table 4.7: Summary of radiometric performance with baseline design*

| | VIS | NIP | NIS |
|---|---|---|---|
| Plate Scale | 0.1 arcsec | 0.3 arcsec | R=500 in 2 pixels |
| Magnitude (AB) | 24.5 | 24 | $5 \ 10^{-16}$ erg cm$^{-2}$ s$^{-1}$ (AB mag 19.1 mag) |
| SNR | 14.3 | 7.1 | 5 (/spectral element) |
| Radiometric aperture | 1.3 arcsec | 0.5 arcsec | 3×5 pixels |
| Sky background | 22.3 | 22.1 (J) | $10^{(-17.75-0.73(\lambda-0.61))}$ erg cm$^{-2}$ s$^{-1}$Å arcsec$^{-2}$ |
| Overlapping frames to reach SNR | 3 out of 4 | 3 out of 4 | 4 |
| Frame duration | 500 s | 170 / 200 / 100 s Y / J / H | 480 s |

*Table 4.8: Summary of field coverage for gaps by baseline dither strategy*

| Frames | VIS (%) | NIP (%) | NIS (%) |
|---|---|---|---|
| 1 | 2.6 | 4 | 0.4 |
| 2 | 1.3 | 1.5 | 8 |
| 3 | 54 | 47 | 39 |
| 4 | 41 | 47 | 53 |
| >3 | 96 | 94 | TBC (depends on loss due to rotation of spectra) |

## 4.6  Field of View gap-filling evaluation

As the dithering is performed at spacecraft level, NIP imposes the size of the dithering step since it has the largest gap compared to VIS. VIS and NIP have requirements addressing the coverage by 3 or by 4 dither frames. Considering a global optimisation of dither pattern for all instruments together, we only consider the requirement on 3 frames, since the SNR is calculated with 3 frames. The proposed sequence of offsets in $x$ and $y$ (long and short focal plane dimensions) is (0,0 ; 100,+40 ; +200, +40 ; +300,+40, arcsec). In this case the estimated coverage of each sky pixel by number of dither frames is listed in Table 4.8.

From Table 4.8 the requirement for completing the gap coverage of 95% is marginally met. In addition the losses of the NIS spectrograph must be addressed more carefully to account for the complete spectrum length (~600 pixels on a 2048 pixel wide detector) that itself rotates between dither frames

## 4.7 DMD slit spectrograph (option)

The DMD spectrograph option description here is focused on performance and feasibility issues. This design has not been reviewed by industry, and budgets, performance and requirements therefore not verified as part of the system assessment. Multi-object spectroscopy (MOS) with multi-slits is the best approach to eliminate the problem of spectral confusion, to optimize the quality and the S/N ratio of the spectra, to reach fainter limiting fluxes and to maximize the scientific return both in cosmology and in legacy science.

The design uses micro-mirror arrays (MMA). A dedicated programmable multi-slit mask cannot be developed for the Euclid timescale and a commercially available component has been identified: the DMD Cinema chip from Texas Instruments to get more than 2 million independent mirrors in a 2048×1080 "pixels" format, with a pitch of 13.68µm (see Figure 4.23). The nominal DMD operational parameters are room temperature, atmospheric pressure and mirrors tilting several hundreds times in a second, while for Euclid, the device might work in vacuum, at low temperature, and each MOS exposure lasts approximately 500 s with mirrors frozen in one state (either ON or OFF) during that duration. ESA has engaged with Visitech and LAM a space evaluation of a DMD chip. Specific test facilities have been developed for this evaluation. Imaging capability for resolving each micro-mirror has also been developed for determining any failure for a single mirror. A dedicated electronics and software permit to freeze any pattern on the device for duration as long as 1500s. Tests in vacuum at low temperature, radiations, vibrations, thermal cycling, and preliminary life tests are under way. No show stoppers have been identified concerning the ability of the DMD chip to fulfill Euclid requirements. MOS-like tests on a specific optical bench are also scheduled[4].

The spectrograph fore-optics must accommodate the micro-mirror tilt along the diagonal of each mirror. The spectrograph must provide the required spectral resolution, i.e. between 200 and 400. The ENIS Consortium has studied a pool of solutions, reflective, refractive and mixed. Among these, the number of arms, of DMDs, of detectors has been varied according to performance and sky coverage. A 4-arms 8 detectors solution has been selected to undergo a deeper analysis for this study. For geometrical reasons (input-output beam respective locations) as well as contrast requirement, a beam of F/3 on the DMD has been chosen. This means a plate scale on the DMD of 0.77" / micro-mirror. For a single DMD, the FOV is then 0.10 deg$^2$; for covering a maximum FOV with the 8 detectors, 4 spectrographs are foreseen with 2 detectors for each one. The total FOV would be 0.4 deg$^2$. The slit size is related to the size of the astronomical objects: we can set one micro-mirror for compact objects, while two micro-mirrors will be used for more extended objects. Spectral resolution for one-mirror slit is 400 and 200 for a two-mirror slit. Other designs with different field coverage have been examined.

Four spectrographs were designed which are folded in a single plane perpendicular to the telescope axis (Figure 4.24). Each spectrograph includes a 3-mirror fore optics (different for each arm), the DMD, a 3-mirror spectrograph, a grism for the beam dispersion and the detector. The optical design is also based only on mirrors with aspherical values within standard values, commercially available from main manufacturers. The optical quality on the DMD and the detector are also reached with this design. The spot diagram is within one micro-mirror in the DMD plane (optical quality of the fore-optics), and around two detector pixels in the detector plane.

The opto-mechanical design of NIS/DMD option is based on a central optical bench with two spectrographs located each side (Figure 4.24). Four flat pick-up mirrors are sending the beams towards the spectrographs; they are located within the NIS volume and before telescope focal plane. This leads to 4 non-adjacent sub-fields in the sky providing a pattern suitable to complete sky coverage achievable with an optimization of the scanning and dithering strategy to be performed in the next phase in case DMD spectroscopy will be selected as the baseline. The bench is linked to the telescope structure, via 3 bipods. Optical bench and structures are based on honey-combs with carbon skins; optics are made with Zerodur and attached with Invar bipods. All detectors are located in the area of the instrument for a global electronics/thermal design of the focal planes. A preliminary DMD board design has been produced, based on actual boards developed for DMD testing. A specific 2-chip assembly, based on H2RG detectors, has been designed. The proximity electronics has also to be specific for driving 2 detectors.

---

[4] A Non-Disclosure Agreement is being negotiated between TI and ESA and between ESA and the Euclid Consortia

*Table 4.9: Summary of DMD spectrograph parameters*

| DMD spectrograph parameters | Value |
|---|---|
| FOV | 4 x 0.1 deg$^2$ = 0.4 deg$^2$ |
| F# on DMD | F/3 |
| Plate scale on DMD | 0.77 arcsec / micro-mirror |
| Plate scale at detector | 2 detector pixel / 1 micro-mirror, 36µm |
| Wavelength band | 0.9 to 1.7 µm |
| Spectral resolution | R = 200 – 400 |

Common interfaces for the slitless and the DMD spectrograph option have been developed for the opto-mechanics, the thermal architecture, the electronics, the data handling, most of AIT/AIV and development plan.

The DMD-spectrograph design presented here was developed to demonstrate its feasibility as an option to the current slitless baseline. In case DMD spectroscopy will be baselined, further optimizations of the survey strategy will be necessary to meet the requirement of maximum mission duration of 5 years. This may be possible by trading-off the scientific performances of the DMD spectroscopic survey with the optical design, total sky coverage, or integration time, in a way fully compatible with the VIS/NIP survey strategy.

## 4.7.1  Electronics Architecture

In the DMD option, four Focal Planes are located in each of four spectrographs, accommodating 2 chips (2048×2048 H2RG chips). Detector architecture is identical to the slitless case and the main features of the electronics are identical with the slitless design, for the FPA electronics and data processing. However, for DMD handling and configuration, specific additional electronics cards are considered, including a DMD formatter board and DMD thermal control unit. Additional computing power must be also installed for on-board processing of the DMD pattern, after a pre-imaging step.

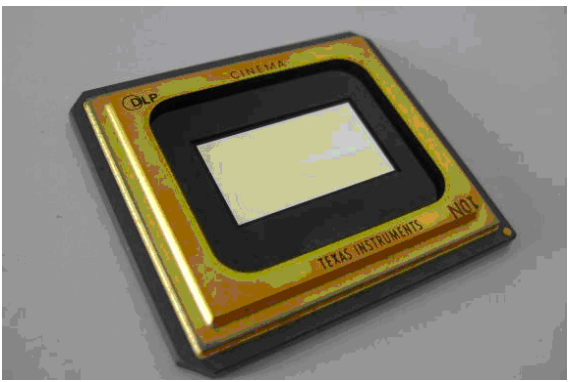

*Figure 4.23: DMD Cinema chip from Texas Instruments (2048 x 1080 micro-mirrors)*

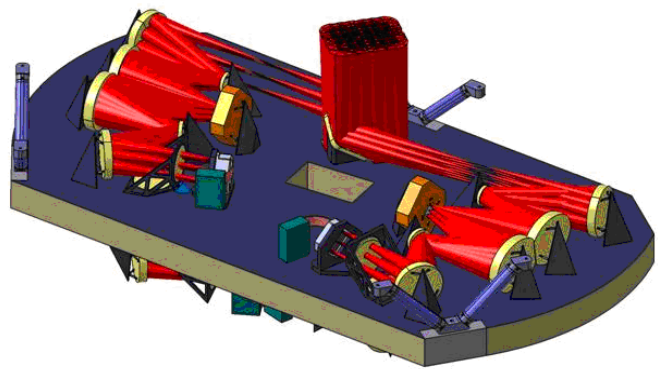

*Figure 4.24: Opto-mechanical designs of the DMD spectrograph option*

## 4.7.2  Thermal Architecture

The Thermal Architecture of the DMD option is likewise similar to the slitless case, with the exception of the need to keep the DMD subsystem at 253 K. This has been thoroughly computed and the result reported in the documenttation package.

## 4.7.4  Mass and Power budget

The current mass budget for the E-NIS DMD has been generated by using the CAD model without margin except for optical bench for which a 10% margin has been taken into account. In addition, an overall system level margin provision of 20% has been taken.

*Table 4.5: Euclid Spectrometer Channels mass budget*

| System | Sub-system | Mass (incl. 20% margin) |
|---|---|---|
| *NIS* | | *128 kg* |
| | Detectors, ASIC +harness | 15.5 |
| | Spectrograph | 41 |
| | DAPU/ICU/harness | 41 |
| | Structure | 30 |

*Table 4.6: Euclid Spectrometer Channels power budget*

| System | Sub-system | Power (incl 20% margin) |
|---|---|---|
| *NIS* | | *110 W* |
| | Detectors, ASIC +harness | 6 |
| | DAPU/ICU | 104 |

## 4.7.5 System Implications

The DMD slit spectrometer case has not been assessed during the industry study, and therefore it has not been possible to consolidate the system implications of this design, compared with the baseline. Items that would need to be addressed include:

- Complete the assessment of DMD environment and radiation qualification aspects, including also the peripheral electronics drivers. Slower than anticipated progress prevents confirmation of TRL status of the DMD at this time.
- Another aspect of TRL confirmation will be to verify optical performance in a lab/breadboard environment particularly with appropriate f-number optical stimulation.
- Verify mass and power budget assumptions, including interfaces
- Verify processing power requirements, for both target acquisition and data processing, and the impacts of any changes to data rate and compression ratios
- Verify wavelength regime and implications for different detector cut-off wavelength (This may affect thermal design, as well as detector procurement cost and yield, with a lower level of TRL than assumed for the baseline H2RG array)
- Verify accommodation requirements in terms of optical and thermal impact on PLM design
- Identify system requirements, for example on AOCS requirement for on-board Absolute Attitude Measurement Knowledge. This is considerably more complicated than the providing these data post facto.
- Verify sky scan law strategy, and ~20-30% prolongation for mission due to the slightly reduced field of view. Verify the compatibility with EIC dithering requirements

# 5    Mission Description

Two major trades were performed at the outset of the Assessment Study. The sky observation mode, i.e. the way the images are obtained during sky observation, could be either continuous scanning or step & stare. NIR detectors cannot be operated in a TDI mode, so that de-rotation is required for the NIR instruments in case of continuous scanning. The resulting optics is complex and incompatible with the available mass and volume envelope. The step & stare mode therefore was baselined. A dithering of the FoV is required to fill the gaps between the detectors and to achieve sub-pixel sampling. Actuating a mirror within the payload was found to be infeasible due to the detector geometry and the large FoV. Spacecraft dithering therefore has been baselined. Analysis of the AOCS performance shows that the impact on the duty cycle for these choices is acceptable.

## 5.1    Mission analysis

We selected a large amplitude orbit around the Sun-Earth Lagrange point 2 (SEL2) because it imposes minimum constraints on the observations and allows scanning of the sky outside a ±30 deg band around the Milky Way within the mission duration. Euclid will be launched from Kourou with a Soyuz ST 2.1-B carrying an increased tank for the Fregat. As for Gaia, Euclid is directly inserted into SEL2 from a circular parking orbit close to the equator. Transfer manoeuvres will correct for launcher dispersion and will perform the fine-targeting to a stable manifold of the free insertion libration orbit. The minimum thrust required is 1N for a manoeuvre duration less than 1day. The launch date and the launch conditions determine the geometry of the operational orbit, and influence the Sun-spacecraft-Earth angle (SSE angle) plus the daily visibility from the ground station. The launch is possible at any day of the year with minor restrictions to avoid eclipses. For a specimen reference date for launch on November 2, 2017 at 05:50:48 UT the SSE angle is limited to 30 deg and the daily visibility is at least 4 hours, considering potential interference with other missions. The total ΔV required for transfer (3 manoeuvres) and station keeping is about 50 m/s adopting 10% margin.

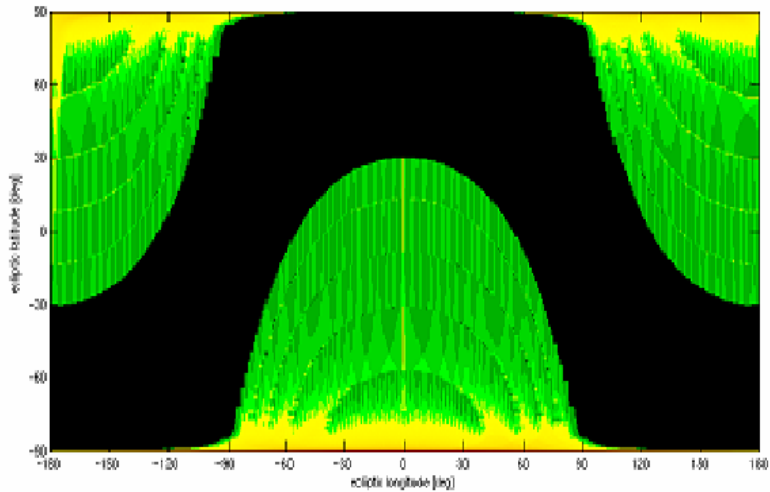

*Figure 5.1: Sky coverage, the dark region depicts the ±30 deg band containing the galactic plane, the colour coding of the coverage scales from single sky coverage (dark green) to multiple coverages (yellow) due to overlap of fields.*

For the operational orbit, the maximum Sun-S/C-Earth angle is roughly about 30 deg, and the in- and out-of-plane orbital periods are both close to 180 days. The frequency of station-keeping manoeuvres is ~30 days.

For the observations' scenario, the scanning strategy is to perform a scan in latitude ("stripe") of maximum possible extent in a given time interval, and then to place the next stripe adjacent to it. The latitude scan is a roll around the spacecraft-Sun direction with the telescope boresight at a right angle to the spacecraft-Sun direction. The design allows an angle of telescope boresight to Sun direction between 90 and 120 deg. The roll in ecliptic latitude is about 15 to 20 deg per stripe. It consists of step and stare steps ("fields") that are 1 deg wide and extend 0.5 deg in roll direction. There is a 2.5% linear overlap between the fields on each side.

Each "step and stare" period, including slew, settling time and observation takes ~2400 s. With a constrained Sun-aspect angle of the spacecraft of order 1 deg and the Earth's motion around the Sun, the average rate of the scan in ecliptic longitude is given by the Earth's mean motion divided by the FOV, i.e. after 0.9 days, an adjacent stripe is observed. After 0.5 year of scanning adjacent stripes, the spacecraft is flipped and a zone in the opposite hemisphere will be observed. This continuous scan strategy is effective, except for the ecliptic poles. These are observed with a special strategy around the equinox periods (Figure 5.1). Consolidation work is ongoing to maximize the scientific return in the context of embedding Deep Survey fields, as well as optimizing the thermal stabilization time between long slews.

*Table 5.2: Mission Summary timeline used for sizing purposes*

| Event | Date |
|-------|------|
| Launch (=L) | 01/11/2017- (assumed launch date) |
| First trajectory correction manoeuvre | **L+2** - 03/11/2017 |
| Additional trajectory control manoeuvres | **L+5** / L+20 |
| | 06/11/2017 / 21/11/2017 |
| Cover ejection | **L+23** TBC - 16/11/2017 |
| Outgassing | 1 week TBC - 17/07/2022 – 01/12/2017 |
| Commissioning | 2 months TBC - 02/12/2017 – 31/01/2018 |
| Science Operations | 4.5 years - 01/02/2018 – 31/01/2023 |
| Orbit correction manoeuvres | Every 30 days |
| End of mission | 31/07/2022 |

# 5.2 Spacecraft

## 5.2.1 AOCS

The VIS channel places stringent attitude stability requirements to ensure optimum PSF quality, especially with respect to its ellipticity. This means that the orthogonal axes of AOCS must perform equally well. A CCD pixel size projection of 0.1 arcsec demands a Relative Pointing Error (RPE) of ~0.025 arcsec (1$\sigma$) over any image accumulation time, which is 500 sec. An Absolute Pointing Error (APE) of 10 arcsec (1$\sigma$) is required to ensure that neighbouring image fields have the required overlap. An Absolute Pointing Knowledge (APK) of 0.1 arcsec (1$\sigma$) ensures that the NIS target locations can be traced accurately to the equivalent field imaged by VIS and NIP.

The pointing acquisition for each field and between dithers is attainable with a standard high accuracy star tracker (e.g. multi head Hydra). However, the pointing stability requirement demands a higher accuracy. A Fine Guidance Sensor (FGS) must be implemented, and must be located in the focal plane of VIS, as this instrument drives the pointing stability. Both industrial studies consider an FGS sensor consisting of CCDs similar to the ones in the VIS imager but operated at a higher cadence. The definition phase must address the complicated interface between SVM and payload for the FGS location, as well as the FGS procurement, AIV, and testing. Some system functionality has to be developed to combine the FGS and IMU system to ensure stable pointing in science phases and to manage transitions between small slew steps and the pointing lock.

## 5.2.2 Propulsion and actuators

Chemical propulsion is used for transfer corrections, and monthly station-keeping manoeuvers. The propellant budget is based on the manoeuvres defined in the ESA Consolidated Mission Analysis document. The total $\Delta$V budget for transfer corrections, station keeping and safe mode is ~68 ms$^{-1}$, leading to a CPS propellant budget of ~85kg, depending on efficiency and unbalanced directions for station-keeping.

Conventional reaction wheels were discounted as the actuator, because the noise budget is too high for the RPE. Cold gas micro-propulsion is the preferred option allowing attitude control, 0.5 deg slew manoeuvers and ~70" dither steps. A set of balanced 1mN thrusters are considered, (delta qualification cf. Gaia of pressure regulator may be necessary) and a budget of ~65kg nitrogen is required. The propellant tanks of standard size define the height of the spacecraft service module. If the TRL of magnetic bearing reaction wheels can be advanced in time in case of a European solution or if it can be US sourced (e.g. by Teledix), then the magnetic bearing reaction wheel solution offers mass saving with respect to cold gas. The wheel noise has to be verified for consistency with the pointing stability budget.

The cold gas option has been calculated for attitude control manoeuvres for 5 years with 100% margin, and attitude step and dither manoeuvres of 0.5 deg and 70″, respectively, including a 100% margin. This leads to an estimate of ~65 kg gas, and similar tank and thruster mass. This compares with 5 (including redundancy) magnetic reaction wheels of 90 kg total, indicating the level of additional mass margin.

## 5.2.3  Communications

For the high data rate of ~850 Gbit/day, a K-band capability is required. Due to the strong dependency of the K-band link margin with the ground station elevation and the atmospheric propagation conditions, the expected link performance requires application of a statistical model, which was not done during this assessment. We therefore assumed the following parameters for the calculation of the Euclid link budget: 95% availability (calculated on yearly average conditions); Elevation range of 20 deg; and a 4h/day maximum pass duration. An ESA 35m ground station antenna is assumed. A two degrees of freedom steerable 40 cm dish on the spacecraft is required for ~65 Mbit/s with 35 W RF power (1.1 dB pointing loss are allocated in link budget for ±0.7 deg mispointing). New designs based on existing technology will be required for a dual X/K feed transponder. System operations for automated retransmission protocols or large transmission margins may be needed to accommodate extreme weather losses. Prototype K-band receiving equipment is already being developed and operational scenarios for K-band ground stations are to be studied. Euclid is forseen to be the first science mission to require K band capability at SEL2, and should bear the cost of these station upgrades.

## 5.2.4  Data handling

Data handling for the detector systems can be located within the relevant payload sections. However, the SVM equipment bay has sufficient volume to house payload-related data handling systems, and could be used as a centralised unit. A key driver for the data handling sub-system (PDHU) is the analysis of the non-destructive readouts of the NIR detector arrays and the provision of processing power for the "follow-up-the-ramp" (FUR) sampling, necessary for noise reduction and cosmic ray glitch removal. In addition, the PDHU could provide data-compression for the science instruments, and windowing/centroiding for the FGS. Preliminary estimates for the FUR sampling are 20Mflops for NIS, compared with existing LEON2FT 100MIPS/25MFLOPS. Alternatively, the Maxwell 750 PowerPC processor is ~10 time faster and provides more margin. This device would be required for target selection in the case of a DMD spectrograph. NIP has ~2.5 times more pixels than NIS implying a similar factor increase in required processing power. It is to be inversigated if the NIP noise performance can relax the FUR sampling requirement. A detailed trade-off for reducing mass by using only 1 central PDHU, versus redundancy and bus complexity has resulted in different architecture choices between the studies, but gave no major technical weaknesses.

The possibilities for data compression have been studied for Euclid. The driving instrument (VIS) has been assessed by using standard RICE compression algorithms on a set of simulated images with a range of noise levels, background zodiacal light levels and cosmic ray frequencies. The average compression factor achieved was ~2.8. Similar exercises were performed for the NIP and NIS data, where less efficient compression factors (2.5 and 1.5) were achieved. We assume for this assessment study that the SVM central data handling will realize these compression factors on the instrument-generated data.

Assuming a data volume of 850 Gbit/day and 3 days of storage capability, the mass memory size should be ~2.6 Tbit. With 25% margin for memory cells degradation and similar margin on overall memory size for additional working space, this gives approximately 4 Tbit of required memory. This can be covered by 3 boards of flash NAND assuming that one board provides 2 Tbit storage capacity and that one additional board covers board failures. Flash memory exhibits a higher degradation than SDRAM. NAND flash is preferred for Euclid because of its much higher storage density and lower power consumption compared to SDRAM. In addition, NAND flash is non volatile and needs no power for storage only.

## 5.2.5 Thermal

The thermal control of the instrument is needed to guarantee the VIS focal plane CCD to be at around 150 K, while the front end proximity electronics and payload Interface Module prefer to be at room temperature. The existing configuration for Gaia seems an adequate solution. An alternative solution calls for a better thermal decoupling of CCD and PEM by longer harness in order to optimise the relative sizes of CCD and PEM radiators. For the infrared focal planes, the Hawaii detector and SIDECAR ASIC operate at around 100 K, and again the IM is at room temperature. For the telescope, initial analysis shows the temperature must be stable to few mK over 500 s in order to meet the ellipticity stability specification (0.02% over 500 s). This partly depends on ultimate choice of mirror and bench materials. Purely passive or active control have been considered, the latter requiring some 50-100W additional power. The PLM can be designed as an isothermal cavity, and arranged for no thermal background to impact onto the NIR focal planes. Alternatively, if the cavity is at a more elevated temperature (150-200K) the last optical element must be cooled and the FPA view baffled to a lower temperature. More detailed trade studies must be carried out to explore these options.

As arranged for other L2 missions (Herschel, Planck) the SVM has to be thermally de-coupled from the PLM with low-conductive mounts and by multi-layer insulation, either to establish different temperature levels at PLM and SVM, and to obtain a significantly better temperature stability at PLM level compared to SVM level. This sunshield protects the payload module from any incident sunlight. The sunshield, the thermal baffles, and the PLM telescope itself have to be designed such that the required PLM temperatures are met by passive means.

*Table 5.3: Estimated Mass Budget*

| Item | PLM including maturity margins (kg) | Item | SVM including maturity margins (kg) |
|---|---|---|---|
| Structure | 185 | AOCS | 83 |
| Telescope parts and I/F | 115 | TT&C | 27 |
| Thermal | 50 | Data Management | 24 |
| VIS + NIP | 361 | Harness | 60 |
| NIS | 119 | Power | 48 |
| Harness | 25 | Mechanisms | 11 |
| | | Thermal | 28 |
| | | Propulsion | 33 |
| | | Structure | 377 |
| **PLM Total** | **855** | **SVM Total** | **691** |
| **Spacecraft Item** | | **Mass (Kg)** | |
| Dry Mass | | 1546 | |
| System Margin 20% | | 309 | |
| Propellant | | 150 | |
| Adapter | | 100 | |
| Estimated Euclid Launch mass | | 2105 kg | |
| Launcher capability | | 2150 kg | |

## 5.2.6 Power

The solar array (GaAs triple junction) consists of body mounted panels. For an end of life power at 100 deg C, the budget of ~1200W requires ~8m$^2$. The battery sizing is not driven by eclipses, because an eclipse-free mission design is feasible during transfer as well as during operational orbit. The sun-facing SVM side is can accommodate an 8 m$^2$ array, leaving margin via populating the less efficient neighbouring and canted side.

## 5.2.7 Overall configuration and budgets

To reach a mechanical support for an optical payload module the SVM primary structure is realized in a shape of a Central Cylinder which enables a load path from the actual launcher interface of 1194 mm to the interface points of the payload with an increased diameter of around 1700 mm. Around the Central Cylinder there are 6 to 8 Equipment Panels radially and equidistantly arranged over circumference. They serve for the mechanical interface to the PLM instrument as well as stiffener for the top and bottom planes. These planes have an outer diameter of ~3.1m. The top plane carries also the sunshield, which protects the PLM instruments from direct Sun light and thermal loads. The primary structure is constructed from stiff Aluminum honeycomb panels with CFRP/Aluminum face-skins to keep it light and to avoid thermo-elastic distortions.

The sunshield is combined with the solar array and shields the PLM from solar irradiance, by ±5 deg in spacecraft azimuth, and +5 to -45 deg in spacecraft elevation. The PLM is covered by a thermal baffle which is highly reflective for reflecting off the residual thermal radiation from the sunshield backside and cooling down the telescope and instruments passively. The shape of the sunshield is adapted from Herschel. Its width is limited by the Soyuz fairing, the height of the sunshield (5000 mm) is driven by the required solar array size of 9 m$^2$ and by the PLM height. A detailed analysis of sunshield surface properties may be needed to quantify thermal stability properties as function of Solar Aspect Angle. A synthesis of both industries' designs is presented in Table 5.3 and Table5.4.

*Table 5.4: Estimated Power Budgets. Tthe instrument budgets include local power converter efficiencies. The solar array is sized for 1560 W at 30° Solar angle, 100°C at EOL = 9m².*

| Unit | Nominal Power (W) |
|---|---|
| VIS + NIP | 267 |
| NIS | 83 |
| Payload Total | 350 |
| AOCS | 95 |
| TT&C (Peak) | 220 |
| DHS/Memory | 55 |
| Propulsion | 10 |
| Power System | 65 |
| Thermal | 150 |
| ***Total (Peak)*** | ***945*** |
| Harness loss (3%) | 28 |
| PCDU loss (3%) | 30 |
| System Margin (20%) | 201 |
| **Total Power** | **1204** |

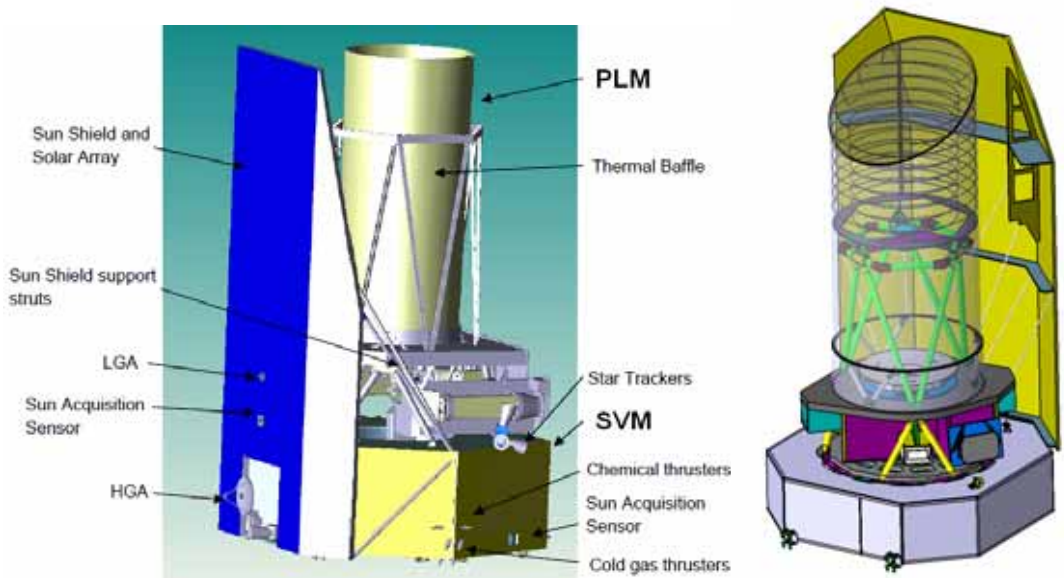

*Figure 5.2 Comparison of EADS (left) and: Thales (right) concept of Euclid Spacecraft*

## 5.3   Payload interfaces

Eventually a contractor should be identified for the payload module and be responsible for interfaces. He shall naturally also be responsible for the telescope. All optical elements up to the instrument interfaces shall be part of that telescope development. The telescope should consist of all items needed to validate the performance and alignment. Therefore the optical bench, dichroic mirror and the tip/tilt mechanism could be part of the telescope.

An Instrument Control Unit (ICU) can be the interface of each instrument to the SVM. Separate instrument DHUs will process FURs data from NIP and NIS. This allows adequate processing power depending on the capabilities of standard units. A centralized approach has been proposed for both DHUs to minimize mass, although increasing interface and AIV complexity significantly. This approach will be further studied in the early Definition Phase and compared with the approach envisaged by the instrument consortia.

The optical interface to VIS is at the Korsch image plane. Essentially the VIS instrument consists of the focal plane, which also includes the FGS CCDs (incl. the associated electronics) and a possible wavefront sensor. The interface of FGS to the spacecraft AOCS sub-system will complicate the VIS focal plane, as well as a potential EMC coupling between sensors operated at different cadence. The conflicting procurement schedules could drive the schedule for adequate testing with appropriate PLM and SVM components. The optical interface to NIP may be the intermediate pupil of the Korsch, where the dichroic mirror is located. The interface to NIS could be the afocal pupil where the grism is located. The grism should be part of NIS to provide full performance verification of the instrument.

Optical interfaces must be validated at cold temperature: The image plane for VIS can be validated through WFE measurements, the pupils for NIP and NIS by geometrical properties and WFE measurements.

We assume that the service module and the payload module will be functionally verified separately and then integrated during the spacecraft AIV campaign. Environmental tests shall be performed for the telescope and each instrument separately. Relying only on spacecraft level environment tests may decrease the costs but unnecessarily increases the risks. This parallel development of instruments, telescope and service module naturally implies some risk of interface problems and this must be mitigated by strict interface control and by early verification of all electrical interfaces with the aid of development models.

## 5.4   AIV and Development Issues

The main development drivers include the schedule, constrained by the 2012 implementation start and ~2018 launch period; the procurement of the payload sensors, and the demanding payload mechanical-thermal stability and operational temperature requirements, which have implications for the AIT programme. A system model philosophy has been proposed, consisting of AVM and PFM models, which is driven by the schedule and cost ceiling for M class missions. For the SVM this is considered acceptable by industry, on account of the test heritage from GAIA/Herschel. The Avionics Model (AVM) or "flat-sat" is necessary for an early test of the electrical interfaces and software between sub-systems and with the warm electronics of the instruments. The SVM (Service Module) and PLM (Payload Module) are well-separated thermally and mechanically, justifying the lack of STM (Structural Thermal) model of the SVM.

The integration of the space segment comprising PLM and SVM shall be under the prime contractor's responsibility. He is also responsible for all tests for verification of the full system functionality under environmental conditions. These tests will include end-to-end tests on the functional chains. To satisfy all procurement options currently envisaged by ESA, the instruments are assumed to be provided by consortia, and therefore are treated separately for AIV and development. This requires a dedicated overall PLM development approach which is compatible with the overall development requirements and which coordinates the development and AIV activities shared between the involved parties.

Technology development activities are proposed already in early phases. These include the optical design for validation of lens support structure and validation of the alignment procedure, detector test campaigns to validate the performance in specific Euclid modes; proximity electronics module (PEM) and interface module (IM or PCU) breadboard development where they are coupled to the detector allows to acquire knowledge of the video chain to improve and refine their specifications; focal plane mechanical and thermal design to validate the detector support structure, thermal shield design, validation of the capability to have detectors at cold temperature with low gradients and electronics at room temperature, and the validation of the alignment procedures. The general AIV flow is expedited by establishing clear interfaces, sub-assemblies and assemblies on various levels.

## 5.4.1  System Level

It is assumed that the service module and the payload module are functionally verified separately and then integrated during the spacecraft AIV campaign. An avionics model (AVM) supports the verification of interfaces with the instruments. The environmental tests are performed for the telescope and each instrument separately. To consider environmental tests only at spacecraft level decreases the costs but is not acceptable as it increases the risks. The parallel development of instruments, telescope and service module naturally implies the risk of interface problems.

The Avionic Verification Model allows early proof of interfaces and for software development. Its purposes include:

• Checking electrical and functional interfaces between the units.
• Verifying the functionality of the avionics subsystems and on-board software including closed loop tests for AOCS functional verifications
• Validation of the On-Board Control Procedures.
• Validation of communication and power interfaces between the payload instrument warm units and the CDMS and EPS subsystems.
• Testing the GSE / SVM interfaces including the EGSE software and verify the EGSE capability to perform the planned tests.
• Validation of the test sequence to be re-used for the PFM/FM test campaign
• EMC Conducted tests on avionics units.

A deliverable form of an instrument Development Model is suitable for the testing of interfaces with AVM

Thermal control may be fully validated on the satellite PFM, where the PFM test sequence duration might allow for detailed thermal control adjustments, as special attention needs to be paid to thermal stability, due to the very stringent performance requirements. Generally the purposes of the PFM include:

• thermal qualification
• mechanical qualification
• functional / performance qualification
• Verification of alignments
• Completion of EMC qualification

## 5.4.2  Payload Level

The identified PLM schedule and cost drivers are:

o Primary mirror: availability of this key element of the telescope starts the AIT sequence.
o Instruments: the entire PLM AIT schedule depends on the schedule of each instrument. Slight delay of one instrument can be compensated if another instrument is delivered earlier, assuming the PLM design allows integration to be made whatever the order. Slippages longer than 1Q will likely have a direct impact on the PLM schedule.

A PFM approach for the PLM is dictated by schedule constraints, but to achieve qualification for the instruments they should be subject to a more full programme including development / engineering-qualification / (proto)flight models, subject to schedule constraints of an early start on DM before mission selection to implementation. For the Telescope, we consider good heritage from previous programs, and the availability of analyses of the thermal environments from Herschel/GAIA allowing us to model reliably the Euclid thermal environment. The PFM programme must also include an extended thermal test of the PLM and telescope.

The following payload modules models may be envisaged for instruments:

• **Development Model**: all focal planes (to verify functional performance and interfaces for data processing, characterize cryogenic performance alone and/or with DM of cryo-optics); for optical sub-systems characterize filters, dichroics, lenses at ambient and cryo-conditions, alignment concept demonstration, PSF enlargement, filter wheel and shutter mechanisms. The corresponding TDAs occur in parallel with DM-level activities at the scientific institutes. At the end of the TDAs, without loss of continuity, the DM activity continues, in parallel with the ITT process, under the responsibi-lity of ESA as regards the interfaces. Some time after the award of the Implementation Phase contract, during the satellite Phase B2, the responsibility of the instrument interfaces is transferred to the satellite contractor. The warm electronics DM can be delivered to be used with the AVM

- **Structural Model**: should be used for telescope and the focal planes structural and thermal design verification. However no specific PLM STM is necessarily needed: The lower cavity will experience thermal tests at an early stage of the PLM integration. If the baseline PLM design is made very modular this allows parallelisation of instruments development until the delivery to PLM. Then each instrument can be integrated separately on the PLM, with no functional constraints. To ease the schedule, ctrictal thermal and structural tests could be demonstrated on the EQM, whose procurement could then begin immediately after DM.

- **Engineering Qual Model**: For VIS a reduced set of CCD/PEM chains, perform electrical command-/control validation, and EGSE to simulate remaining PEMs. For the NIR instruments the EQM should also include optical system, delivered optical bench, mechanisms, together with instrument electronics that may be sited in the SVM, mechanisms (wheels, shutter, etc), PCU, Interface Units

- **Prototype Flight Model** : all

Due to the issue of cryo-optics, extended AIV activities are expected, in particular w.r.t. Thermal/Vacuum testing**.** The optical system can be aligned initially at ambient conditions. The optical mounting has to consider the CTE of the optical elements as well of the mounts and optical bench w.r.t. to shrinkage and deformation from ambient to the operating temperature to be chosen within ~150 - 190K, including tolerancing within a range a few degrees of nominal. The deformations due to the large temperature range from manufacturing and alignment temperature (ambient) to operation temperature might have an impact on the optical performance.

The autonomous sequencing of the observation scenario and care over EMC performance (low noise and cross-coupling of sensors in different cadences) has to be demonstrated at cryogenic conditions. Involvement of the instrument consortia in this complex testing should be mandatory to ensure their EGSE is also deployed as a "pull-through" for the eventual data analysis tasks.

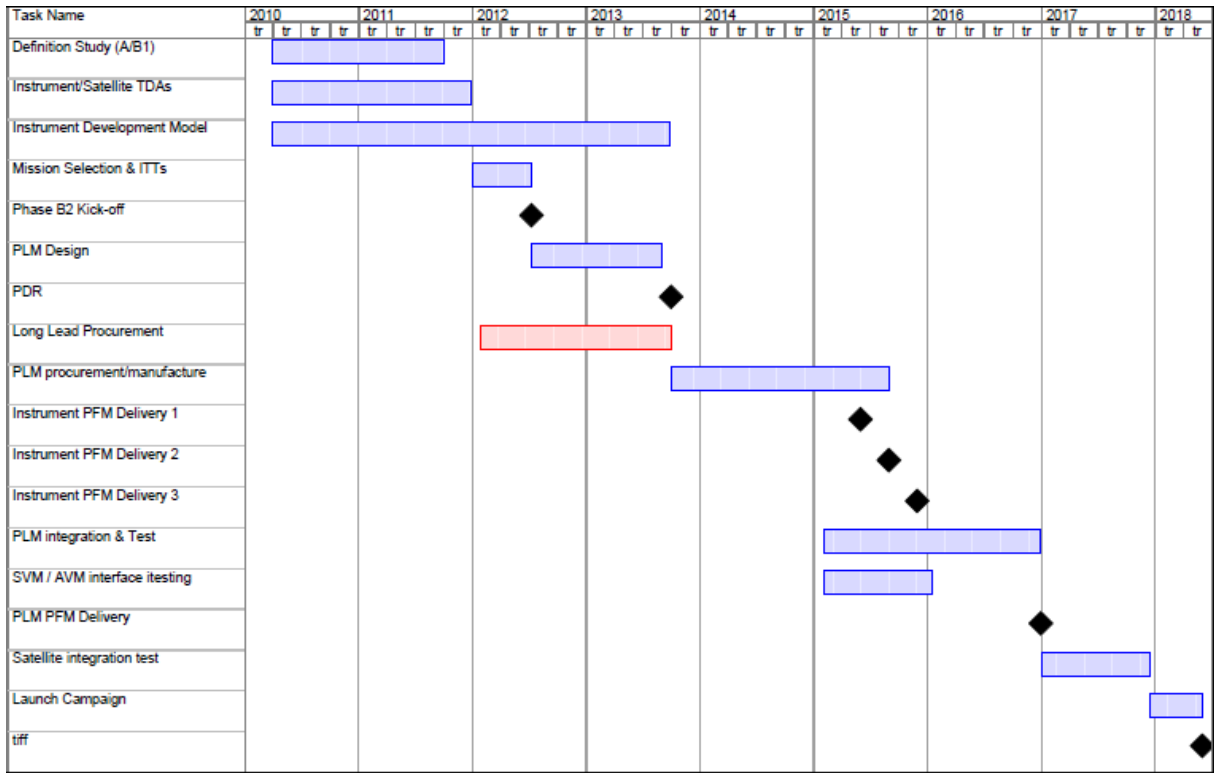

*Figure 5.3: Preliminary schedule for PLM*

# 6 Ground Segment and Data Handling

Euclid will deliver an unprecedented large volume of data for astronomical space missions: e.g. about 4 times more down linked data than Gaia. Relying on long-standing and in-depth experience gained from large ground based and space based all-sky surveys, the instrument consortia in collaboration with ESA can present a feasible framework for the end-to-end handling of the Euclid data. This section addresses the data stream in an overall coordinated fashion, ensuring a most cost-effective and efficient approach. The data stream will be handled by the Ground Segment (GS), which will be designed, implemented and supported jointly by ESA and the instrument consortia. The GS in turn will be propelled by a data handling system (DHS), which amongst other things will maintain all administration of data at various stages of processing, data products and quality controls.

Dedicated teams from the instrument consortia will process the data from their instruments during all phases of the mission. The instrument consortia are expected to provide the Instrument Operations Centres (IOCs), hosting the operations teams and a number of Science Data Centres (SDCs) in charge of the higher level science data products. These IOCs are responsible for the first level standard data products, which involve the data calibration, the removal of the instrumental effects, and production of mosaics and preliminary source catalogues. The SDCs are in charge of further science data processing, the creation of second and higher level data products, and the development of simulation packages to support the development and testing of the operational pipelines.

The Mission Operations Centre (MOC) operates the spacecraft and will deliver the raw scientific data to the GS. ESAC will implement the Euclid Science Operations Centre (SOC) which will act as the central node for the mission planning, will distribute the science and housekeeping telemetry to the IOCs after a first quick quality check, and will be the custodian of the Euclid Legacy Archive. The quick quality check at the SOC will directly feed back to the mission planning by means of rescheduling or re-planning. The SOC will populate and maintain the Euclid Legacy Archive (ELA) and deliver the data products to the astronomical community at large.

The logistic link between the various detached settlements of the GS (the MOC, SOC, IOCs and SDCs) is provided by the DHS. This system shall guarantee cost- effectiveness, avoiding duplications of work and tasks, a challenging requirement for the large data volume of Euclid. The dataflow rate of Euclid is high indeed, but similar to a number of currently operating and future large imaging surveys on the ground (e.g, ESO- VISTA, ESO -VST-OmegaCAM, CFHT MegaCAM) for which there is extensive experience with the data handling at various European institutes. The *datacentric* design of the DHS builds both on this ground based astronomy expertise and also on Gaia and Planck and elaborates on the following more Euclid specific issues:

- Optimal hierarchical data handling infrastructure from SOC to science communities involving quality control at each stage and capitalizing as much as possible on the experience of the European scientific community in the development of data processing systems for ESA missions (e.g. XMM, INTEGRAL, Planck, and Gaia).
- Publication of all relevant Euclid data items into a large Euclid Legacy Archive (ELA) ready for additional studies. The ELA will be propelled by a distributed Euclid Mission Archive (EMA), which is a logical rather than a physical entity, containing all quality controls and intermediate products. The ELA content will range from raw observational data to processed spectra, images and source catalogs etc. provided in a format compatible with the international Virtual Observatory and other networks; data will be accessible through public e-infrastructures and web services.
- Optimal involvement of the Consortia in the quality controls and calibration by the Instrumental Operation Centers, together with additional Science Data Centers connected in a distributed European wide network; this is made possible by exploiting the extensive experience in European collaboration networks such as Euro-VO (EU FP5, FP6 and FP7), and also the expertise on operating large surveys and particularly lensing surveys, such as in the DUAL network (EU-FP7), the Astro-WISE network (EU - FP5), and spectroscopic survey networks (ESO-GOODS, zCOSMOS, APPLES, GRAPES, RAVE, SDSS);

- Provision of the infrastructure for the Consortia and partners to exchange and share data also facilitating redundant processing as a verification of the key science results;
- Provision of a long-term (at least 10 years) and cost efficient solution for the data processing, storage, archiving and dissemination;
- Focus on the more challenging algorithms required for obtaining the specific Euclid results (many CCD imaging algorithms are common practice):
  - Handling cosmics/glitches.
  - Image decomposition of the Grism spectral data optimizing the dynamic range of the observations.
  - Determination of photometric redshifts – (photo-z).
  - PSF modeling techniques for the lensing analysis.
  - Optimal combination of multi-angle and/or multi-filter data.
  - Optimal spectral extraction.
  - Accurate spectroscopic redshift estimate.

The EMA may be distributed, but the consortia have the responsibility of providing integrity, security and the appropriate level of quality control. The EST, via the SOC, will authorize data access.

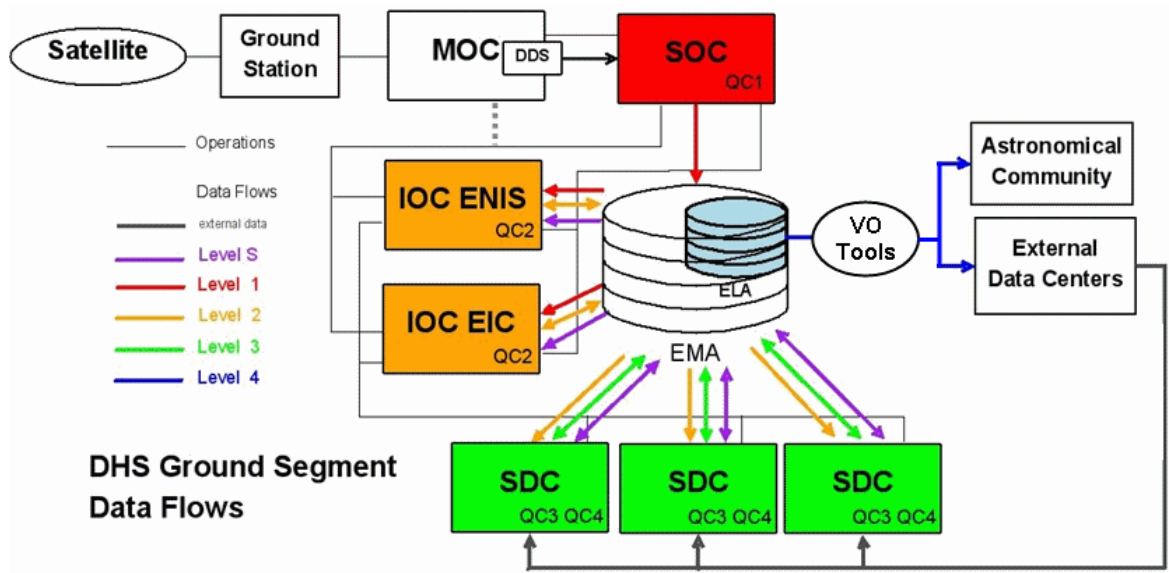

*Fig. 6.1: Envisaged Ground Segment Data Flows. External data from external astronomical data centers (CDF etc.) and the Virtual Observatory (SDSS, 2MASS and other surveys) are used by the Scientific Data Centers to produce pre-mission data and simulation data. All data flows inside the mission (between MOC, IOCs and SDCs) will go through the Euclid Mission Archive supported by the Euclid Archival System (EAS). Components of GS which generate data are highlighted by the corresponding colour, quality controls at different levels (QC1, QC2, QC3 and QC4) are shown. The dashed grey connection between MOC and the IOTs represents the possibility to receive real time data (near-real science and real time HK) to be used for non-routine activities.*

## 6.1  Data Handling System Components and Data Flows

The DHS covers all the operations of the Euclid data processing done by the Ground Segment. The inputs to the DHS are raw data (telemetry) from the satellite, the final product is provided by the Euclid Legacy Archive (ELA). The IOCs and SDCs participate in the common archival infrastructure, the Euclid Mission Archive- (EMA) and populate this with their data products. The SOC will operate the ELA, the public subset of the Euclid Mission Archive and disseminate data products to the astronomical community after a suitable proprietary period. The infrastructure provided by the DHS shall:

- provide a safe interface for the inflow of raw data,
- provide a quality control for the inflow of data and to monitor the instruments and satellite
- provide safe and robust operational interfaces for mission and instrument control,
- store, archive and provide access for dozens of Petabytes of the raw, processed and science-ready data during the mission,

- enable a full data processing chain from the raw image to the science data products,
- reprocess huge chunks of data (sometimes even all raw data) during the mission,
- integrate a number of scientists in a number of geographically remote institutes across Europe, this requirement is especially critical for IOTs and SDCs
- prepare and distribute science data products and provide access to a wide astronomical community.

The Euclid Ground Segment will consist of two blocks: mission control components, controlled by ESA, and instrument-dependent data processing and archiving components controlled by the Consortia. Mission control components are:

(a) Ground Station manned by ESA which will support a daily telemetry communications period (DTCP), expected to be 4 hours during nominal operations, and longer during the Commissioning and Performance Verification phases scheduled before the start of the nominal operations.

(b) Mission Operations Centre (MOC) managed by ESA, which will be in charge of monitoring spacecraft health and safety, monitoring instruments safety, controlling the spacecraft attitude, handling telemetry/telecommands for both spacecraft and instruments; Electrical Ground Support Equipment supports MOC at all phases.

(c) Science Operation Centre (SOC) managed by ESA, which will be in charge of planning the surveys, scheduling the spacecraft slews, scheduling the science observations, monitoring the survey performances, rescheduling, requesting MOC action via predefined procedures and sequences of telecommands. At present we adopt the configuration where the SOC will carry out the production of daily reports and the level 1 data (telemetry) ingestion into the Euclid Mission Archive with first-level quality control (QC1). The SOC will operate the Euclid Legacy Archive.

The instrument-dependent data processing and archiving components of the GS are:

(d) Instrument Operation Centers (IOCs), responsible for: maintenance of the instruments, monitoring of instruments health, instrument trend analysis, production of weekly instrument reports, instrument calibration activities, second-level quality control (on calibrated data, QC2). The IOCs are in charge of the conversion of the raw telemetry frames into science frames and the creation of the Level 2 data products (calibrated mosaics of the survey data). As well as SDCs, IOCs will be responsible for the selection of data items from Level 2 and Level 3 (science ready data products) for Level 4 (final public dissemination) after formal approval by the EST.

(e) Science Data Centres (SDCs) in charge of science data processing and of the creation of Level 3 data products and quality control of these data (QC3), of linking back to the IOCs mission critical issues, such as systematic instrumental errors propagating into the errors of cosmological parameters (QC4). As well, SDCs will provide the mission with simulated and reprocessed external data.

The data flows include:

- MOC-SOC data flow of telemetry and housekeeping via Data Distribution System (DDS)
- SOC-IOCs data flow of Level 1 data
- SDCs-IOCs data flow of Level S (simulations) data
- IOC-IOC data flow of exchange of calibrated (Level 2) data between them
- IOCs-SDCs data flow of Level 2 and Level 3 data for science data processing.

Data flows between GS elements will be specified in suitable Interface Control Documents. Reports which influence the instrument operations are delivered by the SOC (daily) and by the IOCs (weekly) and will be realized with secure and robust interfaces. The Data Distribution System (DDS) propagates telemetry and housekeeping data from MOC to SOC and is a responsibility of ESA.

Time-critical operations, i.e. the capability of identifying instrumental problems and reacting within the DTCP, will involve the MOC, SOC and IOCs. It is understood that, to ensure the proper quality of the data products, feedback from the whole GS is needed: this will be generated in both IOCs and SDCs and fed back to SOC (e.g., for replanning) through the IOCs either within the weekly reports or *ad-hoc*; see Figure 6.2

## 6.2 Euclid Data Levels and Related Processing Levels

The actual data processing is modeled into "data processing levels" and associated "data levels". Data levels represent all data produced at the corresponding data processing level including intermediate data. The criterion for the definition of a specific data level is that it can be implemented separately from the others and forms a closed and complete part of the data processing chain. The baseline is that the data processing is made in consequent steps. However, exceptions like feed back or by-pass by some special pipelines and workflows are not excluded. The list of data levels includes:

- External Data. Reprocessed data from existing missions and Ground based surveys which will be used for calibrations, photo-metric redshifts and simulations before and during the mission.
- Level S (Simulations) data. Pre-launch simulations and modeling done by the Consortia impacting on calibrations and observing strategies.
- Level 1 data. Unpacked and checked telemetry data from the satellite.
- Level 2 data. Calibrated data and intermediate data products produced during the calibrations. Calibrated data are the data with all instrumental signatures removed.
- Level 3 data. Science ready data products.
- Level 4 data. Data ready for public dissemination. These data consist of Level 3 data and part of Level 2 data and form the Euclid Legacy Archive (ELA). The EST will approve the selection of data to be included in the ELA.

Each data level has corresponding quality controls. The interface for the outputs and the quality control results at the various data levels will be implemented by a common distributed Euclid Archiving System, providing the data link between all participants. Here, we describe each data level.

**External Data and Level S (Simulations) Data**: Level S (simulations) data will be produced before the mission starts and during the mission. Simulations will be produced by the Consortia and used for instrument calibrations, covariance, model testing, science data processing and optimizing observing strategies.

The ground-based data from the Dark Energy Survey, PanSTARRS-2 Survey and other surveys will be used for calibrations, quality control tasks and scientific data reduction, specifically for settling photometric redshift techniques. Some Ground based data will have to be (partially) re-processed, e.g. for aperture matching. The mission will use raw and calibrated data of numerous surveys /observatory archives (e.g. ESO, Edinburgh WFAU, TeraPix, NOAO, specialized data centres). A number of interfaces and cross matching tools, such as the Virtual Observatory interfaces connect any component of the GS to these data.

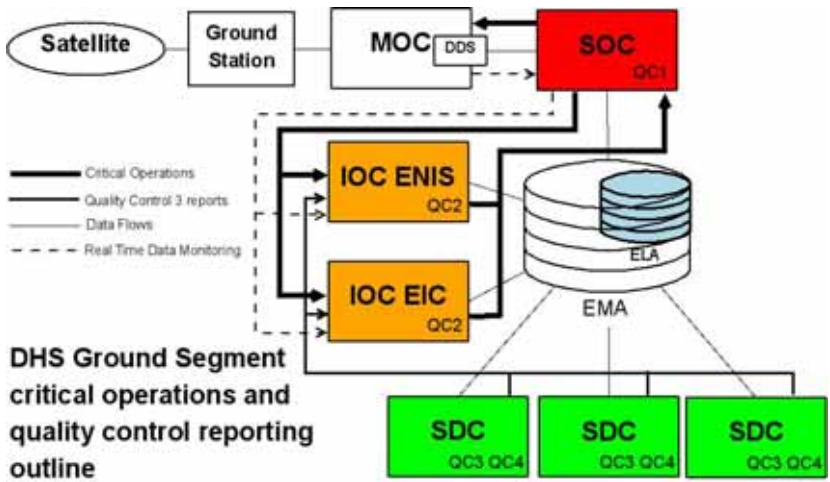

*Figure 6.2: Ground Segment critical operations and quality control reporting outline. The Figure shows components which are participating in the operations of the DHS and connections between these components. SOC performs QC1, IOCs – QC2, SDCs – QC3 and QC4. SDCs report to IOCs on the quality of scientific data reductions, IOCs deliver weekly reports to SOC. SOC reports to MOC on the satellite performance and scheduling of operations and distributes to IOCs new instrumental parameters.*

For nearly all Euclid fields deep ground-based observations will be available. Integration of external data sources into the mission archive will require a check of consistency and homogeneity. In particular, for extended objects the applied apertures will need to be matched. All ground-based imaging data will have to be homogenized with the Euclid imaging survey to make sure the same photometric zeropoints are employed. The ground based surveys currently considered are the Dark Energy Survey and the PanSTARRS-2 survey.

They will produce images and object catalogues. Several other surveys will also be crucial for calibration purposes, e.g., several VISTA and VST surveys will provide optical and infrared data.

The calibration of the photometric redshifts relies on sufficient spectroscopic redshift measurements. All current spectroscopic surveys have made the reduced spectra available. The data resides in observatory archives (e.g. ESO, SDSS, AAT) and can be accessed freely. Planned spectroscopic surveys with the VLT will significantly enhance the database for faint objects necessary for the calibration.

The simulation data volume and reprocessed external data will take 2 PB. The data model of simulation data will mimic main data flows and data processing steps of the Euclid mission, and will be close to the formats and structure of Level 1 - Level 3 data.

**Level 1 Data – Telemetry and Housekeeping**: Level 1 data are telemetry and housekeeping data down linked from the satellite. Level 1 processing will deliver decompressed, edited raw data, checked for correctness and self-consistency. The following observing strategy defines the data volume of Level 1 data.

The Euclid sky coverage strategy is driven by the wide survey requirement to cover 20 000 $deg^2$ of extragalactic sky during the mission lifetime of 5 years. A *field* (which is ~0.5×1.0 $deg^2$) is the area covered during dithered exposures. The extragalactic sky is defined by the regions with galactic longitude |b|>30 degrees. On a daily basis, Euclid will observe *strips*, the strips will be approx. 20 degrees long. On approximately monthly basis a *patch* will be completed, this is an area of about 400 $deg^2$.

The Euclid programme includes observations of dedicated deep field patches of at least 10 $deg^2$ covering a total area of 40 $deg^2$, which are 2 mag (=6.3 times) deeper than the wide survey. Since the deep field employs the same observing mode as the wide survey, the data acquisition rates are similar, thus no additional requirements for the dataflow are set.

Table 4.6 in Section 4.4 gives an overview of the estimated daily data rate by Euclid, which can be at most 850 Gbit. We assume on-board loss-less compression and we adopt conservative compression factors, taking into account realistic glitch rates. The total duration of a field observation, including overheads for slewing/pointing and instrument settings, is assumed to be 2400 s. Consequently 36 fields per day can be observed. All frames are stored as 16 bits per pixel images.

The covering of 22000 (20000 + 10% overlap) $deg^2$ with a 0.5 $deg^2$ field of view leads to 44000 fields. The assumed 2400 s duration of each frame gives about 3.4 years of observing time. For the required minimum mission life time we must add time for slewing, deep fields and calibrations plus maintenance periods. These would account for about 1 more year.

**Level 2 Data – Calibrated Data**: Level 2 data are calibrated data with all instrumental fingerprints removed. The data processing from Level 1 data to Level 2 data and Level 2 data delivery is a responsibility of the IOCs, run by the Consortia. The mission requires an extensive In Orbit Calibration Requirement Document. However, most of the spectroscopic and direct imaging calibrations, such as biasing, read-noise, dark currents and flat fielding are 'standard practice' implemented by many data reduction software (see Section 6.4.1 and 6.4.2).

In ground based surveys of similar extent, during the routine phase less than 5% of the observing time is dedicated to calibrations; this will be less in space, because the majority of the ground based calibrations actually monitor the variations in atmospheric conditions. Thus the data rate for calibration observations is expected to be less than 40 Gbit/day.

For NIS the following calibrations are essential:

(a) Astrometric mapping from the NIP to the NIS field of view. During the commissioning phase this will require offset pointings of a given astrometric field for a direct comparison of the images produced by the two instruments (NIS in imaging mode as well). During the routine phase standard observations are sufficient with presumably only bi-weekly or monthly periodicity.

(b) Wavelength calibration involves the observations of predefined fields with a number of relatively bright targets and well known redshifts with bi-weekly or monthly periodicity. The astrophysical objects that are suitable for these calibrations are compact Planetary Nebulae (i.e. with strong emission lines) and M-dwarf stars. Prior to the start of Euclid mission Gaia will provide many objects with very accurate radial velocity information.

(c) Spectro-photometric calibration is done on a few fields with relatively bright targets with known spectral energy distribution, also with bi-weekly or monthly periodicity.

For NIP and VIS the following calibrations are essential:

1. Photometric calibration is done on a few fields with relatively bright targets with known spectral energy distribution, also with bi-weekly or monthly periodicity.
2. PSF calibration across the whole instrument field-of-view involves the observations of predefined fields with a relatively high density of moderately bright stars, presumably with monthly periodicity.

The periodicity depends on both the instrument stability and reproducibility, and the availability of secondary calibrators.

Calibration data items: for the astrometric mapping each NIS detector will produce one image, while for the wavelength calibration and the spectrophotometric calibration N images will be produced, corresponding to different grism roll angles (or the different wavelength coverage if the solution adopted to reduce spectra cross-contamination should be to sub-divide the total spectral range into N sub-intervals). For the photometric and PSF calibrations each NIP and VIS detector will produce one image.

Science data items: for each survey pointing each NIS, NIP and VIS detector will produce N images, where N is the number of dithering offsets on a survey field.

**Level 3 Data – Science Ready Data:** At Level 3, the Level 2 data will be processed by a number of pipelines into science ready data – mostly catalogs to eventually obtain the various science goals of the mission. Based on the experience of optical image surveys (KIDS, PannSTARS), the data volume of Level 3 data is about four times that of Level 1 and Level 2 mostly due to intermediate images.

**Level 4 Data - Euclid Legacy Archive**: Level 4 data are in fact Level 3 data and a part of Level 2 data (calibrated images) which are authorized by the EST for the public dissemination. Level 4 data form the Euclid Legacy Archive. Level 4 data will be disseminated by SOC under responsibility of ESA as the end product of the mission.

## 6.3   Operations and Interfaces

It is assumed that ESA will support the transfer of data from the MOC to the SOC with redundant transmission lines with a bandwidth allowing the download of all daily data in 6 hours.

There are 3 types of activities which will be performed by DHS:

    a) ***operations*** – the time critical operations which involve the management of the satellite and are done by MOC, SOC and partially IOCs;

    b) ***quality control*** – at the various levels the quality controls monitor the satellite and data processing performance. Quality controls will lead to reports for operations on a daily and weekly basis. Quality controls involve time critical data processing;

    c) ***data processing and archiving*** – actual data processing operations from raw data from the satellite to the science ready end products.

**Operations:**

SOC provides to the IOCs all relevant instrument housekeeping and science telemetry, spacecraft telemetry and attitude information.

IOCs will send to the SOC weekly reports, which will include information on instrument health and performance (including trend analysis). IOCs will furthermore send to the SOC requests for changes in instrumental setup in the form of pre-defined procedures built as sequences of telecommands.

SOC will receive from MOC spacecraft telemetry and attitude information, and any other auxiliary information. SOC will send to MOC pre-defined procedures built as sequences of telecommands to plan the surveys, schedule the spacecraft slews, schedule the observations, and rescheduling;

SOC will make it possible for IOCs to monitor real time telemetry from the satellite with an ability to check response of the satellite to IOC commands

**Quality control:**

Detailed data quality control is essential for both the management of the satellite, the data processing and achieving the science goals. Quality control will be performed at the different levels of data processing. Quick simple quality controls are done by SOC, more elaborate quality controls involving full pipeline reductions and calibrations (e.g. effective tracking stability over different dithers of a single pointing field) are done by IOCs. IOCs report at least weekly to SOC. All these quality control are critical for the success of the mission.

The results of the quality control will be stored in the Euclid Mission Archive so that each participant can be supplied with quality information. *All Quality Control information is shared over all participants*; there are 4 domains, which are described in Table 6.1.

*Table 6.1: Quality control domains*

| Name | Data processing level | Description | Locations |
|------|----------------------|-------------|-----------|
| Quality Control 1 (QC1) | Level 1 | Data integrity (file level) and Quick Look Analysis | SOC |
| Quality Control 2 (QC2) | Level 2 | Instrumental fingerprints removal (instrumental calibration) | IOCs |
| Quality Control 3 (QC3) | Level 3 | Data product quality checkup | SDCs |
| Quality Control 4 (QC4) | Level 4 | Final scientific product quality checkup | SDCs |

**Data processing and archiving:**

1. MOC will propagate to SOC telemetry and housekeeping data via DDS
2. SOC will deliver to IOCs and SDCs observational data via the Mission Archive.
3. IOCs will deliver to SDCs calibrated data via the Mission Archive.
4. IOCs and SDCs will deliver to the Euclid Legacy Archive (ELA) all science data products selected by the EST for public distribution..
5. The scientific community will access ELA through VObs-compliant interfaces.

## 6.4   Applications and Data Processing

Important principles for the development of a cost-efficient and coordinated Euclid DHS are

1. A component based software engineering. This is a modular approach to software development, each module can be developed independently and wrapped in the language adopted as a standard language for the system to form a pipeline or workflow. The concept is currently in use in already working systems (Astro-WISE) and is employed by some future missions (Gaia).
2. A common data model used in the system. This means that each module, application and pipeline will deal with the unified data model for the whole cycle of the data processing from the raw data to the final data product.
3. Persistence of the data model objects (each frame in the data processing chain is described by the common data model and saved in the Euclid Mission Archive along with the parameters used for the data processing).

These principles for the development of the data processing software combined with the Mission Archive allow parallel and independent data processing on different levels of data. This facilitates cross-checking of results by a number of science groups which is crucial for the objective quality control of the data and the validation of the science results, particular the cosmological parameter values.

The scientific groups of the Consortia have a long and successful experience in managing data processing activities for both space-borne and ground based instrumentation similar to the data processing expected for Euclid. Here, we describe the critical components of the Euclid data processing. As well Euclid will provide data for a number of additional science cases including Supernovae and transient events study.

## 6.4.1  Standard Image Pipeline

A number of optical image reduction pipelines have been developed, including among others Astro-WISE for the KIDS survey (Valentijn et al., 2007), Terapix for CFHT-LS (Marmo, 2007), the IPP for PanSTARRS (Heasley 2008) and the HST-COSMOS data reduction pipeline (Leauthaud, 2007). These show that there are multiple solutions that are able to meet Euclids reduction needs.

For lensing studies the reduction involves homogeneous incorporation of external data, PSF estimation and shape measurement levels. CCD effects, for example CTE, will be corrected at the pixel level, followed by correction for bias correction and flat-fielding. The calibrated images will be stacked so that object detection can be performed on a high signal-to-noise image.

## 6.4.2  Standard Spectroscopy Pipeline and Spectroscopic Redshifts Pipeline

Both pipelines will be developed based on existing pipelines. Slitless spectroscopic data have features (frequent overlap of spectra resulting in contamination, spectral resolution dependent on object size, each pixel potentially receiving radiation of any wavelength) which require dedicated software. The reduction package aXe, originally developed for the HST Advanced Camera for Surveys (ACS) grism and prism spectroscopy (Kümmel et al. 2009), is by design instrument independent. It has been successfully applied to Near-Infrared Camera and Multi-Object Spectrometer (NICMOS) slitless data (Freudling et al., 2008) and to data obtained during the ground calibration campaigns of the Wide Field Camera 3 (Kuntschner et al., 2008). Spectral extraction is driven by an object catalogue, allowing the optimization of extraction parameters given an object shape and a quantitative estimate of the contamination. Finally, all spectral bins have an error associated, which is derived via rigorous error processing from the errors associated to the original slitless imaging data. The simulation package aXeSIM simulates slitless spectroscopic data based on the identical configuration and calibration files used by aXe for the extraction. The aXe and aXeSIM software packages have already been deployed by the Euclid team to help in defining the properties of the spectroscopy part of Euclid.

Once the 1D spectrum has been extracted, and all instrument signatures removed, a fully automated redshift measurement technique has to be applied to provide a redshift estimate for every object. This technique will be the natural extension and development of the current EZ software package (Fumana et al., 2008 and Garilli et al., 2009). currently used to obtain redshifts for . EZ has been used by the E-NIS team to assess the redshift measurement success rate during this first study phase. No additional work is required to integrate EZ in the instrument pipeline. Consortia expertise in the building of spectroscopic data reduction pipelines includes the development of VIPGI, EZ, and aXe. VIPGI is the de-facto standard data reduction pipeline for the VIMOS spectrograph at the VLT, already used in the reduction of more than 90,000 spectra obtained from the largest redshift surveys carried out with the VIMOS spectrograph at the VLT. EZ is designed to measure the redshifts of the spectra reduced with VIPGI. These two software packages have been developed at INAF - IASF Milano. High quality photometric redshifts (Section 6.4.3) will also be used in some critical cases in order to cross-check ambiguous or uncertain estimates of spectroscopic redshifts.

## 6.4.3  PSF Matching and Photometric Redshift Determination

The data handling for determining photometric redshifts involves several challenges including:

- Comparison and extension of photometric redshifts codes, eg. HyperZ (Bolzonella et al 2000), BPZ (Benitez 2000), ANNz (Collister & Lahav 2004), Zebra (Feldman et al 2008) and codes implemented for Astro-WISE and PanSTARRS (Bender et al. 2001) with the goal of producing full reliable probability distribution functions.
- In many cases, the external data sets, such as DES and PanSTARRS, will need to be reanalysed to measure the photometry of faint Euclid objects that may not be included in the object catalogues of the external data sets. Objects of low signal/noise, as even upper limits on galaxy fluxes could significantly improve the photo-z accuracy.
- No single ground based survey can cover the 20000 $\deg^2$ of Euclid, matching of these samples is crucial.
- Spectroscopic redshifts sample are also needed for calibration of the full survey.

Many tools for these tasks are already in place, for example the photo-z packages mentioned above and pipeline packages already implemented in Astro-Wise (Valentijn et al. 2006) and IPP/PSPS (Heasley 2008; Saglia 2008) for the VST and PanSTARRS-1 imaging telescopes. These software solutions use the experience collected in the last decade of wide field imaging (such as SLOAN), offering efficient modular tools to derive the end-products of imaging surveys, with the support of powerful databases. While computationally demanding, the experience from COSMOS has shown that it is feasible to combine photometry from diverse sources to achieve accurate redshifts. In fact, COSMOS photo-z's now reach an accuracy of $\sigma(z)=0.01(1+z)$ (Ilbert et al 2009), which is significantly higher than what is needed for Euclid. Following these schemes, the Euclid Photo-z application will be structured in four modules:

1. PSF homogenization,
2. object photometry,
3. star/quasar/galaxy photometric classification and
4. photometric redshift determination.

Object colours can be determined once the PSF has been homogenized. Alternatively, model magnitudes (as in SLOAN) can be also derived by fitting PSF-convolved analytic models (Sersic one-component profiles, disk+bulge models, simultaneous multi-object fitting) to the images. This approach could solve blending issues in crowded fields. As a byproduct, PSF and model photometry provide also a morphological classification into the classes "point-like" or "extended". The current implementation of the PanSTARRS-1 Photometric Classification Server (Saglia 2008), the star/qso/galaxy classification and stellar parameters are based on a support vector machine (Bailer-Jones and Smith 2008), which will also be used for the Gaia mission.

## 6.4.4 Weak Lensing

**General Principles:** The weak lensing experiment of Euclid requires precision measurements of the shapes of galaxies and determination of their photometric redshifts. This is achieved through a combined analysis of the Euclid VIS and NIP data, ground based visible photometry and additional calibration spectra. This process is standard in lensing analysis and several experiments have already developed dedicated pipelines, such as KIDS, CFHTLS, DES, Pan-STARRS and the space-based HST-COSMOS. In particular, the COSMOS approach is similar to that envisioned for Euclid, with galaxy shape measurements from space and redshifts derived from external data sets. The components necessary for the Euclid weak lensing pipeline, therefore, are well known. The specificities for Euclid come from the fact that Euclid aims to perform measurements to an unprecedented accuracy and include:

- The requirement for an integrated analysis approach. These include both the significant cross talk between different analysis levels due to instrument calibration and inter-linkages between measurements. For instance, accurate galaxy shapes require a measure of the spectral energy distribution of the galaxy, which can be estimated as part of the photo-z analysis.
- The data volumes involved are very large, which will require the pipeline to be efficient and highly automated. As an example, many current pipelines ask for user input in refining masks. For Euclid, this is no longer feasible.
- The shape measurement codes existing today have been developed for current (and near future) data. These codes are essential for a successful weak lensing analysis and must be modified to support the level of precision needed for Euclid. We will develop several independent methods.
- At the level of precision that Euclid will operate, all sections of the data analysis pipeline must perform to high accuracy. Quality control and cross-checks have to be integrated in all analysis stages. This will also require simulations at all levels.

**Implementation and Interfacing:** The data analysis for weak lensing will be integrated in Euclid's data handling structure. The performance of VIS and NIP will be monitored by the EIC IOC using a series of quick-look analysis tests. These tests will include monitoring of the system PSF stability, of optical distortion, and of the performance of the CCD and NIR detectors. If needed, the EIC IOC will request the SOC to alter the observation procedure or modify instrument and/or observing parameters.

**Data Processing:** The data processing element of the Euclid weak lensing pipeline consists of basic reduction, homogeneous incorporation of external data, PSF estimation and shape measurement levels. Each level should be interconnected and yet separable, so as to allow for cross checking, modular analysis and tracking of results. Any CCD effect, for example CTE, is corrected at the pixel level, followed by correction for bias correction and flat-fielding. The calibrated images are stacked so that object detection can be performed on a high signal-to-noise image. Objects will be classified as stars or galaxies, for the purpose of PSF characterization. Such a classification will make use of all available photometry and spectra. Each individual exposure is masked for saturated stars, telescope reflections, CCD defects and transient objects, including cosmic rays and other artifacts. This will result in a weight map for each exposure, which incorporates the mask. These basic reduction steps need to be automated and modular. The requirements to data processing facilities will not exceed 0.2 Tflops (based on an extrapolation of currently available pipelines). The PSF is modeled as a function of position and colour for each individual calibrated exposure. The PSF model is based on inputs from the pre-flight hardware tests, optical design simulations and observations of the stars over many exposures. The PSF is also characterised for each co-added image and for each of the VIS and NIP images for cross checks. The PSF is an input for at least two shear measurement pipelines to enable cross checks and redundancy for this critical step. The primary shape measurement is performed on the individual exposures. Catalogue control and systematic cross-checks are performed at this stage, including star-galaxy cross correlations and removal of outliers in the shear catalogue. The PSF and shear measurement levels should take approximately 132 Tflop per year (based on an extrapolation of currently available pipelines). Shear measurement and PSF estimation are both critical steps that require further algorithm development.

***Data Products:*** The deliverable data products associated with the weak lensing analysis include:

- Raw and processed images
- PSF model and optical distortion maps
- Catalogues (including shear, redshift, etc)
- Dark matter mass maps
- Shear correlation functions and covariance errors

## 6.4.5 Large Scale Structure

To limit systematics induced by the analysis we aim to compare results calculated using many different methodologies. Briefly, we will work in Fourier space using the power spectrum $P(k)$, and in configuration space with the correlation function $\xi(r)$. These form a Fourier pair and therefore contain the same information. We need to split measurements in scale, redshift and angle to line-of- sight μ, and fit models as a function of these parameters in order to recover the cosmological information from BAO, redshift-space distortions, and other physcial processes as decribed in Section 2.2.5.

Models will be constructed and tested using a combination of analytic and numerical techniques, such as fitting formulae (e.g. Eisenstein & Hu 1998), perturbation theory (e.g. Crocce & Scoccimarro 2006), and numerical simulations (e.g. CMBFAST, CAMB, N-body simulations). For $\xi(r)$, the standard measurement technique is based on counting galaxy pairs and comparing to the expectation without clustering (Landy & Szalay 1993). In Fourier space, the simplest technique is to place galaxies on a grid, convert into an over-density field, and Fourier transform to measure the power spectrum (Feldman, Kaiser & Peacock 1994). In such an approach it is difficult to split in μ, which can be done by decomposing instead onto a basis composed of Spherical Harmonics and spherical Bessel functions (e.g. Fisher Scharf & Lahav 1993). All of these techniques allow us to optimally weight the galaxies to allow for changing galaxy properties, and optimal weighting schemes have been suggested for a population with varying galaxy density by Feldman, Kaiser & Peacock (1994) and with varying bias by Percival, Verde & Peacock (2004).

The survey mask which leads to the window through which we observe galaxies has to be included by modeling the selection function. In addition to simply quantifying where observations are taken, this needs to include effects such as confusion, obscuration by bright stars, and selection effects due to the slitless spectroscopy. These effects, and the full sampling and analysis procedure adopted for the data will be simulated using mock catalogues. As well as using these standard techniques for measuring 2-pt statistics, we will also consider less-standard methods such as Gibbs sampling and topological analyses based on genus statistics. These measurements will be supported by a program of simulations which, as well as helping to obtain cosmological constraints from the non-linear regime, will also allow us to estimate the multi-variate, possibly non-Gaussian error distribution of the measured data.

## 6.5   The Archival System: Mission- and Legacy Archive

The Euclid data processing will be built using a *datacentric* approach which means that the Euclid Mission Archive (EMA) is at the core of any data processing activity. The EMA is a logical unit which may be distributed/federated/partially replicated over all participants. The data processing requires safety, security, consistency and integrity of the EMA. Each component of the DHS (each scientist operating in DHS) will be provided with a safe and reliable interface to the data according to a set of privileges.

The EMA will hold all data obtained, processed or simulated during the mission along with the reports on the quality of the data. From a logical point of view, the Euclid Legacy Archive is the public subset of the Mission Archive. The EMA is a distributed system with storage facilities located on each node of the DHS (SOC, IOCs, SDCs). Each node stores data items (files) relevant to the data processing done by this node. The EMA integrity is provided by dedicated archival system which uses hardware of SOC, IOCs and SDCs and provides access to the processing environment. The security is provided by the DHS authorization and authentication mechanism assigning privileges to all users. After the data products have passed the quality control and after a proprietary period, which is under the supervision of the EST, the data can be published. The data will be at the disposal of the astronomical community with the full data curation identical to that available to the mission staff, allowing independent scientists to perform detailed analyses and reprocessing of the data. Thus the mission provides the interfaces (including Virtual Observatory interfaces) to the data obtained during the mission. Results in the form of images and catalogues will be published in Virtual Observatory by SOC. The Euclid Legacy Archive will be operated under ESAC responsibility. The archival system shall have a simple baseline, which can evolve and be refined during design and implementation with increasing complexity, up to becoming a fully distributed architecture. This approach is already used in some functioning systems such as Astro-WISE and partly for current and ongoing missions (Planck and Gaia). Data access rights are specified in Chapter 7. PIs in collaboration with SOC will authorize the access privileges levels on all the EMA data items.

*Table 6.2: data processing matrix defines responsibilities of DHS components*

| Data Level | Consumer | Producer | Quality Controller |
|---|---|---|---|
| External Data | SDCs. IOCs | SDCs | SDCs |
| Level S Data | SDCs. IOCs | SDCs | SDCs |
| Level 1 Data | IOCs | SOC | SOC |
| Level 2 Data | SDCs | IOCs | IOCs |
| Level 3 Data | SDCs | SDCs | SDCs |
| Level 4 Data | External users | IOCs. SDCs | care of ESST |

## 6.6   Requirements to Data Processing and Data Storage Facilities

***Data Volume.*** The total data volume produced by the mission will not exceed 5 PB (this includes all stored intermediate data products, simulations and reprocessed external data). The initial raw data will be multiplied by a factor 10 by data processing pipelines and cross-identifications with external data.

***Data Processing Facilities.*** The required computing facilities (i.e., combined computer performance of computing facilities of all participants) should be at least 1 Tflops. This estimation is based on the outline of Euclid data processing (see Section 6.4), the existing image pipelines of Pan-STARRS and Gaia requirements to data processing.

***Data Processing Software.*** Data Processing systems handling an amount of data close to the data volume expected from the Euclid processing and archiving environment exist and are already in use.

The Euclid DHS will not be developed from scratch: participants in the project already operate stable and reliable systems similar to Euclid DHS both in terms of data storage and data flow, and in terms of data processing. No specific risks are envisaged under this aspect.

# 7    Management

## 7.1    Elements of the Euclid Mission

Euclid is a survey mission of nominally 5 years duration, to perform imaging and spectroscopy of galaxies distributed over the entire extra-galactic sky in the visible and near-infrared. This **wide survey** should fulfill all cosmology objectives described in Section 2, but the capabilities of Euclid are such that it can also perform imaging and spectroscopy on dedicated regions on the sky. The completion of a **deep survey** on a smaller area is also a main science requirement. To exploit efficiently the all-sky survey aspects and maximize the capabilities of the spacecraft and instruments, the mission will be developed and operated with two Principal Investigators to oversee instrument development and the related science exploitation.

## 7.2    Industrial Organisation and Payload Procurement

The industrial assessment study has been initiated in September 2008 and was performed by parallel teams led by Astrium GmbH, and Thalès Alenia Space Italy (TAS-I). As for all other Cosmic Vision missions, the industrial Phase A/B1 will be opened for competition early 2010, with the possibility for running two parallel industrial contracts. The final industrial organization will be completed only in Phase B2, mostly through a process of competitive selection and according to the ESA Best Practices for subcontractor selection, by taking into account geographical distribution requirements.

The Euclid telescope system including 1.2m primary mirror, secondary and possible tertiary elements will be provided by the Agency under industrial contract. The remaining payload elements, i.e. the instruments, will be entirely provided by nationally funded institutions. These will be the subject of an ESA Announcement of Opportunity to be issued in July 2010. The activities will also cover processing of the scientific and housekeeping data generated by the payload, and necessary housekeeping data from the spacecraft.

## 7.3    Euclid Schedule

The Definition Phase (A/B1) system study is expected to start in July 2010 for a period of 16 months, with the objective to enable the mission final adoption early 2012 (Table 5.1) It will include two major reviews: the Preliminary Requirements Review (PRR), to be held by the mid-term of the study, and the System Requirements Review (SRR), which will close the Definition Phase. The Technology Development Activities (TDAs) will be initiated as soon as possible after the mission down-selection in February 2010. These activities will run in parallel with the Definition Phase and their intermediate results will be fed into the System Study as necessary. Output from the TDAs which are critical to ensuring the mission feasibility or its development schedule are expected to be available before the decision for the mission final adoption.

*Table 5.1: Definition Phase major objectives*

| |
|---|
| • *Consolidate the spacecraft design using the results of the Technology Development Activities, including AOCS elements, and K band communications hardware* |
| • *Update the system, subsystem and equipment requirements and interface specifications.* |
| • *Assess the preliminary performance of the system.* |
| • *Perform further detailed mission analysis.* |
| • *Issue Requests For Quotations of reusable and modified GAIA and NirSpec units in order to consolidate the cost estimate for phase B2/C/D and be able to start procurement immediately upon kick-off of phase B2.* |

At the PRR, the mission baseline should be well established and documented. It will be critically reviewed, with the aim of confirming the technical and programmatic feasibility of the space segment, and more generally of the overall mission concept. The System Requirements Review will close the Definition Phase by consolidating the overall mission concept for enabling an efficient start of the Implementation Phase, should the mission be finally adopted.

The mission readiness for starting the implementation phase is essentially dependent on the completion of a number of spacecraft-related Technology Development Activities in the critical areas AOCS sensors, actuators and propulsion, to K-band communication adaptations and several on-board sensor and optics technologies. However, the assessment of current status gives a good degree of confidence that this mission will reach by the end of 2011 the maturity level required by the Cosmic Vision schedule. Particular attention will be paid to the establishment of interfaces between industrial telescope provider and the science instruments, to ensure clean thermo-mechanical design which also optimizes the AIV flow improve the schedule margins.

# 7.4 Science Management

## 7.4.1 Responsibilies

After the spacecraft commissioning phase, the ESA Science Operations Department (SOD) assume responsibility of the Euclid mission. ESA's Space Operations Centre (ESOC) will implement the Euclid Mission Operations Centre (MOC), operate the spacecraft and deliver the raw scientific data to the Euclid Science Ground Segment. ESA's Space Astronomy Centre (ESAC) will implement the Euclid Science Operations Centre (SOC), populate and maintain the Euclid Legacy Archive and deliver the data products to the astronomical community at large.

The Euclid Science Team (EST) composed of the PIs and scientists representing the instrument consortia, with the ESA Project Scientist (PS) as the chairman will oversee the preparations and execution of the scientific operations. The members of the EST will monitor and advise on all aspects of Euclid which affect its scientific performance. In particular, they will participate in major project reviews and perform specific tasks needed during the development and operations phases.

Dedicated teams from the instrument consortia will process the data from their instruments during all phases of the mission. The instrument consortia are expected to provide, pay for, handle and manage the Instrument Operations Centres (IOCs), hosting the operations teams and a number of Science Data Centres (SDCs) in charge of the production and validation of higher level science data products, and with the responsibility of the instrument consortia to monitor the operations of the instruments, process the scientific data, and deliver the final science data products. Details about the organisation of the uplink and downlink data processing have been described in detail in Chapter 6.

The IOCs are responsible for the first and second level standard data products, which involve the data calibration, the removal of the instrumental effects, and production of mosaics and preliminary source catalogues. The SDCs are in charge of the science data processing, the creation of third level data products, and the development of simulation packages to support the development of the data processing pipelines and their interfaces.

The SOC will act as the central node for the mission planning, will distribute the science and housekeeping telemetry to the IOCs after a first quality check, and will be the custodian of the Euclid Legacy Archive. The quality check at SOC will directly feed back to the mission planning by means of rescheduling or re-planning. The other important task of the SOC is to manage the end-to-end testing of the pipelines producing the scientific products to ensure that all processes in the Euclid science ground segment are validated and ready before launch.

## 7.4.2 Handling of External Data

To fulfil the Euclid science objectives, the Euclid mission requires complementary datasets obtained via ground-based or space based observations. These datasets are (1) ground based multi-band photometry of galaxies covering the entire wide survey area necessary to achieve the photo-z accuracy, and (2) to complete a sample of at least 100,000 spectra providing d$z$=0.001($z$+1) to calibrate the photo-z determination down to AB=24 mag. The latter is necessary for the baseline case in which Euclid will perform slitless spectroscopy where a significant fraction but not all of these spectra can be obtained. Presently, the Pan-STARRS and the DES surveys have expressed their interest in providing their multi-band photometry data to the Euclid mission. Both projects have been endorsed and the data can be made available before the launch of Euclid.

To collect the calibration spectra, a large observing programme must be conducted which still has to be organised.

The EST will be responsible for the definition of the required external data. They will coordinate the availability of the data either through MoUs with endorsed ground based survey programmes or by facilitating ground or space based observing campaigns. The EST will monitor the completion of these datasets before launch to ensure that the science requirements can be fulfilled within the nominal mission time. The connection of the external data sets to the Euclid Mission Archive and the development of the dedicated processing pipelines, fall within the framework of the SDCs development.

### 7.4.3 Data Rights

To obtain the necessary understanding of the data including all the sources of systematic effects ("systematics") an initial period of 1 year or 5000 deg$^2$ (whichever takes longer) of survey data is required for the initial analysis. After this period the instrument consortia will have 1 year of proprietary period to prepare the data products for release to the general astronomical community. The remaining parts of the wide survey will be released on an annual basis.

## 7.5   Instrument Procurement

Figure 7.1 and 7.2 summarise the hardware activities of the two consortia, including the representatives from different nationalities. We appended some key technical arguments about payload interfaces and AIV in chapter 5 that strongly drive the most likely payload procurement approach. It is assumed that the scientific instruments are procured by national agencies or institutes. The remaining elements of the space segment are procured by ESA. This includes the SVM but also the PLM with the telescope.

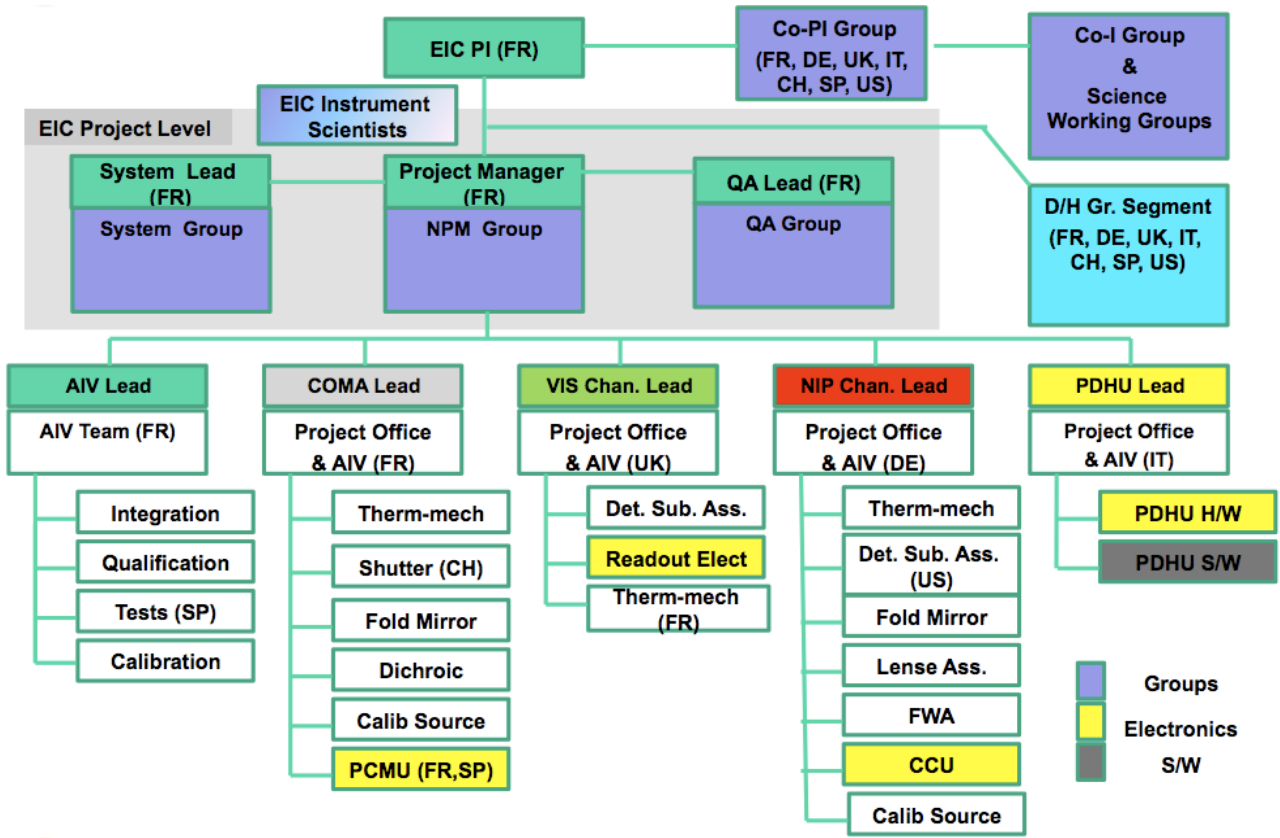

*Figure 7.1: Key activities for EIC with leading member states identified.*

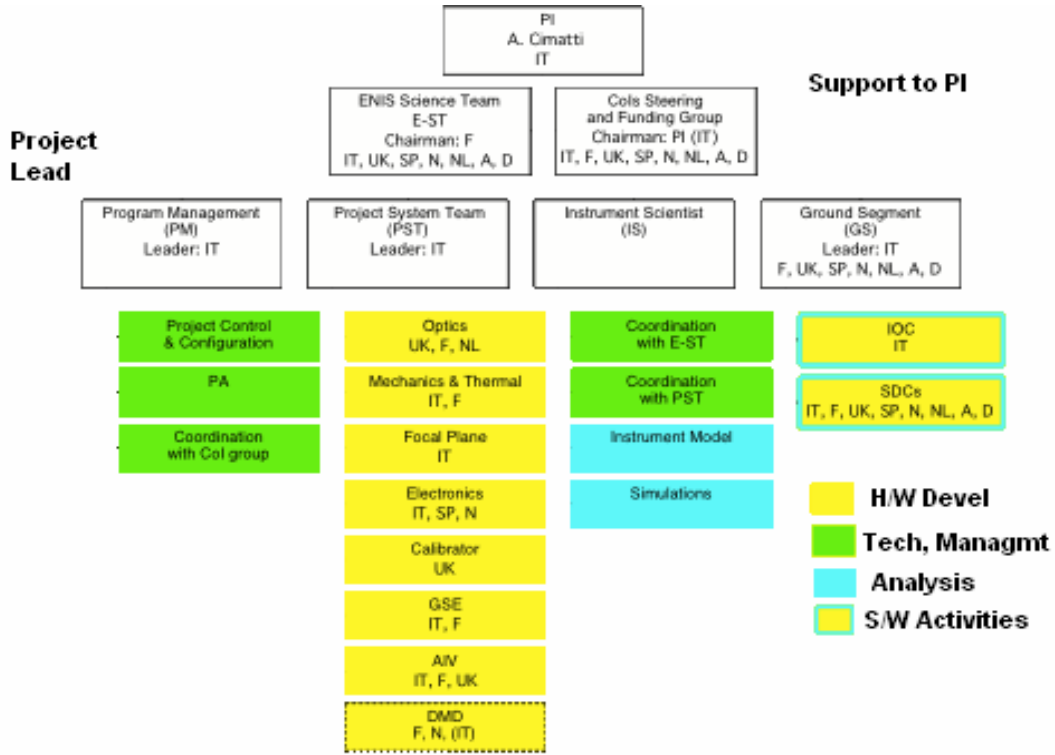

*Figure 7.2: Key activities E-NIS hi-level management structure and WBS. Activities are reported in different colours and contributing countries are indicated (brackets indicates non-equal share).*

The procurement approach has a significant influence on the development and AIV plan and therefore on costs, schedule and risk. Industry has elaborated a development approach and schedule that is ambitious. It needs an early start to the instrument development and qualification programmes.

# Appendix 1 – Comparison of Slitless and DMD Spectroscopy

The first part of this Appendix describes the simulations of spectroscopic observations employing slit (DMD) and slitless spectroscopy. The purpose of such simulations is to guide the instrument development by quantifying the performance of the spectrograph and verifying the science requirements. An important specific goal is to obtain a realistic estimate of the spectroscopic redshift success rate and redshift accuracy. The second part of the Appendix briefly illustrates the advantages of using DMD spectroscopy for the Euclid-NIS spectroscopic science cases.

## A.1.1    Simulations of Euclid-NIS Performances: Overall Approach

In the context of simulations, the success rate can be defined as the fraction of measured spectra with a redshift value within a target error of the true (input) value. Within the constraints of telescope aperture, telescope throughput, instrumental throughput, spectral resolution, exposure time, background level, the surface density and astrophysical parameters of the source galaxies, the success rate can be converted into a total survey redshift yield. In view of the redshift and luminosity range of different galaxy types and the range of emission line properties, only careful sky simulations, which include data dependent properties, such as spectral overlap and selection, can be considered as realistic. Taking the single spectrum radiometry as basis for the signal-to-noise estimate, full simulations of on-sky fields with surface densities following observations and models can be constructed. Furthermore, the resulting spectra and/or measured redshift distributions can be used to perform (mock) measurements of the quantities of interest, enabling improvement of the success rate by adjusting the spectrograph specification, survey strategy and data reduction procedure.

## A.1.2    Spectroscopic radiometry

The full treatment of the spectroscopic radiometry is given in the *E-NIS Radiometric Document*. Here we summarize only the main aspects. The general expression for the signal-to-noise ratio per resolution element of the measurement of a signal C is given by;

$$SNR = \frac{C}{\sqrt{Var(C) + \left(1 + \dfrac{1}{n_B}\right)\left(Var(B) + Var(D)\right)}}$$

where $n_B$ is the ratio of the total number of detector pixels used to measure the background to the number of pixels spanned by the resolution element containing the signal. The quoted SNRs in this document are computed assuming a resolution element of 2×2 (DMD) and 3×5 (slitless) pixels along the spatial and dispersion directions respectively. C and B are source and background signals within the resolution element in counts and their variance is given by:

$$Var(C) = C + C^2 \frac{1}{n_{pix}n_{exp}}\left(\frac{\delta\alpha}{\alpha}\right)^2 \text{, and} \qquad Var(B) = B + B^2 \frac{1}{n_{pix}n_{exp}}\left(\frac{\delta\alpha}{\alpha}\right)^2 ,$$

where $n_{pix}$ is the effective number of pixels spanned by the resolution element over which the signal is extracted, $n_{exp} = t_{exp}/t_{sub}$ is the number of dithered sub-exposures each of duration $t_{sub}$ making up the total integration time $t_{exp}$, and *(δα /α)* is the residual pixel-to-pixel relative rms fluctuations in detector response after flat-fielding. The first variance term is the Poisson noise in the source (background) photon signal; the second term represents the additional noise contribution due to residual uncorrected flat-fielding errors. The background B is composed of the zodiacal light, the scattered light and the thermal background. The variance of the electronic noise (which is negligible in the case of slitless spectroscopy) is given by:

$$Var(D) = n_{exp}n_{pix}(DC \cdot t_{sub} + RON^2 + (DC \cdot t_{sub})^2\left(\frac{\delta c}{c}\right)^2)$$

where DC is the dark current, RON is the read-out noise and *(δc/c)* is the residual pixel-to-pixel relative rms fluctuation in detector response after dark noise subtraction. The first term is due to the fluctuations in detector dark current, which are assumed to obey Poisson statistics, the second term is the net read-out noise achieved per sub-exposure and the last term represents the additional noise introduced by errors in dark current subtraction. Note that in slitless mode the above SNR expression is valid for unresolved sources only.

## A.1.3    End-to-End simulations

The end-to-end NIS simulations comprise three components: *i)* an input source catalogue with spectro-photometric information; *ii)* a module simulating 1D-spectra (for the DMD mode) or 2D dispersed images (in slitless mode) based on the radiometric model, for a given instrumental and observational setup; *iii)* an automated analysis of the extracted spectra to measure the galaxy redshift (and to provide source classification).

The wealth of information on the space distribution of galaxies and their physical properties (e.g. luminosities, colours, star formation rates, morphologies) available from a variety of large multi-band imaging and spectroscopic surveys (e.g. GOODS, VVDS, COSMOS) can be used to create mock galaxy catalogues and sky simulations which closely reproduce the surface density and redshift distribution of galaxies of different types at varying fluxes, their spectral energy distributions and sizes.

Specifically, we assign to each galaxy in the mock catalogue a spectral template extracted from a theoretical or observational library, maintaining a realistic variety of galaxy types. Then, the spectral template is rescaled so to have the galaxy redshift, magnitude, Hα line flux and equivalent width. Finally, using the radiometric model outlined above, the appropriate noise is added to the spectrum. In the DMD case, a 1D simulated spectrum for each input galaxy is created. In the slitless case, the aXeSIM software (Kümmel el al. 2007, 2009, see below) is used to simulate a sky field image and its corresponding dispersed image from which 1D spectra are extracted and analysed. An example of simulated DMD and slitless spectra using the radiometric model is shown in Figure A.1.

***Redshift measurement:*** The redshift is measured from 1D-spectra using the EZ software (Fumana et al. 2008), a program developed at IASF-Milano for automatic redshift measurement that has been widely used for the ground-based VVDS and zCosmos surveys with the VLT. Based on detected spectral features, EZ estimates in each spectrum the emission line redshift and a flag related to the reliability of the redshift measurement (based on the number of detected lines and their S/N). In absence of emission lines, EZ uses the cross-correlation of a set of templates to determine the redshift. By comparing the redshift obtained by EZ, with the one of the input catalogue, we can assess the redshift measurement success rate as a function of parameters of interest (e.g. redshift, magnitude, emission line flux, galaxy type, etc.) for a given observational setup (e.g. exposure time, sub-exposure strategy, etc).

## A.1.4    DMD Spectroscopy Simulations

***Input catalogue and sampling:*** For the DMD spectroscopy simulations we have used the catalogue obtained by Jouvel and collaborators based on observations of the Cosmos field (see Jouvel et al. 2009). This catalogue provides the realistic angular and redshift distribution of galaxies of different types selected in the near-infrared. The spectrograph collects light only from objects upon which a micro-mirror has been "opened". In multi-object spectroscopy, it is important that the spectra of different targets do not overlap either in the spatial or in the dispersion direction. As a result, the fraction of targets which satisfy such geometrical constraints (i.e. the "sampling") depends on galaxy surface density and on spectral length (which in turn depends on spectral resolution and wavelength coverage). Using the SPOC algorithm described in Bottini et al., 2005, the expected sampling has been computed assuming a minimum slit separation of 2 pixels (1 micro-mirror), the dispersion direction along the longer FoV size, and not taking into account any gap between detectors. With the DMD spectrograph configuration and for a limiting magnitude of $H_{AB} = 22$, the resulting sampling is 35%, i.e. about one third of the objects will be targeted randomly.

***Creation of simulated Spectra and Results***: Spectral templates are rescaled to the H-band magnitude of each object and then degraded to the expected S/N ratio using a custom code developed by the NIS consortium. Based on the DMD spectrograph and detector parameters, and on the analysis of a few thousands simulated spectra, the global spectroscopic redshift success rate to H=22 (AB) is extremely good, being always >90% for emission line galaxies (Sa, Sb, Sc, Irr, Starbursts) and >70% for early-type galaxies (E/S0). Given the typical fraction of galaxy types to H=22 (~20% early-type galaxies), this results in an overall success rate of >85%. This results in a total yield of ~1.5x10$^8$ galaxy redshifts from the 20,000 deg$^2$ survey to H=22. We note here that EZ is not completely optimized for absorption line galaxies (i.e. E/S0) and that the success rate for E/S0 galaxies is a strict lower limit that will improve with further optimization of the software tools. We verified that the input underlying redshift distribution is very well recovered. Finally, when the redshift is successfully measured, the error on the measurement (defined as the rms of the difference between the real and the measured redshift) is ~ 5×10$^{-4}$, i.e. consistent with the requirement of the ENIS redshift accuracy.

## A.1.5    Slitless simulations

Slitless spectroscopy is characterized by a significantly higher background compared to the multi-slit case (since integrated over the entire spectral range), contamination/confusion due to overlapping spectra, and a lower effective spectral resolution for resolved objects of increasing size.

On-sky simulations become in this case absolutely essential to give any realistic estimate of the redshift success rate as a function of line and continuum flux, redshift and galaxy type (see Section 2.5 on "Additional Science" for more details).

***Input catalogue:*** Targets of the Euclid slitless survey are essentially emission-line (i.e. star-forming) galaxies at *z*~0.5-2, with a negligible contribution from early-type galaxies whose redshift is very difficult to measure in the slitless case. The input catalogue is based on empirical count predictions of Hα emitters in combination with a K-selected COSMOS sample which includes photometric redshifts and size information from HST. Redshift distribution of Hα emitters at varying Hα fluxes were derived from a luminosity function model at *z*<2 based on HST-NICMOS slitless and narrow-band NIR surveys (Geach et al. 2009). All emission line objects were assigned Starburst and Late Spiral spectral templates, by reproducing the observed ratio of 1.5:4. The Hα flux was assigned to each object in such a way to produce the realistic distribution of the rest-frame equivalent width for the two galaxy types (with averages around 150 and 65 Å respectively). In slitless mode, the spectrograph collects light from all the objects in the field of all magnitudes, including early-type galaxies, without strong emission lines, and stars. These objects would produce a non-negligible contamination in the slitless spectra. Thus, in addition to the input catalog of emission line galaxies, we have included in the simulations also 20% of early-type galaxies, a fraction appropriate for this type of galaxies in near-infrared selected samples, and stars. The number of stars in the simulations was the number expected at high Galactic latitude (~ 3000 stars/deg$^2$ down to H(AB) = 20 mag for |*b*| ~ 60 deg). Each star was assigned a spectrum according to their types, using the Pickles library. We note that at high Galactic latitudes (|*b*|>45 deg) the contribution of stars to the total number of objects in a FoV is of the order of few percent, but the contamination increases for decreasing b (at |b|~30° the number of stars being a factor of 2 – 3 higher). Note that in slitless mode, the "sampling" is 100% by definition.

***Creation of simulated slitless data***: Slitless simulations were performed using the aXeSIM software (see Kümmel el al. 2007, 2009), developed by ST-ECF to support the slitless modes of HST. For a given set of E-NIS slitless spectrograph parameters, using an input catalogue of sources containing H-band magnitudes, morphological parameters and associated spectral templates, aXeSIM simulates direct and dispersed images (see Figure A.2) and then performs optimal extraction of 1D spectra from the 2D spectra of all objects in the FoV (Rosati et al. 2009)). Slitless simulations have been made using three observational configurations.

(1) "Full spectrum": the full spectral range (1–2 micron) is covered in all four dithers (450 s each).

(2) "Roll angles": the full spectral range (1–2 micron) is covered in all four dithers, but the dispersion direction is rotated in each dither (0, 45, 90, 135 deg) to mitigate spectral contamination and confusion.

(3) "Multi-filter": only ¼ of the spectral range is observed in each dither and the final spectrum is created by joining the four "pieces" (sub-spectra). In each exposure, the spectrum is shorter making the probability of overlapping with another spectrum is lower and also the zodiacal background (which is the integral of the zodiacal spectrum within the wavelength range) is much lower. As a consequence, the S/N ratio of emission lines remains equivalent to that of the "full spectrum" case, despite the shorter integration time.

Figure A.3 shows a blowup of a region of one of the "full spectrum" simulations, where a number of sources and their corresponding 2D spectra can be identified to appreciate the extent of spectral confusion. Note that due to the spectral "confusion" problem, one generally needs a higher signal-to-noise than in slit spectroscopy (typically 5σ) in order to detect an emission line. Automated detection of Hα lines is possible in slitless mode with Euclid down to line fluxes of $3 \times 10^{-16}$ erg cm$^{-2}$s$^{-1}$ but only for compact sources in absence of confusion, which occurs very rarely. This explains why redshifts are efficiently measured only for sources with SNR well above the canonical value of five (see below).

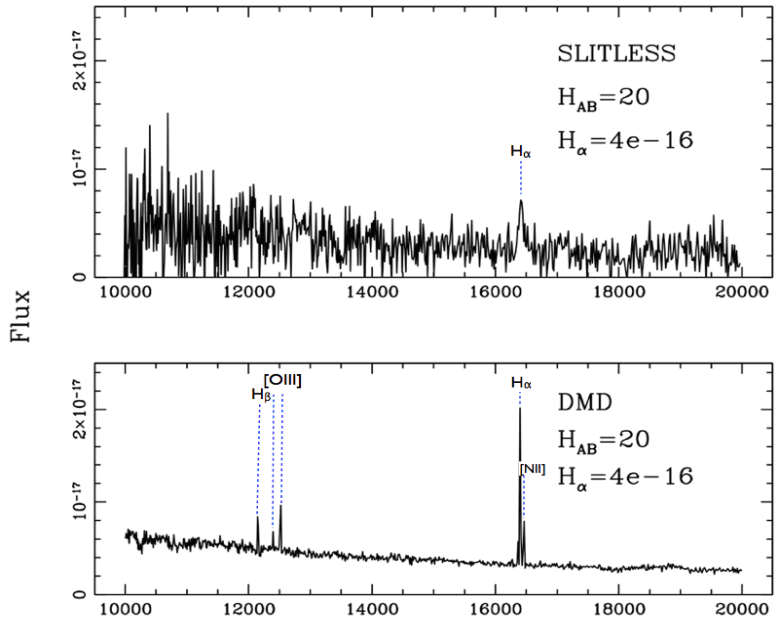

*Figure A.1.1: Simulated extracted spectra (flux in erg cm$^{-1}$s$^{-1}$, wavelength in Angstrom) of an unresolved emission line galaxy at $z=1.5$ as observed in slitless and DMD mode. The Hα flux ($4 \times 10^{-16}$ erg cm$^{-1}$s$^{-1}$, S/N ≈ 7) and continuum magnitude $H_{AB}=20$ (EW =40Å) are typical of a large fraction of the sources expected near to the flux limit of the slitless survey. Note the Hα-6563 and [NII]6583 lines will be just resolved in the slitless case in case of higher S/N.*

## A.1.6      Simulation Results

For each configuration, simulations were executed on ~5 deg$^2$ to produce a statistical sample of thousands extracted spectra and then the redshift measured with the EZ automated algorithm. The catalog of measured redshifts is then analyzed to obtain a reliable estimate of the redshift success rate. In this process, we assumed the availability of photometric redshifts with an accuracy ($\Delta z/z=0.05$) to filter out stars and galaxies at $z<0.5$ and help to discriminate between largely discordant redshifts in case of line misidentification. We note that the current simulation workflow, although sophisticated, still contains simplified assumptions, such as the more-than-optimal extraction with a a-priori knowledge of position and source size. On the other hand, the EZ software which was developed for ground-based optical surveys can be further improved for the Euclid slitless mode (e.g. taking advantage of a physical model of the spectrograph). As a result, the derived redshift success rates and redshift yields for the entire survey should be considered a realistic estimate.

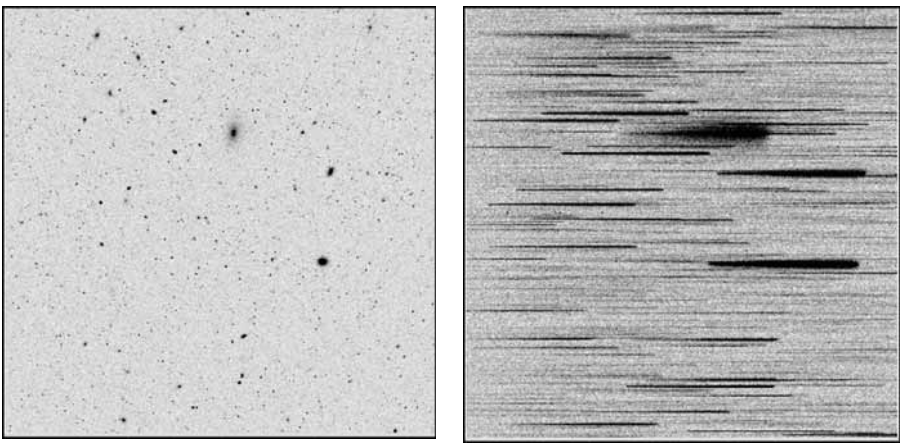

*Figure A.1.2: Example of a simulated direct image (0.5 deg$^2$) (left) and its dispersed counterpart (right) obtained with the aXeSIM software.*

A summary of the effective success rate in redshift determination as a function of Hα flux and redshift is shown in Figure A.4. Depending on the flux (but also on redshift), the efficiency ranges between 30 and 60%. By convolving these curves with the distribution of Hα emitters per unit redshift and sky area, d$N$/d$z(z,F)$ (Figure 2.23), one obtains for the 20,000 deg$^2$ survey a yield of $\approx 6.5 \times 10^7$ galaxies with robust redshift determination, with an accuracy of $\sigma_z = 10^{-3}$, in the "single spectrum" mode. A similar analysis of the "multi-filter" configuration suggests a success rate improvement of $\approx 10\%$, with the "multi-roll" approach somewhat in the middle range. With these configuration strategies, we then estimate the redshift yield of the slitless survey to be $N_z \sim 7 \times 10^7$. These yields, together with the estimated redshift accuracy, the consistency with the scientific requirements on the accuracy of the dark energy parameters that can be derived from the BAO and redshift-space distorsions measurements.

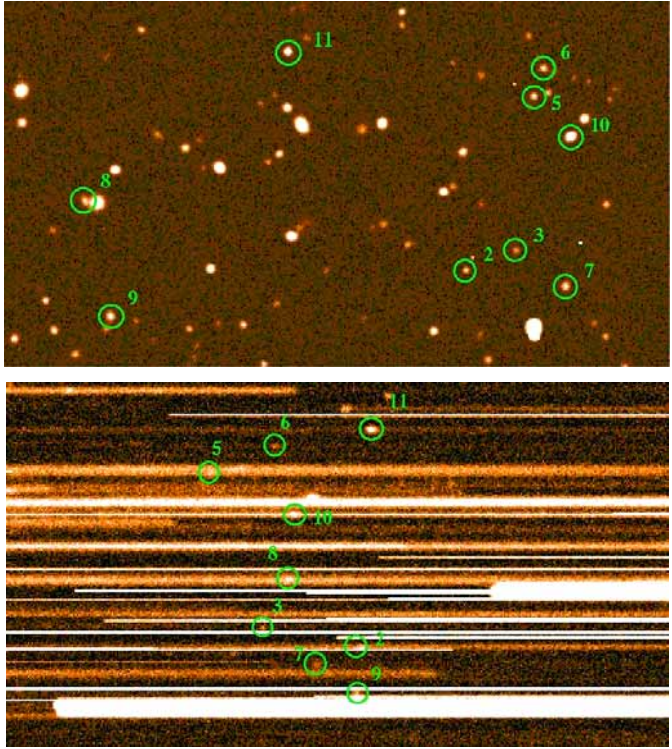

*Figure A.1.3 Blow-up of a region in the direct (top) and dispersed (bottom) images. The green circles indicate selected objects in the direct image and the corresponding Hα emission lines in the grism image. For reference, the Hα fluxes of objects no.6,3 are 7.8 and 5.9× 10⁻¹⁶ erg/cm²/s. Redshifts for objects 3, 5, 10 were not recovered by the automated procedure due to confusion*

## A.1.7 Scientific Performances: Slitless vs DMD spectroscopy

The well-known general advantage of slit (DMD) spectroscopy (either in space or from the ground) is the decrease of the sky background through the use of a small aperture (slit) matched to the typical angular size of the galaxy targets (e.g. 1×1 arcsec in the case of galaxies at $z$>0.5 as in Euclid surveys). In the case of Euclid-NIS-slitless, the dominant source of sky background, the zodiacal light, integrated along the entire spectral direction instead of a few pixels as for slit spectroscopy, makes the sky background a factor of $\approx$100 stronger than for DMD spectroscopy. Thus, DMD spectroscopy strongly mitigates the sky background allowing the detection of most of the galaxy flux, but at the same time minimizing the contribution of the sky background. For this reason, slitless spectroscopy has a disadvantage with respect to slit (DMD) spectroscopy due to (i) shallower limiting magnitudes by 2-3 mag depending on the source size, (ii) limited sensitivity to the emission lines of star-forming galaxies, (iii) detection of only Hα in the majority of galaxies and (iv) the problem of "spectral confusion" due to the overlap of spectra of different objects. Instead, thanks to the deeper limiting fluxes (H=22, or even H=23 mag for a 1.5m telescope), slit spectroscopy based on a DMD-based spectrograph allows:

(1) a clean and unbiased selection function for galaxies (in the case of DMD spectroscopy, the sample would be selected by randomly observing 35% of all the objects down to a limiting magnitude H(AB)=22 instead of searching for emission lines as with slitless spectroscopy)
(2) the detection of all galaxy types (star-forming and E/S0)
(3) the detection of several spectral features in each spectrum (absorptions and emission lines)
(4) the detection of much fainter objects over a wider redshift range
(5) to collect larger samples of galaxies and redshifts (>10$^8$)\
(6) to recover the science cases penalized or completely lost with slitless spectroscopy.
(7) to perform a truly spectroscopic survey instead of a "redshift" done with slitless spectroscopy.

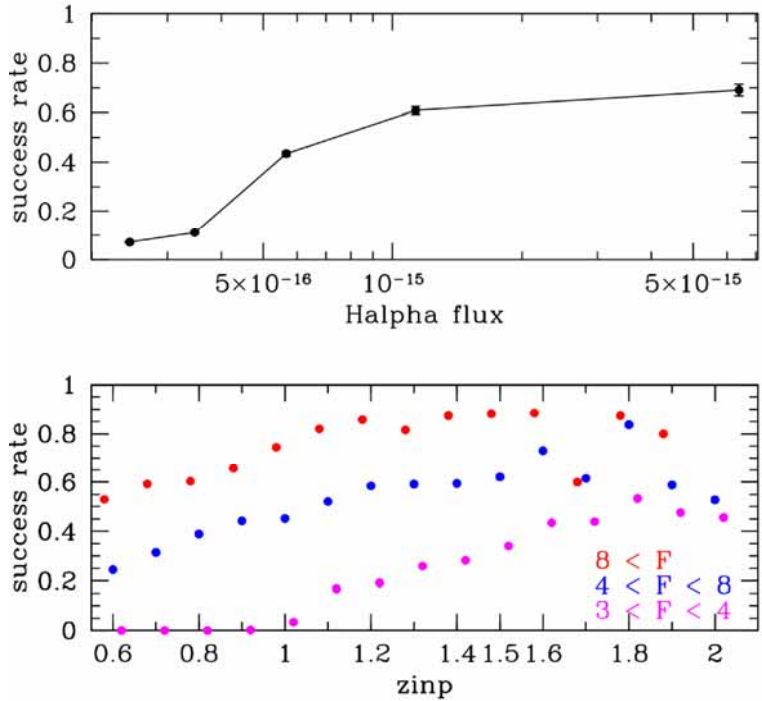

*Figure A.1.4: Effective redshift success rate as a function of Hα flux (integrated over all redshifts) (top) and as a function of redshift for different flux bins (F in units of $10^{-16}$ cgs) (bottom), as determined from sky simulations of the Slitless survey ("full spectrum" configuration, see text). By convolving these redshift success rate, $\varepsilon(F,z)$, with the redshift and number counts distribution $dN/dz(F,z)$ of Hα emitters (see Fig. 2.23) the number of robust redshifts from the 20,000 deg² survey can be estimated.*

**Cosmological Science Cases with Spectroscopic Data**: Due to the improved statistics on clustering and power spectrum estimates (larger number of redshifts and suitable redshift distributions for pure continuum flux-limited samples selected down to H(AB) ≈ 22-23), the gain provided by DMD spectroscopy is substantial for all the "spectroscopic" cosmological science cases. Also, the capability to select simultaneously both star-forming and elliptical (luminous red) galaxies is a clear advantage in order to cross-check the results using galaxy populations with very different bias factors. Observing different populations will also allow us to use techniques that decrease cosmic variance errors when measuring the growth of structure from redshift-space distortions (e.g. McDonald & Seljak 2009). Observing different populations will also help us to use large-scale clustering measurements to measure primordial non-Gaussianity (e.g. Dalal et al. 2008).

Compared to the baseline of the Wide Survey done with slitless spectroscopy (20,000 deg², $F > 4 \times 10^{-16}$ erg s$^{-1}$ cm$^{-2}$, ε =0.35), a survey of the same sky area done with the optional DMD-based spectroscopy to H=22 (AB) (35% sampling) would allow to significantly increase the accuracy of the measurement of both BAO [H(z) and $D(z)$] and redshift-space distorsions (growth factor), resulting in an increase in the DETF dark energy Figure of Merit (FoM) of galaxy clustering (plus Planck) by a factor of 2-2.5. This is a significant improvement: the DETF defined different stages in DE projects based on a factor of 3 improvement in the FoM. Further significant improvements could be in principle be allowed by reaching deeper limiting magnitudes (e.g. H=23) with a larger telescope (e.g. 1.5 m diameter), or increasing the sampling rate of galaxy targets by optimizing the spectral resolution (i.e. the "length" of the spectra on the detector array), or increasing the integration time. Besides BAO and redshift-space distorsions, the other major gain allowed by DMD-based spectroscopy in the cosmological context would be for distant galaxy clusters thanks to the sensitivity to E/S0 galaxies and the resulting larger samples (see below).

**Distant clusters of galaxies**: The main advantages of DMD spectroscopy are: (i) about 1.5 orders of magnitude more clusters at z>0.5 for which accurate dynamical masses can be measured (see Fig. 2.24); (ii) clusters identified out to much higher redshifts (newly indentified by E-NIS or selected from previous cluster surveys); iii) the homogeneous sampling of the whole cluster galaxy population, i.e. passively evolving as well as star-forming galaxies. In particular, a larger number of well-sampled clusters and the availability of redshifts for passive galaxies would improve both the cluster mass calibration and the determination of cluster mass profiles. In fact, star-forming galaxies in nearby clusters are known to have larger σ$_v$ than passive galaxies. Therefore, a more robust determination of $M_{200}$ and $M(r)$ can be obtained if both populations can be used as tracers of the cluster dynamics. Simultaneous sampling of the passive and star-forming cluster galaxy populations would also provide a better understanding of galaxy evolution in clusters. In addition, one could study the build-up of the color-magnitude sequence by sampling the population of passively-evolving cluster galaxies.

**Physics and evolution of galaxies**: The deeper limiting fluxes of the emission lines and continuum, and the capability to detect both absorptions lines and several emission lines (i.e. not only Hα) in normal galaxies, would allow to address an enormous variety of science cases on the formation, evolution and physical properties of all galaxy types to z ≈ 2+ in the Wide Survey and at 2<z<7 in the Deep Survey in a way analogous to what SDSS achieved at low redshifts. For instance, a coherent study of the evolution of stellar populations in galaxies would only be possible with DMD spectroscopy. The strong optical-ultraviolet spectral features of galaxies that are primary redshift and diagnostic tracers fall, in the redshift range z~0.5−3, in the near-IR. At higher redshifts, DMD-based spectroscopy would allow to obtain redshifts and diagnostics of the young stellar populations and the interstellar medium using rest-frame ultraviolet lines.

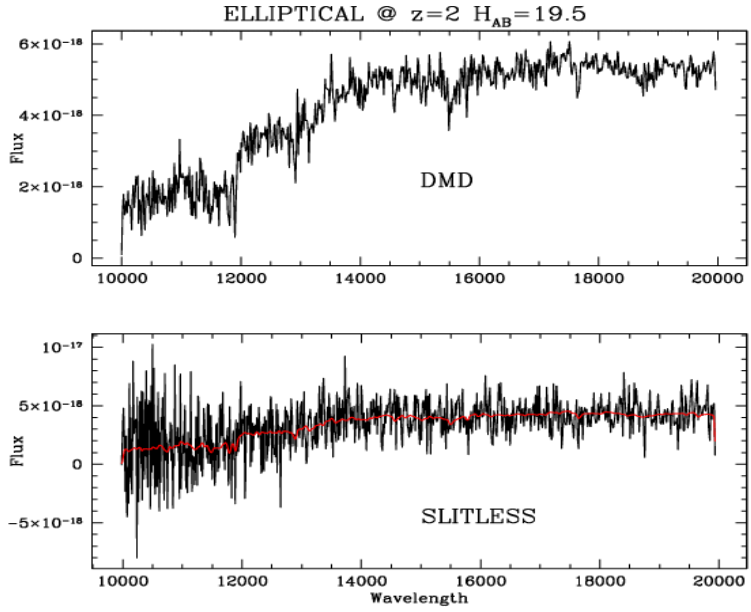

*Figure A.1.5 – Comparison between the simulated spectrum of an elliptical galaxy with H=19.5 at z=2 obtained with DMD (top) and slitless spectroscopy. The H-band magnitude of this galaxy is representative of the high-luminosity tail of the elliptical luminosity function at z ≈ 1.5-2, and corresponds to the continuum limiting magnitude for E-NIS slitless spectroscopy.*

**Formation and evolution of early-type galaxies (ETGs)**: ETGs (also known in the recent literature as Luminous Red Galaxies, LRGs, e.g. Eisenstein et al 2005) play a crucial role in cosmology as probes of the galaxy mass assembly, the evolution of large scale structure and the co-evolution with their central supermassive black holes. ETGs have spectra typical of old passive stellar populations with absorption lines, continuum breaks and no (or very weak) emission lines. ETGs are the most penalized galaxies with slitless spectroscopy (see the example in Fig. A.5): due to the shallow limiting magnitude (H<19.0-19.5 mag), only a few hundreds ETGs are expected in 1 deg$^{-2}$, and their redshift distribution will be limited to z < 1, hence making particularly difficult to measure ETG redshifts because their primary spectral features (e.g. D4000 break and Ca II H&K absorption lines) do not enter in the E-NIS observed spectral range (1-2 micron) for z < 1. Instead, thanks to the deeper limiting magnitudes and the detection of absorption lines, DMD spectroscopy would extend the identification and study of these galaxies over the critical redshift range of 1<z<3, i.e. from the passive population at z ≈ 1-2 to their star-forming progenitors at z >2 and to perform detailed studies on their evolutionary and physical properties using the high-quality continuum and absorption line spectra. This would also nicely complement the sample of ETGs at z<1 obtained by the SDSS-III BOSS survey.

**Identification and study of high-z galaxies:** In the Deep Survey, DMD spectroscopy enables identification of the most luminous galaxies at 7<z<10+ through Lyman-break continuum and/or Lyα emission by accumulating the required integration time on pre-selected targets. For instance, for H<25.5 in 10 deg$^2$ there will be up to ≈500(≈100) Lyman-break galaxies at z >7(>8) depending on the evolution of the luminosity function. As the most massive objects at these redshifts are extremely rare, the large FoV of E-NIS gives an enormous advantage and would be a perfect complement to JWST.

**Identification of QSOs at very high redshifts**: In the Wide Survey, with a limit of H$_{AB}$=22 mag and 1/3 sampling, in 20,000 deg$^2$ E-NIS is expected to find up to 40 QSOs with z≥7 and 0-10 with z ≥9 depending on the evolution of the QSO luminosity function. The spectra will have a S/N ratio on the continuum suitable to perform IGM studies and estimate the SMBH masses. The expected number of *z*>7 QSOs would nearly double in case of a deeper survey down to H=23 (e.g. with a 1.5m diameter telescope).

**Supernovae**: Supernovae used in cosmology need spectroscopy for typing but also to measure spectral features. This is needed to test in for systematic effects such as redshift evolution. Then, following SN in the deep survey with DMD, which is ~4 times more sensitive than the slitless option, can allow for the measurement of spectral properties to control such issues. It can be also very helpful to have the host galaxy properties, which will be difficult to do from the ground in an efficient way and is not possible with slitless.

**The origin of the Near Infrared background (NIRB)**: DMD spectroscopy of the sky background itself would also obtain the definitive spectrum of the so far elusive cosmic NIRB.

**More extended studies of our Galaxy**: DMD-slit spectroscopy would allow one to obtain deeper observations by 2-3 mag for a better analysis of the distribution of the ultracool dwarf population in the disk of the Galaxy and will indeed contribute to our understanding of the Galactic structure. The surveyed space volume would increase by a factor of 1 dex for Y dwarfs, and 2 dex for L-T dwarfs. Besides the larger number of detections, these observations would lead to the study of the scale height perpendicular to the Galactic Plane of the free-floating, least massive population of the Milky Way. Because of the extemely faint nature of these objects, such study can be achieved only with dedicated, deep near-infrared spectroscopic experiments capable of unambiguous detections and of covering a large area of the sky in a reasonable amount of time. DMD spectroscopy would also allow to identify and characterize the planetary-mass objects (<~0.015 $M_{sol}$) of the nearest star-forming regions of the Galaxy, representing a step forward in our understanding of the formation and evolution of the least-massive population of the Milky Way.

*Table A.1.1: Summary on spectroscopic additional science cases: DMD versus slitless spectroscopy*

| Science case (survey)➔ | Galaxies physics | Galaxies evolution | High-z gals | High-z QSOs | Clusters (z<1) | Clusters (z>1) | E/S0 (z<1) | E/S0 (z>1) | Our Galaxy |
|---|---|---|---|---|---|---|---|---|---|
| DMD Wide | YES | YES | NO | YES | YES | YES | YES | YES | YES |
| Slitless Wide | NO | Limited and biased (only starforming galaxies | NO | YES | YES | NO | NO | Marginal | YES (but more limited) |
| DMD Deep | YES | YES | YES | NO | YES | YES | YES | YES | YES |
| Slitless Deep | NO | Limited and biased | marginal | NO | YES | NO | YES | Marginal | YES (but more limited) |

# Appendix 2 – Euclid Imaging Consortium Simulations

The Euclid Imaging Consortium has performed simulations at each stage of the performance monitoring and prediction chain. A comparison of the key results with our requirements is provided in Section 2.4.1 here we set out additional details of our simulations. An overview of the entire simulations process is provided in Figure A2.1 and we give details of individual components in the subsections below.

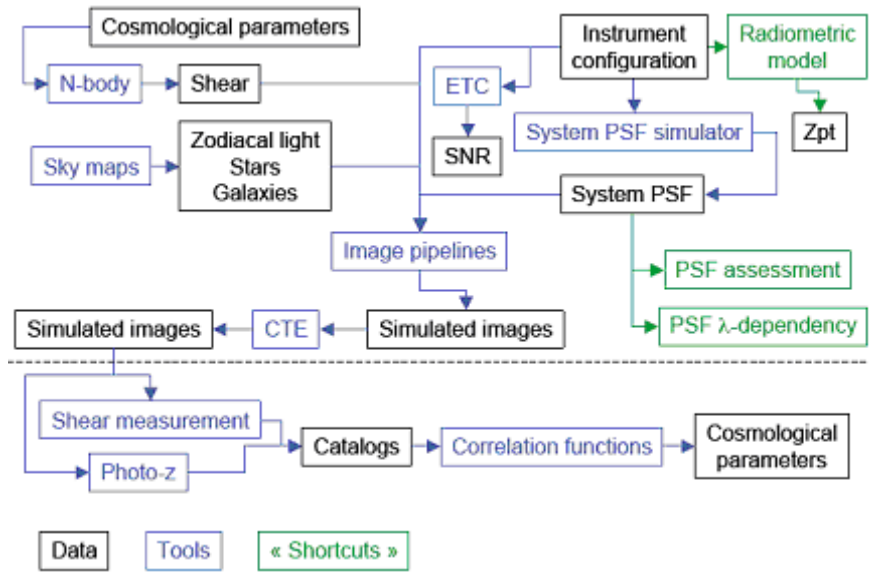

*Figure A2.1: Overview of the EIC simulations tools.*

## A.2.1 Image Simulations

We assess the number of galaxies usable for lensing by making simulated images from two independent pipelines. These pipelines have been analysed using three independent image analysis pipelines.

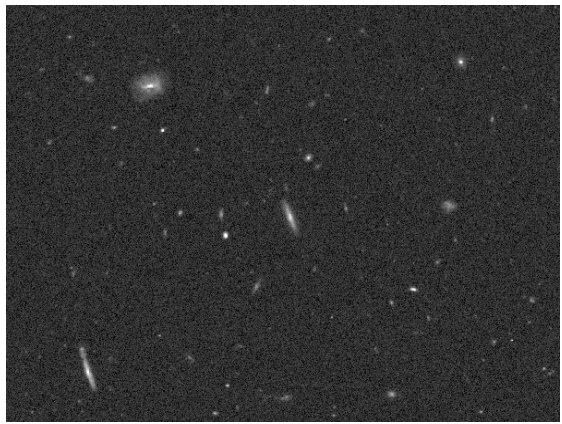

*Figure A2.2: A sample simulated Euclid observation from the* SkyLens *simulations.*

The basic *SkyLens* simulation package is described in detail in Meneghetti et al. (2008), although for the purposes of Phase A substantial improvement were implemented. The *simage* simulation software is decribed in Massey & Réfrégier (2005) and Dobke et al. (2009).

Both simulation pipelines use galaxies taken from the Hubble-Ultra-Deep-Field (Beckwith et al, 2004), each galaxy is decomposed into a shapelet model (the shapelet formalism varies between the simulations,

Melchior et al., 2007 for SkyLens; Massey & Réfrégier, 2008 for *simage*). This modeling is perfomed in the *B, V, i* and *z* bands for *SkyLens* and in addition the J and H bands for *simage*. The HUDF ensures a realistic sample of galaxy morphologies to a limiting magnitude similar to expected Euclid performance. The shapelet models allow many simulation operations to be performed analytically; as a result galaxies can be "painted" on a virtual sky at any desired resolution.

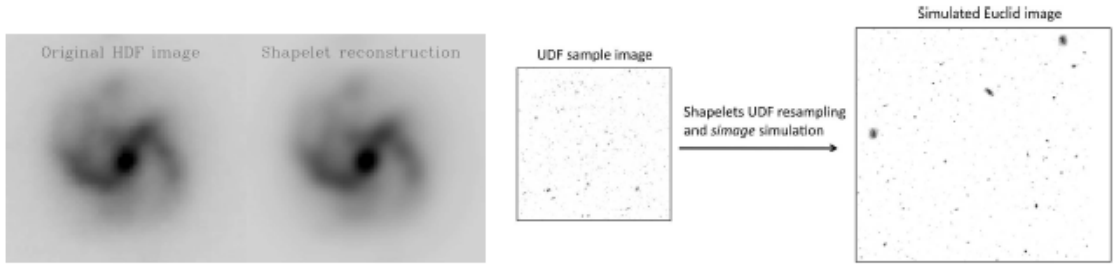

*Figure A2.3: Left: An example of a spiral galaxy (from the Hubble Deep Field) modelled using shapelets simulation code (Massey & Réfrégier, 2005). Right: Output image from* simage *pipeline based on a nominal Euclid configuration (Dobke et al., 2009).*

While the decomposed shapelet galaxy catalogues are created from the UDF, the magnitude distribution of the resultant simulated images are normalised to a variety of existing galaxy surveys, detailed below, rather than just drawing randomly from the UDF galaxy magnitudes. For example in *simage* the distribution of galaxies is described by $N = B \, 10^{Am}$ where *A* and *B* are normalization factors, and m is the galaxy magnitude. The form of this expression and the values of the normalisation factors A and B are taken from existing COSMOS survey analysis between magnitudes 21 and 26 (Leauthaud et al., 2007). COSMOS is used because it is the largest existing HST survey and furthermore it is closest to the Euclid specifications. Below magnitude 21 and above magnitude 26 the number counts are fit to various survey sources as compiled in Metcalfe et al. (2001). The form is continuous at these transition points and results in a realistic number count relation for the simulated images at low, mid, and high magnitudes. For a nominal Euclid configuration Figure A2.2 displays a typical simulated output for which analysis exceeds the requirement of 30 galaxies per square arcminute.

After the simulated galaxy populations are created both independent simulation pipelines create a virtual observations by creating a virtual telescope. These are created using information such as mirror diameter, field of view, PSF throughput, pixel scale, noise properties, pointing coordinates and CCD specifications. The final simulated images contain realistic galaxy, telescope and detector effects.

The final result is images that contain galaxies, observation strategies and instrument effects that match expected Euclid performance to a high accuracy.

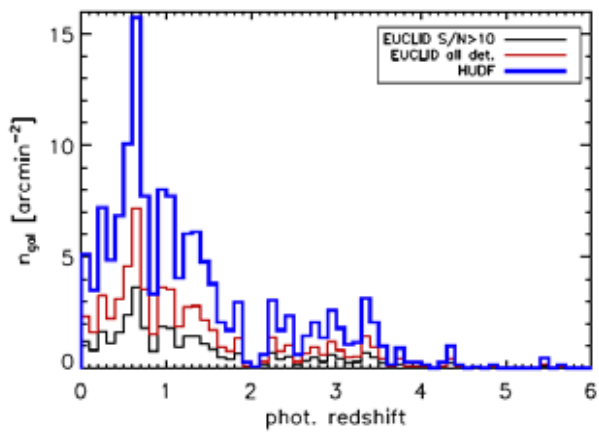

*Figure A2.4: The redshift distribution of the sources in the Euclid images. The lowest (black) line shows the counts per redshift bin for galaxies with S/N>10 and size(FWHM) > 1.25 PSF(FWHM). The middle (brown) line corresponds to all galaxies detected in the image. The thick upper (blue) line shows counts for the HUDF, which is the input to the Italian simulations.*

The simulations from the two simulation pipelines were analysed by two further independent weak lensing image analysis packages. The simulation pipelines themselves used SExtractor to find the number of detectable galaxies down to a given signal-to-noise ratio. The values quoted in Section 2.4.1 correspond to a SExtractor signal-to-noise cut of 10. The *lens*fit (Miller et al., 2007; Kitching et al., 2008) and im2hape (Bridle, 2000) weak lensing pipelines both performed a weak lensing shear measurement on images created using both simulation pipelines and calculated the number of galaxies for which shears were successfully measured, in addition to an effective number of galaxies, The effective number in each pipeline were calculated by including a weighting scheme that reflects the quality of the shear measurement; *lens*fit measures uses a Bayesian weighting scheme, im2shape uses a measured error on the shear.

In this document we have taken a conservative approach in which we assume systematic effects will place the main limit on the number of usable galaxies. In many cases the effective number of galaxies exceeds in the requirement and goal. In addition if galaxies of lower signal to noise can be included then the galaxies number density increases – whether such galaxies can be used for weak lensing is linked to the achievable photometric redshift accuracy at these signal-to-noise levels, this is discussed later.

## A.2.2    Exposure Time Calculator (ETC)

The Euclid exposure time calculator has been created with web interface. This can calculator the exposure time needed to observe any object (star, galaxy – spiral or elliptical) using the optical or IR instruments to any magnitude using either F606_W or F814_W HST filters. This calculator is the front end for a full simulator for optical telescopes adapted from tools developed for the Large Binocular Telescope. The calculator assumes a specific realistic quantum efficiency of the detector and includes Zodical background noise. Both the read-out noise, dark current and wavelength range are free parameters. Galaxies are assumed to have a half light radius of 0.4'' and the PSF is assumed to be 0.23'' at 8500 A.

## A.2.3    Photometric Redshifts Simulations

The photometric redshift requirements for Euclid have been carried out independently by two teams. The concept of photometric redshifts is illustrated in Figure A2.5. The dotted blue curve and solid blue curve show the same galaxy at two different redshifts. The ratio of luminosity received in the different bands will be different for the two objects. Expecially for the higher redshift objects the Euclid infrared bands are crucial for accurate photometric redshifts. The complementarity between the ground and space based observations are clear from the positions of the different observing bands.

**UCL team:** The UCL simulations are described in more detail in Abdalla et al. (2008) and Banerji (2009). The supervised artificial neural network photometric code ANNz (Collister & Lahav 2004) is a training set method which has been shown to produce competitive results compared to other training set methods available (Abdalla et al. 2009). ANNz requires a training set which is the data used to optimise a cost function with respect to the free parameters ('weights'). If the data is noisy, a validation set is also required in order to prevent over-fitting. The remaining freedom left in a neural network analysis is the architecture of the network. A network with architecture N:2N:2N:1 (i.e. which has N inputs, two hidden layers with 2N nodes each and only one output estimating the redshift, and where only adjacent layers are interconnected) has been shown to work well on photometric data (Collister & Lahav 2004) where N is the number of different photometric bands available.

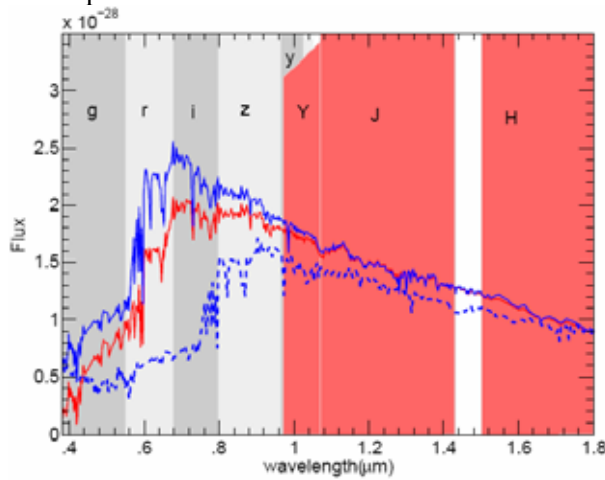

*Figure A2.5: Spectra of different galaxies at different redshifts. The shaded regions are the wavelength range covered by the Pan-STARRS-like (grey) and Euclid NIR (red) simulation. The solid blue curve and the red curve are spectra of different galaxies at same redshift. The blue dotted curve has the same SED as the solid blue curve but has a different redshift. Flux is in units of erg/sec/angstrom.*

The mock catalogues were simulated as described in Abdalla et al. (2008). In order to simulate these catalogues, the GOODS-N spectroscopic sample (Cowie; Wirth) was used. This is a flux limited sample with R < 24.5 mag and redshifts $z$<4. The broadband photometry available for this sample in multiple wavebands was used to generate a series of galaxy templates using a method similar to Budavari et al. but assuming a prior set of templates consisting of the observed CWW templates and the starburst galaxies of Kinney et al. Any reddening is removed from the photometry before template construction and the new templates used to calculate a best-fit SED value and reddening for each object assuming the reddening law of Calzetti et al (1997) as well as a correction for the IGM absorption according to the prescription of Madau et al.

The model assumed for the luminosity function evolution is taken to be an interpolation between the local r-band luminosity function at redshift zero and the Steidel et al. (1999) luminosity function at redshift 3. This is then used to estimate the RIZ magnitude and redshift distribution from which galaxies are drawn using a Monte Carlo technique. The fluxes for each galaxy are then calculated based on the redshift, SED type, reddening and filter profiles specified after normalising to the EIC RIZ filter. Gaussian noise is added to the fluxes before calculating magnitudes and errors for the noisy sample in order to produce the final catalogue.

We have carried out a number of studies for Euclid using these tools including a trade-off study on the near infrared filter configuration, an investigation of the necessary depths of ground-based observations and a trade-off between ground-based and space infrared filters. These are described in detail in Banerji (2009).

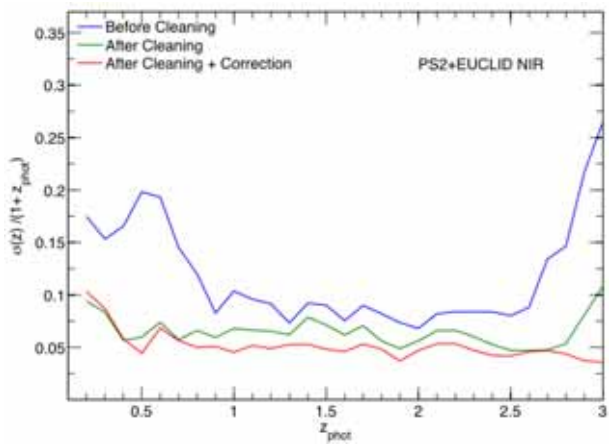

*Figure A2.6: General performance of Pan-STARRS-2 + Euclid NIR survey. This will be the same as the performance for DES. Here 13% of objects are rejected after cleaning the photo-z catalogues for outliers and 5σ outliers are reduced below 0.25%.*

**Zurich team:**

The Zurich team use a template fitting approach is used in which the available photometry is compared with that expected with a set of templates at different redshifts i.e. with two free parameters ($T$,$z$). In practice, a large set of templates are required since the difference in colour between adjacent templates at a given redshift must be smaller than the typical photometric errors. The strength of this method is that it utilizes the knowledge of how the redshifts affect the colours of a particular galaxy. It can in principle therefore be applied at any redshift independent of any spectroscopic redshift that may or may not be available. Template fitting is usually coupled with internal self-calibration of photometric zero point offsets or effective wavelength of filters, using any systematic offset between the photometry and the best fit template of the galaxies to determine this. Extinction in the galactic foreground or in the galaxy itself can be accommodated with the addition of two more parameters ($A_V$, $G$ & $A_V$, $z$ ) in that fit, using our physical knowledge of the effects of dust.

Photo-z code ZEBRA (Feldmann et al 2006) is used to produce photo-z s in maximum likelihood mode. ZEBRA gives the best fit redshift and template type together with their confidence limits estimated from constant $\chi^2$ boundaries. ZEBRA also outputs the likelihood functions for individual galaxies in various formats etc. Further information is available in ZEBRA user manual[5].

In order to simulate the catalogues for this work, we use the COSMOS mock catalogues derived by Kitzbichler & White (2007). To match the templates with the colours of the objects in the catalogue, we take these templates and fit the best fit SED, knowing the redshift of the objects. The fluxes for the galaxies are calculated and converted to magnitudes based on their redshifts, SED type, and filter profiles. This mock is the basic ideal mock without error. We make a cut of $I_{AB} \leq 24.5$ and this sample serves as the ideal mock for our work. We use a random sub sample from combining all the 24 COSMOS mocks together in order to mimic a survey over a large area in the sky. The sample is converted back to flux and Gaussian noise is added to the fluxes according to the different filter sensitivity and survey configuration considered. Thus a realistic mock with the perturbed photometry in flux and error estimations for each filter is created, which should suffice for a Euclid like survey. The details of the depth are given in Bordoloi et al. (2009).

---

[5]can be retrieved from http://www.exp-astro.phys.ethz.ch/ZEBRA/

The maximum likelihood output of ZEBRA are cleaned for outliers and likelihood functions are corrected as described in detail in Bordoloi et al. (2009) and the sum of the likelihood functions are used to characterize the tomographic bins (Figure A2.7).

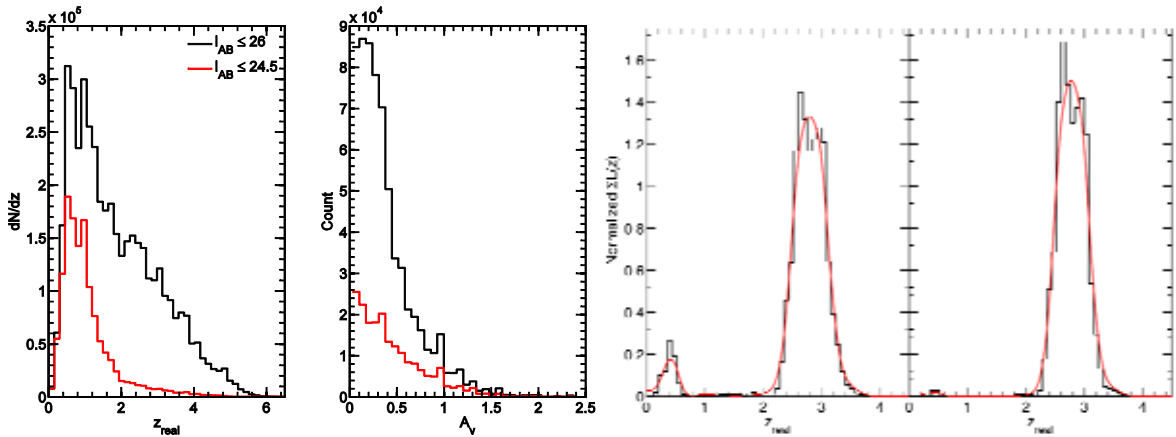

*Figure A2.7: The leftmost panel gives the distribution characterizing one of the mock catalogues used in the simulation. The distribution of number density of objects as a function of redshift for the full simulation is given in black ($I_{AB} \leq 26$ mag) and the red line is after a cut of $I_{AB} \leq 24.5$ mag. The next left panel gives the amount of reddening applied to the templates in the simulation. The black histograms in the two right-most panels show the underlying distribution of galaxy in a particular tomographic bin (with and without cleaning). The red curve shows the reconstruction of this distribution using the likelihood information coming from the photo-z measurement code.*

## A.2.4 PSF Characterisation

We divide the PSF Characterisation into two separate parts. First we investigate the broad-band shape of the PSF and its impact on shear measurement. Second we assess the impact of wavelength dependence of the PSF on shear measurements.

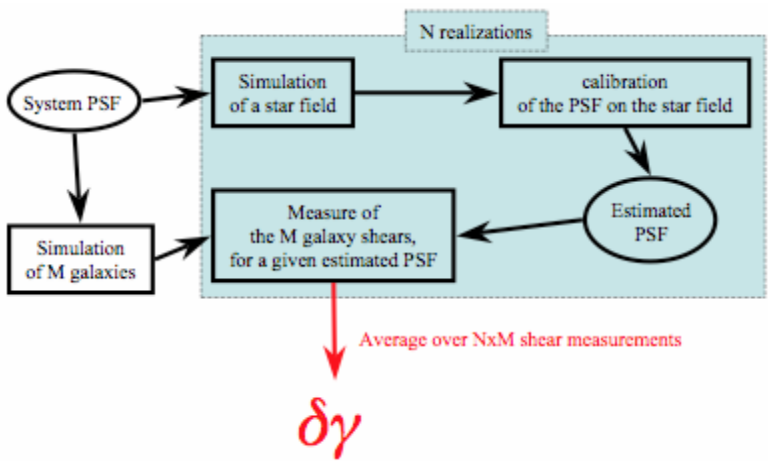

*Figure A2.8: The simulation to estimate the bias on shear measurements, for a given system PSF and galaxy population. The system PSF is used as input to simulate star fields and galaxies. The star fields are used to simulate the PSF calibration, which lead to an estimated PSF. This PSF is then used to measure galaxy shears. The bias on shear measurements comes from the slight difference between the estimated and true PSF. The procedures highlighted in blue are repeated a large number of times.*

**PSF shape assessment:** An overview of the PSF shape assessment simulations is shown in Figure A2.8. The simulated Euclid PSFs are created by simulating the optical throughput of the instrument design. The pixilation and convolution are performed using a full ray tracing processes. The flux is then integrated inside pixels, in which the dithering strategy, the Intra-Pixel Sensitivity Variations and the CTE degradations have all been taken into account.

The system PSF varied with few arbitrary chosen degrees of freedom (e.g. a rotation, a dilation/compression and a distortion) and these these degrees of freedom are used to simulate star field, galaxies (modeled using an exponential profile) are then added and convolved with the system PSF.

In this idealised case the galaxy shears, for a given estimated PSF, are measured by fitting a 2D-exponential profile on each galaxy. Note that fitted ellipticities are not equal to the true ones, because of the slight difference between the estimated PSF and the truth. These measured shear values are compared to the input values, and are then translated into the requirements that appear in Section 2.4.

This flexibility of these simulations mean that all detector effects such as Intra-Pixel Sensitivity Variations, CTE degradations, the complexity and sparsity of the PSF (Paulin-Hendrinksson et al., 2008) and telescope design can be included and the effect on the weak lensing measurement assessed. These investigations have lead to requirement on the PSF properties, as well as requirements on detector effects

**PSF Wavelength Dependence:** Stars are conventionally used to estimate the instrument PSF during the lifetime of an experiement. However in general the PSF depends on the spectral energy distribution of the object, and therefore will not be the same for stars and galaxies. In reality a pre-launch telescope model in addition to stars will be used to characterise the PSF wavelength dependence, the accuracy of this procedure will depend on the stability of the wavelength dependence with telescope properties, and the number of stars available to calibrate. In the case of HST, a PSF model taken from the telescope design is routinely used in conjunction with calibration from any stars in the field to assess the telescope configuration in a given observation.

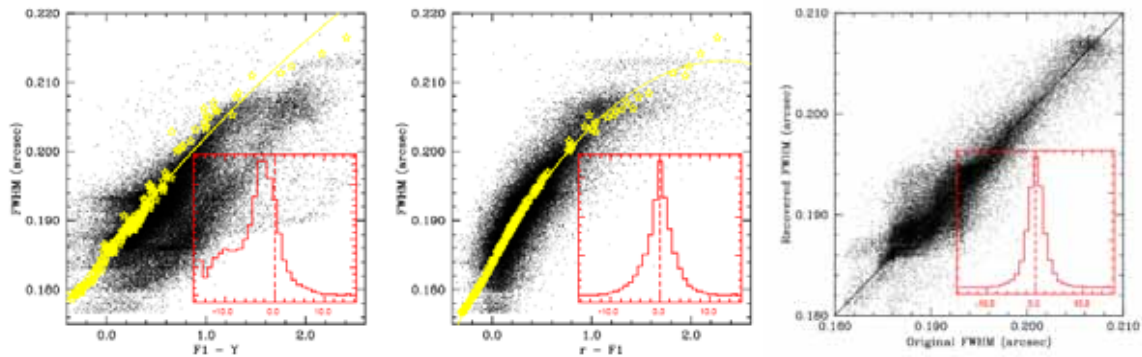

*Figure A2.9: PSF FWHM for galaxies (black dots) and stars (yellow stars) as a function of object colour. The left hand panel shows calibration using space-data alone and the middle panel shows the result of using a colour composed of the wide Euclid optical filter and a ground based r band filter such as that used for photometric redshift estimates. The insets show the dispersion about the best fit first or second order polynomial to the stars (yellow line). We see the results using r-F1 are significantly better than using F1-Y. The right hand panel shows the results of our template fitting method.*

Cypriano et al. (2009) have simulated this effect, and investigated the impact on cosmic shear measurements. These simulations use realistic galaxy and stellar populations to quantify the amount by which the PSF FWHM is likely to be mis-estimated. Galaxy SEDs are generated using mock catalogues designed to simulate the distribution of redshifts, colours and magnitudes of galaxies in GOODS-N (Cowie et al. 2004; Wirth 2004). Template spectra are taken from Kinney et al. (1996) and intermediate types obtained by linear interpolation of these templates; for further details see Abdalla et al. (2007). The PSF sizes for stars are estimated using stellar SEDs from the Bruzual-Persson-Gunn-Stryke Spectrophotometric Atlas. Fluxes are then obtained for each object in the mock catalogues for the Euclid on-board filters F1 and Y up to the limiting magnitudes 24.5 and 22.80 respectively (AB magnitudes, 10σ detections). In addition, fluxes are measured in the *g, r, i, z* and *y* filters up to the magnitudes 24.45, 23.85, 23.05, 22.45 and 20.95 mag. These configurations are chosen to simulate data from future ground based cosmology surveys such as DES and Pan-STARRS, which could be used to correct optimally for the wavelength dependence of the PSF.

The simulations use a template fitting method to predict the observed PSF FWHM of a galaxy by using all the colours available. Redshifts, specral type and reddening for simulated galaxies were estimated using the ANNz (Collister & Lahav 2007) photometric redshift pipeline making full use of the multi-colour information. With this information the SED of each object was computed and used as a model of the wavelength dependence to predict the PSF FWHM for this galaxy. The simulations use the exact PSF model for each sample of stars and propagate the noisy and potentially biased galaxy SED estimates through to PSF biases and cosmological parameter biases. The comparison between this predicted PSF FWHM and the truth for the exact galaxy SED and redshift can be seen in the right hand panel of Fig A2.9. As discussed in Section 2.4 this method results in residual biases in cosmological parameter estimation that meet the Euclid requirements.

## A.2.5      Detector Performance Simulations

A full analysis technique for looking at CTE effects on galaxy shapes has been demonstrated by Rhodes et al (2009). This uses a combination of laboratory tests on irradiated CCDs and a software model developed to mimic the process of CCD readout. The model has been tested against and calibrated upon data from the Hubble Space Telescope (Massey et al., 2009) and laboratory-controlled data (Dawson et al 2008, Rhodes et al 2009). The CTE model's parameters include the density and characteristic release time of multiple charge trap species. These were predicted for Euclid by irradiating several CCDs in the LBNL 88-Inch Cyclotron with a (large) known flux of 12.5 MeV protons, and measuring the image degradation was measured in subsequent first pixel response (FPR) / Extended Pixel Edge Response (EPER) tests behind point sources created by 55Fe radiation. The effect on galaxy shape measurement was demonstrated to be linearly dependent upon trap density (by testing both the increased degradation over time and as a function of on-chip distance from the readout register), so the effect at any given point in the Euclid mission can be deduced by rescaling the trap density measured in the irradiated CCD.

The software to mimic CCD readout can also be used in an inverse mode to undo the charge trailing caused on board Euclid, and thus mitigate the adverse effects on the scientific objectives. This has been demonstrated to reduce the effects of CTE by a factor of 10 on extant HST data (Massey et al 2009) using only science data to calibrate the model. These tests suggested that an even better improvement should be possible, with minimal overhead during the mission to better calibrate the readout model using on-board EPER/FPR and pocket pumping tests.

The conclusion is that a coherent solution has been demonstrated in which a combination of radiation-tolerant hardware and software post-processing can reduce the effect of CTE on galaxy shapes to 10% of the total shape measurement error budget (Rhodes et al 2009). Further work will continue during the Definition Phase to further reduce this load by investigating additional hardware solutions such as increasing the number of readout registers, charge injection, and changing the charge clocking time.

## A.2.6      Shape Measurement Method

The GRavitational lEnsing Accuracy Testing (GREAT) challenges are a controlled set of shape measurement simulations designed to closely match the statistical size of a Euclid-like cosmic shear survey. These simulations form a roadmap that is designed to all improvement and calibration for existing shape measurement methods and to encourage the development of new algorithms. GREAT 2008 (Bridle et al., 2009) began these challenges by creating simulations that match the zeroth order shape measurement goal of measuring a constant shear given a known and constant PSF. In only 6 months of operation, the accuracy of shape measurement techniques improved by over a factor of 2, with the Euclid requirement on the systematic variance being achieved under some observing conditions. GREAT 2010 (Kitching et al., in prep) will increase the complexity by including variable shear and PSF fields as well as determination of the PSF itself as part of the challenge

# Acronyms

| | |
|---|---|
| 2dGRS | 2 degree Galaxy Redshift Survey |
| ACS | (HST) Advanced Camera For Surveys |
| AGN | Active Galactic Nucleus |
| AIT | Assembly, Integration and Testing |
| AIV | Assembly, Integration and Verification |
| ALMA | Atacama Large Millimetre/Submillimetre Array |
| ACMSS | Attitude Control Management System |
| AOCS | Attitude and Orbit Control System |
| APE | Absolute Pointing Error |
| APPLES | ACS Pure Parallel Lyman Alpha Emission Survey |
| ASIC | Application Specific Integrated Circuit |
| astro-WISE | Astronomical Wide-Field Imaging System |
| AVM | Avionics Model |
| aXe | software to extract spectra from slitless spectroscopy |
| BAO | Baryonic Acoustic Oscillations |
| BOSS | Baryon Oscillation Spectroscopic Survey |
| CCAT | Cornell Caltech Atacama Telescope |
| CCD | Charge Coupled Device |
| CCSDS | Consultative Comittee on Space Data System |
| CDM | Cold Dark Matter |
| CEA | Commissariat a L'énergie Atomique |
| CMB | Cosmic Microwave Background |
| CMBFAST | CMB Polarisation and power spectra extraction |
| COMA | Common Opto-Mechanical Assembly |
| COSMOS | Cosmological Evolution Survey |
| CPL | Chevallier, Polarski and Linder |
| CTE | Charge Transfer Efficiency |
| CTI | Charge Transfer Inefficiency |
| DE | Dark Energy |
| DES | Dark Energy Survey |
| DETF | (NASA) Dark Energy Task Force |
| DGP | Dvali, Gabadadze, Porrati |
| DHS | Data Handling System |
| DM | Development Model |
| DMD | Digital Micromirror Device |
| DMD | Digital Micro-Mirror Device |
| D-SRE-PJ | Project division of ESA's SRE Directorate |
| DTCP | Daily Telemetry Communications Period |
| ELA | Euclid Legacy Archive |
| ELT | (European) Extremely Large Telescope |
| EMA | Euclid Mission Archive |
| EMC | Electro-Magnetic Compatibility |
| EPER | Extended Pixel Edge Response |
| eROSITA | Extended Röntgen Survey Imaging Telescope Array |
| ESA | European Space Agency |
| ESAC | European Space Astronomy Centre |
| ESO | European Southern Observatory |
| EST | Euclid Science Team |
| ETC | Exposure Time Calculator |
| EZ | Easy-Z (Redshift Determination) |
| FGS | Fine Guidance Sensor |
| FoM | Figure Of Merit |
| FoV | Field-Of-View |
| FPA | Focal Plane Array |
| FPR | First Pixel Response |
| FRLW | Friedmann, Lemaitre, Robertson And Walker |
| FUR | Follow-Up-The-Ramp |
| FWA | Filter Wheel Assembly |
| FWHM | Full Width Half Maximum |
| GOODS | Great Observatories Origins Deep Survey |
| GR | General Relativity |
| GRAPES | Grism ACS Program for Extra-Galactic Science |
| GREAT | Gravitational Lensing Accuracy Testing |
| GSTP | General Support Technology Programme |
| HETDEX | Hobby-Eberly Telescope Dark Energy Experiment |
| HST | Hubble Space Telescope |
| HUDF | Hubble Ultra Deep Field |
| IM | Interface Module |
| IMU | Inertial Mesurement Unit |
| IOT | Instrument Operations Team |
| ISW | Integrated Sachs Wolfe (effect) |
| JWST | James Webb Space Telescope |
| LAM | Laboratoire Astrophysique de Marseille |
| LOFAR | Low Frequency Array For Radio Astronomy |
| LOS | Line of Sight |
| LSST | Large Synoptic Survey Telescope |
| MLI | Multilayer Insulator |
| MMA | Micro Mirror Array |
| MOC | Mission Operations Centre |
| MOEMS | Micro-Optical Electro-Mechanical Systems |
| MOS | Multi Object Spectroscopy |
| MoU | Memorandum of Understanding |
| MTF | Modulation Transfer Function |
| NASA | National Aeronautics and Space Agency |
| NFW | Navarro, Frenk and White |
| NIP | Euclid Near-Infrared Imaging Photometer |
| NIR | Near Infrared |
| NIS | Near Infared Spectrometer |
| PanSTARRS | Panoramic Survey Telescope Rapid Response System |
| PCDU | Power Conditioning & Distribution Unit |
| PDHU | Payload Data Handling Unit |
| PEM | Proximity Electronics Module |
| PLM | Payload Module |
| PMCU | Power Management Controller Unit |
| PSF | Point Spread Function |
| PSU | Power Supply Unit |
| PV | Performance Verification |
| QC | Quality Check |
| QSO | Quasi-Stellar Object |
| RAVE | Radial Velocity Experiment |
| ROE | Read Out Electronics |
| RPE | Relative Pointing Error |
| RSSD | (ESA) Research and Science Support Department |
| S/C | Spacecraft |
| SAA | South Atlantic Anomaly |
| SDC | Science Data Centre |
| SDRAM | Synchronous Dynamic Random Access Memory |
| SDSS | Sloan Digital Sky Survey |
| SEL2 | Sun-Earth Lagrange point 2 |
| SFR | Star-Formation Rate |
| SKA | Square Kilometre Array |
| SIDECAR | System for Image Digitisation, Enhancement, Control and Retrieval Application |
| SNe | Supernovae |
| SNLS | Supernovae Legacy Survey |
| SNR | Signal to Noise Ratio |
| SOC | Science Operations Centre |
| SPRT | (ESA) Science Programme Review Team |
| SPT | South Pole Telescope |
| SRE | (ESA Directorate) Science and Robotic Exploration |
| SRE-O | (ESA) Science Operations Department |
| SSE | Sun-Spacecraft-Earth angle |
| STEP | Shear Testing Programme |
| SVM | Service Module |
| SZ | Sunyaev-Zeldovich |
| TBC | To Be Confirmed |
| TBD | To Be Done / To Be Decided |
| TDA | (ESA) Technology Development Activity |
| TRL | Technology Readiness Level |
| UDF | (HST) Ultra-Deep Field |
| US | United States of America |
| UT | Universal Time |
| VIMOS | Visible Multi-Object Spectrograph |
| VIPERS | VIMOS Public Extra-Galactic Redshift Survey |
| VIPGI | VIMOS Interactive Pipeline Graphical Interface |
| VIS | Euclid Visible Instrument |
| VISTA | Visible and Infrared Survey Telescope for Astronomy |
| VLT | (ESO) Very Large Telescope |
| VO | Virtual Observatory |
| VST | VLT Survey Telescope |
| VVDS | VIMOS VLT Deep Survey |
| WFE | Wave Front Error |
| WFPC | Wide Field and Planetary Camera |
| WFXT | Wide Field X-Ray Telescope |
| WISE | (NASA) Wide Field Infrared Survey Explorer |
| WL | Weak Lensing |
| WMAP | Wilkinson Microwave-Anisotropy Probe |